\documentclass[10pt]{article}
\usepackage{geometry}              
\usepackage{graphicx}
\usepackage{amssymb}
\usepackage{url}

\newcommand{\figurecontent}[1]{#1}    

 \newcommand{\comment}[1]{}   

\newcommand{\mktall}[1]{}
\renewcommand{\mktall}[1]{\mbox{\rule[-1ex]{0ex}{2ex}{#1}}}

\newcommand{\I}{\mathrm{i}}

\def\sign{\mathop{\rm sign}\nolimits}
\def\det{\mathop{\rm det}\nolimits}
\def\tr{\mathop{\rm tr}\nolimits}

\newcommand{\Eqn}[1]{Eq.~(\ref{#1})}
\newcommand{\Fig}[1]{Fig.~(\ref{#1})}

 \newcommand{\Hat}[1]{\hat{\bf #1}}
 \newcommand{\Vec}[1]{{\bf #1}}

\newcommand{\mathReal}{\mathbb{R}}

\newcommand{\Sphere}[1]{\mathbf{S}^{#1}}

\newcommand{\SO}[1]{\mathbf{SO}({#1})}

\newcommand{\opt}{\mbox{\small opt}}
\newcommand{\ID}{\mbox{\scriptsize ID}}
 \renewcommand{\t}{\mbox{\footnotesize \,t}}   
\newcommand{\td}[1]{\stackrel{\scriptscriptstyle \sim}{{#1}} }

\def\ShowFiles{0}

\begin{document}                  




\title{
      The Quaternion-Based Spatial Coordinate\\ and Orientation Frame
      Alignment Problems}

\author{Andrew J.\ Hanson\\
Luddy School of Informatics, Computing, and Engineering\\
Indiana University, Bloomington, Indiana, 47405, USA}
\date{}

 \maketitle

\begin{abstract}
   We review the general problem of finding a global rotation that
  transforms a given set of points and/or coordinate frames (the ``test'' data) into the
  best possible alignment with a corresponding set (the ``reference" data).  
  For 3D point data, this ``orthogonal Procrustes problem''  is often phrased in terms of
  minimizing a root-mean-square deviation or RMSD corresponding to a
  Euclidean  distance measure relating the two sets of matched coordinates.  We focus
  on quaternion eigensystem methods that have been exploited to solve this problem
  for at least five decades in several different bodies of scientific literature where they 
  were discovered independently.  While numerical methods 
  for the eigenvalue solutions dominate 
  much of this  literature,  it has long been realized that the quaternion-based 
  RMSD optimization problem can  also  be solved using
  exact algebraic expressions  based on the form of the quartic equation
  solution published by Cardano in 1545;  we focus on these
  exact solutions to expose the structure of the entire eigensystem 
  for the traditional 3D spatial
  alignment problem.  We then explore the structure of the less-studied
   orientation  data context, investigating 
  how  quaternion methods  can be extended to
  solve the corresponding 3D quaternion orientation frame alignment (QFA) problem, noting
  the interesting equivalence of this problem to the rotation-averaging problem,
  which also has been the subject of  independent literature threads.
  We conclude with a brief discussion of the combined 3D translation-orientation
   data alignment problem.
   Appendices are devoted to a tutorial on quaternion frames, a related quaternion
   technique for extracting quaternions from rotation matrices, and    
   a review   of quaternion rotation-averaging methods relevant to the orientation-frame
   alignment problem.   Supplementary
  Material sections cover  novel extensions of quaternion methods to
  the 4D Euclidean point alignment and  4D orientation-frame alignment problems,
  some miscellaneous topics, and additional details of the quartic algebraic eigenvalue problem.  
\end{abstract}


\section{Context}

Aligning matched sets of spatial point data is a universal problem that occurs in a
wide variety of applications.  In addition, generic objects  such as  protein residues,
parts of   composite object models,   satellites,   cameras, or   camera-calibrating reference
objects are not only located at points
in three-dimensional space, but may also need 3D orientation frames to describe
them effectively for certain applications.   We are therefore  led to consider both the Euclidean
translation alignment problem and the orientation-frame alignment problem on the same footing.

Our purpose in this article is to review, and in some cases to refine, clarify, and extend,
the possible quaternion-based approaches to the optimal alignment problem for
matched sets of translated and/or rotated objects in 3D space, which could be
referred to in its most generic sense as the "Generalized Orthogonal Procrustes
Problem"~\cite{Golub-vanLoan-MatrixComp}.
  We also devote some attention to identifying the surprising breadth
of domains and literature where the various approaches, including particularly
quaternion-based methods, have appeared; in fact the number of times in which
quaternion-related methods have been described independently without
cross-disciplinary references is rather interesting, and exposes some
challenging issues that scientists, including the author, have faced in coping
 with the wide dispersion of both historical and modern scientific literature relevant
 to these subjects.

We present our study on two levels.  The first level, the present main article,
is devoted to a description of the 3D spatial and orientation alignment
problems, emphasizing quaternion methods, with an historical perspective and a
moderate level of technical detail that strives to be accessible.  The second
level, comprising the Supplementary Material, treats novel extensions of the
quaternion method to the 4D spatial and orientation alignment problems, along with
many other technical topics, including analysis of algebraic quartic eigenvalue
solutions and numerical studies of the applicability of certain common approximations
and methods.


In the following, we first review the diverse bodies of literature regarding the
extraction of 3D rotations that optimally align matched pairs of Euclidean point data
sets.  It is important for us to remark that we have repeatedly become aware of additional
literature in the course of this work, and it is entirely possible that other worthy references
have been overlooked: if so, we apologize for any oversights, and hope that the
literature that we have found to review will provide an adequate context for the interested reader.
We then introduce our own preferred version of the quaternion approach to the
spatial alignment problem, often described as the root mean square deviation
(RMSD) minimization problem, and we will  adopt that terminology when convenient; 
our intent is to consolidate a range of distinct variants
in the literature into one uniform treatment, and, given the wide variations in
symbolic notation and terminology, here we will adopt terms and conventions that work well
for us personally.  Following a technical
introduction to quaternions, we treat the quaternion-based 3D spatial alignment
problem itself.  Next we introduce the quaternion approach to the 
3D orientation frame alignment (QFA) problem in a way
that parallels the 3D spatial problem, and note its equivalence to  quaternion frame
averaging methods.  We conclude with a brief
analysis of the  6-degree-of-freedom problem, combining the 3D spatial and 3D
orientation-frame measures. Appendices include treatments of the basics of quaternion orientation
frames, an elegant method that extracts a quaternion from a numerical 3D rotation
matrix, and the generalization of that method to compute averages of rotations.


\section{ Summary of Spatial Alignment Problems, Known Solutions, and Historical Contexts}

{\bf The Problem, Standard Solutions, and the Quaternion Method.} The fundamental problem we will be 
concerned with arises when we are given a
well-behaved $D\times D$ matrix $E$ and we wish to find the optimal
$D$-dimensional proper orthogonal matrix $R_{\opt}$ that maximizes the measure
$\tr(R \cdot E)$.  This is equivalent to the RMSD problem, which 
seeks a global rotation $R$ that rotates an ordered set of point test data   $X$  
in such a way as
  to minimize the squared Euclidean differences relative to a matched reference set $Y$.
We will find below that $E$ corresponds to the
cross-covariance matrix of the pair $(X,Y)$ of $N$ columns of $D$-dimensional
vectors, namely $E=X \cdot Y^{\t}$, though we will look at cases where $E$ could
have almost any origin.

One solution to this problem in any dimension $D$  uses the decomposition of the
general matrix $E$ into an orthogonal matrix $U$ and a symmetric matrix $S$
that takes the form
$E = U\cdot  S = U \cdot (E^{\t} \cdot E)^{1/2}$, 
giving $R_{\opt} =U^{-1} =  (E^{\t} \cdot E)^{1/2} \cdot E^{-1}$; note that there exist
 several equivalent forms (see, e.g., \cite{Green-matsoln-1952,HornHN1988}).
General solutions may also be found using
singular-value-decomposition (SVD) methods, starting with the decomposition $E = U\cdot S
\cdot V^{\t}$,  where $S$ is now diagonal and $U$ and $V$ are orthogonal matrices, 
 to give the result $R_{\opt}= V \cdot D \cdot U^{\t}$, where $D$
is the identity matrix up to  a possible sign in one element (see, e.g.,
\cite{Kabsch1976,KabschErratum1978,Golub-vanLoan-MatrixComp,Markley1988}).

In addition to these general methods based on traditional linear algebra approaches, a
significant literature exists for three dimensions that exploits the
relationship between 3D rotation matrices and quaternions, and rephrases the
task of finding $R_{\opt}$ as a \emph{quaternion eigensystem} problem.  This
approach notes that, using the quadratic quaternion form $R(q)$ for the rotation
matrix, one can rewrite $\tr(R \cdot E)\to q\cdot M(E) \cdot q$, where  
  the \emph{profile matrix} 
$M(E)$ is a traceless, symmetric $4\times 4$ matrix
consisting of linear combinations of the elements of the $3 \times 3$ matrix
$E$.  Finding the largest eigenvalue $\epsilon_{\opt}$ of $M(E)$ determines  
the optimal quaternion eigenvector $q_{\opt}$ and
thus the solution $R_{\opt} = R(q_{\opt})$.  The quaternion framework will be
our main topic here.

\qquad

{\bf Historical Literature Overview.}   Although our focus is the quaternion eigensystem
context,  we first note that one of the original approaches to the RMSD task exploited
 the singular-value-decomposition  directly to obtain an optimal rotation matrix.  This solution
 appears to date at least from 1966 in Sch\"{o}nemann's thesis
 \cite{Schonemann-Procrustes1966} and possibly \cite{CliffSVD1966} later in the same journal issue; Sch\"{o}nemann's work is chosen for citation, for example, in the earliest editions of Golub
 and van Loan~\cite{Golub-vanLoan-MatrixComp}.  Applications of the SVD to alignment
 in the aerospace  literature appear, for example, in the context of Wahba's problem~\cite{wiki:Wahbas_problem,Wahba1965}, and are used explicitly, e.g.,  in~\cite{Markley1988}, 
 while the introduction of the SVD for the alignment problem in molecular chemistry generally is attributed to  Kabsch~\cite{wiki:Kabsch_algorithm,Kabsch1976}.

We believe that the quaternion
 eigenvalue approach itself was first noticed around 1968 by Davenport \cite{Davenport1968}
 in the context of  Wahba's problem,  rediscovered in 1983 by Hebert and
 Faugeras~\cite{HebertThesis1983,FaugerasHebert1983,FaugerasHebert1986}  in the
 context of machine vision, and
 then found independently a third time in 1986 by Horn~\cite{Horn1987}.  

An alternative quaternion-free approach by \cite{HornHN1988} with the optimal rotation of the
form  $ R_{\opt} =(E^{\t} \cdot E)^{1/2} \cdot E^{-1}$  appeared in 1988, but this basic form
was apparently known elsewhere as early as 1952 \cite{Green-matsoln-1952,Gibson-matsoln-1960} .

\begin{figure}
 \figurecontent{ \centering
 \includegraphics[width=6.0in]{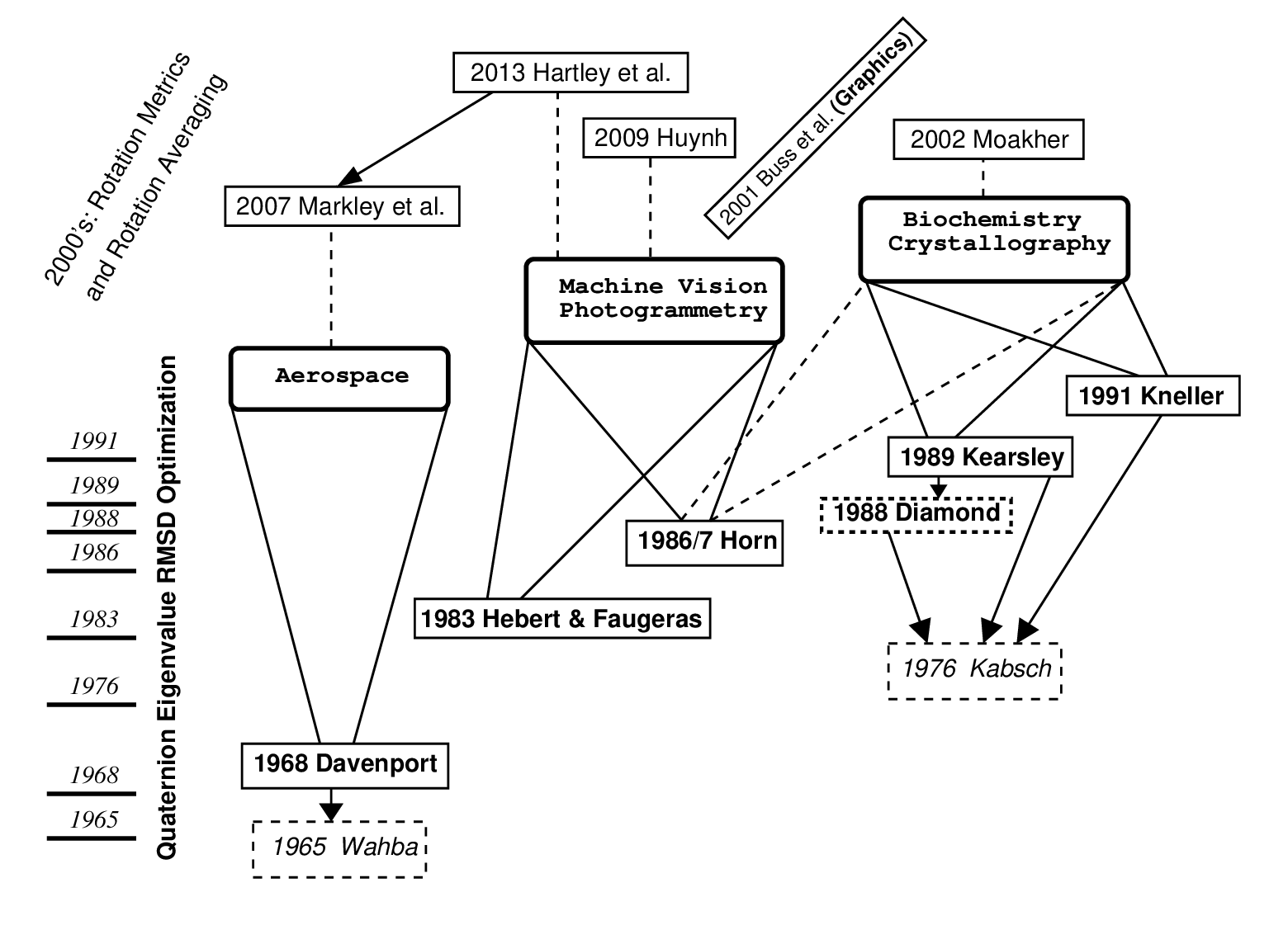} }
   \vspace*{.1in}
\caption{\ifnum\ShowFiles=1 {\bf QuatEigTimesPlusRot.eps.} \fi
 The quaternion eigensystem method for computing the optimal
 rotation matching two spatial data sets was discovered independently
 and published without cross-references  in  at least three
 distinct literatures.  Downward arrows point to the introduction of the
 abstract problem, and upward rays indicate domains of publications
 specifically citing the quaternion method.  Horn eventually appeared 
 routinely in the crystallography citations, and reviews such as  \cite{Flower1999} 
  introduced multiple cross-field citations.  Several fields have included activity
  on quaternion-related rotation metrics and rotation averaging with varying degrees of 
  cross-field awareness.
 }
\label{QuatTimes.fig}
\end{figure}

Much of the recent activity has occurred in the context of the molecular alignment problem, starting
from a basic framework put forth by Kabsch~\cite{Kabsch1976,KabschErratum1978}.
So far as we can determine, the  matrix eigenvalue approach to
 molecular alignment was introduced in 1988 without actually mentioning quaternions by name in 
 Diamond~\cite{RDiamond1988}, and refined to  specifically incorporate quaternion methods in 1989 by
 Kearsley~\cite{Kearsley1989}.  In 1991 Kneller~\cite{Kneller1991} independently
 described a version of the quaternion-eigenvalue-based approach that is widely
 cited as well.   A concise and useful review can be found in Flower~\cite{Flower1999},
 in which  the contributions of 
 Sch\"{o}nemann, Faugeras and Hebert, Horn, Diamond, and Kearsley    are  
  acknowledged and  all cited in the same place.  A graphical summary of the discovery
  chronology in various domains is given in \Fig{QuatTimes.fig}. 
 Most of these  treatments mention using numerical methods to find the optimal
 eigenvalue, though several references, starting with ~\cite{Horn1987},  point out that 
 16th century algebraic methods
  for solving the quartic polynomial characteristic equation, discussed in the next paragraph,
 could also be used to determine the eigenvalues. 
  In our treatment we will study the explicit form of  these algebraic solutions  for  the 3D problem
   (and also for 4D in the Supplementary Material), taking advantage of several threads
   of the literature.    
   
   \qquad 
   
   {\bf Historical Notes on the Quartic.}
   The  actual solution to the quartic equation, and thus the solution of the characteristic polynomial of
   the 4D eigensystem of interest to us,  was first published in 1545 by Gerolamo
 Cardano \cite{ArsMagnaCardano1545} in his  book \emph{Ars Magna}.  The intellectual 
  history of this fact is controversial and narrated with varying emphasis in diverse sources.
  It seems generally agreed upon that Cardano's student Lodovico Ferrari was the first to discover the basic  method  for solving the quartic in 1540, but his technique was incomplete as it only reduced the problem to the cubic equation, 
  for which no solution was publicly known at that time, and that apparently prevented him from 
  publishing it.   The complication appears to be that Cardano had actually learned of a method for
  solving the cubic already in 1539 from Niccol\`{o} Fontana Tartaglia (legendarily in the form of a poem), but had been  sworn to secrecy, and so could not reveal   the final explicit step needed to complete Ferrari's implicit
  solution.
   Where it gets controversial is that at some point between 1539 and 1545, Cardano learned that
    Scipione del Ferro  had found the  same cubic solution  as the one of Tartaglia that he had sworn
    not to reveal, and furthermore that del Ferro  had discovered his solution before Tartaglia did. 
   Cardano interpreted that fact as releasing him from his oath of secrecy (which Tartaglia did not
   appreciate),   allowing him to publish the complete
    solution to the quartic, incorporating the cubic solution into Ferrari's result.    Sources  claiming that Cardano ``stole'' Ferrari's solution may perhaps be exaggerated, since Ferrari did not have access 
    to the cubic equations, and Cardano did not conceal  his sources;  exactly who 
    ``solved'' the quartic is   thus philosophically complicated, 
    but Cardano does seem to be the one who combined the multiple threads 
    needed to express the equations  as a single complete formula.

    Other interesting observations were 
 made later, for example, by Descartes in 1637~\cite{Descartes-onRoots-1637}, and
  in 1733 by Euler~\cite{Euler-onRoots-1733,Euler-Bell-1733}.  
 For further descriptions, one may consult, e.g.,
 \cite{AbramowitzStegun1970} and \cite{BoyerMerzabach1991}, as well as the narratives in
 Weisstein~\cite{mathworld:CubicFormula,mathworld:QuarticEquation}.  
 Additional amusing pedagogical investigations
 of the historical solutions may be found in several expositions by
 Nickalls~\cite{NIckalls1993,NIckalls2009}.

 \qquad 
 
 {\bf Further Literature.} A very informative treatment of the features of the quaternion eigenvalue solutions
 has been given by Coutsias,  Seok, and Dill in 2004, and expanded in 2019~\cite{CoutsiasSeokDill2004,CoutsiasWester2019}.   
 Coutsias et al.\  not only  take on a thorough 
 review of the quaternion RMSD method,  but also derive the complete relationship
 between the  linear algebra of the SVD method and  the quaternion eigenvalue system;
 furthermore,  they  exhaustively enumerate the special cases involving
 mirror geometries and degenerate eigenvalues that may appear rarely, but must be dealt with on occasion.    Efficiency is also an area of potential interest, and Theobald et al.\ in 
 \cite{Theobald2005,TheobaldAgrafiotisLiu2010} argue that among the many variants of
 numerical methods that have been used to compute the optimal quaternion eigenvalues,
  Horn's original proposal to use Newton's method
 directly on the characteristic equations of the relevant eigenvalue systems may well be the best approach. 
 
 There is  also a rich literature dealing with distance measures among representations of
 rotation frames themselves, some dealing directly with the properties of  distances computed
 with rotation matrices or quaternions, e.g., Huynh~\cite{Huynh:2009:MRC}, and others
 combining discussion of the distance measures with associated applications such as
 rotation averaging or finding ``rotational centers of mass,'' e.g.,\ 
 \cite{BrownWorsey1992,ParkRavani:1997,BussFillmore2001,Moakher2002,MarkleyEtal-AvgingQuats2007,HartleyEtalRotAvg2013}.   The specific computations explored in
 our section on optimal alignment of matched pairs of orientation frames make 
 extensive use of the quaternion-based and rotation-based measures discussed in
 these treatments.   
 In the Appendices,  we review the details of some of these orientation-frame-based
 applications.

 \qquad


\section{Introduction}


We explore the problem of finding global rotations that optimally
align pairs of corresponding lists of spatial and/or orientation data.
This issue is significant in diverse application domains.  Among these
are aligning spacecraft (see, e.g.,
\cite{Wahba1965,Davenport1968,Markley1988,MarkleyMortari2000}), obtaining
correspondence of registration points in 3D model matching (see, e.g.,  
\cite{FaugerasHebert1983,FaugerasHebert1986}), matching structures in aerial imagery (see, e.g.,
\cite{Horn1987,HornHN1988,HuangBlosteinMargerum1986,ArunHuangBlostein1987,Umeyama1991,Zhang-cam-cal-PAMI2000}),
and alignment of matched molecular and biochemical structures (see, e.g., \linebreak
\cite{Kabsch1976,KabschErratum1978,maclachlan1982,Lesk1986,RDiamond1988,Kearsley1989},\linebreak
\cite{Kearsley1990,Kneller1991,CoutsiasSeokDill2004,Theobald2005,TheobaldAgrafiotisLiu2010},
\linebreak\cite{CoutsiasWester2019}). 
There are several alternative approaches that in principle produce the same optimal
global rotation to solve a given alignment problem, and the SVD and 
$ (E^{\t} \cdot E)^{1/2} \cdot E^{-1}$ methods apply to any dimension.
Here we critically examine the quaternion eigensystem decomposition
approach to studying the rotation matrices appearing in the RMSD
optimization formulas for the 3D and 4D spatial alignment problems,
along with the extensions to the 3D and 4D orientation-frame alignment problems.
Starting from the exact quartic algebraic solutions to the eigensystems
arising in these optimization problems, we direct attention to the  elegant
algebraic forms of the eigenvalue solutions appropriate for these applications.
(For brevity, the more complicated 4D treatment is deferred to the Supplementary Material.)
  
Our extension of the quaternion approach to \emph{orientation data} exploits
the fact that 3D  orientation frames can {\it themselves\/} be
expressed as quaternions, e.g., amino acid 3D orientation frames
written as quaternions (see \cite{HansonThakurQuatProt2012}), and we will refer to the
corresponding ``quaternion frame alignment''  task as the QFA problem.  Various proximity
measures for such orientation data have been explored in the
literature (see, e.g.,~\cite{ParkRavani:1997,Moakher2002,Huynh:2009:MRC,Hugginsdj2014Measures}),
and the general consensus is that the most rigorous measure minimizes
the sums of squares of geodesic arc-lengths between pairs of quaternions.  This
ideal QFA proximity measure is highly nonlinear compared to the analogous
spatial RMSD measure, but fortunately there is an often-justifiable
linearization, the chord angular distance measure; we present 
several alternative solutions exploiting this approximation that closely
parallel our spatial RMSD formulation.  In addition, we analyze 
 the problem of optimally aligning  \emph{combined} 
  3D spatial and quaternion 3D-frame-triad data, e.g., for
molecules with composite structure.  Such rotational-translational
measures have appeared mainly in the molecular entropy
literature~\cite{Hugginsdj2014,FogolariEtAl:2016}, where, after some
confusion, it was recognized that the spatial and rotational measures
are dimensionally incompatible, and either they must be optimized
independently, or an arbitrary context-dependent
dimensional constant must appear in any combined measure for the
RMSD+QFA problem.

In the following, we organize our thoughts by first summarizing the
fundamentals of quaternions, which will be our main computational
tool.  We next introduce the spatial measures that underlie the alignment
problem, then examine the quaternion approach to the 3D problem,
together with a class of exact algebraic solutions that can be used
as an alternative to the traditional numerical methods.  Our quaternion approach to
the 3D orientation-frame triad alignment problem is presented next, along
with a discussion of the combined spatial-rotational problem.
  Appendices provide  a tutorial on
the quaternion orientation-frame methodology,  an alternative formulation
 of the RMSD optimization equations, 
 and a summary of Bar-Itzhack's
method~\cite{BarItzhack2000} for obtaining the corresponding
quaternion from a numerical 3D rotation matrix,  along with a treatment of the
closely related quaternion-based rotation averaging problem.

 In the Supplementary Material, we extend all of our 3D results to 4D space, including 
4D spatial RMSD alignment and 4D orientation-based QFA  methods employing
double quaternions  and their relationship
to the singular value decomposition, and also a Bar-Itzhack method for finding a pair of quaternions
corresponding to a numerical 4D rotation matrix.   Other Sections of the 
Supplementary Material explore properties of the RMSD problem for 2D data and
evaluate the accuracy of our 3D orientation frame alignment
approximations, as well as studying and evaluating the properties of combined
measures for spatial and orientation-frame data in 3D.    An appendix is devoted
to further details of the quartic equations and forms of the algebraic solutions related to our
eigenvalue problems.

\section{Foundations of Quaternions}
\label{QuatFoundations.sec}


For the purposes of this paper, we take a quaternion to be a point $q
= (q_0, q_1, q_2, q_3)= (q_0,\,\Vec{q})$ in 4D Euclidean space with
unit norm, $q \cdot q = 1$, and so geometrically it is a point on the
unit 3-sphere $\Sphere{3}$ (see, e.g.,
\cite{HansonQuatBook:2006} for further details about
quaternions).  The first term, $q_{0}$, plays the role of a real
number, and the last three terms, denoted as a 3D vector $\Vec{q}$, play the role of a
generalized imaginary number, and so are treated differently from the
first:   in particular the conjugation operation is taken to be
$\bar{q} = (q_0, - \Vec{q})$.  Quaternions possess a multiplication
operation denoted by $\star$ and defined as follows:
\begin{equation} 
 q \star p =    Q(q)  \cdot p =  \left[ \begin{array}{cccc}
   q_0& -q_1& -q_2& -q_3 \\
   q_1& q_0& -q_3& q_2  \\
   q_2& q_3& q_0& -q_1\\
   q_3& -q_2& q_1& q_0
  \end{array} \right] \cdot\left[ \begin{array}{c}
  p_0\\p_1\\p_2\\p_3 \end{array} \right]  = 
( q_0 p_0 - \Vec{q}\cdot \Vec{p},\; q_0 \Vec{p}+ 
   p_0 \Vec{q} + \Vec{q}\times\Vec{p} ) \ ,
\label{qprod.eq} 
\end{equation}
where the orthonormal matrix $Q(q)$ expresses a form of quaternion multiplication that can
be useful.  Note that the orthonormality of $Q(q)$ means that quaternion multiplication
of $p$ by $q$ \emph{literally} produces a rotation of $p$ in 4D Euclidean space.

Choosing exactly one of the three imaginary components 
in both $q$ and $p$ to be nonzero 
gives back the classic complex algebra
$(q_0 +{\I} q_1)(p_0 + {\I} p_1)= \left( q_0 p_0 - q_1 p_1\right) + 
   {\I} \left(q_0 p_1  + p_0 q_1  \right)$, 
so there are three copies of the complex numbers embedded in the
quaternion algebra; the difference is that in general the final term
$\Vec{q}\times\Vec{p}$ changes sign if one reverses the order, making
the quaternion product order-dependent, unlike the complex product.
Nevertheless, like complex numbers, the quaternion algebra satisfies the nontrivial
``multiplicative norm'' relation 
\begin{equation}
 \|q\|\,\|p\| = \| q \star p\| \ ,
 \label{multiplicativeNorm.eq}
 \end{equation}
 where  $\|q\|^2 =  q\cdot  q =\Re{(q \star \bar{q} )}$,  i.e., quaternions are one of the four 
 possible Hurwitz algebras (real, complex, quaternion, and octonion).
 
 Quaternion  triple products obey generalizations of the 3D vector identities
 $A\cdot (B\times C) = B\cdot (C \times A) = C\cdot(A\times B)$, along with 
 $A\times B = - B\times A$.  The corresponding quaternion
 identities,  which we will need
 in Section \ref{3dframes.sec}, are
 \begin{equation}
 \begin{array}{c@{\ = \ }c@{\ = \ }c}
 r \cdot (q \star p) & q\cdot (r \star \bar{p}) & \bar{r} \cdot (\bar{p} \star \bar{q}) \\
 \end{array}  ,
 \label{quatTriples.eq}
 \end{equation}
 where the complex conjugate entries are the natural consequences of the sign changes
occurring only in the 3D part.
  
It can be shown that  quadratically conjugating a vector $\Vec{x}=(x,y,z)$, written as
a purely ``imaginary'' quaternion $(0,\Vec{x})$ (with only a 3D part), 
by quaternion multiplication is isomorphic to
the construction of a 3D Euclidean rotation $R(q)$ generating all
possible elements of the special orthogonal group $\SO{3}$.  If we compute
\begin{equation}
  q \star (c,\, x,\, y,\, z) \star \bar{q} = (c,\, R(q) \cdot \Vec{x}) 
  \ , \label{qqR3.eq} 
\end{equation} 
we see that only the purely imaginary part is affected, whether or not
the arbitrary real constant $c=0$.
The result of collecting coefficients of the vector term is a  \emph{proper}  orthonormal
3D rotation matrix quadratic in the quaternion elements that takes the form
\begin{equation}
 \left.  \begin{array}{rcl}
R_{ij}(q)    &= & \delta_{ij}\left({q_{0}}^2 - {\Vec{q}}^2\right) + 2 q_{i} q_{j} 
   - 2 \epsilon_{ijk}q_{0} q_{k}\\[0.15in]
  R(q) & =&  \left[ 
 \begin{array}{ccc}
 {q_0}^2+{q_1}^2-{q_2}^2 - {q_3}^2 & 2 q_1 q_2  -2 q_0 q_3 
& 2 q_1 q_3 +2 q_0 q_2   \\
 2 q_1 q_2 + 2 q_0 q_3  &  {q_0}^2-{q_1}^2 + {q_2}^2 - {q_3}^2   
      &  2 q_2 q_3 - 2 q_0 q_1  \\
 2 q_1 q_3 - 2 q_0 q_2  &  2 q_2 q_3 + 2 q_0 q_1 
 & {q_0}^2 - {q_1}^2 - {q_2}^2 + {q_3}^2
\end{array}
\right]  
\end{array} \right\} \ ,
\label{qrot.eq}
\end{equation}
with determinant $\det R(q) = (q\cdot q)^3 = +1$.
The formula for $R(q)$ is technically a two-to-one mapping from
quaternion space to the 3D rotation group because $R(q) = R(-q)$;
changing the sign of the quaternion preserves the rotation matrix.
Note also that the identity quaternion $q_{\ID} =
(1,0,0,0) \equiv q \star \bar{q}\,$ corresponds to the identity rotation
matrix, as does  $ - q_{\ID} = (-1,0,0,0)$. 
The $3\times 3$ matrix $R(q)$ is fundamental not only to the
quaternion formulation of the spatial RMSD alignment problem, but will
also be essential to the QFA orientation-frame problem because the {\it
  columns\/} of $R(q)$ are exactly the needed 
quaternion representation of the {\it frame triad \/} describing the
orientation of a body in 3D space, i.e., the columns are the vectors
of the frame's local $x$, $y$, and $z$ axes relative to an initial
identity frame.

Multiplying a quaternion $p$ by the quaternion $q$ to get a new
quaternion $p' = q\star p$ simply {\it rotates\/} the 3D frame
corresponding to $p$ by the matrix \Eqn{qrot.eq} written in terms of
$q$.  This has non-trivial implications for 3D rotation matrices, for 
which  quaternion multiplication
corresponds exactly to multiplication of two \emph{independent}
$3\times 3$ orthogonal rotation matrices, and we find that 
\begin{equation}
R(q\star p) = R(q)\cdot R(p) \ .
\label{Rqp.eq}
\end{equation}
 This collapse of repeated rotation matrices to a single
 rotation matrix with multiplied quaternion arguments 
 can be continued indefinitely.

 If we choose the following specific 3-variable
parameterization of the quaternion $q$ preserving $q\cdot q = 1$,
\begin{equation}
q = \left(\cos(\theta/2),\,\hat{n}_{1} \sin(\theta/2),\,
     \hat{n}_{2} \sin(\theta/2),\, \hat{n}_{3} \sin(\theta/2) \right)
\label{qqnth.eq}
\end{equation}
(with $\Hat{n}\cdot \Hat{n}=1$), then $R(q) = R(\theta,\Hat{n})$ is
precisely the ``axis-angle'' 3D spatial rotation by an angle $\theta$
leaving the direction $\Hat{n}$ fixed, so $\Hat{n}$ is the lone real
eigenvector of $R(q)$.

\quad

\noindent{\bf The Slerp.} Relationships among quaternions can be studied using the \emph{slerp},
or ``spherical linear interpolation''~\cite{shoemake-1985-quat,JuppKent1987}, that
smoothly parameterizes the points on the shortest geodesic quaternion path
between two constant (unit) quaternions, $q_{0}$ and $q_{1}$, as
\begin{eqnarray} 
\mbox{\it slerp\/}(q_{0},q_{1},s) \equiv q(s)[q_{0},q_{1}] & = &
q_{0}\frac{\sin ((1-s)\phi)}{\sin \phi} +  
q_{1}\frac{\sin (s\: \phi)}{\sin \phi} \ .
\label{slerpdef.eq}
\end{eqnarray}
Here $\cos \phi = q_{0} \cdot q_{1}$ defines the angle $\phi$ between
the two given quaternions, while $q(s=0) = q_{0}$ and $q(s=1) =
q_{1}$.  The "long" geodesic can be obtained for $1 \le s \le 2 \pi/\phi$.
For small $\phi$, this reduces to the standard linear
interpolation $(1-s)\, q_{0} + s\, q_{1}$.  The unit norm 
is preserved,   $q(s)\cdot q(s) =1$ for all $s$,  
so $q(s)$ is always a valid quaternion and $R(q(s))$ defined by
\Eqn{qrot.eq} is always a valid 3D rotation matrix.  We note that one can
formally write \Eqn{slerpdef.eq}   as an exponential of the form
$ q_{0} \star \left(\bar{q}_{0}\star q_{1}\right)^{s}$, but since
this requires computing a logarithm and an exponential whose most efficient
reduction to a practical computer program is \Eqn{slerpdef.eq}, this is 
mostly of pedagogical interest.

In the following we will make little further use of the quaternion's
algebraic properties, but we will extensively exploit \Eqn{qrot.eq} to
formulate elegant approaches to RMSD problems, along with employing
\Eqn{slerpdef.eq} to study the behavior of our data under smooth
variations of rotation matrices.

\vspace*{0.1in}\noindent{\bf Remark on 4D.} Our fundamental formula \Eqn{qrot.eq} 
can be extended to four Euclidean
dimensions by choosing two {\it distinct\/} quaternions in
\Eqn{qqR3.eq}, producing a 4D Euclidean rotation matrix.  Analogously
to 3D, the columns of this matrix correspond to the axes of a 4D
Euclidean orientation frame.   The nontrivial details of the quaternion approach to  aligning
both 4D spatial and 4D orientation-frame data are given in the Supplementary
Material.

\section{Reviewing the 3D Spatial Alignment RMSD Problem}
\label{3DRMSDalign.sec}

We now review the basic ideas of spatial data alignment, and then
specialize to 3D  (see, e.g.,
\cite{Wahba1965,Davenport1968,Markley1988,MarkleyMortari2000,Kabsch1976,KabschErratum1978,maclachlan1982,Lesk1986,FaugerasHebert1983,Horn1987,HuangBlosteinMargerum1986,ArunHuangBlostein1987,RDiamond1988,Kearsley1989,Kearsley1990,Umeyama1991,Kneller1991,CoutsiasSeokDill2004,Theobald2005}).
We will then employ quaternion methods to reduce
the 3D spatial alignment problem to the task of finding the optimal
quaternion eigenvalue of a certain $4\times 4$ matrix.  This is the
approach we have discussed in the introduction, and it
can be  solved using numerical or algebraic eigensystem methods.
In a subsequent section, we will explore in particular the classical
 quartic equation solutions for the
exact algebraic form of the entire four-part eigensystem,
whose optimal eigenvalue and its quaternion eigenvector produce
the optimal global rotation solving the 3D spatial alignment problem.

\qquad

\noindent{\bf Aligning Matched Data Sets in Euclidean Space.}
We begin with the general least-squares form of the RMSD problem, which
is solved by minimizing the optimization measure over the space of rotations,
which we will convert to an optimization over the space of unit quaternions.
We take as input one data array with $N$
columns of D-dimensional points $\{ y_{k}\}$ as the {\it reference\/}
structure, and a second array of $N$ columns of \emph{matched} points $\{ x_{k}\}$ as
the {\it test\/} structure.  Our task is to rotate the latter in space by a global
$\SO{D}$ rotation matrix $R_{D}$ to achieve the minimum value of the
cumulative quadratic distance measure
\begin{equation}   
 {\mathbf S}_{D}  = 
   \sum_{k=1}^{N} \| R_{D} \cdot x_{k} -  y_{k}\|^{2} \ .
\label{RMSD-basic.eq}
\end{equation}
We assume, as is customary, that any overall translational components have been
eliminated by displacing both data sets to their centers of mass 
(see, e.g., \cite{FaugerasHebert1983,CoutsiasSeokDill2004}).  
When this measure is minimized with respect to the
rotation $R_{D}$, the optimal $R_{D}$ will rotate the  test set $\{x_{k}\}$
to be as close as possible to the reference set $\{y_{k}\}$.  Here we
will focus on  3D data sets (and, in the Supplementary Material, 4D data sets) because
those are the dimensions that are easily adaptable to our targeted
quaternion approach.  In 3D, our least squares measure \Eqn{RMSD-basic.eq} 
can be converted directly into
a quaternion optimization problem using the method  of Hebert and Faugeras
detailed in Appendix \ref{fullRMSD3D.app}. 

\qquad

 {\it Remark:} Clifford algebras may support alternative methods as well
as other approaches to higher dimensions (see, 
e.g.,~\cite{HavelNajfeld-GeomAlg1994,CliffordAvg2005}).

\qquad 

\noindent{\bf Converting from Least-Squares Minimization to Cross-Term Maximization.}
  We choose from here onward to focus on an
equivalent method based on expanding the measure given in \Eqn{RMSD-basic.eq},
removing the constant terms, and recasting
the RMSD least squares minimization problem as the task of maximizing the surviving
cross-term expression.  This takes the general form
\begin{equation} 
 \Delta_{D} = \sum_{k=1}^{N} \left(R_{D} \cdot x_{k}\right) \cdot y_{k} =
 \sum_{a=1,b=1}^{D} [R_{D}]_{ba} E_{ab} \ = \  \tr R_{D} \cdot E ,
\label{RMdef.eq}
\end{equation}
 \centerline{where}
\begin{equation} 
E_{ab} = \sum_{k=1}^{N} \, [x_{k}]_{\textstyle_a} \:[y_{k}]_{\textstyle_b} 
     = \left[ \Vec{X}   \cdot \Vec{Y}^{\t} \right]_{ab} 
\label{Edef.eq}
\end{equation}
is the \emph{cross-covariance matrix} of the data,  $[x_{k}]$ denotes the $k$th column of $\Vec{X}$,
and the range of the indices $(a,b)$  is the spatial dimension $D$.


\qquad 

\noindent{\bf Quaternion Transformation of the 3D Cross-Term Form.}  
We now restrict our attention to the \emph{3D cross-term form} 
of  \Eqn{RMdef.eq} with 
pairs of 3D point data related by a proper rotation.  The key step is to substitute
\Eqn{qrot.eq} for $R(q)$ into \Eqn{RMdef.eq}, and pull out the terms corresponding
to pairs of components of the quaternions $q$.  In this way
the 3D expression is transformed
into the $4 \times 4$  matrix  $M(E)$
sandwiched between two identical quaternions (not a conjugate pair),
of the form
\begin{equation}
 \Delta(q)\,  = \,  \tr R(q) \cdot E  \, = \, (q_0,q_1,q_2,q_3) \cdot M(E) \cdot
 (q_0,q_1,q_2,q_3)^{\t} \equiv q \cdot M(E) \cdot q \ .
\label{qM3q.eq}
\end{equation}
 Here $M(E)$ is  the traceless, symmetric  $4\times 4$ matrix  
\begin{equation} 
     M(E) \!  = \!
\left[ \begin{array}{cccc}
 \!\!   E_{xx} + E_{yy} + E_{zz} & E_{yz} - E_{zy} & E_{zx} - E_{xz} &
                     E_{xy} - E_{yx} \!  \\
 \!   E_{yz} - E_{zy} &  E_{xx} - E_{yy} - E_{zz} & E_{xy} + E_{yx} & 
                     E_{zx} +E_{xz}  \! \\
 \!   E_{zx} - E_{xz} &  E_{xy} + E_{yx} & - E_{xx} + E_{yy} - E_{zz} & 
          E_{yz} + E_{zy}  \!  \\
  \!   E_{xy} - E_{yx} &  E_{zx} +E_{xz}  &  E_{yz} + E_{zy} & 
              - E_{xx} - E_{yy} + E_{zz}  \!\!
                     \end{array} \right] \ .
       \label{basicHorn.eq}              
\end{equation} 
built from our original $3\times 3$ cross-covariance matrix $E$ defined by \Eqn{Edef.eq}. 
We will refer to $M(E)$ from here on as the \emph{profile matrix}, as it essentially
reveals a different viewpoint of the optimization function and its relationship to the 
 matrix $E$.  Note that in some literature, matrices related
to the cross-covariance matrix $E$ may be referred to as  ``attitude profile matrices,''  and one also may
see the term ``key matrix'' referring to $M(E)$.

 
 
The bottom line is that if one decomposes \Eqn{basicHorn.eq} into its
eigensystem, the measure \Eqn{qM3q.eq} is maximized when the
unit-length quaternion vector $q$ is the eigenvector of $M(E)$'s largest
eigenvalue~\cite{Davenport1968,FaugerasHebert1983,Horn1987,RDiamond1988,Kearsley1989,Kneller1991}.
The RMSD optimal-rotation problem thus reduces to finding the maximal
eigenvalue $\epsilon_{\opt}$ of $M(E)$ (which we emphasize depends only on the
numerical data).   Plugging the corresponding eigenvector $q_{\opt}$
into \Eqn{qrot.eq}, we obtain the rotation matrix $R(q_{\opt})$ that
solves the  problem.  The resulting proximity measure relating
$\{x_{k}\}$ and $\{y_{k}\}$ is simply
\begin{equation}  \left. \begin{array}{rcl}
 \Delta_{\opt} & = & q_{\opt} \cdot M(E) \cdot q_{\opt}\\
  & = & q_{\opt} \cdot \left(\epsilon_{\opt}\, q_{\opt} \right)\\
  & = &  \epsilon_{\opt}   \\ \end{array} \right\} \ ,
\label{DeltaOptIsEig.eq}
\end{equation}
and does not require us to actually compute $q_{\opt}$ or
$R(q_{\opt})$ explicitly if all we want to do is compare various test data sets to a
reference structure.

\qquad

\begin{quote}
\noindent{\it \large Note.}  In the interests of conceptual and notational simplicity, we
have made a number of assumptions.  For one thing, in declaring that \Eqn{qrot.eq}
describes our sought-for rotation matrix, we have presumed that the optimal rotation matrix will always
be a proper rotation, with $\det{R} = +1$.  Also, as mentioned, we have omitted any general
translation problems, assuming that there is a way to translate each data set to an appropriate center, e.g., by subtracting the center of mass.  The global translation optimization process is treated
 in~\cite{FaugerasHebert1986,CoutsiasSeokDill2004}, and discussions of center-of-mass alignment, scaling, and point weighting are given in much of the original literature, see e.g.,  \cite{Horn1987,CoutsiasSeokDill2004,Theobald2005}.  Finally, in real problems,  structures such as molecules may appear in mirror-image or enantiomer form, and such issues were introduced
early on by Kabsch~\cite{Kabsch1976,KabschErratum1978}.   There can also be particular symmetries, or very close approximations to symmetries, that can make some of our natural assumptions about the good behavior of the profile matrix invalid, and many of these issues, including ways to treat degenerate cases, have been carefully studied,
see, e.g.,   \cite{CoutsiasSeokDill2004,CoutsiasWester2019}.   
The latter authors also point out that if a particular data set $M(E)$ produces a negative smallest eigenvalue
$\epsilon_{4}$ such that $| \epsilon_{4} | > \epsilon_{\opt}$, this can be a sign of a reflected match,
and the \emph{negative} rotation matrix $R_{\opt} = - R(q(\epsilon_{4}))$ may
 actually produce the best alignment.  These considerations
 may  be essential in some applications, and readers are referred to the original literature
 for details.
 \end{quote}

\qquad

\begin{figure}
\figurecontent{
\centering
 \includegraphics[width=2.6in]{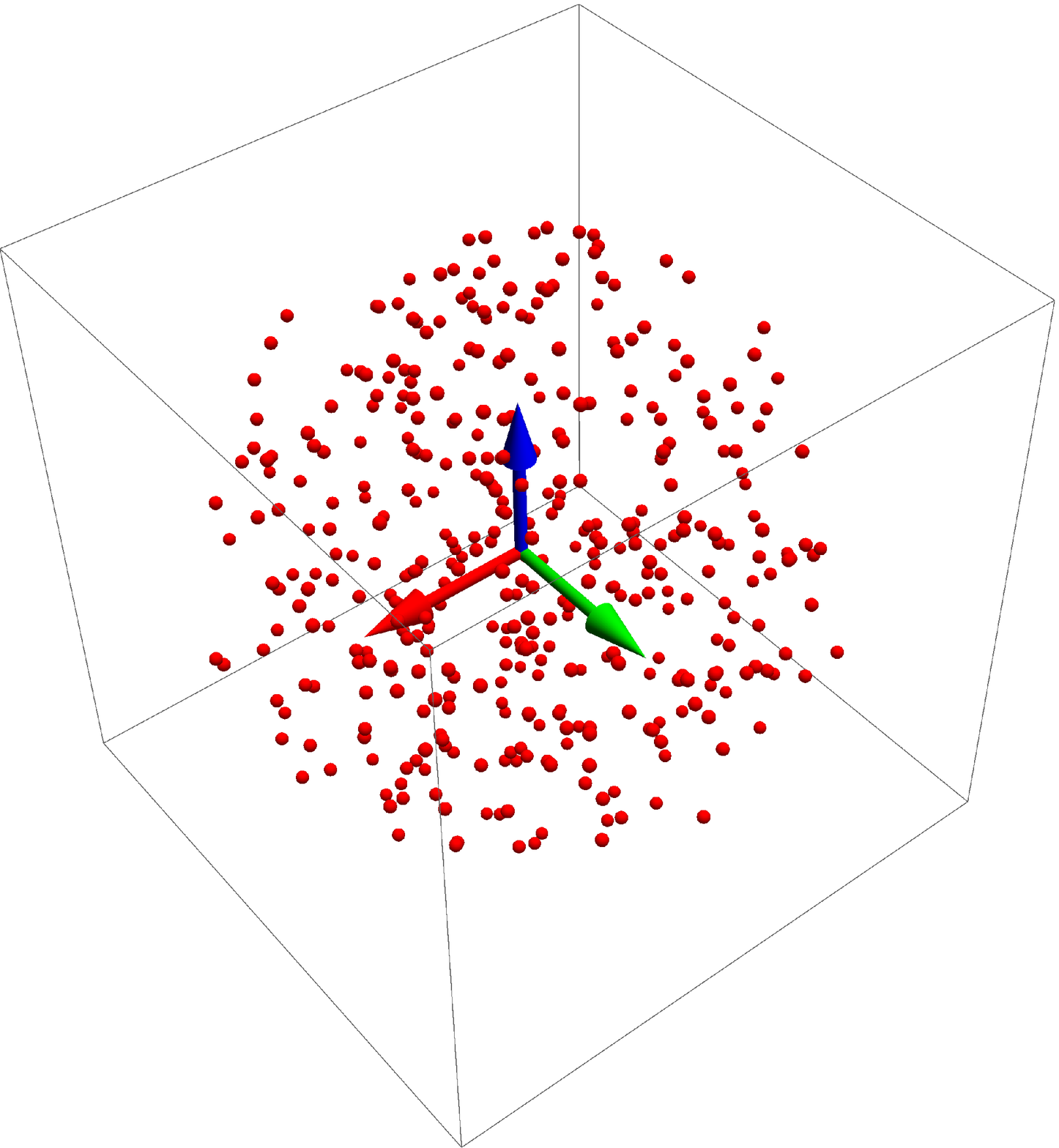}
\hspace{.25in}
\includegraphics[width=2.6in]{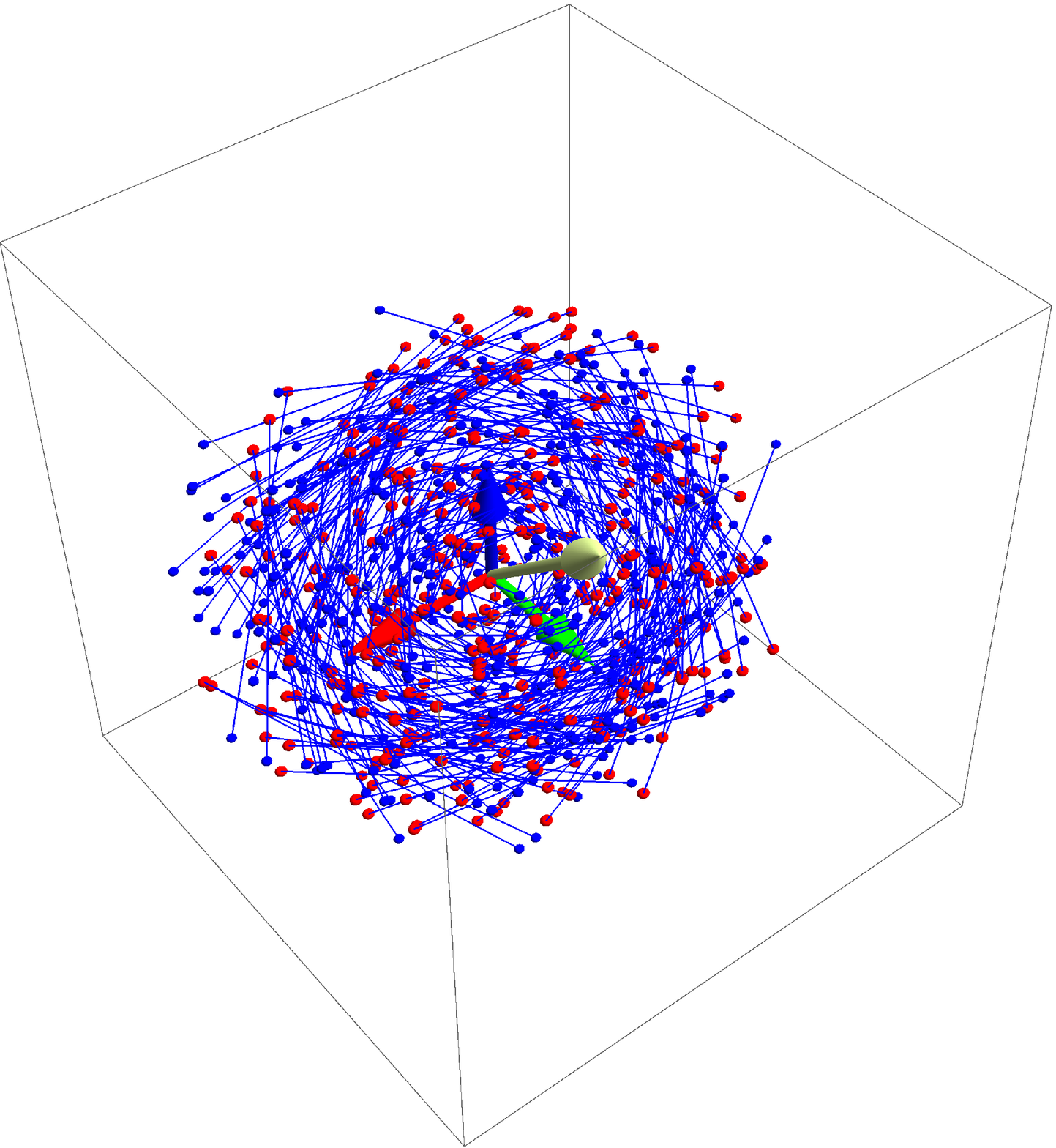}
\centerline {\hspace{1.5in} (A) \hfill (B) \hspace{1.5in} }
\centering
{\ }\\[0.2in]
 \includegraphics[width=2.8in]{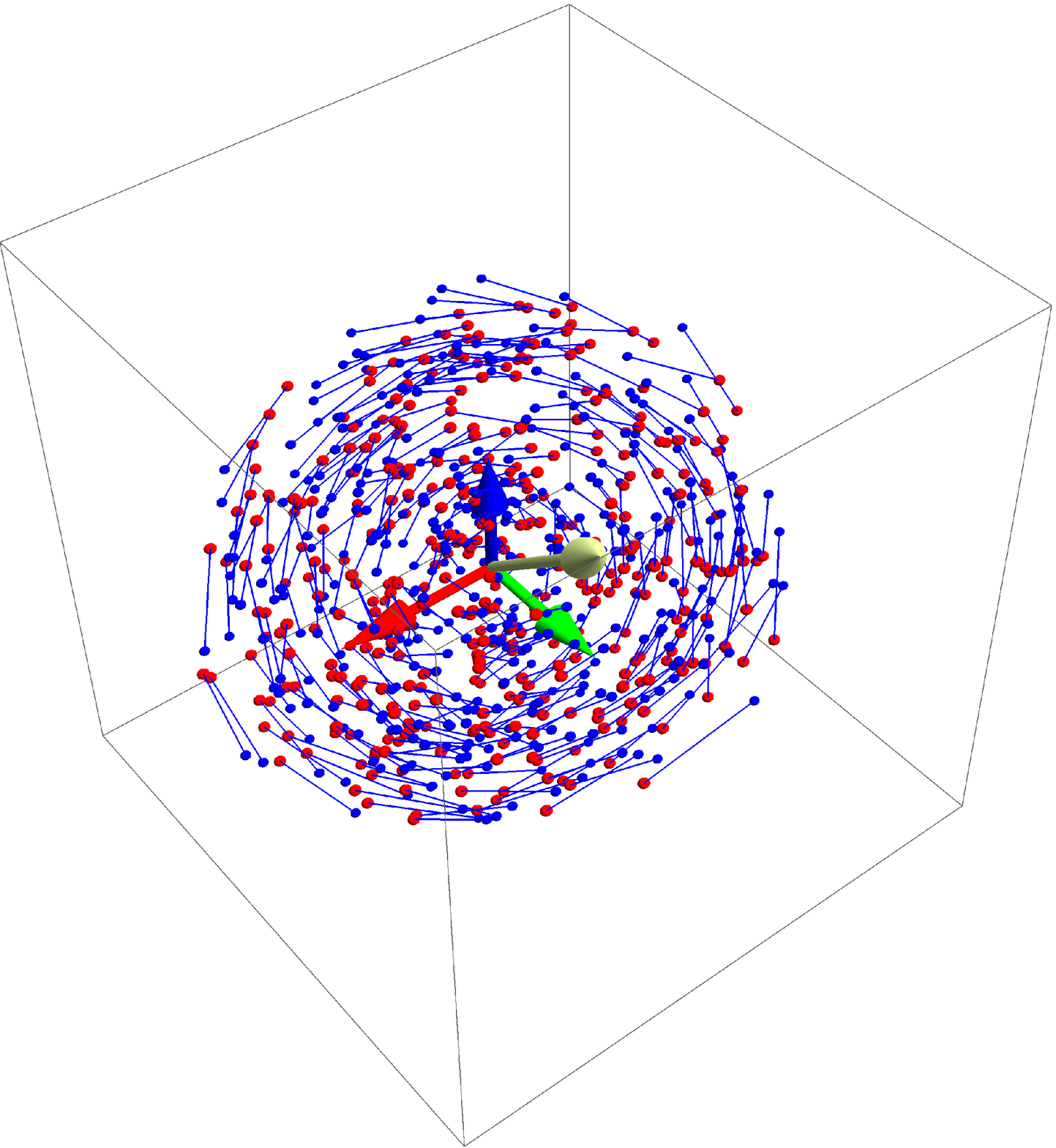}
\hspace{.25in}
\includegraphics[width=2.8in]{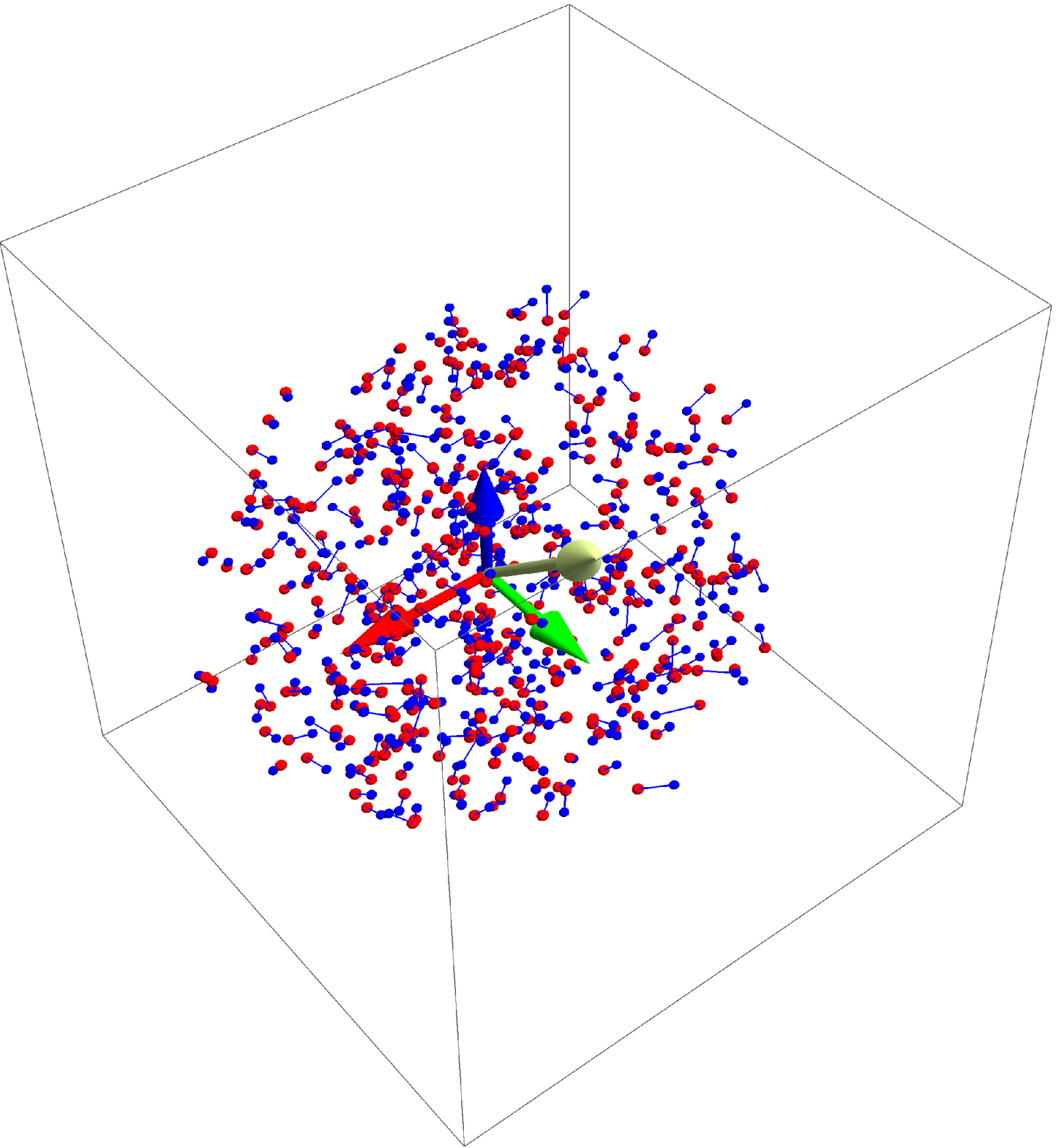}
\centerline {\hspace{1.5in} (C) \hfill (D) \hspace{1.5in} }
}
\caption{\ifnum\ShowFiles=1  {\bf fig3DRef.eps, fig3D2TgtQ.eps,
fig3DRotMatch3abQ.eps, fig3DRotMatch3bQ.eps.} \fi
 (A) A typical 3D spatial reference data set. (B)  The reference data in red
  alongside the test data in blue, 
 with blue lines representing the Euclidean distances connecting each test
data point with its corresponding reference point.
 (C) The partial alignment at $s=0.75$.  (D) The optimal alignment for
this data set at $s=1.0$. The yellow arrow is the axis of rotation
specified by the optimal quaternion's spatial components.}
\label{fig3DRefRotTgt.fig}
\end{figure}

\begin{figure}
\figurecontent{
\centering
 \includegraphics[width=2.6in]{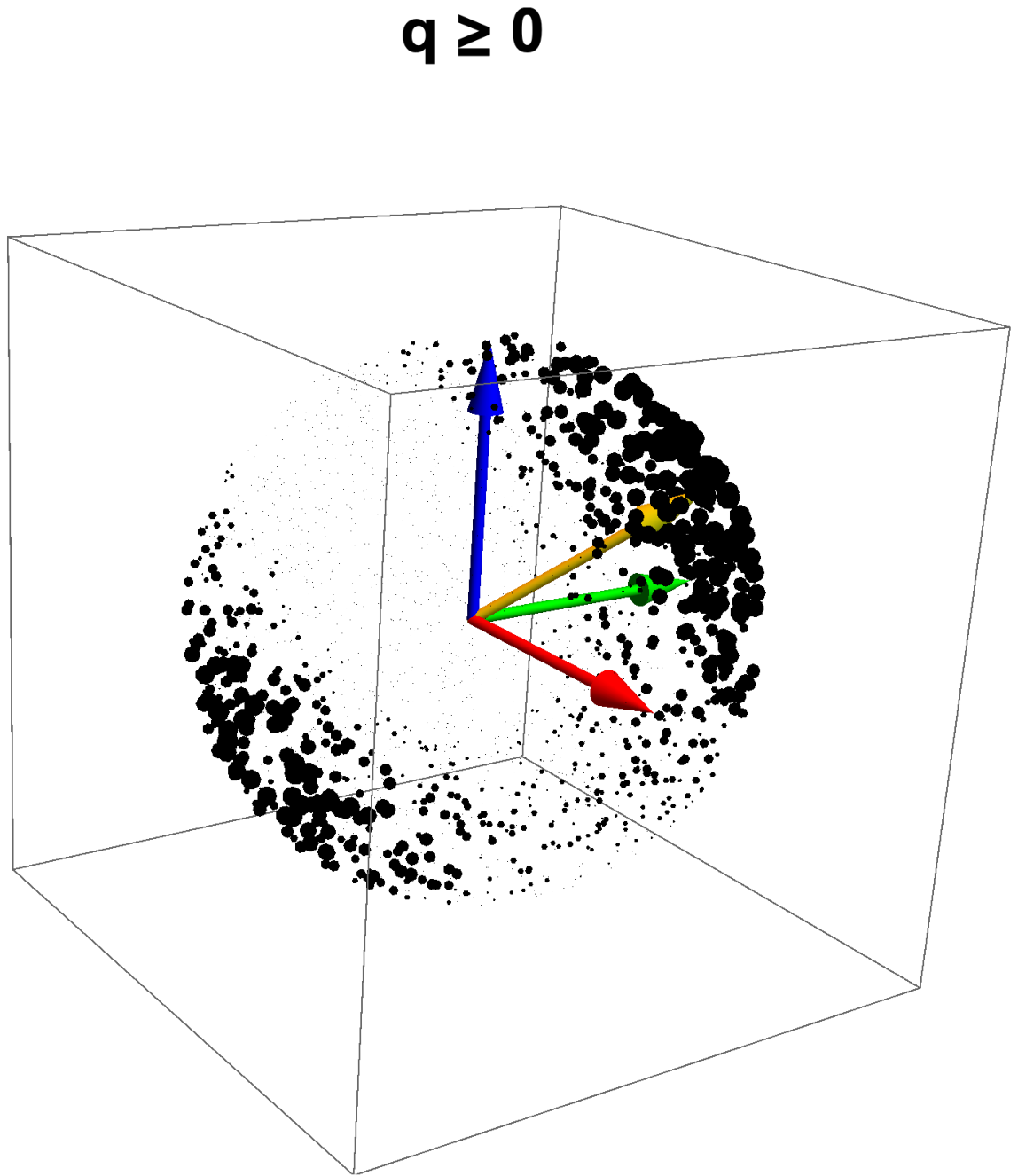} 
 \hspace{0.35in}
  \includegraphics[width=2.6in]{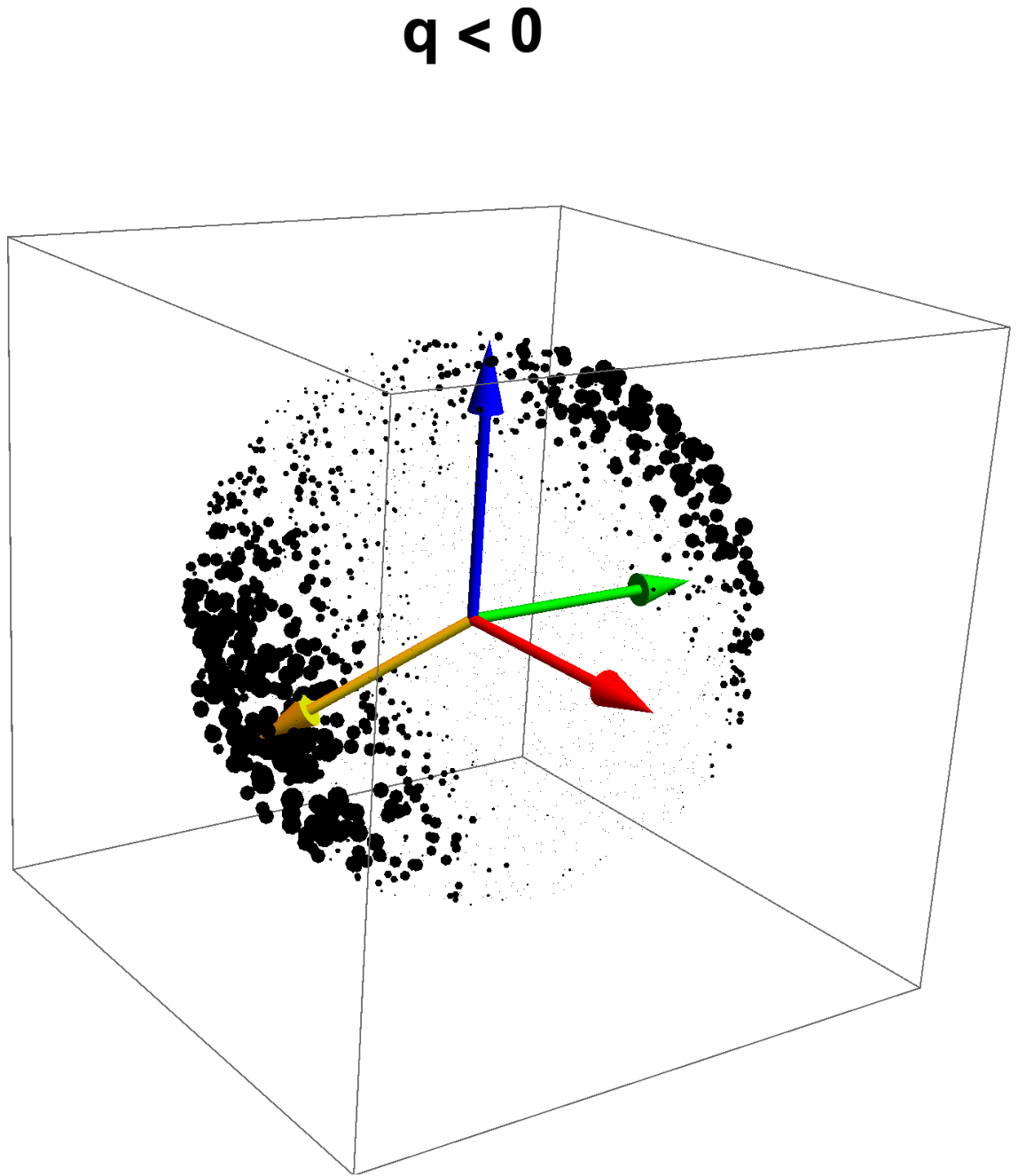} 
  \centerline {\hspace{1.5in} (A) \hfill (B) \hspace{1.5in} }
  }
 \caption{\ifnum\ShowFiles=1  {\bf  fig3DQDeltaPlotA.eps, fig3DQDeltaPlotB.eps.} \fi
 The values of $\Delta(q) = \tr R(q)\cdot E = q \cdot M(E)\cdot q\,$ represented by
 the sizes of the dots placed randomly in the  "northern" and "southern" 3D solid balls
 spanning the entire hypersphere $\Sphere{3}$ with (A) containing the 
 $q_{0} \ge 0$ sector  and {B} containing the $q_{0}<0$ sector.
   We display the data dots at the locations of their
 spatial quaternion components $\Vec{q} = (q_1,q_2,q_3)$,
  and we know that  $q_0= \pm \sqrt{1 - \Vec{q}\cdot\Vec{q}} $
  so the $\Vec{q}$ data uniquely specify the full quaternion.
 Since $R(q)=R(-q)$, the points in \emph{each} ball
 actually represent all possible unique rotation matrices.  The spatial component
 of the maximal eigenvector
 is shown by the yellow arrows, which clearly end  in the middle of the maximum
 values of $\Delta(q)$.  Note that, in the quaternion context, diametrically opposite
 points on the spherical surface are identical rotations, so the cluster of larger
 dots at the upper right of (A) is, in the entire sphere,  representing the
 same data as the "diametrically opposite" lower left  cluster in (B),
 both surrounding  the tips of their own yellow arrows.  The smaller dots
 at the upper right of (B) are contiguous with the upper right region of (A), forming
 a single cloud centered on $\Vec{q}_{\opt}$, and similarly for the lower left of (A) and the
 lower left of (B).  The whole figure contains   two distinct clusters of dots
  (related by  $q \to -q$) centered around  $\pm \Vec{q}_{\opt}$.}
 \label{fig3DQDelta.fig}
 \end{figure}

\noindent{\bf Illustrative Example.} We can visualize the transition from
the initial data $\Delta(q_{\ID}) = \tr{E}$ to the optimal alignment
$\Delta(q_{\opt}) = \epsilon_{\opt}$ by exploiting the geodesic
interpolation \Eqn{slerpdef.eq} from the identity quaternion $q_{\ID}$
to $q_{\opt}$ given by
\begin{equation}
q(s) = \mbox{\it slerp}(q_{\ID},q_{\opt},s) \ ,
\label{IDoptSlerp.eq}
\end{equation} 
and applying the resulting rotation matrix $R(q(s))$ to the test
data, ending with $R(q_{\opt})$ showing the best alignment of the 
two data sets.  In \Fig{fig3DRefRotTgt.fig}, we show a sample
reference data set in red, a sample test data set in blue connected to
the reference data set by blue lines, an intermediate partial
alignment, and finally the optimally aligned pair, respectively.
The yellow arrow is the \emph{spatial part of the quaternion solution},
proportional to the eigenvector $\Hat{n}$ (fixed axis) of the optimal 3D rotation matrix
$R(q) = R(\theta,\,\Hat{n})$, and
whose length is $\sin(\theta/2)$, sine of half the rotation
angle needed to perform the optimal alignment of the test data with
the reference data.  In \Fig{fig3DQDelta.fig}, we visualize the optimization process
in an alternative way, showing  random samples of 
$q=(q_{0},\Vec{q})$ in $\Sphere{3}$, separated into the ``northern hemisphere''
 3D unit-radius ball  in (A) with $q_{0}\ge 0$, and
 the ``southern hemisphere''
 3D unit-radius ball  in (B) with $q_{0} < 0$. (This is like representing 
 the Earth as two flattened disks, one showing everything above the equator
  and one showing everything below  the equator; the distance from the equatorial
  plane is implied by the location in the disk, with the maximum at the centers,
  the north and south poles.)
 Either solid ball contains one unique quaternion 
  for every possible choice of $R(q)$.  The values
of $\Delta(q) = \tr R(q)\cdot E\,$ are shown as scaled dots located
at their corresponding spatial (``real'') quaternion points  $\Vec{q}$ in the solid balls.
The yellow arrows, equivalent negatives of each other, show 
the spatial part $\Vec{q}_{\opt}$
of the optimal quaternion $q_{\opt}$, and
the tips of the arrows clearly fall  in the middle of the mirror pair of clusters 
of the largest values of $\Delta(q)$.  Note that the lower left dots in (A)
 continue smoothly into the larger lower left dots in (B), which
 is the center of the optimal quaternion in (B). 
 Further details of such methods of displaying
quaternions are provided in Appendix \ref{IntroQuatFrames.app}
(see also \cite{HansonQuatBook:2006}).

\section{Algebraic Solution of the Eigensystem for 3D \\ Spatial Alignment}
\label{eigensystem.sec}


At this point, one can simply use the traditional  \emph{numerical} methods to
solve \Eqn{qM3q.eq} for the 
maximal eigenvalue $\epsilon_{\opt}$ of $M(E)$ and its eigenvector
$q_{\opt}$, thus solving the 3D spatial alignment problem of \Eqn{RMdef.eq}.
Alternatively, we can also exploit \emph{symbolic} methods to study the properties of the 
eigensystems of  $4 \times 4$ matrices $M$  algebraically  to provide deeper insights into 
the structure of the problem, and that is the subject of this Section.

\quad

Theoretically, the algebraic form of our eigensystem is a textbook problem following from
the 16th-century-era solution of the quartic algebraic equation in, e.g.,\ \cite{AbramowitzStegun1970}.  
Our objective here is to explore this textbook solution in the specific context
of its application to eigensystems of  $4\times 4$ matrices and its  behavior
relative to the properties of such matrices.  The real, symmetric, traceless profile  matrix
$M(E)$ in \Eqn{basicHorn.eq}
appearing in the 3D spatial RMSD optimization problem must necessarily possess
only real eigenvalues, and the properties of $M(E)$ permit some particular
simplifications in the algebraic solutions that we will discuss.  The quaternion RMSD literature
varies widely in the details of its treatment of the algebraic solutions, ranging from no discussion at all,
to Horn, who mentions the possibility but does not explore it, to the work of Coutsias 
et al.~\cite{CoutsiasSeokDill2004,CoutsiasWester2019}, who present an exhaustive treatment, in addition 
to working out the exact details of the correspondence between the SVD eigensystem and
the quaternion eigensystem, both of which in principle embody the algebraic solution to the RMSD optimization
problem.  In addition to the treatment of Coutsias et al., other approaches similar to the
one we will study  are due to 
 Euler~\cite{Euler-onRoots-1733,Euler-Bell-1733}, as well as  a series of papers on the quartic by 
Nickalls~\cite{NIckalls1993,NIckalls2009}.

\qquad

\noindent{\bf Eigenvalue Expressions.}
We begin by writing down the eigenvalue expansion of the profile matrix,
\begin{equation}
\det [M - e I_{4}]  \; = \; 
     e^4 + e^3 p_1 + e^2 p_2 + e p_3 + p_4 \; =0\;  \label{3DeigEqne.eq}\ ,
\end{equation}
where $e$ denotes a generic eigenvalue,  $I_{4}$ is the 4D
identity matrix, and  the $p_{k}$ are homogeneous polynomials of degree $k$ in the elements of $M$.
For the special case  of a traceless, symmetric profile matrix $M(E)$ defined by \Eqn{basicHorn.eq},
the $p_{k}(E)$ coefficients simplify and can be expressed  numerically as the following 
 functions  either of $M$ or of $E$:
\begin{eqnarray}
p_{1}(E) & = & - \tr [ M ] \  = 0   \label{p1-3D.eq} \\
p_{2}(E) &   =& -\frac{1}{2} \tr [ M \cdot M ]  \ = \   -2 \tr [ E \cdot  {E}^{\t} ] \nonumber \\
 & = & -2
   \left(E_{xx}^2+E_{xy}^2+E_{xz}^2+E_{yx}^2+E_{yy}^2+E_{yz}^2+E_{zx}^2
     +E_{zy}^2+E_{zz}^2\right) 
          \label{p2-3D.eq}     \\
p_{3}(E) &   =  & -\frac{1}{3} \tr \left[ M\cdot M \cdot M\right] 
\  = \  -8 \det [ E ] \nonumber \\ 
  &  = &   8 \left(E_{xx} \,E_{yz} \,E_{zy} +  E_{yy}\, E_{xz} \,E_{zx}+   E_{zz} \,E_{xy} \,E_{yx}   \right)   \nonumber\\
      && \mbox{}   - 8 \left( E_{xx}\, E_{yy}\, E_{zz}+E_{xy}\,  E_{yz}\, E_{zx} + E_{xz}\, E_{zy}\, E_{yx}  \right)   \label{p3-3D.eq}  \\
p_{4}(E) & =& \det [ M ]   \ = \    
                         2  \tr [  E \cdot  {E}^{\t} \cdot  E \cdot  {E}^{\t} ] - \left( \tr [  E \cdot  {E}^{\t}  ]\right)^2 
     \label{p4-3D.eq}  \ .
\end{eqnarray}
Interestingly, the polynomial $M(E)$ is arranged so that $-p_{2}(E)/2$ is the (squared) 
Fr\"{o}benius norm of $E$, and $-p_{3}(E)/8$ is its determinant.
Our task now is to express the four eigenvalues   $e = \epsilon_{k}(p_1,p_2,p_3,p_4)$,
 $k=1,\ldots,4$, usefully in 
 terms of the matrix  elements, and also to find their eigenvectors;  we are of course particularly  interested in the  maximal eigenvalue $\epsilon_{\opt}$.

\qquad

\noindent{\bf Approaches to Algebraic Solutions.\,}  Equation (\ref{3DeigEqne.eq}) can be solved directly 
using  the quartic equations published by Cardano in 1545 (see, e.g., \cite{AbramowitzStegun1970,mathworld:QuarticEquation,ArsMagnaCardano1545}),
which are incorporated into the Mathematica function  
\begin{equation}
\mathtt{Solve[myQuarticEqn[e] == 0, e, Quartics}\, \rightarrow\, \mathtt{True]}
   \label{SolveQuarticT0.eq}  \end{equation}
that immediately returns a suitable algebraic formula.  At this point we defer
detailed discussion of the textbook solution to the Supplementary Material,
and instead focus on a particularly symmetric version of the solution and the
form it takes for the eigenvalue problem for traceless, symmetric $4\times 4$
matrices such as our profile matrices $M(E)$.   For this purpose, we
 look for an alternative solution by considering the following traceless ($p_{1}=0$)
Ansatz: 
\begin{equation}  \left.
\begin{array}{rclp{0.25in}rcl}
\epsilon_{1}(p) & \stackrel{\mbox{?}}{ = } &  \sqrt{X(p)} + \sqrt{Y(p)} + \sqrt{Z(p)} &&
\epsilon_{2}(p) &\stackrel{\mbox{?}}{ = } &    \sqrt{X(p)} - \sqrt{Y(p)}  -  \sqrt{Z(p)} )\\[0.2in]
\epsilon_{3}(p) & \stackrel{\mbox{?}}{ = } &    \sqrt{X(p)} + \sqrt{Y(p)} - \sqrt{Z(p)}&&
\epsilon_{4}(p) &\stackrel{\mbox{?}}{ = } &   \sqrt{X(p)} - \sqrt{Y(p)} +  \sqrt{Z(p)} \\
\end{array}  \hspace{0.25in}\right\} \label{eps1to4.eq} \ .
\end{equation}
This form  emphasizes some  additional explicit
symmetry that we will see is connected to the role of cube roots
 in the quartic algebraic solutions (see, e.g, \cite{CoutsiasWester2019}) .
   We can turn it into an equation for $\epsilon_{k}(p)$
 to be solved in terms of the matrix parameters $p_{k}(E)$ as follows:
 First we  eliminate $e$ using $(e- \epsilon_{1})(e- \epsilon_{2})(e- \epsilon_{3})(e- \epsilon_{4})=0$ 
   to express the matrix data expressions $p_{k}$ directly in terms of
   totally symmetric polynomials
of the eigenvalues in the form \cite{AbramowitzStegun1970}
\begin{equation} 
\left. \begin{array}{rcl}
p_1  &  = &-\epsilon_{1} - \epsilon_{2} -  \epsilon_{3}
    -  \epsilon_{4}  \\[0.05in]
p_2 & = &\epsilon_{1} \epsilon_{2} + \epsilon_{1}  \epsilon_{3} +
    \epsilon_{2}  \epsilon_{3} + \epsilon_{1}  \epsilon_{4} +
    \epsilon_{2}  \epsilon_{4} +  \epsilon_{3}  \epsilon_{4}
     \\[0.05in]
p_3  & = & -\epsilon_{1} \epsilon_{2}  \epsilon_{3} -
    \epsilon_{1} \epsilon_{2}  \epsilon_{4} - \epsilon_{1} 
    \epsilon_{3}  \epsilon_{4} - \epsilon_{2}  \epsilon_{3} 
    \epsilon_{4}    \\[0.05in]
p_4    & = &   \epsilon_{1} \epsilon_{2}  \epsilon_{3} \epsilon_{4} 
\end{array}  \ \ \right\} \ .
\label{pNasEps.eq}
\end{equation}
Next we substitute our expression \Eqn{eps1to4.eq}  for   the
$\epsilon_{k}$  in terms of the $\{X,Y,Z\}$ functions into
\Eqn{pNasEps.eq}, yielding a completely different alternative to \Eqn{3DeigEqne.eq}
that \emph{also}  will solve the 3D RMSD eigenvalue problem if we can invert it
to express $\{X(p),\,Y(p),\,Z)p)\}$ in terms of the data $p_{k}(E)$ as presented in
\Eqn{p4-3D.eq}:
\begin{equation} 
\left. \begin{array}{rcl}
p_{1} & = & 0   \\
p_2 & = &  - 2\,(X + Y + Z)  \\
p_3  & = &  -8\,  \sqrt{X\, Y\, Z}   \\ 
p_4 & = &  X^2 + Y^2 + Z^2   -2 \left( Y Z  + Z X  + X Y \right) 
\end{array} \right\} \ .
 \label{pkOfxyzT0.eq}
\end{equation} 
We already see the critical property in $p_{3}$ that, while $p_{3}$ itself
has a deterministic sign from the matrix data, the possibly
variable signs
of the square roots in \Eqn{eps1to4.eq}  have to be constrained so their
product $  \sqrt{X\, Y\, Z}$ agrees with the sign of $p_{3}$.
 Manipulating the quartic equation solutions that we can obtain by
applying the library function \Eqn{SolveQuarticT0.eq} to \Eqn{pkOfxyzT0.eq}, 
and restricting our domain to real traceless, symmetric matrices (and hence  real eigenvalues),
we find solutions
for  $X(p)$, $Y(p)$, and $Z(p)$  of the following form:
\begin{equation}
  \begin{array}{rcl}   
F_{ f}(0,p_2,p_3,p_4) & =& {\displaystyle+ \frac{1}{6}}
\left(\rule{0in}{1.2em} r(p) \cos_{ f}(p) -p_{2}  \right) \\[0.05in]  
\end{array} \label{xyzsolnMain.eq} \\[0.1in] 
\end{equation}
      \centerline{\hspace{-1.0in}\mbox{where the $\cos_{f}(p)$ terms differ only by a cube root phase:}}
\begin{equation}
\begin{array}{c@{\hspace{.07in}}c@{\hspace{.07in}}c}
  \cos_{\textstyle x}(p) \! = \! \cos\left( { \displaystyle \frac{\arg{(a + \I b)}}{3}} \right), & 
  \cos_{\textstyle y}(p) \! = \! \cos\left(  {\displaystyle \frac{\arg{(a + \I b)}}{3} -  \frac{2 \pi}{3} }\right), &
  \cos_{\textstyle z}(p) \! = \! \cos\left(  {\displaystyle\frac{\arg{(a + \I b)}}{3} +
    \frac{2 \pi}{3} }\right) \ . \end{array} \label{xyzcosMain.eq} 
\end{equation}
Here    $\arg(a+\I b) = \mathrm{atan2}(b,a)$ in the C mathematics
library, or $\mathrm{ArcTan}[a,b]$ in Mathematica, 
$F_{f}(p)$ with $f=(x,y,z)$ corresponds to $X(p)$, $Y(p)$, or $Z(p)$, and the utility
functions appearing in the equations for our traceless  $p_1=0$ case are
\begin{equation} 
\left. \begin{array}{rcl}
r^{2}(0, p_2,p_3,p_4) &=&  
   {p_{2}}^{2}   + 12 p_{4} \; = \;\sqrt[3]{a^2 + b^2} \;= \;(a+\I b)^{1/3}  (a-\I b)^{1/3}  \\[0.075in]
a(0, p_2,p_3,p_4) & = &  {p_{2}}^{3} + 
  {\textstyle \frac{1}{2} } \left(  27   {p_3}^2   - 72 {p_2}   {p_4} \right)   \\[0.075in]
b^{2}( 0,p_2,p_3,p_4) & = &  r^{6}(p) - a^{2}(p) \\[0.07in]
 & = &  {\displaystyle \frac{27}{4} }\left(16 p_4 {p_2}^4 - 4 {p_3}^2 {p_2}^3 - 
128 {p_4}^2 {p_2}^2 + 144 {p_3}^2 p_4 p_2 - 27 {p_3}^4 + 
   256 {p_4}^3 \right) 
\end{array} \right\} \ . 
\label{4DxyzsolnabrFuns2.eq}
\end{equation}
 The function $b^{2}(p)$ has the essential
property that, for real solutions to the cubic, which imply the required real solutions to
our eigenvalue equations  \cite{AbramowitzStegun1970}, we must have $b^{2}(p) \ge 0$.
That essential property allowed us to convert the bare solution into
terms involving $\{(a+ \I b)^{1/3}, (a - \I b)^{1/3}\}$  whose sums form  the manifestly
 real cube-root-related cosine terms in \Eqn{xyzcosMain.eq}.
 
 \qquad
 
 \noindent{\bf Final Eigenvalue Algorithm.\,}
While Eqs.~(\ref{xyzsolnMain.eq}) and (\ref{xyzcosMain.eq}) are well-defined,
  square-roots must be taken to finish the computation of the eigenvalues postulated
in \Eqn{eps1to4.eq}.  In our special case of symmetric, traceless matrices such as $M(E)$,
we can always choose the signs of the first two square roots to be positive, but the
sign of the $\sqrt{Z}$ term is non-trivial, and in fact is the sign of $\det[E]$.
The form of the solution in Eqs.~(\ref{eps1to4.eq}) and (\ref{xyzsolnMain.eq}) 
that works specifically for all traceless symmetric matrices such as  $M(E)$
is  given by our equations for   $p_{k}(E)$   in Eqs.~(\ref{p1-3D.eq}--\ref{p4-3D.eq}),
along with Eqs.~(\ref{xyzsolnMain.eq}),  (\ref{xyzcosMain.eq}), and
(\ref{4DxyzsolnabrFuns2.eq}) \emph{provided} we modify \Eqn{eps1to4.eq}
using $\sigma(p) = \!\sign{( \det[E])} \!= \sign{( -p_{3})}$ as follows: 
\begin{equation}  \left.
\begin{array}{rclp{0.05in}rcl}
\epsilon_{1}(p) & =&  \sqrt{X(p)} + \sqrt{Y(p)} + \sigma(p) \sqrt{Z(p)} &&
\epsilon_{2}(p) & =&    \sqrt{X(p)} - \sqrt{Y(p)}  -  \sigma(p) \sqrt{Z(p)} )\\[0.2in]
\epsilon_{3}(p) & =&    \sqrt{X(p)} + \sqrt{Y(p)} -  \sigma(p) \sqrt{Z(p)}&&
\epsilon_{4}(p) & =&   \sqrt{X(p)} - \sqrt{Y(p)} +   \sigma(p) \sqrt{Z(p)} \\
\end{array}  \hspace{0.15in}\right\} \label{eps1to4S.eq} \ .
\end{equation}

The particular order of the numerical eigenvalues in our chosen form
of the solution \Eqn{eps1to4S.eq} is found in regular cases 
to be uniformly nonincreasing in numerical
order for  our $M(E)$ matrices, 
so $\epsilon_{1}(p)$ is always the leading eigenvalue.   This is our preferred 
symbolic version of the solution to  the 3D RMSD problem defined by $M(E)$.

\qquad

\begin{quote} {\large\it Note:\,} 
We have experimentally confirmed the numerical behavior of  \Eqn{xyzsolnMain.eq}  
in \Eqn{eps1to4S.eq} 
with 1,000,000  randomly generated sets of 3D  cross-covariance matrices $E$, along with
 the corresponding profile matrices $M(E)$,   producing numerical values of 
  $p_{k}$ inserted into the equations for $X(p)$, $Y(p)$, and $Z(p)$.   We 
 confirmed that the sign of $\sigma(p)$ varied randomly, and found that
 the algebraically computed values of $\epsilon_{k}(p)$) corresponded to the standard numerical
 eigenvalues of the matrices $M(E)$ in all cases, to within expected variations due to numerical
 evaluation behavior and expected occasional instabilities.  In particular, we
 found a maximum per-eigenvalue discrepancy of about $10^{-13}$  for the \emph{algebraic} methods relative
 to the standard \emph{numerical} eigenvalue methods, and a median difference of  $10^{-15}$, in the context of machine precision of about $10^{-16}$.  (Why did we do this?  Because  we had
 earlier versions of the algebraic formulas that produced anomalies due to inconsistent
 phase choices in the roots, and felt  it worthwhile to perform a practical check on the numerical
 behavior of our final version of the solutions.) \end{quote}

\qquad

\noindent{\bf Eigenvectors for 3D Data.}
The eigenvector formulas corresponding to $\epsilon_{k}$ can be
generically computed by solving  any three rows of
 \begin{equation}
 \left[M(E) \cdot v - e \,v\right] = \left[ A \right]= 0
 \label{adjugateDet.eq}
 \end{equation}
 for the elements of  $v$, e.g., 
$v=(1,\,v_1,\,v_2,\,v_3)$, as a function of some eigenvalue $e$
(of course, one must account for special cases, e.g., if some subspace
of $M(E)$ is already diagonal).  The desired unit quaternion for the
optimization problem can then be obtained from the normalized eigenvector
\begin{equation} 
q(e,E) = \frac{v}{\| v \|} \ .
\label{EigV3.eq}
\end{equation}
Note that this can often have $q_{0} < 0$, and that whenever
the problem in question depends on the sign of $q_{0}$, such as a
\emph{slerp} starting at $q_{\ID}$, one should choose the sign
of \Eqn{EigV3.eq} appropriately;  some applications may also
require an element of statistical randomness, in which case one might
randomly pick a sign for $q_{0}$.   

As noted by Liu et al.~\cite{TheobaldAgrafiotisLiu2010},
a very clear way of computing the eigenvectors for a given eigenvalue
is to exploit the fact that the determinant of \Eqn{adjugateDet.eq} must vanish,
that is $\det[A ]=0$;  one simply exploits the fact that the columns of the adjugate
matrix $\alpha_{ij}$ (the transpose of the matrix of cofactors of the matrix $[A]$ )  
produce its inverse by means of creating multiple  copies of the determinant.  That is, 
 \begin{equation}
 \label{EigVadj.eq}
 \sum_{c=1}^{4} A_{ac} \alpha_{cb} = \delta_{ab} \det [A] \equiv 0, 
 \end{equation}
so we can just compute \emph{any} column of the adjugate via the appropriate
set of subdeterminants and, in the absence of singularities, that
will be an eigenvector (since any of the four columns can be eigenvectors,
 if one fails, just try another).

In the general well-behaved case, the
form of $v$ in the eigenvector solution for any eigenvalue $e =
\epsilon_{k}$ may be  explicitly computed to give the corresponding
quaternion (among several equivalent alternative expressions)  as 
\begin{eqnarray}
q(e,E) & =& \frac{1}{\|v\|} \times
\left[ \begin{array}{c}
2 A B C + A^2 e_{x} + B^2 e_{y} + C^2 e_{z} - e_{x} e_{y} e_{z}\\[0.1in]
A( a A -  b B - c C)   - c B e_{y} - b C e_{z} - a\, e_{y} e_{z}\\[0.1in]
 B(b B  - c C - a A)   - a C e_{z} - c A e_{x} - b\, e_{z} e_{x}\\[0.1in]
 C(c C  - a A - b B )  - b A e_{x} - a B e_{y} - c\, e_{x} e_{y}\\
\end{array} \right] \ ,
\label{EigQvec3.eq}
\end{eqnarray}
where for convenience we define $\{ e_{x} = (e-x+y+z),\, e_{y} = (e +x
- y+z),\, e_{z} = (e+x+y-z) \}$ with $x=E_{xx}$, cyclic, $a = E_{yz} -
E_{zy}$, cyclic, and $ A = E_{yz} + E_{zy}$, cyclic.
We substitute the maximal eigenvector
$q_{\opt} = q(\epsilon_{1},E)$ into \Eqn{qrot.eq} to give the
sought-for optimal 3D rotation matrix
$R(q_{\opt})$ that solves the RMSD problem with 
$ \Delta(q_{\opt}) = \epsilon_{1}$, as we noted in
\Eqn{DeltaOptIsEig.eq}. 

\begin{quote}
{\bf Remark:}  Yet another approach to computing eigenvectors that, surprisingly, 
\emph{almost} entirely avoids any reference to the original matrix, but needs only
 its eigenvalues and minor eigenvalues, has  recently been rescued  from relative obscurity~\cite{EigvecsVals2019}.  (The authors uncovered a long list of
 non-cross-citing literature mentioning the result dating back at least to 1934.)
 If, for a real, symmetric $4\times 4$ matrix $M$ 
  we label the set of four eigenvectors $v_{i}$ by the index $i$ and the components of 
 any single such four-vector by $a$, the   squares of each of
 the sixteen corresponding components take the form
 \begin{equation}
  \left( [v_{i}]_{\textstyle_a} \right)^{2} = \frac
 {\displaystyle  \prod_{j=1}^{3}\left(  \lambda_{i}(M) - \lambda_{j}(\mu_{a})   \right)  }
  {\displaystyle    \prod_{k=1;k \neq  i}^{4}\left(    \lambda_{i}(M) - \lambda_{k}(M)   \right) } \ .
 \label{EigVecVals.eq}
\end{equation}
Here the $\mu_{a}$ are the $3\times 3$ minors obtained by removing the $a$th row and column of $M$,  and the $\lambda_{j}(\mu_{a})$ comprise the list of $3$ eigenvalues of each of
these minors.   Attempting to obtain the eigenvectors by taking square roots is of course
hampered by the nondeterministic sign; however, since the eigenvalues $\lambda_{i}(M)$
are known, and the overall sign of each eigenvector $v_{i}$ is arbitrary, one needs  
to check at most eight sign combinations to find the one for which 
$M \cdot v_{i} = \lambda_{i}(M) \; v_{i} $, solving the problem.  Note that the general
formula extends to Hermitian matrices of any dimension.
\end{quote}

\section{The 3D Orientation Frame Alignment Problem}
\label{3dframes.sec}

We turn next to the orientation-frame problem, assuming that the data
are like lists of  orientations of roller coaster cars, or lists of
residue orientations in a protein, ordered pairwise in some way, but
\emph{without} specifically considering any
spatial location or nearest-neighbor ordering information.  In
$D$-dimensional space, the {\it columns\/} of any $\SO{D}$ orthonormal
$D\times D$ rotation matrix $R_{D}$ are what we mean by an orientation
frame, since these columns are the directions pointed to by the axes
of the identity matrix after rotating something from its defining
identity frame to a new attitude; note that no spatial location
information whatever is contained in $R_{D}$, though one may wish to
choose a local center for each frame if the construction
involves coordinates such as amino acid  atom
locations (see, e.g., \cite{HansonThakurQuatProt2012}).

In 2D, 3D, and 4D, there exist two-to-one quadratic maps from the
topological spaces $\Sphere{1}$, $\Sphere{3}$, and $\Sphere{3}\times
\Sphere{3}$ to the rotation matrices $R_{2}$, $R_{3}$, and $R_{4}$.
These are the quaternion-related objects that we will use to obtain
elegant representations of the frame data-alignment problem.  In 2D,
a frame data element can be expressed as a complex phase, while in 3D the
frame is a unit quaternion (see
\cite{HansonQuatBook:2006,HansonThakurQuatProt2012}).  In 4D (see
the Supplementary Material), the frame is described by a \emph{pair} of unit
quaternions.



\begin{quote}
\noindent{\it \large Note.}  Readers unfamiliar with the use of complex numbers
and  quaternions to obtain elegant representations 
of 2D and 3D orientation frames are encouraged to review the tutorial in Appendix
\ref{IntroQuatFrames.app}.
\end{quote}

\qquad

\noindent{\bf Overview.} We focus now on the problem of
aligning corresponding sets of 3D orientation frames, just as we
already studied the alignment of  sets of 3D spatial coordinates by
performing an optimal rotation.  
There will be more than one feasible method.  We might assume we could just define
the quaternion-frame-alignment or 
``QFA'' problem by converting any list of frame orientation matrices to
quaternions (see~\cite{HansonQuatBook:2006,HansonThakurQuatProt2012}
and also Appendix~\ref{baritzhack.app}), and writing down the quaternion
equivalents of the RMSD treatment in \Eqn{RMSD-basic.eq} and
\Eqn{RMdef.eq}.  However, unlike the linear Euclidean problem,
the preferred quaternion optimization function technically requires a \emph{non-linear} minimization
of the squared sums of geodesic arc-lengths connecting the points on the quaternion
hypersphere $\Sphere{3}$.  The   task of formulating this ideal problem as well as studying
alternative approximations is the subject of its
own branch of the literature, often known as the \emph{quaternionic barycenter} problem or the 
\emph{quaternion averaging} problem  (see, e.g.,   \cite{BrownWorsey1992,BussFillmore2001,Moakher2002,MarkleyEtal-AvgingQuats2007,Huynh:2009:MRC,HartleyEtalRotAvg2013} and also   Appendix \ref{quatbarycenter.app}).   We will focus on $L_{2}$ norms (the aformentioned sums of squares of arc-lengths), although
 alternative approaches to the rotation averaging problem, 
  such as employing  $L_{1}$ norms and using the Weiszfeld algorithm to find the
   optimal rotation numerically, have been advocated, e.g., by \cite{HartleyWeiszfeldRotAvg2011}.
The computation of optimally aligning rotations, based on plausible  exact or approximate measures relating collections of  corresponding pairs of (quaternionic) orientation frames, is now our task.

Choices for the forms of the measures encoding the distance between orientation frames
have been widely discussed, see, e.g., \cite{ParkRavani:1997,Moakher2002,MarkleyEtal-AvgingQuats2007,Huynh:2009:MRC,HartleyWeiszfeldRotAvg2011,HartleyEtalRotAvg2013,Hugginsdj2014Measures}.
Since we are dealing primarily with quaternions, we will start with two measures
dealing directly with the quaternion geometry, the  geodesic arc length and the chord length,
and later on examine some advantages of starting with quaternion-sign-independent
rotation-matrix forms.

\begin{figure}[h!]
 \figurecontent{ \centering
 \includegraphics[width=2.5in]{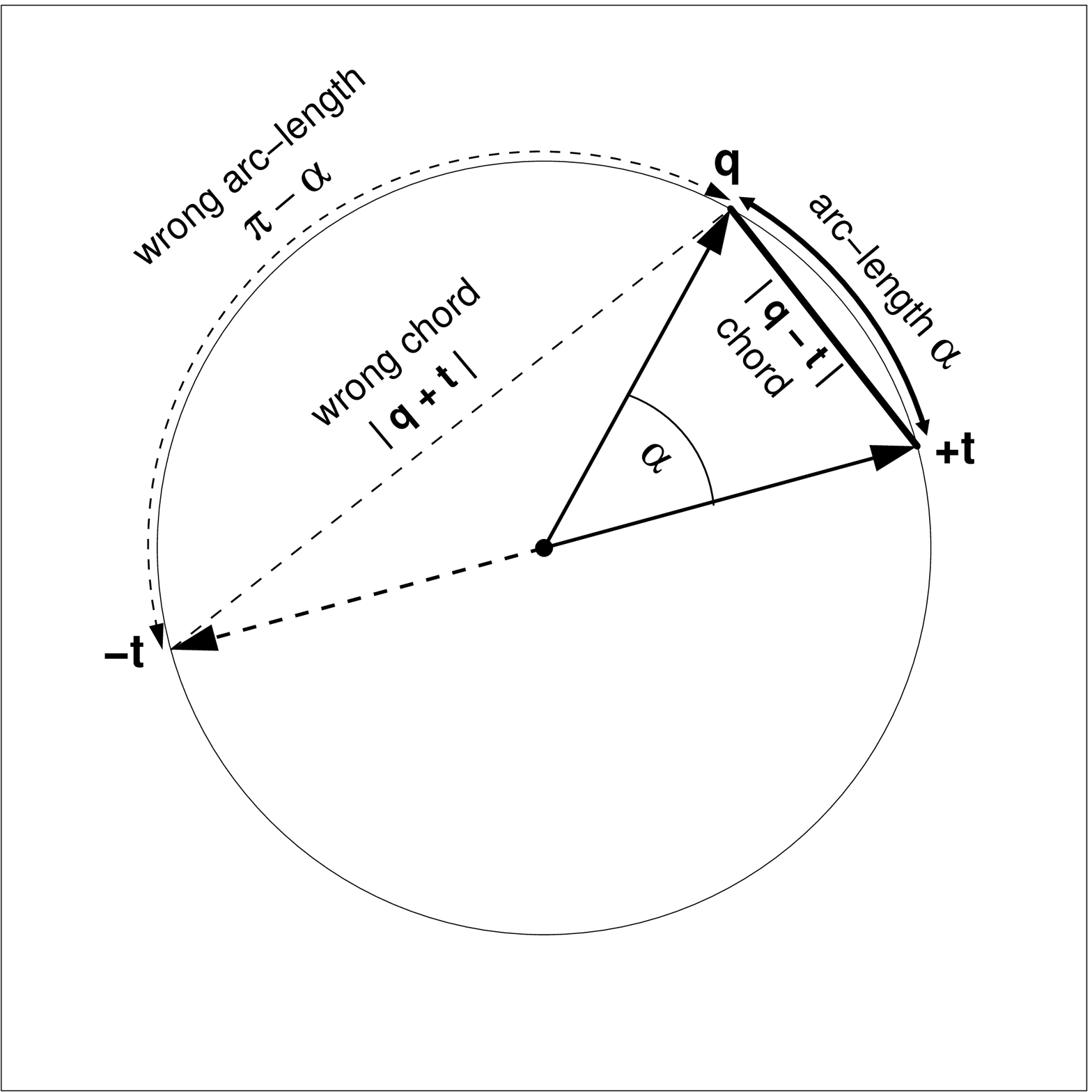} 
 \hspace{0.25in}
  \includegraphics[width=2.5in]{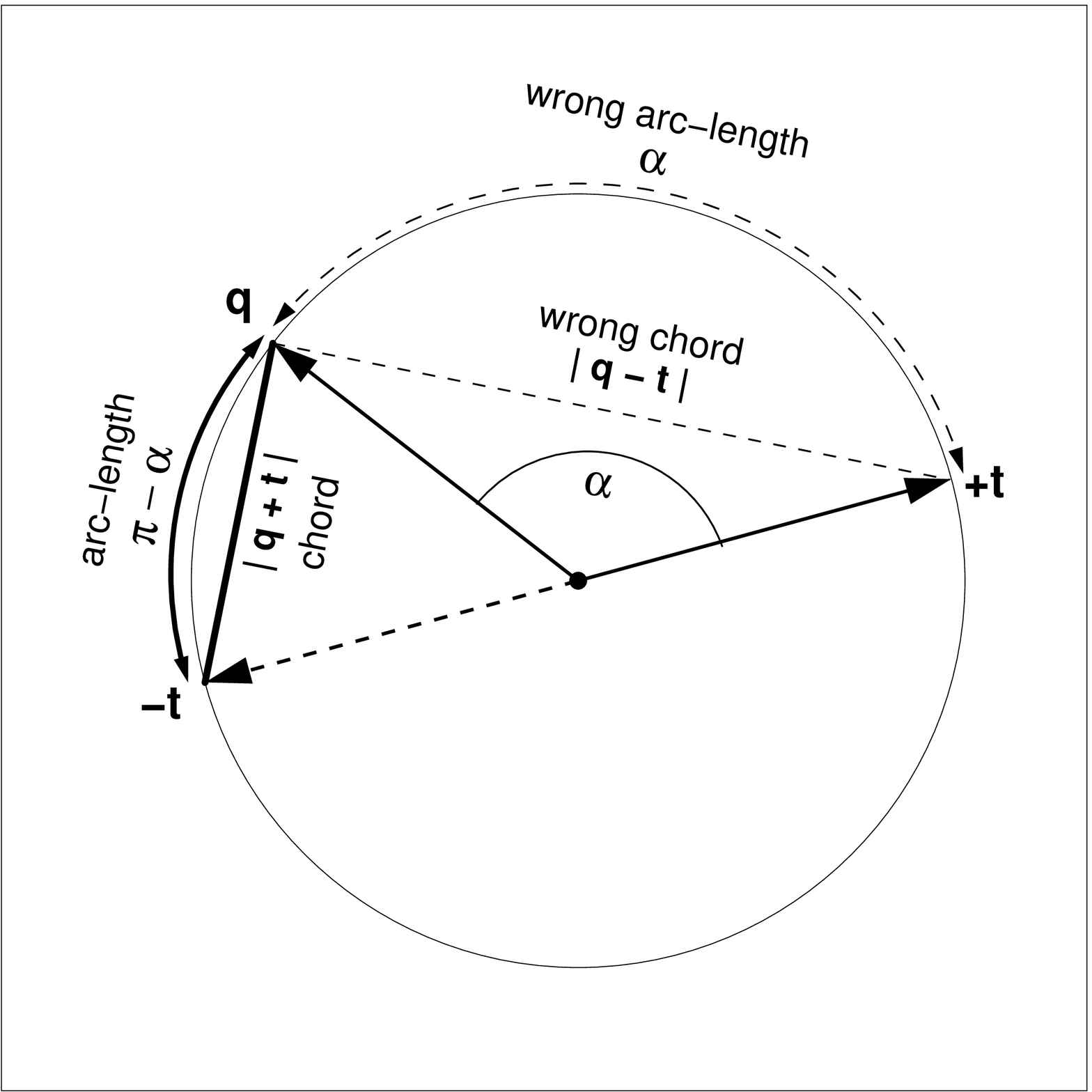} }
  \centerline {\hspace{1.5in} (A) \hfill (B) \hspace{1.5in} }
   \vspace*{.1in}
\caption{\ifnum\ShowFiles=1 {\bf arcvschordA.eps, arcvschordB.eps.} \fi
Geometric context involved in choosing a \emph{quaternion  distance} that will result in
the correct \emph{average rotation matrix} when the quaternion
measures are optimized.
 Because  the quaternion vectors represented by $t$ and $-t$
  give the same rotation matrix, one must choose $|\cos \alpha |$
  or the \emph{minima}, that is   $\min\left( \alpha,\, 
  \pi - \alpha \right)$  or $\min\left(\|q-t\|,\,\|q+t \| \right)$ ,
  of the alternative distance measures to get
  the \emph{correct} items in the arc-length or chord measure summations.
  (A) and (B) represent the cases when the first or second choice should be made, respectively.
  }
\label{arc-chord.fig}
\end{figure}

 \paragraph{3D Geodesic Arc Length Distance.}  First, we recall that the matrix
 \Eqn{qrot.eq} has three orthonormal columns that define a quadratic map from
 the quaternion three-sphere $\Sphere{3}$, a smooth connected Riemannian manifold,
 to a 3D orientation frame.  The squared geodesic arc-length distance between  two quaternions 
 lying  on the three sphere $\Sphere{3}$ is  
 generally agreed upon as the measure of orientation-frame proximity whose
 properties are the closest in principle to the ordinary squared
 Euclidean distance measure \Eqn{RMSD-basic.eq}
  between points~\cite{Huynh:2009:MRC}, and we will adopt this
 measure as our starting point.  We begin by writing down a frame-frame distance measure between two unit quaternions $q_{1}$ and $q_{2}$, corresponding precisely to two orientation
 frames defined by the columns of $R(q_{1})$ and $R(q_{2})$.  We define  the geodesic
 arc length as an angle $\alpha$  on the hypersphere  $\Sphere{3}$ computed geometrically from
 $q_{1} \cdot q_{2}= \cos \alpha$.  As pointed out by 
   \cite{Huynh:2009:MRC,HartleyEtalRotAvg2013}, the  geodesic arc length between
 a test quaternion $q_{1}$ and a data-point quaternion $q_{2}$ of ambiguous sign (since $R(+q_{2}) = R(-q_{2})$)
 can take two values, and we want the minimum value.  Furthermore,
 to work on a spherical manifold instead of a plane, we need basically
 to cluster the ambiguous points in a deterministic way.  
 Starting with the bare angle between two quaternions on $\Sphere{3}$,
 $\alpha = \arccos(q_{1} \cdot q_{2})$, where we recall that $\alpha\ge 0$,
  we define a \emph{pseudometric} \cite{Huynh:2009:MRC}  for the geodesic arc-length 
 distance as
 \begin{equation}
d_{\mbox{geodesic}}(q_{1},q_{2}) = \min (\alpha,\, \pi - \alpha): 
   \  \ 0 \le  d_{\mbox{geodesic}}(q_{1},q_{2}) \le \frac{\pi}{2} \ ,
\label{dgeo.eq}
\end{equation}
illustrated in   \Fig{arc-chord.fig}.
An efficient implementation of this is to take 
 \begin{equation}
d_{\mbox{geodesic}}(q_{1},q_{2}) = \arccos( | q_{1} \cdot q_{2} |) \ .
\label{dgeocos.eq}
\end{equation}
 We now seek to define an ideal 
minimizing $L_2$ orientation frame measure, comparable to our minimizing Euclidean RMSD measure,  but constructed
from geodesic arc-lengths on the quaternion hypersphere instead of Euclidean distances in space.
Thus to compare  a test quaternion-frame data set $\{p_{k}\}$  to a reference data set $\{ r_{k}\}$,
 we propose the geodesic-based least squares measure
 \begin{equation} 
{\mathbf S}_{\mbox{geodesic}} =  \sum_{k=1}^{N}
  \left( \arccos{  \left| \left( q \star p_{k}\right) \cdot  r_{k}
      \right| } \right)^2   =  \sum_{k=1}^{N}
  \left( \arccos{  \left| q \cdot \left(   r_{k} \star \bar{p}_{k}\right)   \right| } \right)^2 
      \ , 
\label{QFA-basic-geodesic}
\end{equation}
where we have used the identities of \Eqn{quatTriples.eq}.
 When $q=q_{\ID}$, the individual
measures correspond to \Eqn{dgeocos.eq}, and otherwise
  ``$q \star p_{k}$''    is the exact analog of ``$R(q) \cdot x_{k}$''  in 
 \Eqn{RMSD-basic.eq},  and
  denotes the quaternion rotation $q$ acting on
the entire set $\{p_{k}\}$ to rotate it to a new orientation that we
want to align optimally with the reference frames $\{r_{k}\}$.  Analogously,
for points on a sphere, the arccosine of an inner product is equivalent
to a distance between points in Euclidean space.

\begin{quote} {\it Remark:}  For improved numerical behavior in the computation
of the quaternion inner product angle between two quaternions,
  one may prefer to convert the arccosine to an arctangent form, 
  $\alpha = \arctan(dx,dy)=\arctan(\cos \alpha,|\sin\alpha |))$ (remember the C math
  library uses the opposite argument order $\mathrm{atan2}(dy,dx)$), 
 with the parameters
\[\cos(\alpha)=| \Re(q_{1} \star {q_{2}}^{-1}) |=  |q_{1} \cdot q_{2}| , \hspace{.25in}
 |\sin(\alpha)|= \left\| \Im(q_{1} \star {q_{2}}^{-1}) \right\| = \left\| -{[q_{1}]}_{0} \Vec{q}_{2} +
  { [q_{2}]}_{0} \Vec{q}_{1}  - \Vec{q}_{1} \times \Vec{q}_{2}  \right\| \  ,
 \]
which is somewhat more stable.
\end{quote}


{\bf Adopting the Solvable Chord Measure.}
Unfortunately, the geodesic arc-length measure does not fit into the linear algebra approach  that we were able to use to obtain exact solutions for the Euclidean data alignment problem treated so far. Thus we are led to investigate instead a very close approximation to $d_{\mbox{geodesic}}(q_{1},q_{2}) $ that \emph{does} correspond closely to the Euclidean data case and does, with some contingencies, admit exact solutions.  This approximate measure is the \emph{chord distance}, whose individual distance terms analogous to \Eqn{dgeocos.eq} take 
the form of a closely related pseudometric \cite{Huynh:2009:MRC,HartleyEtalRotAvg2013},
 \begin{equation}
d_{\mbox{chord}}(q_{1},q_{2}) = \min (\|q_{1} - q_{2} \|, \, \|q_{1} + q_{2} \|): 
   \  \ 0 \le  d_{\mbox{chord}}(q_{1},q_{2}) \le \sqrt{2} \ .
\label{dchord.eq}
\end{equation}
 We compare the geometric origins for \Eqn{dgeocos.eq} and \Eqn{dchord.eq} in \Fig{arc-chord.fig}.
 Note that the crossover point between the two expressions in \Eqn{dchord.eq} is at $\pi/2$, so the
 hypotenuse of the right isosceles triangle at that point has length $\sqrt{2}$.
 \quad
 
 The solvable approximate optimization function analogous to $\|R\cdot x - y \|^{2}$ that we will
 now explore for the quaternion-frame alignment problem will thus take the form
 that must be minimized as
 \begin{equation} 
{\mathbf S}_{\mbox{chord}} =  \sum_{k=1}^{N}
\left( \min (\| (q \star  p_{k} ) - r_{k} \|, \, \|(q \star  p_{k} )+ r_{k} \|)\right)^{2} \ .
\label{initchord.eq}
\end{equation} 
 
  We can convert the sign ambiguity in \Eqn{initchord.eq}
to a deterministic form like \Eqn{dgeocos.eq} by observing, with the help of \Fig{arc-chord.fig}, that 
\begin{equation}
\| q_{1} - q_{2}\|^2 = 2 -2 q_{1} \cdot q_{2}, \hspace{.5in} \| q_{1} +q_{2}\|^2 = 2 +2 q_{1} \cdot q_{2}  \ .
\label{initchordsqrs.eq}
\end{equation}
Clearly $( 2- 2|q_{1} \cdot q_{2}| )$ is always the smallest of the two values.  Thus minimizing 
\Eqn{initchord.eq} amounts to maximizing the now-familiar cross-term form, which we can  write as
\begin{equation}
\left.\begin{array}{rcl}
\Delta_{\mbox{chord}}(q) & = &    \sum_{k=1}^{N}  |(q \star  p_{k} ) \cdot r_{k} | \\[0.1in]
  & = &   
    \sum_{k=1}^{N}  | q  \cdot ( r_{k} \star  \bar{p}_{k} ) | \\[0.1in]
     & = &  \sum_{k=1}^{N}  | q  \cdot t_{k} |
 \end{array} \right\} \ .
 \label{basicQFA.eq}
 \end{equation}
 Here we have used the identity $(q \star  p ) \cdot r  = q  \cdot (r \star  \bar{p}) $ 
 from \Eqn{quatTriples.eq}
 and defined the quaternion displacement or "attitude error" 
 \cite{MarkleyEtal-AvgingQuats2007}
 \begin{equation}\label{deftk.eq}
 t_{k} = r_{k} \star \bar{p}_{k} \ .
 \end{equation}
 Note that we could have derived the same result using \Eqn{multiplicativeNorm.eq} to show that
 $\|q \star p - r\| = \| q \star p - r\| \|p \| = \| q - r \star \bar{p} \|$.
 
 There are several ways to proceed to our final result at this point.  The simplest is to
 pick a neighborhood in which we will choose the samples of $q$ that include our expected
 optimal quaternion, and adjust the sign of each data value $t_{k}$ to $\td{t}_{k}$ by
 the transformation
 \begin{equation}
  \td{t}_{k} =  t_{k}\, \sign(q\cdot t_{k})  \ \ \to  |q\cdot t_{k}|  = q \cdot \td{t}_{k} \ .
\label {tTotilde.eq}
\end{equation}
The neighborhood of $q$ matters because, as argued by \cite{HartleyEtalRotAvg2013},
even though the allowed range of 3D rotation angles is $\theta \in (-\pi ,  \pi)$  (or 
quaternion sphere angles $\alpha \in (-\pi/2, \pi/2)$), convexity of the optimization
problem cannot be guaranteed for collections outside local regions centered on 
some $\theta_{0}$ of size
 $\theta_{0} \in (-\pi/2 ,  \pi/2)$  (or $\alpha_{0}  \in (-\pi/4, \pi/4)$): beyond this range,
local basins may exist that allow the mapping \Eqn{tTotilde.eq} to produce distinct
local variations in the assignments of the $\{\td{t}_{k}\}$ and
in the solutions for $q_{\opt}$.  Within considerations of such constraints, \Eqn{tTotilde.eq}
now allows us to
take the summation outside the absolute value, and write the quaternion-frame optimization
problem in terms of maximizing the cross-term expression
 \begin{equation}
\left.\begin{array}{rcl}
\Delta_{\mbox{chord}}(q) & = & {\displaystyle \sum_{k=1}^{N} }\, q \cdot \td{t}_{k}   \\[0.2in]
  & = &   q \cdot V(t)
   \end{array} \right\}
 \label{solnQFA.eq}
 \end{equation}
 where $V=   \sum_{k=1}^{N}\td{t}_{k}$ is the analog of the Euclidean RMSD profile 
 matrix $M$.   However, since this is \emph{linear} in $q$, we have the remarkable result that,
 as noted in the treatment of \cite{HartleyEtalRotAvg2013} regarding
the quaternion $L_{2}$ chordal-distance norm, the solution is immediate, being simply
 \begin{equation}
 q_{\opt} = \frac{V}{\|V \|} \ ,
\label{VsolnQFA.eq}
\end{equation}
since that immediately maximizes the value of $\Delta_{\mbox{chord}}(q)$
in \Eqn{solnQFA.eq}.  This gives the   maximal value of the measure as 
\begin{equation} 
\Delta_{\mbox{chord}}(q_{\opt}) = \|V\| \ ,
\label{DeltaForVlinear.eq}
\end{equation}
and thus $ \|V\|$ is the exact orientation frame analog of the spatial
RMSD maximal eigenvalue $\epsilon_{\opt}$, except it is far
easier to compute.

\qquad

\begin{figure}
\figurecontent{
\centering
 \includegraphics[width=2.6in]{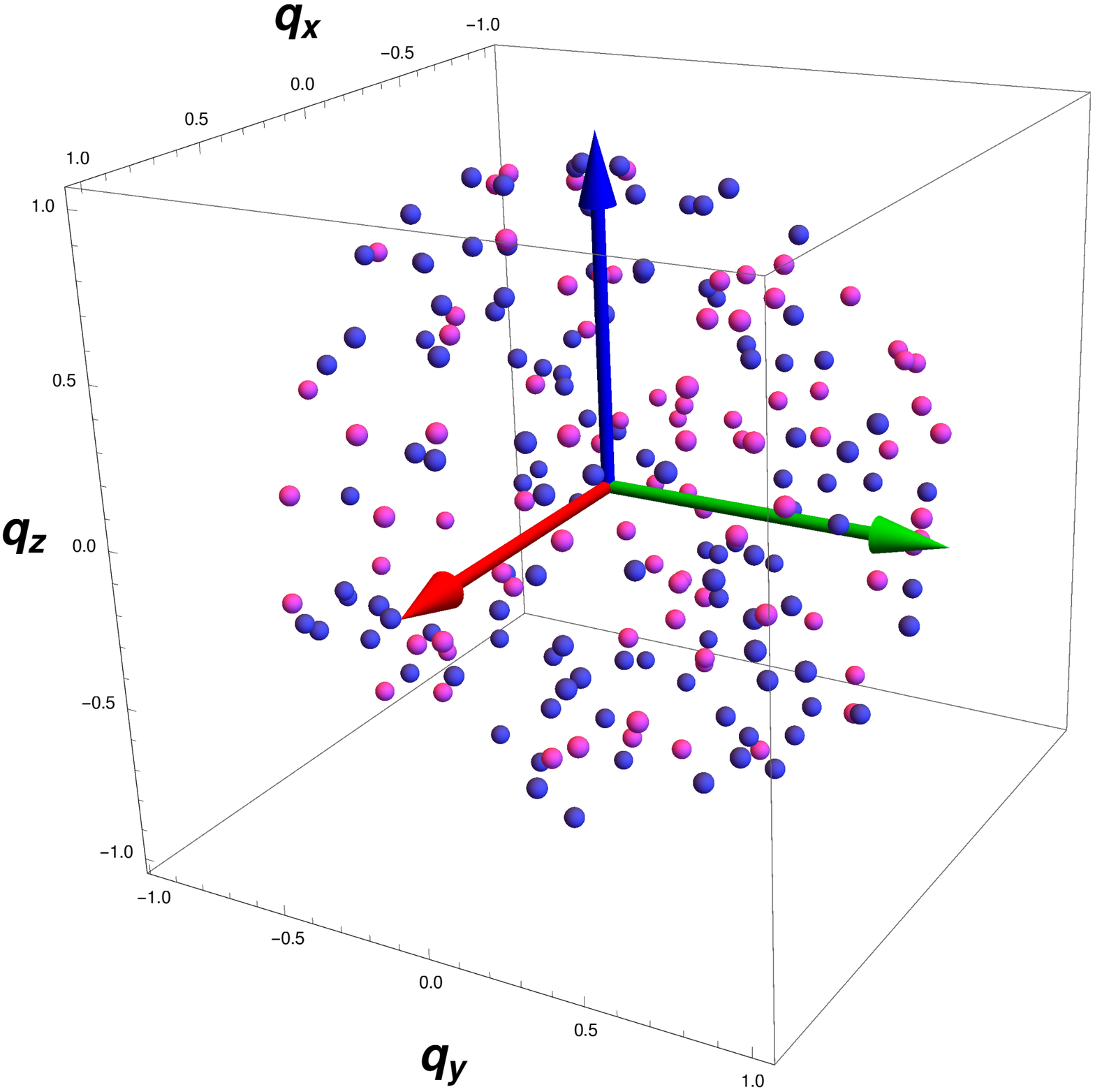}
\hspace{0.35in} 
\includegraphics[width=2.8in]{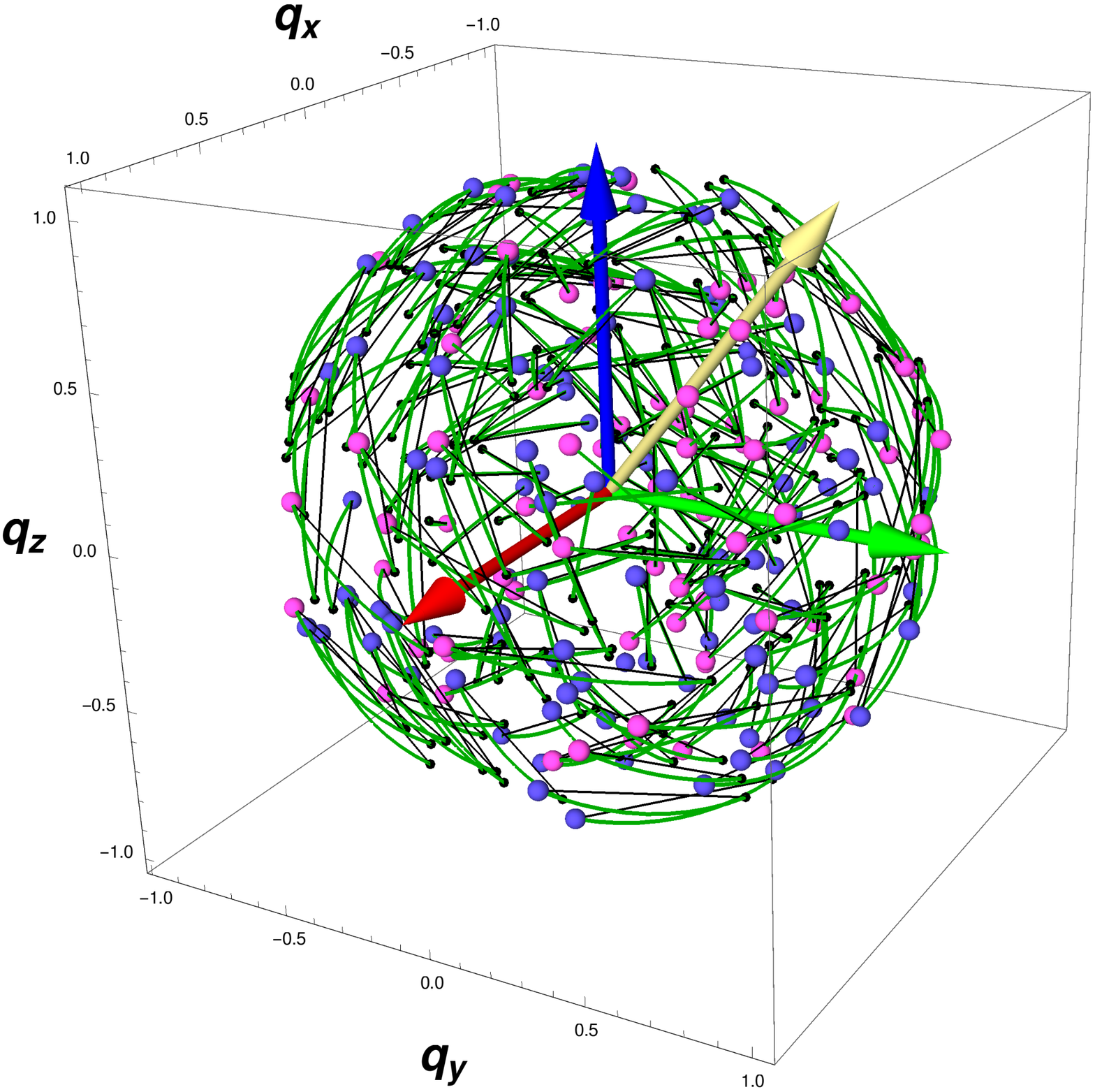} 
\centerline {\hspace{1.5in} (A) \hfill (B) \hspace{1.5in} }
\centering
 \includegraphics[width=2.6in]{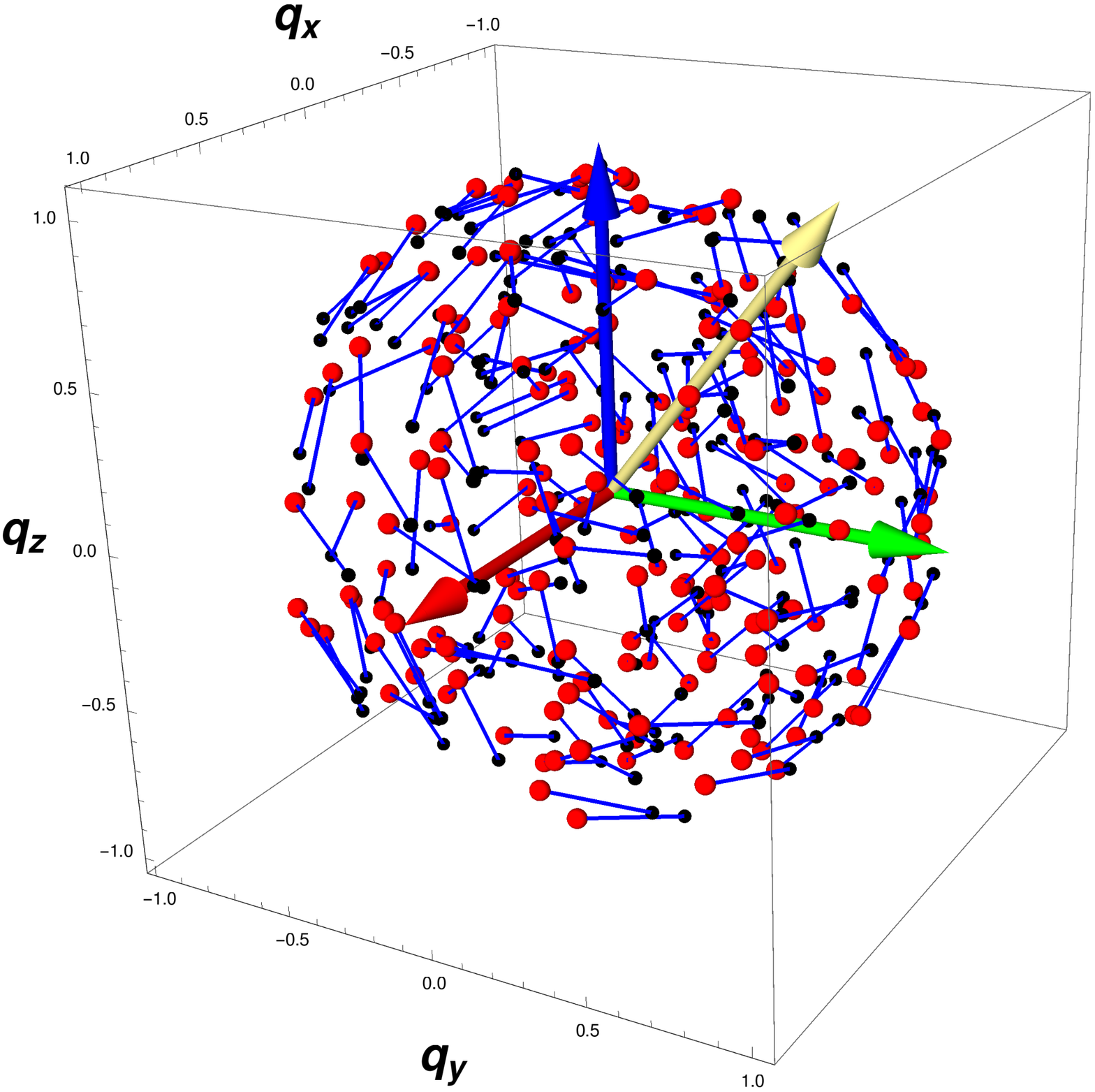}
\hspace{.35in}
 \includegraphics[width=2.6in]{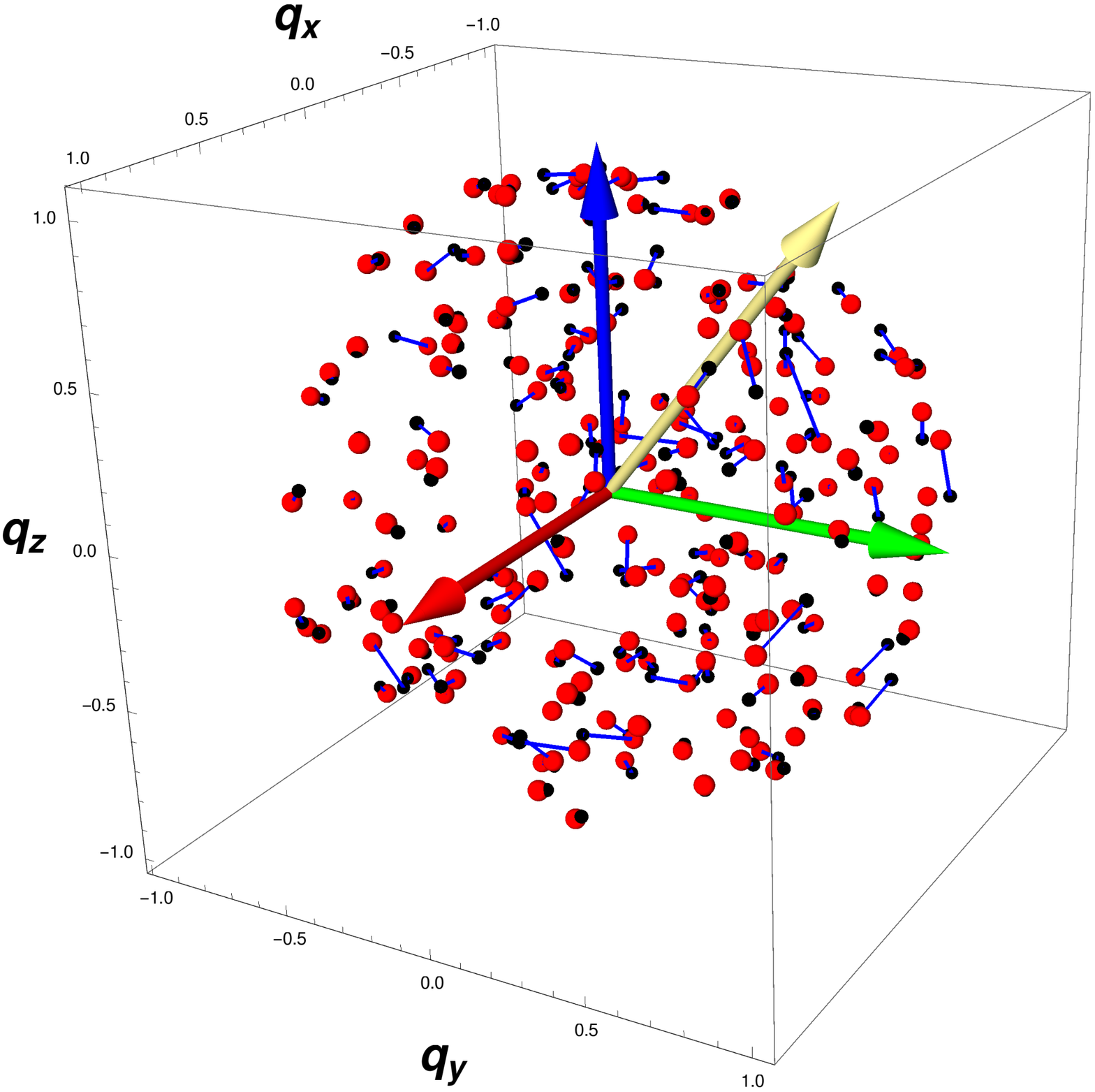}
\centerline {\hspace{1.5in} (C) \hfill (D) \hspace{1.5in} }
}
\caption{\ifnum\ShowFiles=1 {\bf fig3DFrmRef2.eps,
    fig3DFrmArcChord3Q.eps, fig3FRotMatch4BQ.eps, fig3FRotMatch5BQ.eps. } \fi
 3D components of a quaternion orientation data set. 
(A) A quaternion reference set, color coded by the sign of $q_0$.
(B) \emph{Exact} quaternion arc-length distances (green) vs chord
distances (black) between  the test and reference points.
 (C) Part-way from starting state to the aligned state, at $s=0.5$. (D) The final
best alignment at $s=1.0$. The yellow arrow is the direction of the
quaternion eigenvector; when scaled, the length is the sine of half
the optimal rotation angle. }
\label{fig3DFrmMatch.fig}
\end{figure}

\noindent{\bf Illustrative Example.}
Using the quaternion display method described in Appendix
\ref{IntroQuatFrames.app} and illustrated in
\Fig{fig3DRefFrms.fig}, we present in \Fig{fig3DFrmMatch.fig}(A) a
representative quaternion frame reference data set, then in (B) the
relationship of the arc and chord distances for each point in a set
of arc and chord distances (see \Fig{arc-chord.fig}) for each point
pair in the quaternion space.  In \Fig{fig3DFrmMatch.fig}(C,D), we
show the results of the quaternion-frame alignment process using
conceptually the same \emph{slerp} of \Eqn{IDoptSlerp.eq} to
transition from the raw state at $q(s=0)= q_{\ID}$ to $q(s=0.5)$ for
(C) and $q(s=1.0)=q_{\opt}$ for (D).  The yellow arrow is the axis
of rotation specified by the spatial part of the optimal quaternion.

\begin{figure}
\figurecontent{
\centering
 \includegraphics[width=2.6in]{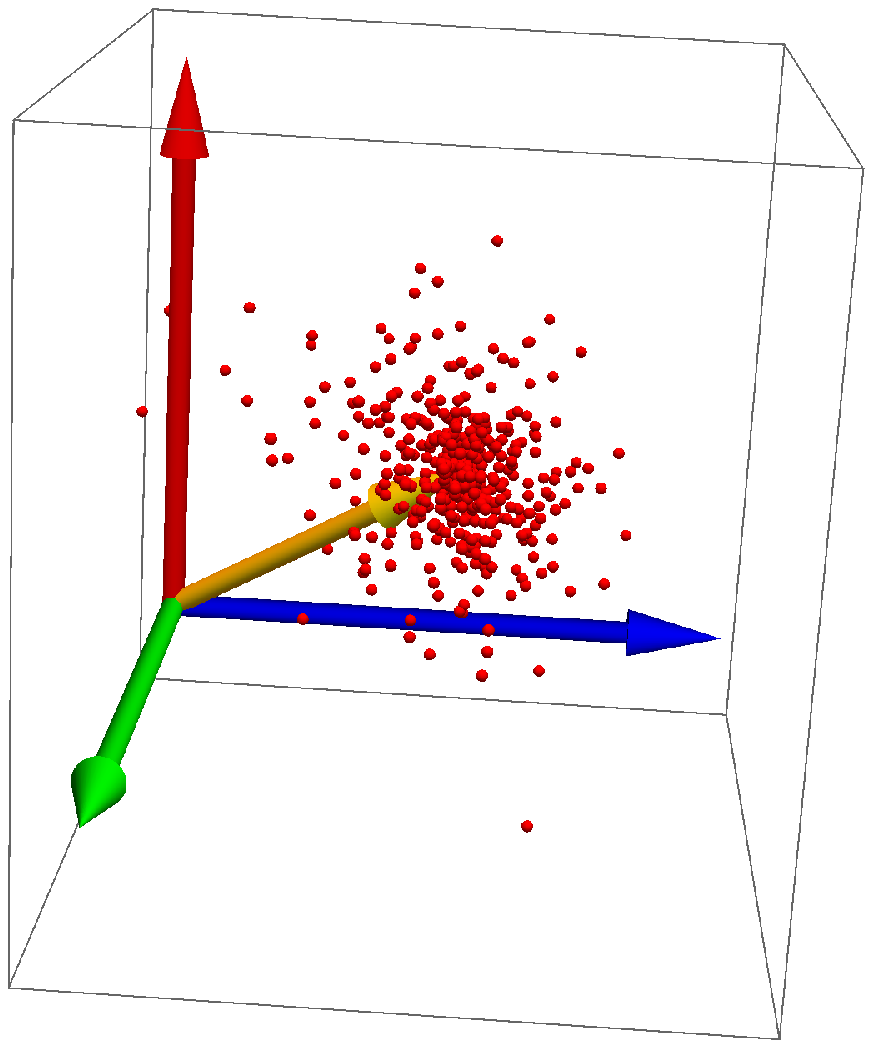}
 \hspace{0.35in} 
 \includegraphics[width=2.6in]{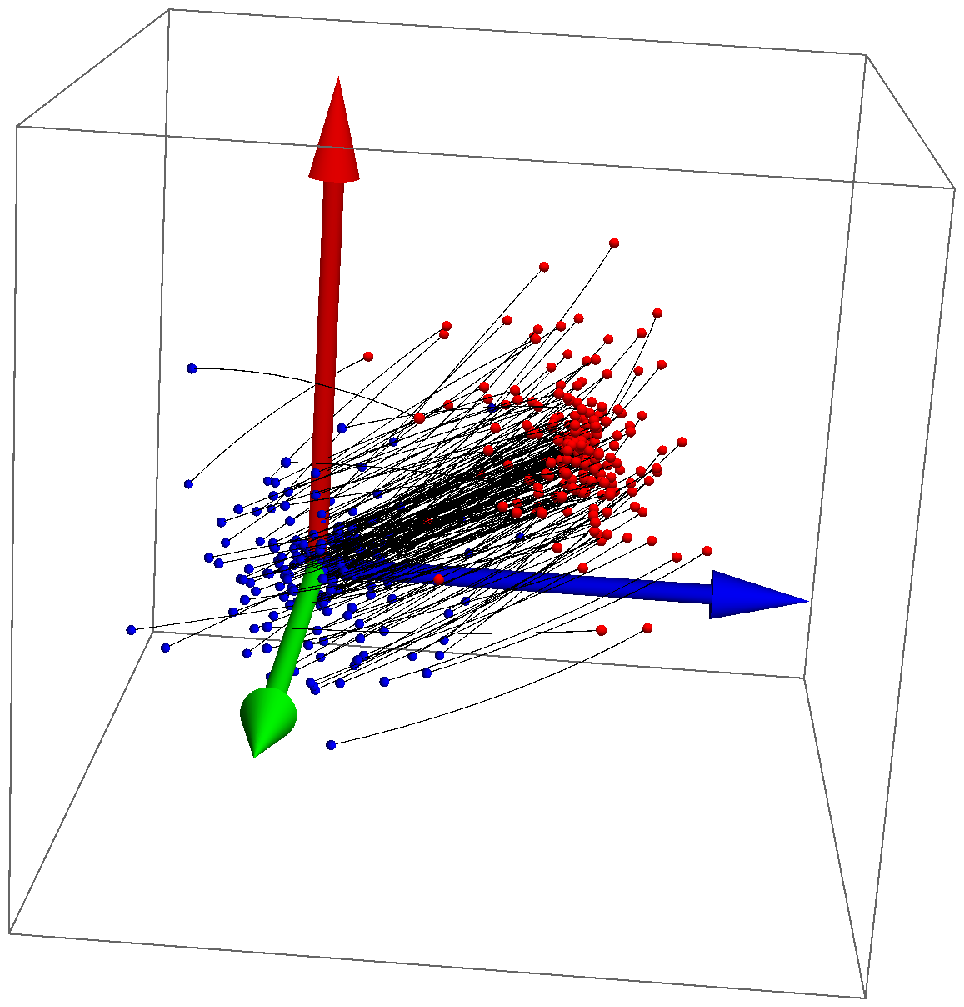}
\centerline{\hspace{1.5in} (A) \hfill (B) \hspace{1.5in} }
\vspace{-.1in}
 \includegraphics[width=2.6in]{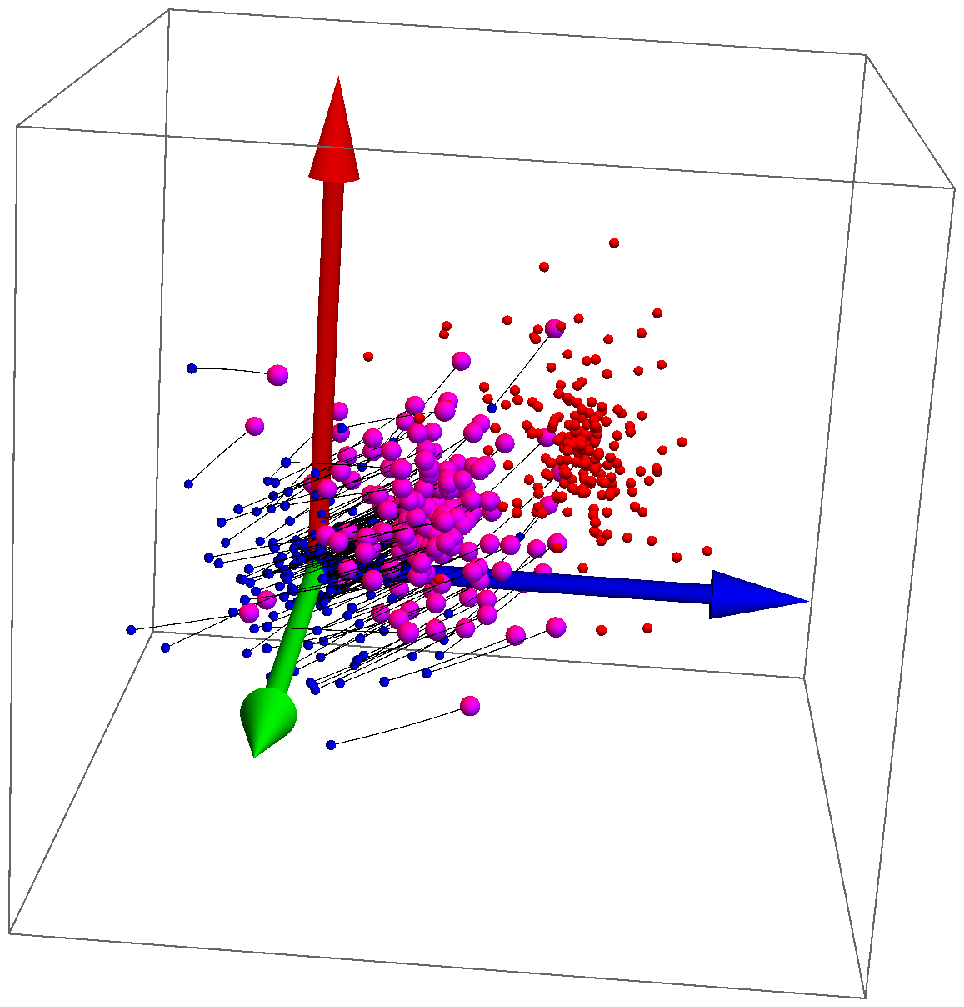}
 \hspace{0.35in} 
 \includegraphics[width=2.6in]{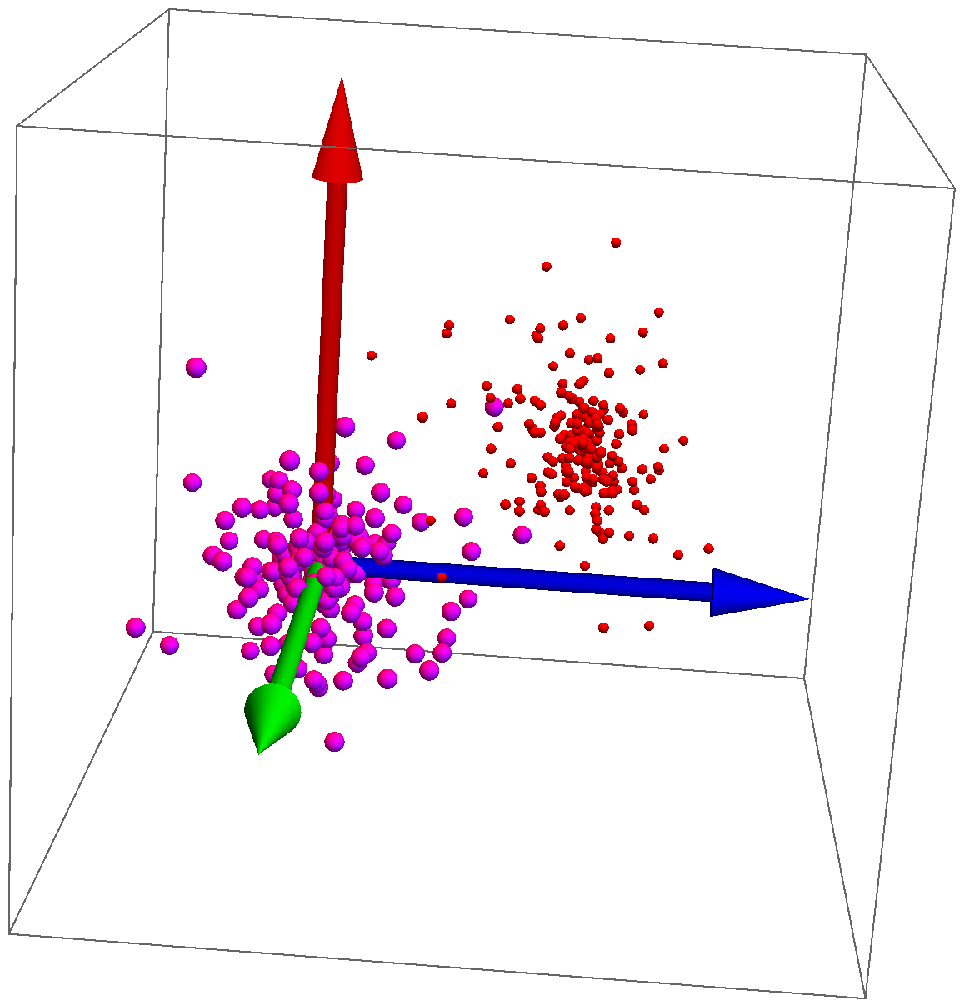}
 \centerline {\hspace{1.5in} (C) \hfill (D) \hspace{1.5in} }
}\vspace{-.25in}
\caption{\ifnum\ShowFiles=1 {\bf InterpFrmDiffsToCtr0.eps, InterpFrmDiffsToCtr.eps, InterpFrmDiffsToCtr2.eps, InterpFrmDiffsToCtr3.eps.} \fi
3D components of the rotation-average transformation of the quaternion orientation data set,
 with each point denoting the \emph{displacement} between each pair of frames as
 a single quaternion, corresponding to the rotation taking the test frame to the reference frame.  
 (A) The cluster of points $t_{k} = r_{k}\star \bar{p}_{k}\to \td{t}_{k}$ derived from the frame 
 matching problem using just the curved arcs in \Fig{fig3DFrmMatch.fig}(B).
    If there were no alignment errors introduced in the simulation,
  these would all be a single point. The yellow arrow is the quaternion solution 
  to the chord-distance centroid
 of this cluster, and is identical to the optimal  quaternion rotation transforming
 the test data to have the minimal chord measure relative to the reference data.
 (B) Choosing a less cluttered subset of the data in (A),
  we display the geodesic paths from the initial quaternion 
 displacements $\td{t}_{k}$ to the origin-centered set with minimal chord-measure distance relative to the origin.
 This is the result of applying the inverse of the quaternion $q_{\opt}$ to each $\td{t}_{k}$.
 Note that the paths are curved geodesics lying properly within the quaternion sphere.
 (C,D )  Rotating the cluster using a slerp between the 
 quaternion barycenter of the initial misaligned data and the  optimally aligned
 position, which is centered at the origin.
 }
\label{fig3DFrmMatchRAalt.fig}
\end{figure}

\begin{figure}
\figurecontent{
\centering
 \includegraphics[width=2.8in]{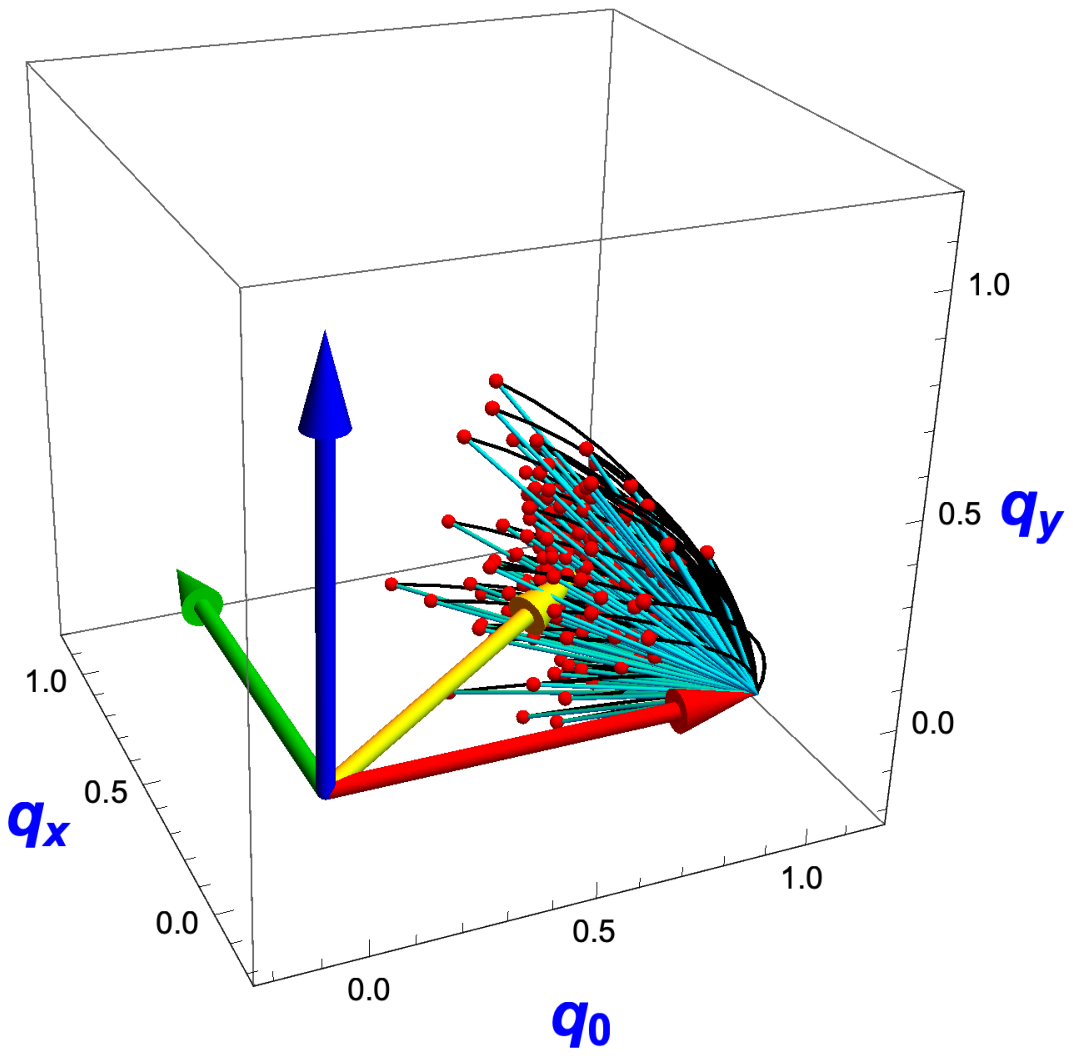}
 \hspace{0.35in} 
 \includegraphics[width=2.6in]{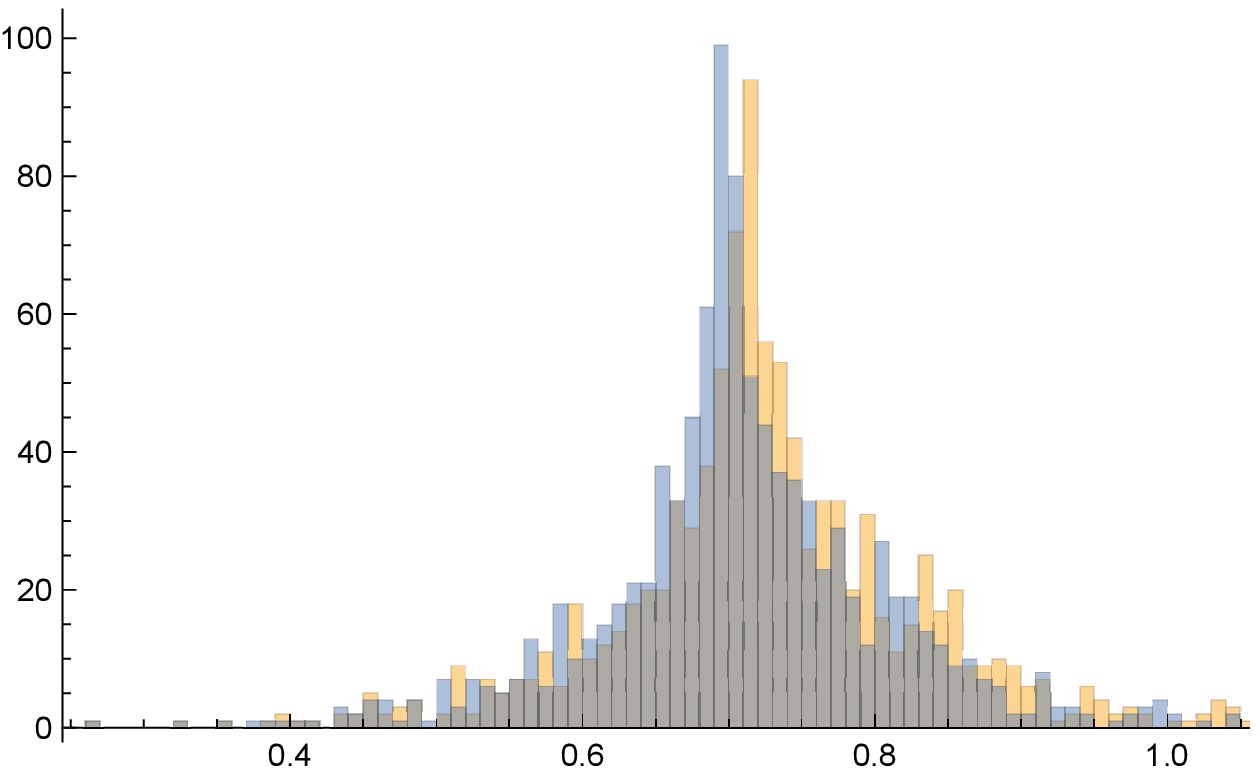}}
 \centerline {\hspace{1.5in} (A) \hfill (B) \hspace{1.5in} }
 \caption{\ifnum\ShowFiles=1  {\bf  arcVsChordRAPlot.eps, arcVsChordFrmHisto.eps.} \fi
 (A) Projecting the geodesic vs chord distances from the origin to sampled points
 in a set of frame-displacement data  $t_{k} = r_{k}\star \bar{p}_{k} \to \td{t}_{k}$.  Since the $\Vec{q}$ spatial
 quaternion paths  project to a straight line from the origin, we use the $(q_{0},q_{1},q_{2})$
 coordinates instead of our standard $\Vec{q}$ coordinates to expose the curvature 
  in the arc-length distances to the origin.
   (B) Comparing the distribution of arccosine values of the rigorous geodesic arc-length cost
 function and vs the chord-length method, sampled with a uniform distribution
 of random quaternions over $\Sphere{3}$.  The arc-length method has a different
 distribution, as expected, and produces a very slightly better barycenter.
 However, the   optimal quaternions for the arc-length vs chord-length measure
 for this simulated data set differ by less than a hundredth of a degree, so drawing
 the positions of the two distinct optimal quaternions would not reveal
 any difference in  image   (A).
 }
   \label{arcVsChordHisto.fig}
 \end{figure}

 The rotation-averaging visualization of the optimization process,
 though it has exactly the same optimal quaternion, is quite different,
 since all the quaternion data collapse to a list of single small quaternions
 $t = r\star\bar{p}$.  As illustrated in  \Fig{fig3DFrmMatchRAalt.fig},
 with compatible sign choices, the $\td{t}_{k}$'s cluster around the optimal
 quaternion, which is clearly consistent with being the barycenter 
 of the quaternion differences, intuitively the place to which all the
 quaternion frames need to be rotated to optimally coincide.
 As before, the yellow arrow is the axis
of rotation specified by the spatial part of the optimal quaternion.
Next, \Fig{arcVsChordHisto.fig} addresses the  question of how
the rigorous arc-length measure is related to the chord-length 
measure that can be treated using the same methods as the
spatial RMSD optimization.  In parallel to  \Fig{fig3DFrmMatch.fig}(B),
 \Fig{arcVsChordHisto.fig}(A) shows essentially the same comparison
 for the $\td{t}_{k}$ quaternion-displacement version of the same data.
In  \Fig{arcVsChordHisto.fig}(B), we show the histograms of the chord distances to a sample point,
the origin in this case, vs the arc-length or geodesic distances.
They obviously differ, but in fact for plausible simulations, 
the arc-length numerical optimal quaternion
barycenter differs from the chord-length counterpart by less than
one hundredth of a degree.   These issues are studied in more detail
 in the Supplementary Material.  
 
 Next, in \Fig{fig3DQDeltaRA.fig},
 we display the values of $\Delta_{\mbox{chord}} = q \cdot V$ that
 parallel the RMSD version in \Fig{fig3DQDelta.fig}.  The dots show
 the size of the cost $\Delta(q)$ at randomly sampled points across
 the entire $\Sphere{3}$, with $q_{0} \ge0$ in (A) and $q_{0}<0$ in (B).
 We have all the signs of the $\td{t}_{k}$ chosen to be centered
 in an appropriate local neighborhood, and so, unlike the quadratic
 Euclidean RMSD case, there is only one value for $q_{\opt}$ which
 is in the direction of $V$.   Finally, in \Fig{checkQFAbasins.fig} we
 present an intuitive sketch of the convexity constraints for the QFA
 optimization related to \cite{HartleyEtalRotAvg2013}.  We start with
 a set of data in (A) (with both $(q,-q)$ partners), that consists of
 three local clouds that can be smoothly deformed from dispersed to
 coinciding locations.  (B) and (C) both contain a uniform sample of
 quaternion sample points $q$ spread over all of quaternion space,
 shown as magenta dots,
 with positive and negative $q_{0}$ plotted on top of each other.
 Then each sample $q$ is used to compute \emph{one} set of
 mappings $t_{k} \to \td{t}_{k}$, and the \emph{one} value of  $q_{\opt} = V(\td{t})/\|  V \|$
 that results.  The black arrows show the relation of  $q_{\opt}$ to each
 original sample $q$, effectively showing us their \emph{votes} for
 the best quaternion average.  (B) has the clusters positioned far enough apart
 that we can clearly see that there are several basins of attraction, with no
 unique solution for  $q_{\opt}$, while in (C), we have interpolated the
 three clusters to lie in the same local neighborhood, roughly in a ball
 of quaternion radius $\alpha < \pi/4$, and we see that almost all of
 the black arrows vote for one unique $q_{\opt}$ or its equivalent negative.
 This seems to be a useful exercise to gain intuition about the nature
 of the basins of attraction for the quaternion averaging problem that
 is essential for quaternion frame alignment.

 \qquad
 
\begin{figure}
\figurecontent{
\centering
 \includegraphics[width=2.6in]{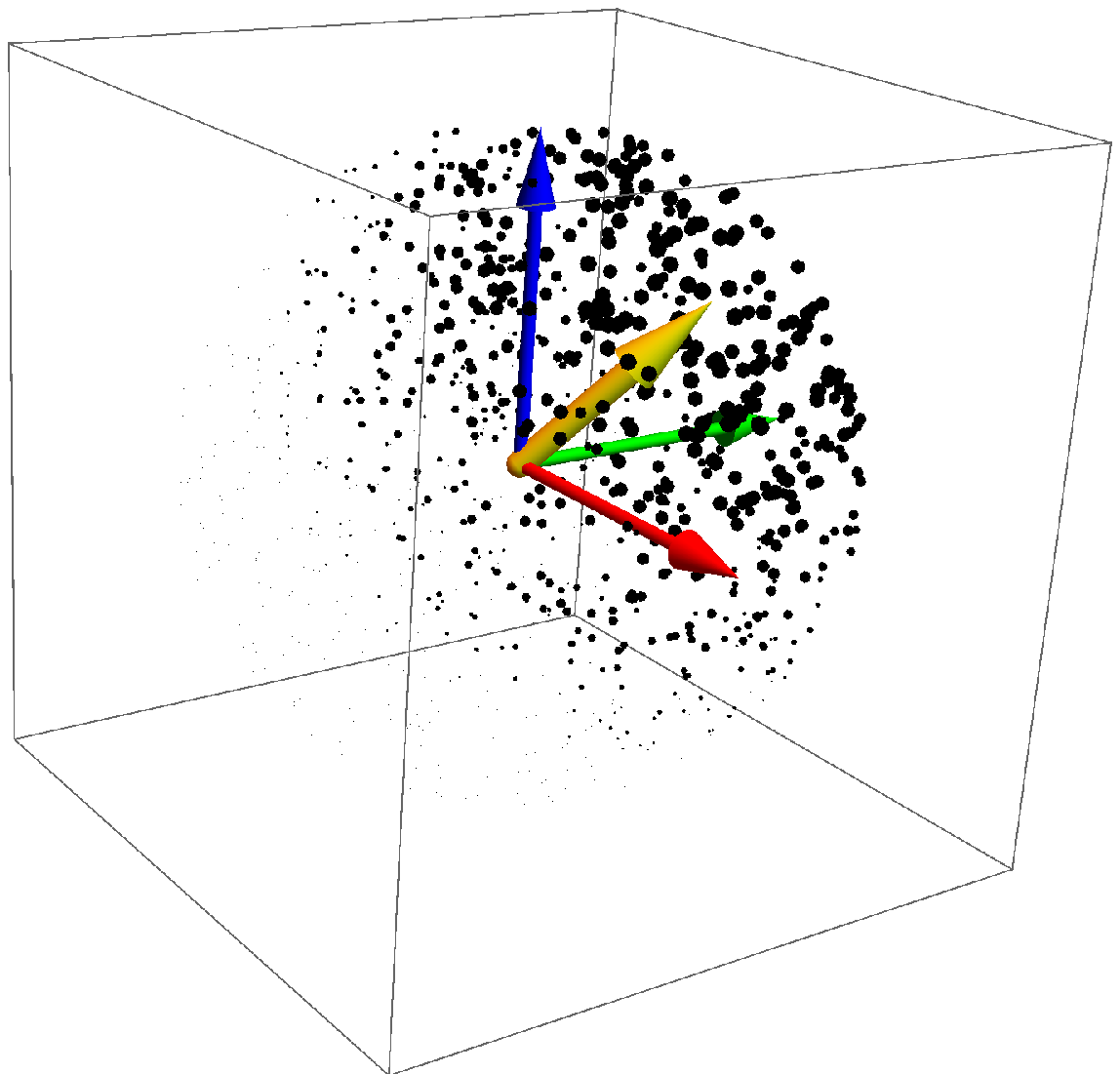}
 \hspace{0.35in} 
 \includegraphics[width=2.6in]{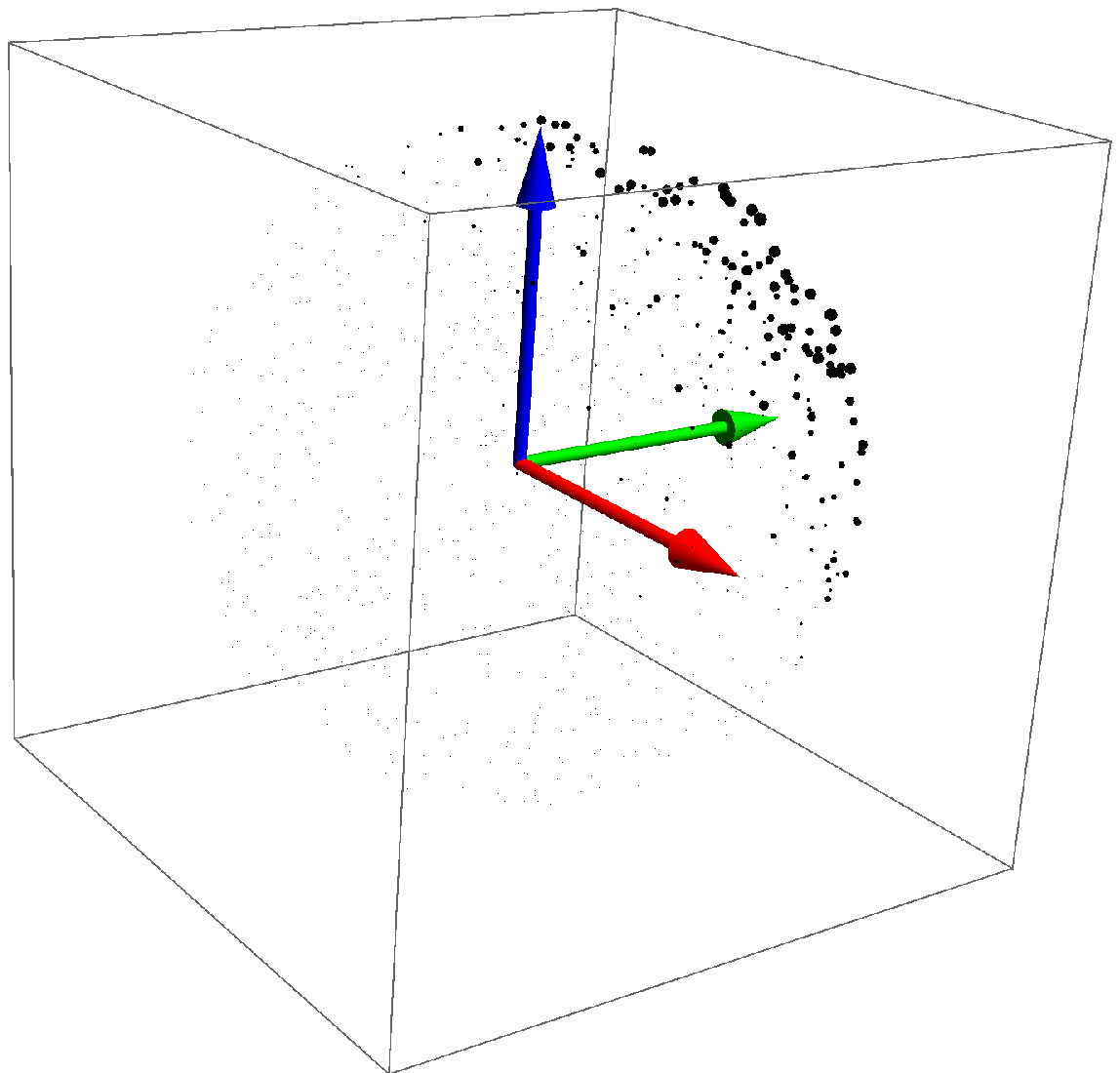}}
 \centerline {\hspace{1.5in} (A) \hfill (B) \hspace{1.5in} }
 \caption{\ifnum\ShowFiles=1  {\bf  fig3DQFrameDeltaPlotA.eps, fig3DQFrameDeltaPlotB.eps.} \fi
 The values of $\Delta = q \cdot V$ represented by
 the sizes of the dots placed  at a random
 distribution of quaternion points.  We display the data dots at the locations of their
 spatial quaternion components $\Vec{q} = (q_1,q_2,q_3)$.
  (A) is the northern hemisphere of $\Sphere{3}$, with $q_{0}\ge 0$,  
 (B) is the southern hemisphere, with $q_{0}<0$, and we implicitly know
 that the value of $q_{0}$ is $\pm \sqrt{1-{q_{0}}^2}$.
 The points in these two solid balls represent the entire space of quaternions,
 and it is important to note that, even though $R(q)=R(-q)$  so each ball alone
 actually represents all possible unique rotation matrices, our cost function
 covers the \emph{entire} space of quaternions, so $q$ and $-q$  are distinct.
 The spatial component  of the maximal eigenvector
 is shown by the yellow arrow, which clearly ends in the middle of the maximum
 values of $\Delta$.  The small cloud at the edge of (B) is simply the rest of the
 complete cloud around the tip of the yellow arrow as $q_{0}$ passes through 
 the ``equator'' at $q_{0} = 0$, going 
 from a small positive value at the edge of (A) to a small negative value 
 at the edge of (B). }
 \label{fig3DQDeltaRA.fig}
 \end{figure}

\begin{figure}
\figurecontent{ \vspace*{-0.32in}
\centering
 \includegraphics[width=3.2in]{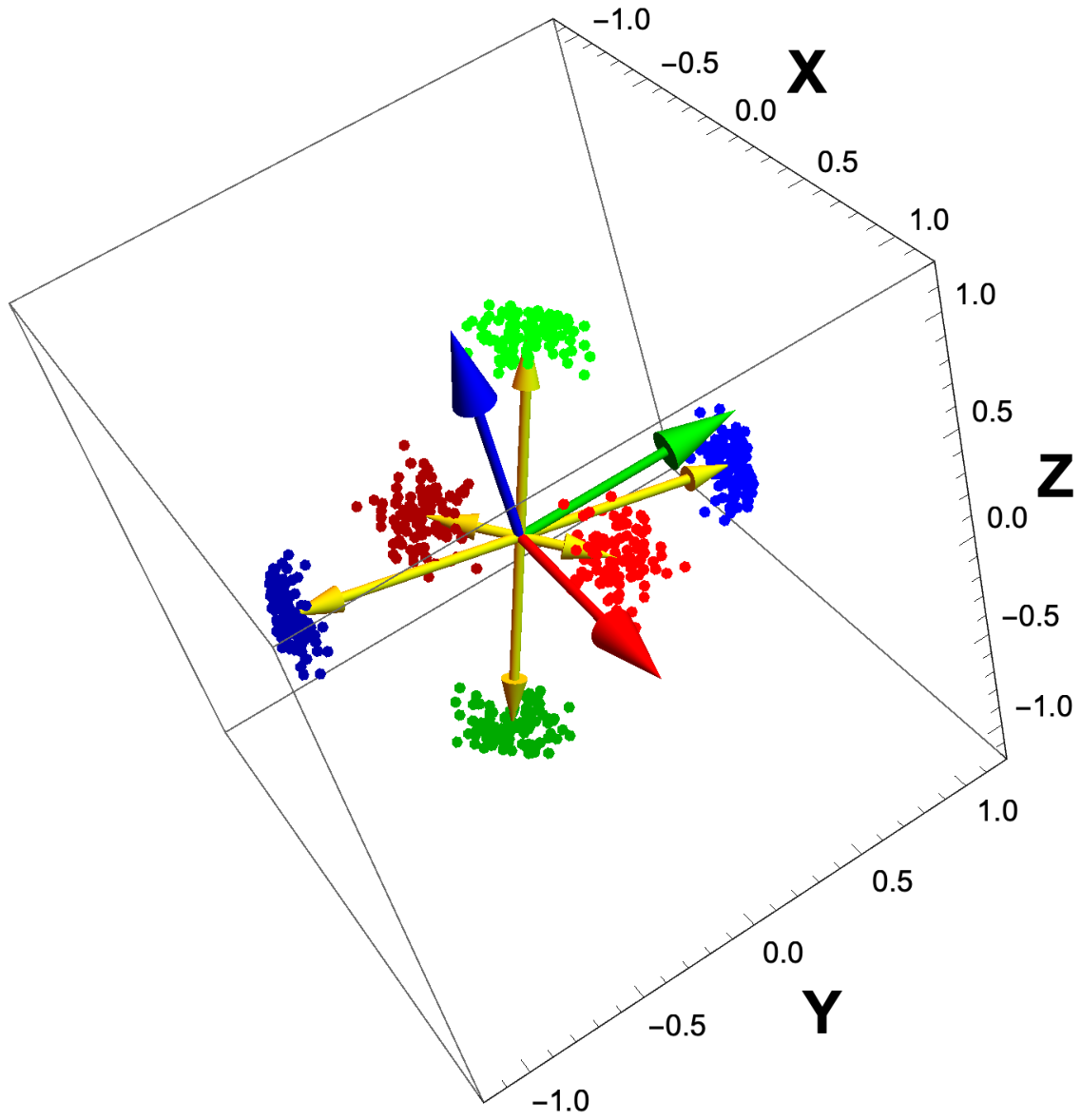}
 \centerline {\hspace{1.0in} \hfill (A) \hfill  \hspace{1.0in} }
 \centering
 \includegraphics[width=2.4in]{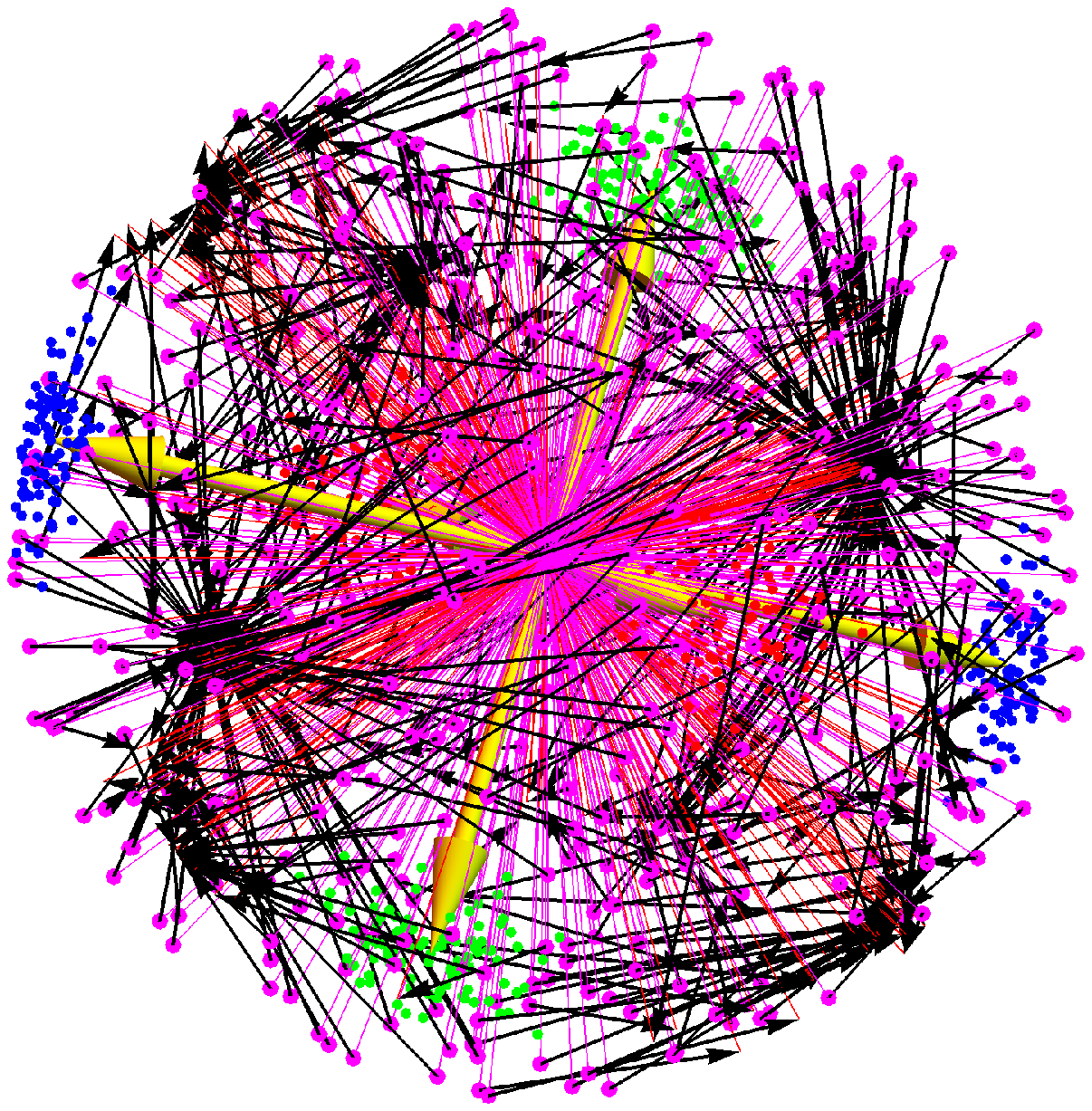}
 \hspace{0.4in} 
 \includegraphics[width=2.4in]{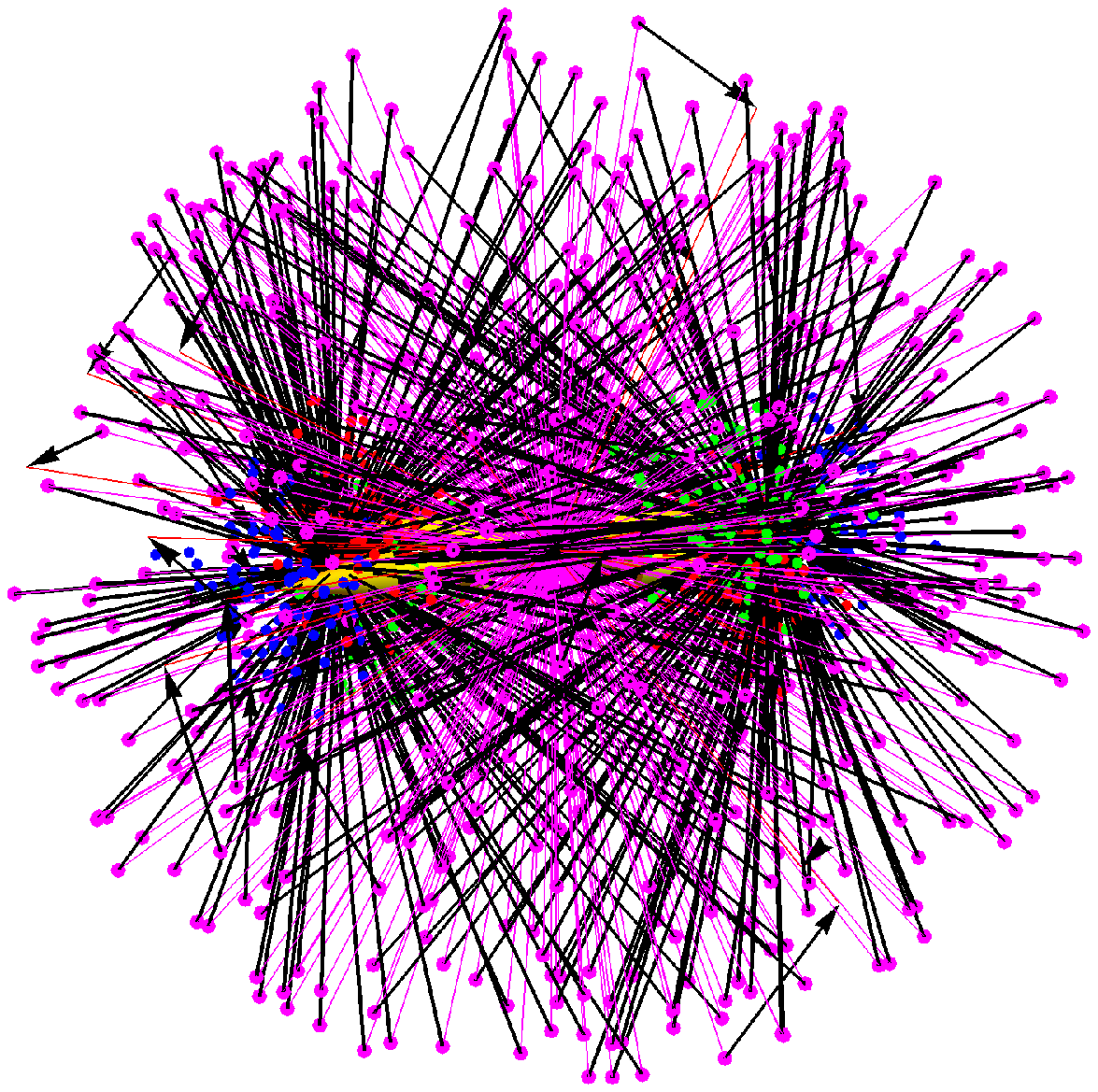}  }
 \centerline {\hspace{1.5in}   (B) \hfill (C)\hspace{1.5in} }\vspace{-.1in}
 \caption{ \small \ifnum\ShowFiles=1  {\bf  basin00.eps, basin1.eps, basin2.eps.} \fi
 The behavior of the basins of attraction for the $t_{k} \to \td{t}_{k}$ map
 is shown here, starting in (A) with the $(q,-q)$ pairs for three movable
 clusters of quaternion frame data, each having a well-defined \emph{local}
 quaternion average $q_{\opt} = V(\td{t})/\|  V \|$ shown as the yellow
 arrows with their $q \to -q$ equivalents.  Next we merge all three samples
 into one data set that can be smoothly interpolated between the data
 being outside the $\alpha = \pi/4$ safe zone to all being together within that
 geometric boundary in quaternion space. (B) shows the results of taking
 500 uniform samples of $q$ and computing the set $ \{\td{t}_{k}\}$  for
 \emph{each} sample $q$,  placed at the magenta dots, and \emph{then}
 computing the resulting $q_{\opt}$; the black arrows follow the line from
 the sample point to the resultant $q_{\opt}$.  Clearly in (B), where the
 clusters are in their initial widely dispersed configuration, the black
 arrows (the ``votes'' for the best $q_{\opt})$  collect in several different
 basins of attraction, signifying the absence of a global solution.  We
 then interpolate all the clusters close to each other, and show the new
 results of the voting in (C).  Now almost all of the samplings of the
 full quaternion space converge to point their arrows densely to the
 two opposite values of $q_{\opt}$, and there is just one effective
 basin of attraction.
 }
 \label{checkQFAbasins.fig}
 \end{figure}

\qquad

\noindent{\bf  Alternative Matrix Forms of the Linear Vector Chord Distance.}
If the signs of the quaternions representing orientation frames are well-behaved,
and the frame problem is our only concern, 
Eqs.~(\ref{solnQFA.eq}) and (\ref{VsolnQFA.eq}) provide
a simple solution to  finding the optimal global rotation.  If we are
anticipating wanting to combine a spatial profile matrix $M(E)$ with
an orientation problem in a single $4\times 4$ matrix, or we have
problems defining a consistent quaternion sign, there are two further
choices of orientation frame measure we may consider.  

\qquad

\noindent{\bf\small  (1) Matrix Form of the Linear Vector Chord Distance.}  The first
option uses the fact  that the square of \Eqn{solnQFA.eq} will yield the same
extremal solution, so we can choose a measure of the form
\begin{eqnarray} 
\Delta_{\mbox{chord-sq}} & = & (q\cdot V)(q \cdot V) \nonumber \\
 & = & \sum_{a=0,b=0}^{3} q_{a}\; V_{a} V_{b}\; q_{b}  \nonumber\\
& = & q \cdot \Omega \cdot q  \ ,
\label{qVsq.eq}
\end{eqnarray}
where $\Omega_{ab} = V_{a} V_{b}$ is a $4\times 4$ rank one symmetric matrix
with $\det \Omega = 0$, and $\tr\Omega = \sum_{a} {V_{a}}^2 \neq 0$.
The eigensystem of $\Omega$  is just defined by the eigenvalue $\|V\|^2$,
and combination with the spatial eigensystem can be achieved either numerically or
algebraically.  The sign issues for the sampled data remain unchanged since 
they appear inside the sums defining $V$.   
This form will acquire more importance in the 4D case.

\qquad

\noindent{\bf \small  (2) Fixing Sign Problem with Quadratic Rotation Matrix Chord
  Distance.}  Our second  approach has a very natural way to 
   \emph{eliminate sign dependence altogether} from the quaternion chord distance
   method, and  has a   close relationship to $\Delta_{\mbox{chord}}$.  
This measure is constructed  starting from a  minimized  Fr\"{o}benius norm of the form
 (this approach is used by \cite{Sarlette2009}; see also, e.g.,
\cite{Huynh:2009:MRC}, as well as \cite{Moakher2002,MarkleyEtal-AvgingQuats2007,HartleyEtalRotAvg2013})
\[ \|  R(q)\cdot  R(p_{k}) -  R (r_{k})  \|^{2}_{\mbox{Frob}} \ , \]
and then reducing to the cross-term as usual.  The cross-term measure to be maximized,
in terms of $3\times 3$   (quaternion-sign-independent)  rotation matrices, then becomes
\begin{eqnarray} 
\Delta_{\mbox{\scriptsize RRR}} & = &  \sum_{k=1}^{N} \tr \left[ R(q)\cdot
  R(p_{k})\cdot  {R^{-1}}(r_{k}) \right]     \, = \,  \sum_{k=1}^{N} \tr \left[ R(q\star p_{k}\star \bar{r}_{k}) \right] 
   \label{RRRsumdef.eq} \nonumber \\
  & = & \sum_{k=1}^{N} \tr \left[ R(q)\cdot
  R(p_{k}\star \bar{r}_{k}) \right]
  \, = \, \sum_{k=1}^{N} \tr \left[ R(q)\cdot
  R^{-1}(r_{k}\star \bar{p}_{k}) \right]  \label{RRRsumdef2.eq} 
   \ ,
\label{RRRUdev.eq}
\end{eqnarray}
where $\bar{r}$ denotes the complex conjugate or inverse quaternion.
We can verify that this is a chord-distance by noting that each relevant $R\cdot R\cdot R$ term
reduces to the square of  an individual chord distance  appearing in  $\Delta_{\mbox{chord}}$:
\begin{equation} 
\left.
\begin{array}{rcl}
 {\displaystyle  \sum_{k=1}^{N} } \tr \left[ R(q)\cdot
  R(p_{k})\cdot  {R}(\bar{r}_{k}) \right] 
  & = &   {\displaystyle  \sum_{k=1}^{N} }\left(4  \left( (q \star p_{k})\cdot r_{k}\right)^2  -
    (q\cdot q)(p_{k}\cdot p_{k})(r_{k}\cdot r_{k})\right)\\
     & = &   {\displaystyle   \sum_{k=1}^{N} } \left(4   \left(q \cdot (r_{k} \star \bar{p}_{k})\right) ^2   - 1 \right)\\
     & = &  4 \,\, {\displaystyle  \sum_{a,b}} q_{a} \left(  {\displaystyle   \sum_{k=1}^{N} } [t_{k}]_{\textstyle_a} \:[t_{k}]_{\textstyle_b} 
      \right) q_{b}  - N\\
      & = &  4\,\, q \cdot A(t = r  \star \bar{p}) \cdot q - N  \ . 
     \end{array} \right\}
\label{VVeqRRR.eq}
\end{equation}
Here the non-conjugated ordinary $r$ on the right-hand side is not 
a typographical error, and the $4\times4$ matrix $A(t)$ is the alternative (equivalent) 
profile matrix that was introduced by \cite{MarkleyEtal-AvgingQuats2007,HartleyEtalRotAvg2013}
for the chord-based quaternion-averaging problem.  We can therefore use either
the measure $\Delta_{\mbox{\scriptsize RRR}}$ or
\begin{equation}
\Delta_{\mbox{\scriptsize A}} =  q \cdot A(t) \cdot q
\label{qAqmeasure.eq}
\end{equation}
with $A _{ab} = \sum_{k=1}^{N} [t_{k}]_{\textstyle_a} \:[t_{k}]_{\textstyle_b}$ as our rotation-matrix-based  sign-insensitive chord-distance optimization measure. 
Exactly like our usual spatial measure, these measures must be \emph{maximized}
to find the optimal $q$.  It is, however, important to emphasize that the optimal
quaternion will \emph{differ} for the   $\Delta_{\mbox{\scriptsize chord}}$ , 
 $\Delta_{\mbox{\scriptsize chord-sq}}$ , and  $\Delta_{\mbox{\scriptsize RRR}} \sim   \Delta_{\mbox{\scriptsize A}}$   measures, though they will normally be very similar (see discussion
 in the Supplementary Material). 

 We now recognize that the sign-insensitive measures are all very closely
 related to our original spatial RMSD problem, and all can be solved by
 finding the optimal quaternion eigenvector $q_{\opt}$ of a $4\times 4$ matrix.
 The procedure for $\Delta_{\mbox{\scriptsize chord-sq}}$ and $\Delta_{\mbox{\scriptsize A}}$
 follows immediately, but it is useful to work out the options for $\Delta_{\mbox{\scriptsize RRR}}$
 in a little more detail.   Defining
$ T_{k} =  R(p_{k})\cdot  {R^{-1}}(r_{k}) = R(p_{k}\star \bar{r}_{k})= R^{-1}(t_{k})$,
we can write our optimization measure as
\begin{equation} 
\Delta_{\mbox{\scriptsize RRR}} =  \sum_{k=1}^{N} \tr \left( R(q) \cdot
  T_{k} \right) 
 =  \sum_{a=1,b=1}^{3} R_{ba}(q) T_{ab}  =
 \sum_{a=0,b=0}^{3} q_{a} \cdot U_{ab}(p,r) \cdot q_{b} = 
  q \cdot U(p,r) \cdot q \ ,
\label{trRS.eq}
\end{equation}
where the frame-based cross-covariance matrix is simply $T_{ab} =
\sum_{k=1}^{N} {[T_{k}]}_{ab}$  and $U(p,r)=U(T)$ has the same relation
 to $T$ as $M(E)$ has to $E$ in \Eqn{basicHorn.eq}.
\quad

To compute the necessary $4\times 4$ numerical
profile matrix $U$, one need only substitute the appropriate 3D frame
triads or their corresponding quaternions for the $k$th frame pair and
sum over $k$ .  Since the orientation-frame profile matrix $U(p,r)$ is
symmetric and traceless just like the Euclidean profile matrix $M$,
the same solution methods for the optimal quaternion rotation
$q_{\opt}$ will work without alteration in this case, which is
probably the preferable method for the general problem.

\qquad

\noindent{\bf Evaluation.}  The details of evaluating the properties
of our quaternion-frame alignment algorithms, including comparison of
the chord approximation to the arc-length measure, 
are available in the Supplementary Material.  The top-level result is
that, even for quite large rotational differences, the mean difference
between the optimal quaternion using the numerical arc-length measure and 
the optimal quaternion using the chord
approximation for any of the three methods is on the order of small
fractions of a degree for the random data  distributions that
we examined.




\section{The 3D Combined Point+Frame Alignment Problem}


Since we now have precise alignment procedures for both 3D spatial
coordinates and 3D frame triad data (using the exact measure for the
former and the approximate chord measure for the latter), we can
consider the full 6 degree-of-freedom alignment problem for combined
data from a single structure.  As always, this problem can be solved either
by numerical eigenvalue methods or 
in closed algebraic form using the eigensystem formulation of the
both alignment problems  presented in the previous
Sections.  While there are clearly appropriate domains of this type,
e.g., any protein structure in the PDB database can be converted to a
list of residue centers and their local frame
triads~\cite{HansonThakurQuatProt2012}, little is known at this time
about the potential value of combined alignment.  To establish the most
complete possible picture, we now proceed to describe the details of
our solution to the alignment problem for combined translational and
rotational data, but we remark at the outset that the results of the
combined system are not obviously very illuminating. 

The most straightforward approach to the combined 6DOF measure is to
equalize the scales of our spatial $M(E)$ profile matrix and our
orientation-frame $U(S)$ profile matrix by imposing a unit-eigenvalue
normalization, and then simply to perform a linear interpolation
modified by a dimensional constant $\sigma$ to adjust the relative importance
of the orientation-frame portion:
\begin{eqnarray} 
 \Delta_{xf}(t,\sigma) & = & 
 q \cdot \left[ (1-t) \frac{M(E)}{\epsilon_{x}} +  t \; \sigma
 \frac{U(S)}{\epsilon_{f}} \right] \cdot q \ .
\label{ScaledXQMdef.eq}
\end{eqnarray}
  Because of the dimensional incompatibility of
$\Delta_{x}$ and $\Delta_{f}$,  we  treat the ratio
\[ \lambda^2 = \frac{t \sigma}{1-t} \]
as a dimensional weight such as that adopted by Fogolari et
al.~\cite{FogolariEtAl:2016} in their entropy calculations, so if $t$
is dimensionless, then $\sigma$ carries the dimensional scale
information. 

Given the composite profile matrix of \Eqn{ScaledXQMdef.eq}, we can now 
extract our optimal rotation solution  by computing the maximal
eigenvalue as usual, either numerically or algebraically (though we
may need the extension to the non-vanishing trace case examined
in the Supplementary Material for some choices of $U$).
   The result is a parameterized eigensystem
\begin{equation} 
\left. \begin{array}{c}
\epsilon_{\opt}(t,\sigma) \\
q_{\opt}(t,\sigma) \end{array} \right\} 
\label{SDNeValVec.eq}
\end{equation}
yielding the optimal values $R(q_{\opt}(t,\sigma))$, 
$\Delta_{xf}=\epsilon_{\opt}(t,\sigma)$ based on the data
$\{E,S\}$ no matter what we 
take as the values of the two variables $(t,\sigma)$.

\qquad

\noindent{\bf A Simplified Composite Measure.}  However, upon 
inspection of \Eqn{ScaledXQMdef.eq}, one wonders what happens
if we simply use the  \emph{slerp} defined in   \Eqn{slerpdef.eq} to 
interpolate between the \emph{separate} spatial and orientation-frame optimal
quaternions.  While the eigenvalues
that correspond to the two   scaled terms $M/\epsilon_{x}$ and
$U/\epsilon_{f}$ in \Eqn{ScaledXQMdef.eq} are both unity, and 
thus differ from the eigenvalues of $M$ and $U$, 
 the individual normalized eigenvectors $q_{x:{\opt}}$ and
$q_{f:{\opt}}$ are the same.  Thus, if we are happy with simply using
a hand-tuned fraction of the combination of the two corresponding
rotations, we can just choose a composite rotation $R(q(t))$ specified
by 
\begin{equation}
q(t) = \mbox{\it slerp} (q_{x:{\opt}},q_{f:{\opt}},t) \ .
\label{simple6DOFslerp.eq}
\end{equation}
to study the composite 6DOF alignment problem.  In fact,
as detailed in the Supplementary Material, if we simply
plug this $q(t)$ into \Eqn{ScaledXQMdef.eq} for any $t$ (and
$\sigma=1$), we find \emph{negligible differences} between the
quaternions   $q(t)$ and
$q_{\opt}(t,1)$ as a function of $t$.  We suggest in addition
that any particular effect of $\sigma \neq 1$ could be achieved at
some value of $t$ in the interpolation.   We thus conclude that, for all
practical purposes, we might as well use \Eqn{simple6DOFslerp.eq} with
the parameter $t$ adjusted to achieve the objective of \Eqn{ScaledXQMdef.eq}
to study composite translational and rotational  alignment similarities.

\section{Conclusion}


Our objective has been to explore quaternion-based treatments of the
RMSD data-comparison  \linebreak
 problem  as developed in the work of
Davenport~\cite{Davenport1968}, Faugeras and Hebert \linebreak
\cite{FaugerasHebert1983},
Horn~\cite{Horn1987},
Diamond~\cite{RDiamond1988}, Kearsley~\cite{Kearsley1989}, and
Kneller~\cite{Kneller1991}, among others, and to publicize the exact
algebraic solutions, as well as extending the method to handle wider problems.
We studied the intrinsic properties of the RMSD problem for comparing
spatial and orientation-frame data in quaternion-accessible domains, and we examined the
nature of the solutions for the eigensystems of the 3D 
spatial RMSD problem, as well as  the corresponding 3D quaternion
orientation-frame alignment problem (QFA).  Extensions of both the translation and 
rotation alignment problems and their solutions 
to 4D are  detailed in the Supplementary Material.  We also examined
solutions for the combined 3D spatial and orientation-frame RMSD
problem, arguing that a simple quaternion interpolation between the
two individual solutions may well be sufficient for most purposes.



\clearpage

\appendix


\section{\vspace{0.1in} The 3D Euclidean Space Least Squares\\  Matching Function}
\label{fullRMSD3D.app}

This appendix works out the details of the long-form least squares distance
measure for the 3D Euclidean alignment problem using the method
of Hebert and Faugeras~\cite{FaugerasHebert1983,HebertThesis1983,FaugerasHebert1986}.  
Starting with the 3D Euclidean minimizing distance measure \Eqn{RMSD-basic.eq},
we can exploit   \Eqn{qrot.eq}  for  $R(q)$, along
with \Eqn{multiplicativeNorm.eq}, to produce an alternative quaternion eigenvalue
problem whose \emph{minimal} eigenvalue determines the eigenvector $q_{\opt}$
specifying the matrix that rotates the test data into closest correspondence
with the reference data.

  Adopting the convenient notation
$\Vec{x} = (0,x_{1},x_{2},x_{3})$ for a pure imaginary quaternion, we  employ the following steps:
\begin{equation} 
\begin{array}{rcl}  
{\mathbf S}_{3}  & = &
   \sum_{k=1}^{N} \| R_{3} (q)\cdot x_{k} -  y_{k}\|^{2} \\[0.05in]
    & = &   \sum_{k=1}^{N}   \| q \star \Vec{x}_{k} \star \bar{q} \, - \, \Vec{y}_{k} \|^{2}  
    \,  =   \, \sum_{k=1}^{N}   \| q \star \Vec{x}_{k} \star \bar{q} \, - \, \Vec{y}_{k} \|^{2} \|q\|^{2} \\[.1in]
    &  =  &   
            \sum_{k=1}^{N}  \| q \star \Vec{x}_{k}    \, - \, \Vec{y}_{k}\star q \|^{2} 
                      \hspace{.2in} {\mbox{by \Eqn{multiplicativeNorm.eq}}} \\[0.05in]
     & = &   \sum_{k=1}^{N}  \| A(\Vec{x}_{k},\Vec{y}_{k}) \cdot q \|^2 
     \, = \,   \sum_{k=1}^{N}  q \cdot {A_{k}}^{\t}  \cdot {A_{k}} \cdot q  \\[0.05in]
     & = &   \sum_{k=1}^{N}  q  \cdot B_{k}  \cdot q    \, = \, q \cdot B \cdot q \ .
     \end{array}
          \label{RMSD-F-Hebert.eq}
\end{equation}
Here we may write, for each $k$, the matrix $A(\Vec{x}_{k},\Vec{y}_{k}))$  as
\begin{equation}
\begin{array}{c}
   A _{k}= \left[ \begin{array}{cccc} 
   0 & -a_1 & -a_2 & -a_3 \\ 
   a_1 & 0 & s_3 & -s_2 \\
    a_2 & -s_3 & 0 & s_1 \\ 
   a_3 & s_2 & -s_1 & 0  \\[0.05in]
     \end{array} \right]_{\textstyle k} \\[0.3in]
     \mbox{where, with ``$a$" for ``antisymmetric'' and ``$s$" for ``symmetric,''} \\[0.05in]
     a_{\{1,2,3\} }  = \{ x_{1} - y_{1}, \,  x_{2}  - y_{2}, \, x_{3} - y_{3} \} \\
     s_{\{1,2,3\} } =  \{ x_{1} + y_{1},\,  x_{2} + y_{2},\,  x_{3} + y_{3} \} \\
 \end{array}  
\label{Adef.eq}
\end{equation}
and, again for each $k$, 
\begin{equation}
B_{k} \, = \, {A_{k}} ^{\t}  \cdot A_{k} \,  =  \,
 \left[ \begin{array}{cccc} 
       {a_1}^2+{a_2}^2+{a_3}^2 & a_3  s_2 - a_2  s_3 & a_1  s_3 - a_3  s_1 & a_2 s_1 - a_1 s_2 \\
      a_3 s_2-a_2 s_3 & {a_1}^2+{s_2}^2+{s_3}^2 & a_1 a_2-s_1 s_2 & a_1 a_3-s_1 s_3 \\
     a_1 s_3-a_3 s_1 & a_1 a_2-s_1 s_2 & {a_2}^2+{s_1}^2+{s_3}^2 & a_2 a_3-s_2 s_3 \\
    a_2 s_1-a_1 s_2 & a_1 a_3-s_1 s_3 & a_2 a_3-s_2 s_3 & {a_3}^2+{s_1}^2+{s_2}^2 \\ 
      \end{array} \right]_{\textstyle k}  \ ,
\label{Bdef.eq}
\end{equation}
and $B = \sum_{k=1}^{N} B_{k}$,
Since, using the full squared-difference minimization measure \Eqn{RMSD-basic.eq} requires the global minimal value, the solution for the optimal quaternion in \Eqn{RMSD-F-Hebert.eq} is the eigenvector of the \emph{minimal} eigenvalue of $B$ in \Eqn{Bdef.eq}.   This is the approach used by Faugeras and 
Hebert in the earliest application of the quaternion method to scene alignment of which we are aware.  
While it is important to be aware of this alternative method, in the main text,
 we have found it more useful, 
to focus on the alternate form exploiting only the non-constant 
cross-term appearing in \Eqn{RMSD-basic.eq}, as does most of the recent molecular structure literature.  The cross-term requires the determination of the \emph{maximal}
eigenvalue rather than the \emph{minimal} eigenvalue of the corresponding data matrix.
Direct numerical calculation verifies that, though the minimal eigenvalue of \Eqn{Bdef.eq} differs
from the maximal eigenvalue of the cross-term approach, the exact same optimal eigenvector is obtained, a result that can presumably be proven algebraically but that we will not need to pursue here.

\qquad

\section{\vspace{0.1in}Introduction to Quaternion Orientation Frames}

\label{IntroQuatFrames.app}



\begin{figure}
\figurecontent{
\centering
 \includegraphics[width=2.7in]{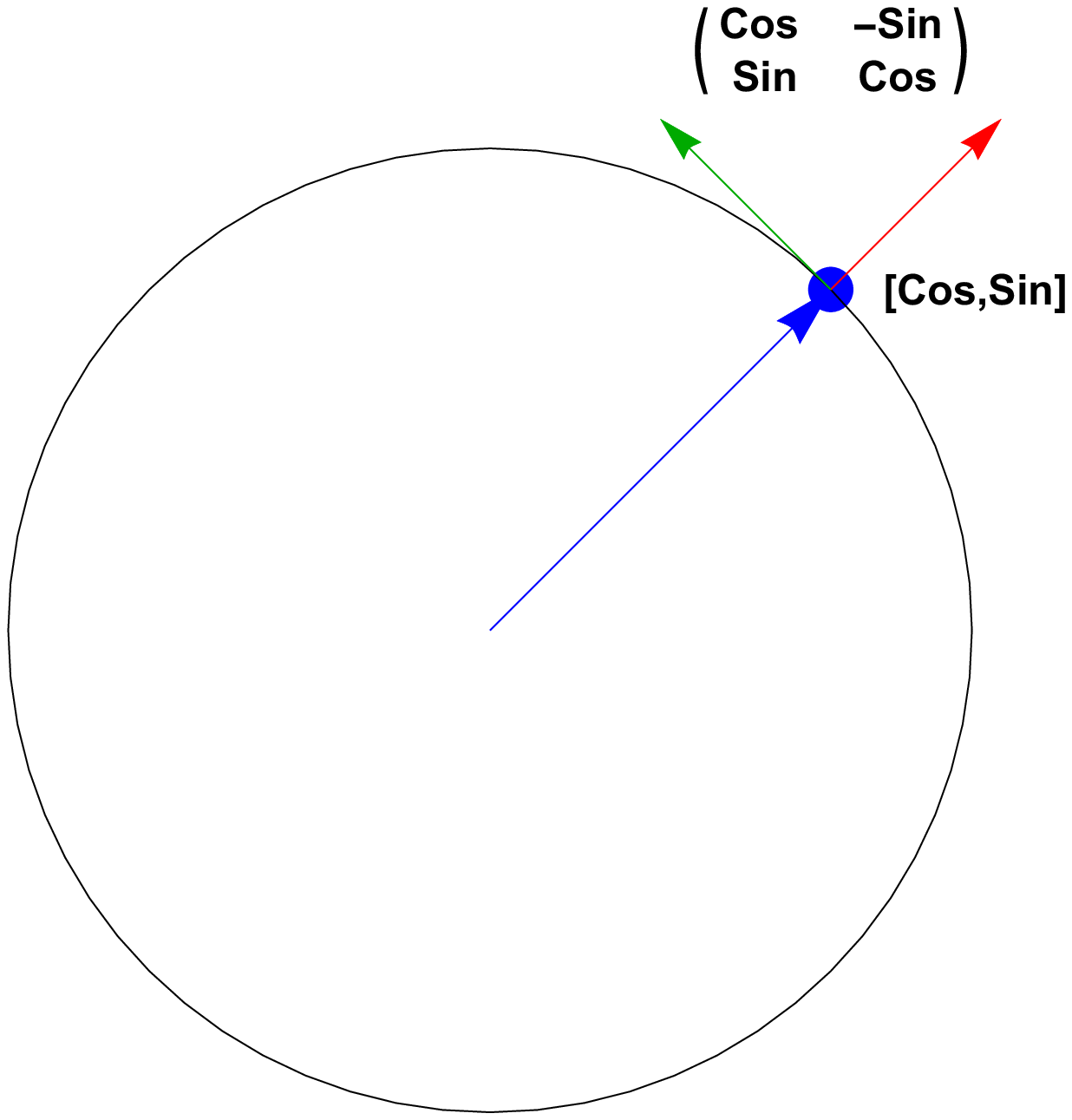}
\hspace{0.1in}
 \includegraphics[width=2.7in]{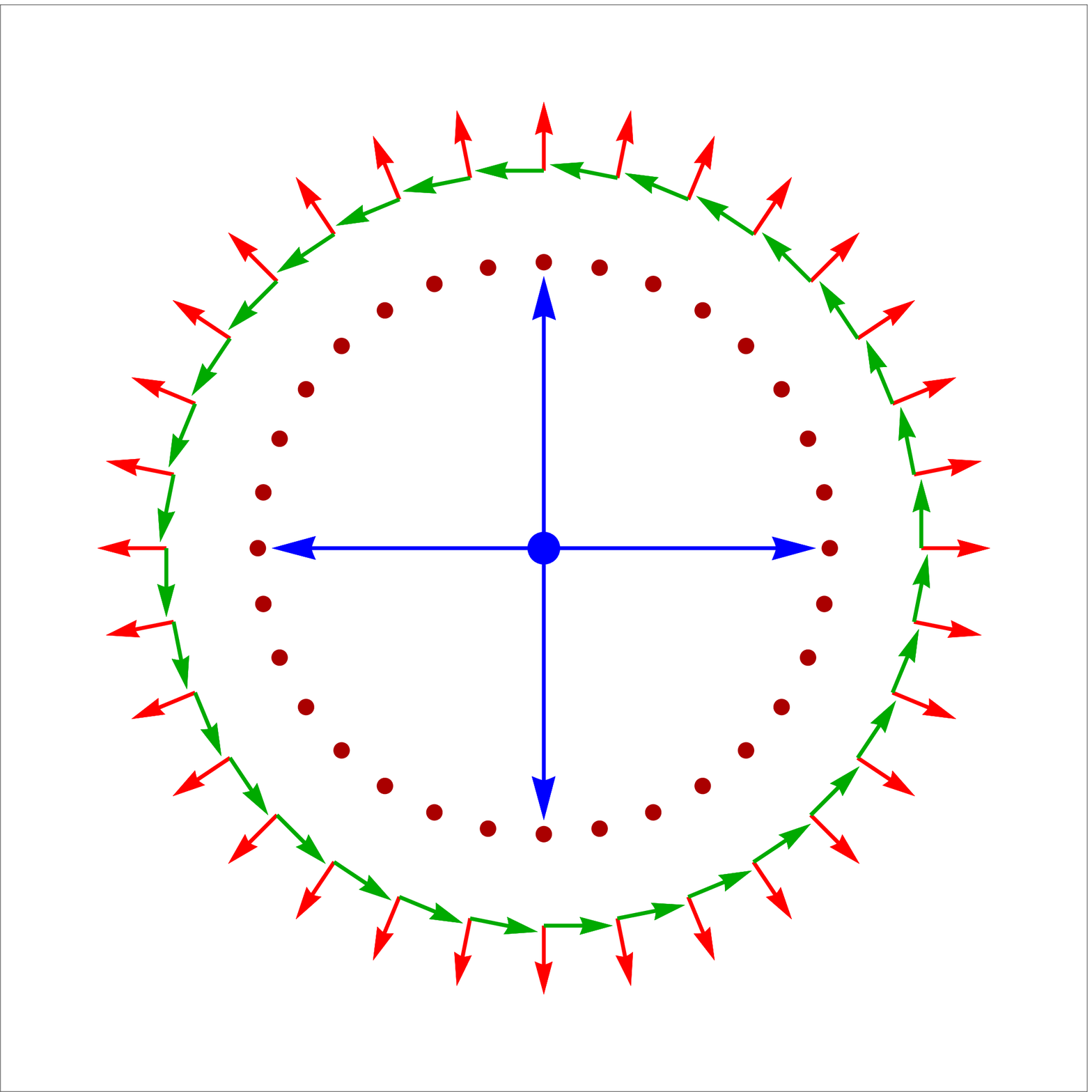}
\centerline {\hspace{1.4in} (A) \hfill (B) \hspace{1.4in} }
\vspace{-0.3in} }
\caption{\ifnum\ShowFiles=1 {\bf One2DFrame.eps, lotsof2DFrames.eps.}\fi
 (A) Any standard 2D
  coordinate frame corresponds to the columns of an ordinary rotation
  matrix, and is associated to the point $(\cos \theta, \sin\theta)$
  on a unit circle.  (B) The standard 2D coordinate frames associated
  with a sampling of the entire circle of points $(\cos \theta,
  \sin\theta)$.}
\label{OneAndLots2DFrames.fig}
\end{figure}

\noindent{\bf What is a Quaternion Frame?}  We will first present a bit
of intuition about coordinate frames that may help some readers with
our terminology.  If we take the special case of a quaternion
representing a rotation in the 2D $(x,y)$ plane, the 3D rotation
matrix \Eqn{qrot.eq} reduces to the standard right-handed 2D rotation
\begin{equation} 
R_{2}(\theta) = \left[\begin{array}{cc} \cos \theta &-\sin\theta\\
\sin \theta &\cos \theta \end{array} \right] \ .
\label{rot2d.eq}
\end{equation}
As shown in Fig~\ref{OneAndLots2DFrames.fig}(A), we can use $\theta$ to define a
unit direction in the complex plane defined by $z = \exp{\I \theta}$,
and then the \emph{columns} of the matrix $R_2(\theta)$ naturally
correspond to a unique associated 2D coordinate frame diad;  an
entire collection of points $z$ and their corresponding frame diads are
depicted in Fig.~\ref{OneAndLots2DFrames.fig}(B).

\begin{figure}
\figurecontent{
\centering
\includegraphics[width=2.7in]{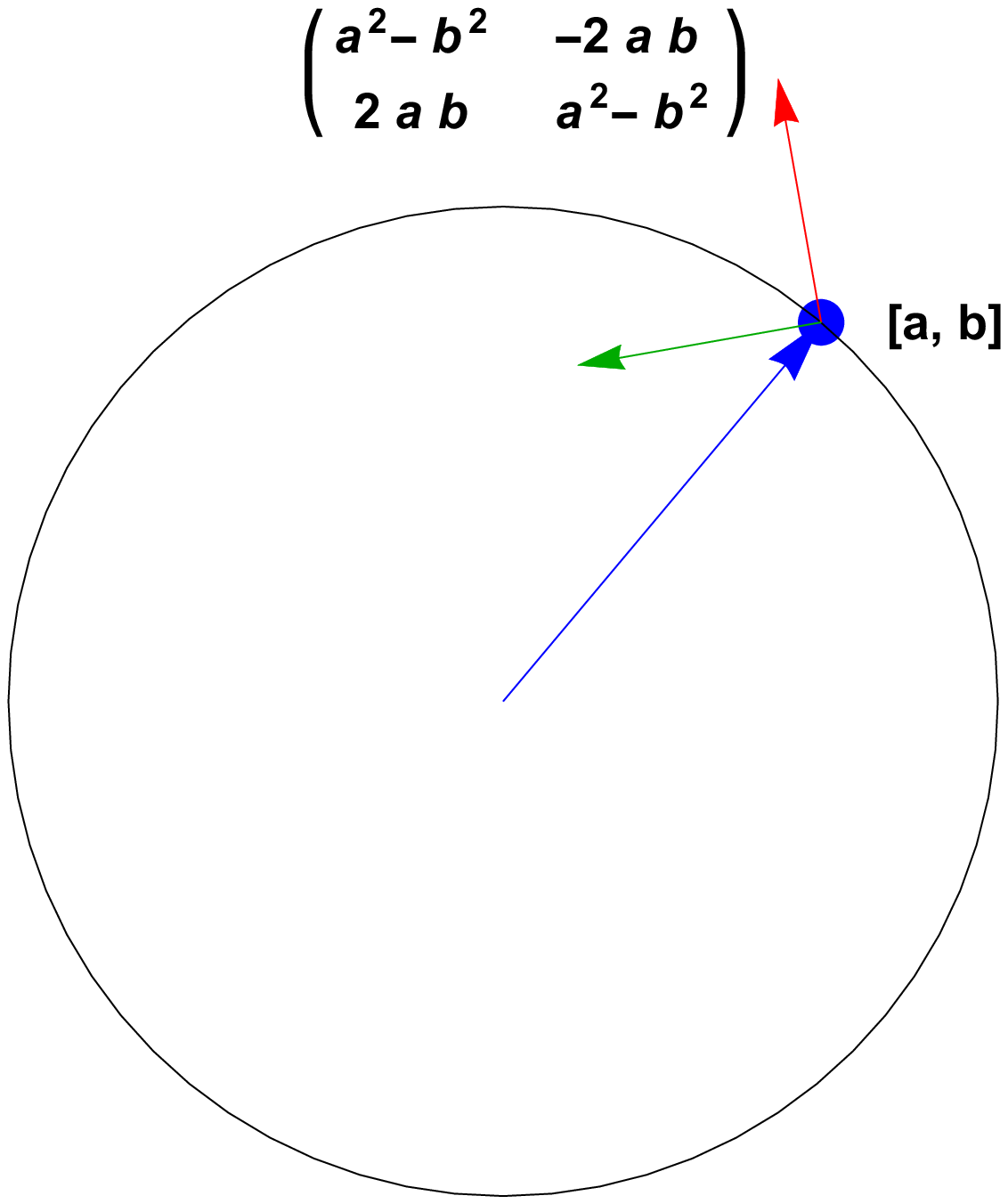}
\hspace{.1in}
\includegraphics[width=2.7in]{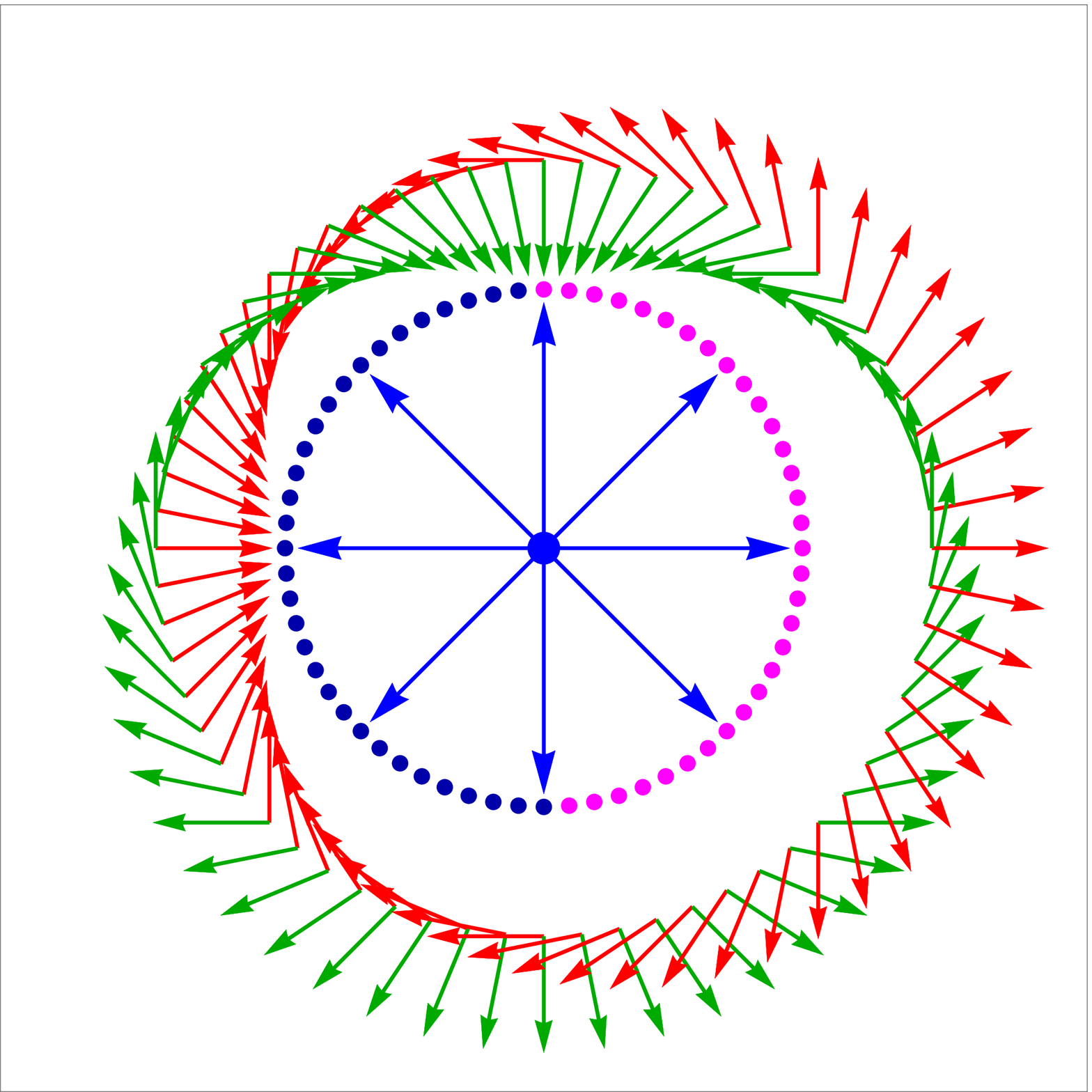}
\centerline {\hspace{1.4in} (A) \hfill (B) \hspace{1.4in} }
\vspace{-0.3in} }
\caption{\ifnum\ShowFiles=1 {\bf One2DabFrame.eps,  lotsof2DabFrames.eps.} \fi
 (A) The quaternion  point $(a,b)$, in 
  contrast, corresponds via the double-angle formula to coordinate
  frames that rotate twice as rapidly as $(a,b)$ progresses around the
  unit circle that is a simplified version of quaternion space. 
(B) The set of 2D frames associated with the entire circle of
  quaternion points $(a,b)$; each diametrically opposite point
  corresponds to an identical frame.  For later use in displaying full
  quaternions, we show how color coding can be used to encode the sign
  of one of the coordinates on the circle. }
\label{OneAndLots2DabFrames.fig}
\end{figure}

Starting from this context, we can get a clear intuitive picture of
what we mean by a ``quaternion frame'' before diving into the
quaternion RMSD problem.  The essential step is to look again at
\Eqn{qrot.eq} for $n_{x} =1$, and write the corresponding quaternion
as $(a,b,0,0)$ with $a^2+b^2=1$, so this is a ``2D quaternion,'' and is
indistinguishable from a complex phase like $z = \exp{\I \theta}$  that we just
introduced.  There is one significant difference, however, and that is
that \Eqn{qrot.eq} shows us that $R_{2}(\theta)$ takes a new form,
quadratic in $a$ and $b$,
\begin{equation} 
R_{2}(a,b) = \left[\begin{array}{cc} a^2-b^2 &-2 a b\\
2 a b &a^2-b^2 \end{array} \right] \ .
\label{rot2ab.eq}
\end{equation}
Using either the formula \Eqn{qqnth.eq} for $q(\theta,\Hat{n})$ or just
exploiting the trigonometric double angle formulas, we see that
\Eqn{rot2d.eq} and \Eqn{rot2ab.eq} correspond and that
\begin{eqnarray} 
(a,b) &  =& \left(\cos(\theta/2),\sin (\theta/2)\right)\\
 u & =& (a + \I\, b) =  \sqrt{z} = e^{\I\theta/2} \ .
\label{abToTheta.eq}
\end{eqnarray}
Our simplified 2D quaternion thus describes the \emph{square root} of
the usual Euclidean frame given by the columns of
$R_{2}(\theta)$. Thus the pair $(a,b)$ (the reduced quaternion) itself
corresponds to a frame.  In Fig.~\ref{OneAndLots2DabFrames.fig}(A), we show how a
given ``quaternion frame,'' i.e., the columns of $R_{2}(a,b)$,
corresponds to a point $u = a + \I\, b$ in the complex plane.
Diametrically \emph{opposite} points $(a,b)$ and $(-a,-b)$ now
correspond to the \emph{same} frame!
Fig.~\ref{OneAndLots2DabFrames.fig}(B) 
shows the corresponding frames for a large collection of points
$(a,b)$ in the complex plane, and we see the new and unfamiliar
feature that the frames make \emph{two} full rotations on the complex
circle instead of just one as in Fig.~\ref{OneAndLots2DFrames.fig}(B).

This is what we have to keep in mind as we now pass to using a full
quaternion to represent an arbitrary 3D frame triad via \Eqn{qrot.eq}.
The last step is to notice that in
Fig~\ref{OneAndLots2DabFrames.fig}(B) we 
can represent the set of frames in one half of the complex circle,
$a\ge 0 $ shown in magenta, as distinct from those in the other half,
$a<0$ shown in dark blue; for any value of $b$, the vertical axis,
there is a \emph{pair} of $a$'s with opposite signs and colors.
  In the quaternion case, we can display
quaternion frames inside one single sphere, like displaying only the
$b$ coordinates in Fig~\ref{OneAndLots2DabFrames.fig}(B) projected to
the vertical axis, realizing
that if we know the sign-correlated coloring, we can determine both
the magnitude of the dependent variable $a = \pm \sqrt{1-b^2}$ as well
as its sign.  The same holds true in the general case: if we display
only a quaternion's 3-vector part $\Vec{q}=(q_x,q_y,q_z)$ along with a color
specifying the sign of $q_{0}$, we implicitly know both the magnitude
and sign of $q_{0} = \pm \sqrt{1 - {q_{x}}^2- {q_{y}}^2- {q_{z}}^2}$,
and such a 3D plot therefore accurately depicts \emph{any} quaternion.
Another alternative employed in the main text is to use \emph{two}
solid balls, one a ``northern hemisphere" for the $q_{0} \ge 0$ components
and the other a ``southern hemisphere'' for the $q_{0} < 0 $ components.
Each may be useful in different contexts.

\qquad

\begin{figure}
\figurecontent{
\centerline{
\includegraphics[width=3in]{fig3DFrmRef2.eps}}
\centerline {\hspace{3in} (A) \hfill }
\centerline{
 \includegraphics[width=3in]{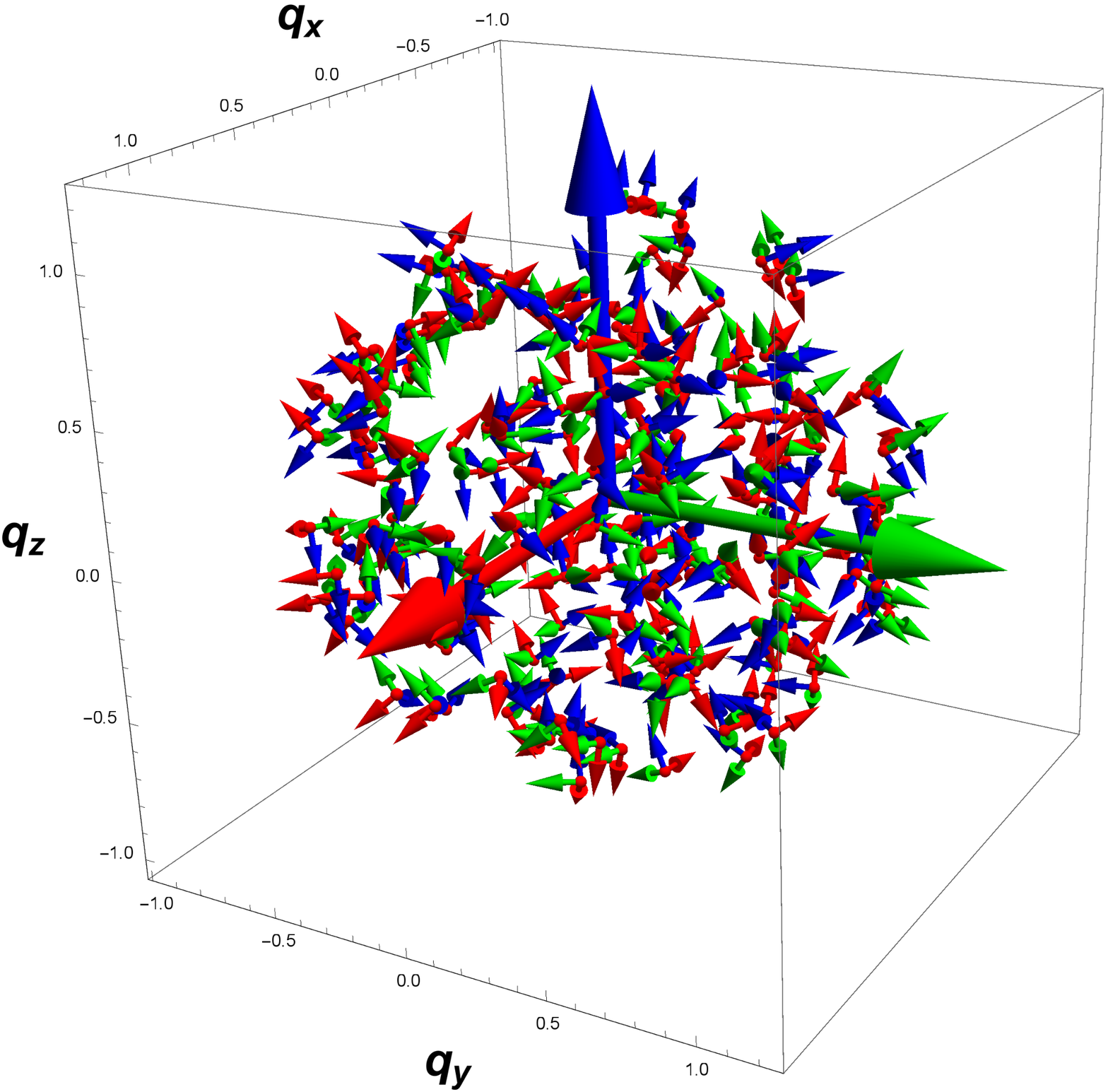}}
\centerline{\hspace{3in} (B) \hfill }
} \caption{\ifnum\ShowFiles=1 {\bf fig3DFrmRef2.eps, fig3DFrmAxes1.eps.} \fi
  (A) The 3D
  portions of the quaternion reference-frame data $q =
  (q_{0},q_{x},q_{y},q_{z})$, using different colors for $q_{0} \ge 0$
  and $q_{0}<0$ in the unseen direction.  Since $|q_{0}| =
  \sqrt{{q_{x}}^2 + {q_{y}}^2 +{q_{z}}^2}$, the complete quaternion
  can in principle be determined from the 3D display.  (B) The
   3D orientation frame triads for each reference  point
   $(q_{0},q_{x},q_{y},q_{z})$ displayed at their associated
   $\Vec{q}=(q_x,q_y,q_z)$.
}
\label{fig3DRefFrms.fig}
\end{figure}

\noindent{\bf Example.} We illustrate all this in
Fig~\ref{fig3DRefFrms.fig}(A), which shows a typical collection of
quaternion reference-frame data displaying only the $\Vec{q}$
components of $(q_0, \Vec{q})$; the $q_{0} \ge 0$ data are mixed with the $q_{0}<0$ data,
but are distinguished by their color coding.
In Fig~\ref{fig3DRefFrms.fig}(B), we show the frame triads resulting
from applying \Eqn{qrot.eq} to each quaternion point and plotting the
result at the associated point $\Vec{q}$ in the display.

\section{\vspace{0.1in}On Obtaining Quaternions from Rotation Matrices}

\label{baritzhack.app}
 

The quaternion RMSD profile matrix method can be used to implement
 a singularity-free algorithm to 
obtain the (sign-ambiguous) quaternions corresponding to  numerical
3D and 4D rotation matrices.  There are many existing approaches to the
3D problem in the literature (see, e.g., \cite{Shepperd1978},
\cite{ShusterNatanson1993}, or Section 16.1 of
\cite{HansonQuatBook:2006}).  In contrast to these
approaches, Bar-Itzhack~\cite{BarItzhack2000} has observed,
 in essence, that if we simply replace the data matrix $E_{ab}$ by a
numerical 3D orthogonal rotation matrix $R$, the numeric quaternion $q$ that 
corresponds to   $R_{\mbox{numeric}}=R(q)$, as defined by
\Eqn{qrot.eq}, can be found by solving our familiar maximal
quaternion eigenvalue problem.  The initially unknown optimal matrix 
(technically its quaternion) computed by
maximizing the similarity measure is equivalent to a single-element quaternion
barycenter problem, and the construction is designed to yield a best approximation
to $R$ itself in quaternion form.  To see this, take
$S(r)$ to be the sought-for optimal rotation matrix, with its own quaternion $r$,
that must maximize the Bar-Itzhack measure.  We start with the 
Fr\"{o}benius measure describing the match of two rotation matrices corresponding
to the quaternion $r$ for the unknown quaternion and the numeric matrix
$R$ containing  the known
$3\times 3$ rotation matrix data:
\begin{eqnarray*}
  {\mathbf S}_{\mbox{\scriptsize BI}} & =&    \|S(r)-R\|^{2}_{\mbox{Frob}} 
    \ = \  \tr \left( [S(r) - R] \cdot[ S^{\t}(r) - R^{\t}] \right)  \\
  & = & \tr \left( I_{3} + I_{3}  - 2  \left( S(r) \cdot R^{\t} \right) \right)  \\
   & = & \mbox{const} - 2   \tr S(r) \cdot  R^{\t}   \ . 
  \label{FrobBI.eq}
\end{eqnarray*}
Pulling out the cross-term as usual and converting to a maximization
problem over the unknown quaternion $r$, we arrive at
\begin{equation}
\Delta_{\mbox{\scriptsize BI}} = \tr{S(r) \cdot R^{\t} } = r \cdot K(R) \cdot r \ ,
\label{deltaBI.eq}
\end{equation}
where $R$ is (approximately) an orthogonal matrix of numerical data, and 
$K(R)$ is  analogous to the profile matrix $M(E)$.
Now $S$ is an abstract rotation matrix,   and $R$ is supposed to be a good numerical 
approximation to   a  rotation matrix,  and thus  the
product $T=S\cdot R^{\t}$ should also be a good approximation to an $\SO{3}$ rotation matrix;
hence that product itself corresponds closely to some axis $\Hat{n}$
and angle $\theta$, where (supposing we knew $R$'s exact quaternion $q$)
\[ \tr {S(r)  \cdot R^{\t}(q)} = \tr T(r \star \bar{q})= \tr T(\theta,\Hat{n}) = 1 + 2 \cos{\theta} \ . \]
The maximum is obviously 
close to  $T$ being the identity matrix, with the ideal value
at $\theta = 0$,   corresponding to $S \approx R$.
Thus if we find the maximal quaternion eigenvalue  $\epsilon_{\opt}$ of
the profile matrix $K(R)$ in \Eqn{deltaBI.eq},  our  closest solution is  well-represented by 
the  corresponding normalized eigenvector $r_{\opt}$,
\begin{equation}
q = r_{\opt}  \ .
\label{deltaBISoln.eq}
\end{equation}
This numerical solution for $q$ will correspond to the targeted numerical rotation
matrix, solving the problem.  To complete the details of the computation,
 we replace the elements
$E_{ab}$ in \Eqn{basicHorn.eq} by a general orthonormal rotation
matrix with columns $\Vec{X}= (x_1,x_2,x_3)$, $\Vec{Y}$, and $\Vec{Z}$,
 scaling by $1/3$, thus obtaining the special $4 \times
4$ profile matrix $K$ whose elements in terms of a known numerical
matrix $R = \left[\Vec{X} | \Vec{Y} | \Vec{Z} \right]$ (transposed in the algebraic
expression for $K$ due to the $R^{\t}$)
 are
\begin{equation}
K(R) = \frac{1}{3} \left[
\begin{array}{cccc}
 x_1+y_2+z_3 &  y_3 - z_2 & z_1 - x_3 & x_2 - y_1 \\
  y_3 - z_2 & x_1-y_2-z_3 & x_2+y_1 & x_3+z_1 \\
  z_1 - x_3 & x_2+y_1 & -x_1+y_2-z_3 & y_3+z_2 \\
  x_2 - y_1 & x_3+z_1 & y_3+z_2 & -x_1-y_2+z_3 \\
\end{array}
\right] \ .
 \label{theKRmatrix.eq}
\end{equation}

Determining the algebraic eigensystem of \Eqn{theKRmatrix.eq} is a 
nontrivial task.
However, as we know, any orthogonal 3D rotation matrix $R(q)$, or
equivalently, $R^{\t}(q) = R(\bar{q})$,  can also be ideally
expressed in terms of quaternions via \Eqn{qrot.eq}, and this
yields an alternate useful algebraic form
\begin{eqnarray}
\lefteqn{K(q) = } \nonumber \\
&& \hspace{-.3in}\frac{1}{3} \left[
 \begin{array}{cccc}
\textstyle \hspace{-.075in}3 {q_{0}}^2-{q_{1}}^2-{q_{2}}^2-{q_{3}}^2 &
      4 q_{0} q_{1} &
      4 q_{0} q_{2} &
      4 q_{0} q_{3} \hspace*{-.1in} \\ 
  4 q_{0} q_{1} &
\textstyle  \hspace{-.075in} -{q_ {0}}^2 + 3 {q_{1}}^2-{q_{2}}^2-{q_{3}}^2 &
     4 q_{1} q_{2} & 4 q_{1} q_{3} \hspace*{-.1in}\\ 
  4 q_{0} q_{2} & 4 q_{1} q_{2} &
\textstyle   \hspace{-.075in} -{q_{0}}^2-{q_{1}}^2 + 3 {q_{2}}^2-{q_{3}}^2 &
    4 q_{2} q_{3}  \hspace*{-.1in}\\  
  4 q_{0} q_{3} & 4 q_{1} q_{3} & 4 q_{2} q_{3} &
   \textstyle  \hspace{-.075in} -{q_{0}}^2-{q_{1}}^2-{q_{2}}^2 + 3 {q_{3}}^2   \hspace*{-.1in}\\   
\end{array}
\right]  \ 
\label{theKqmatrix.eq}
\end{eqnarray}
This equation then allows us to quickly prove that $K$ has the correct
properties to solve for the appropriate quaternion corresponding to $R$.
First we note that  the coefficients $p_n$ of the
eigensystem are simply constants,
\[ \begin{array}{c@{\hspace{.25in}}c@{\hspace{.25in}}c@{\hspace{.25in}}c}
p_1 = 0 & p_2 = -\frac{2}{3} & p_3 = - \frac{8}{27} &
 p_4 = - \frac{1}{27} 
\end{array} \ . \]
Computing the eigenvalues and eigenvectors using the symbolic
quaternion form,  we see that the eigenvalues are constant, with
maximal eigenvalue exactly one, and the eigenvectors are almost
trivial, with the maximal  eigenvector being   the
quaternion $q$ that corresponds to the (numerical) rotation matrix:
\begin{eqnarray}
\epsilon & = & \{ 1, \; -\frac{1}{3},\; -\frac{1}{3},\; -\frac{1}{3}
\} \label{Keigvals.eq} \\
r & = &\left\{ 
  \left[\begin{array}{c} q_0 \\   q_1 \\   q_2 \\   q_3 \end{array}\right], \;
    \left[ \begin{array}{c} -q_1 \\  q_0 \\  0 \\ 0 \end{array}\right], \;
   \left[ \begin{array}{c} -q_2 \\  0  \\  q_0 \\  0 \end{array}\right], \;
   \left[ \begin{array}{c} -q_3 \\  0 \\ 0 \\  q_0 \end{array}\right]
        \right\} \ . 
   \label{Keigvecs.eq}
\end{eqnarray}
The first column is the quaternion $r_{\opt}$, with
 $ \Delta_{\mbox{\scriptsize BI}}  (r_{\opt}) = 1$.  (This would be 3 if we
 had not divided by 3 in the definition of $K$.)
 
 \emph{Alternate version.}  From the quaternion barycenter work of Markley 
 et al.~\cite{MarkleyEtal-AvgingQuats2007} and the natural form of the quaternion-extraction 
 problem in 4D in the Supplementary Material, we know that \Eqn{theKqmatrix.eq} actually
 has a much simpler form with the same unit eigenvalue and natural quaternion eigenvector.
 If we simply take \Eqn{theKqmatrix.eq}  multiplied by 3, add the constant term
 $ I_{4} =    ( {q_{0}}^2 + {q_{1}}^2 +  {q_{2}}^2 +   {q_{3}}^2 ) I_{4}$ , and divide by 4,
  we get a more compact quaternion form of the matrix, namely
\begin{eqnarray}
K '(q) & = &  \left[
 \begin{array}{cccc}
  {q_{0}}^2  &
    q_{0} q_{1} &
       q_{0} q_{2} &    q_{0} q_{3}  \\ 
    q_{0} q_{1} &
   {q_{1}}^2  &
       q_{1} q_{2} &   q_{1} q_{3} \\ 
   q_{0} q_{2} &  q_{1} q_{2} &
    {q_{2}}^2 &
      q_{2} q_{3}    \\  
    q_{0} q_{3} &  q_{1} q_{3} &   q_{2} q_{3} &   
    {q_{3}}^2   \\   
\end{array}
\right]  \  .
\label{theBCsimplematrix.eq}
\end{eqnarray}
This has  vanishing determinant and trace $\, \tr K' = 1= -p_1$, with
all other $p_{k}$ coefficients vanishing, and leading eigensystem
identical to \Eqn{theKqmatrix.eq}: 
\begin{eqnarray} \label{Ksimpleeigvals.eq}
\epsilon & = & \{ 1, \;0 ,\;0,\; 0 \}  \\
r& = &\left\{ 
  \left[\begin{array}{c} q_0 \\   q_1 \\   q_2 \\   q_3 \end{array}\right], \;
    \left[ \begin{array}{c} -q_1 \\  q_0 \\  0 \\ 0 \end{array}\right], \;
   \left[ \begin{array}{c} -q_2 \\  0  \\  q_0 \\  0 \end{array}\right], \;
   \left[ \begin{array}{c} -q_3 \\  0 \\ 0 \\  q_0 \end{array}\right]
        \right\} \ . 
   \label{Ksimpleeigvecs.eq}
\end{eqnarray}
As elegant as this is, in practice, our numerical input data are from the
$3\times 3$ matrix $R$ itself, and not the quaternions, so we will almost
always just use those numbers in \Eqn{theKRmatrix.eq} to solve the problem.

\qquad

 {\bf Completing the solution.}  In typical applications, \emph{the solution is immediate, requiring only trivial algebra}.  The maximal eigenvalue is always known in advance to be unity for any valid rotation matrix, so we need only  to compute the eigenvector from the
numerical matrix \Eqn{theKRmatrix.eq} with unit eigenvalue.  We simply
compute  any column
of the adjugate matrix of  $[K(R) - I_{4}]$,  or  solve the equivalent  linear
equations of the form
\begin{equation}
\begin{array}{c p{0.5in}c}
\left( K(R)- 1  * I_{4} \right) \cdot 
\left[ \begin{array}{c} 1\\v_1\\v_2\\v_3 \\ 
\end{array} \right]  = 0  &&
q = r_{\opt} =   \mbox{normalize\ }\left[ \begin{array}{c} 1\\v_1\\v_2\\v_3 \\ 
\end{array} \right] \\
\end{array} \ .
\label{TheBIsolnQuat.eq}
\end{equation}
As always, one may need to check for degenerate special cases. 

\quad

{\bf Non-ideal cases.} It is important to note, as emphasized by Bar-Itzhack, that if there are
\emph{significant errors} in the numerical matrix $R$, then the actual non-unit 
maximal eigenvalue of $K(R)$ 
can be computed numerically or algebraically as usual, and then that eigenvalue's
eigenvector  
determines the \emph{closest}  normalized quaternion to the errorful rotation
matrix, which can be very useful since such a quaternion always produces
a valid rotation matrix.  

 In any case,  \emph{up to an overall sign},  $r_{\opt}$ is the 
desired numerical quaternion $q$  corresponding to the
target numerical rotation matrix $R= R(q)$  .
 In some circumstances, one is looking for a
uniform statistical distribution of quaternions, in which case the
overall sign of $q$ should be chosen \emph{randomly}.  

 The Bar-Itzhack approach solves the
problem of extracting the quaternion of an arbitrary numerical 3D rotation
matrix in a fashion that involves no singularities and only trivial
testing for special cases, thus essentially making the traditional
methods obsolete.  The extension of Bar-Itzhack's method
to the case of 4D rotations is provided in the Supplementary
Material. 

\qquad


\section{On Defining the Quaternion Barycenter}

\label{quatbarycenter.app}
 
 
 The notion of a \emph{Riemannian Barycenter} is generally
 associated with the work of  \cite{GroveKarcherRuh1974},
 and may also be referred to as the \emph{Karcher mean}~\cite{KarcherMean1977},
 defined as the point that minimizes the sum of  squared geodesic distances from
 the elements of a collection of fixed points on a manifold.  The general class of
 such optimization problems has also been studied, e.g., by \cite{Manton2004}.
  We are interested here in the case of quaternions, which we know are points on the 
 spherical 3-manifold $\Sphere{3}$ defined by the unit-quaternion subspace of ${\mathReal}^{4}$ 
 restricted to  $q\cdot q = 1$ for any point $q$ in ${\mathReal}^{4}$.   This subject
 has been investigated by a number of authors, with Brown and Worsey~\cite{BrownWorsey1992} 
 discussing the problems with this computation in 1992, and Buss and Fillmore~\cite{BussFillmore2001} 
 proposing a solution applicable to computer graphics 3D orientation interpolation problems 
 in 2001, inspired to some extent by Shoemake's 1985 introduction of the quaternion \emph{slerp} as a way to
 perform geodesic orientation interpolations in 3D using  \Eqn{qrot.eq} for $R(q)$.  
 There are a variety of  methods and studies related to the quaternion barycenter problem.  In 2002, Moahker published a rigorous account on averaging in  the group of rotations \cite{Moakher2002},   while
 subsequent treatments included the   2007  work by Markley et 
 al.~\cite{MarkleyEtal-AvgingQuats2007}, focusing on aerospace and astronomy applications,
 and  the comprehensive review in 2013 by Hartley at al.~\cite{HartleyEtalRotAvg2013}, aimed
 in particular at the machine vision and robotics community, with additional  attention to
 conjugate rotation averaging (the ``hand-eye calibration'' problem in robotics), and multiple rotation
 averaging.  While we have focused on measures starting from sums of squares that lead to
 closed form optimization problems,
 \cite{HartleyWeiszfeldRotAvg2011} have carefully studied the utility of the corresponding $L_{1}$
 norm and the iterative  Weiszfeld algorithm for finding its optimal solution numerically.
 
 The task at hand is basically to extend the Bars-Itzhack algorithm to an entire collection of frames instead of a single  rotation.   We need to find an optimal rotation matrix $R(q)$ that corresponds to the quaternion point
 closest to the geodesic center of an \emph{unordered} set of reference data.   We already know
 that  the case of the ``barycenter'' of
 a single orientation frame is solved by the Bars-Itzhack algorithm of Appendix \ref{baritzhack.app}, which  finds
 the quaternion closest to a single item of rotation matrix data  (the quaternion barycenter of a single rotation  is itself).
   For two items of data, $R_{1}= R(q_{1})$ and
 $R_{2}=R(q_{2})$, the quaternion  barycenter is determined by the \emph{slerp} interpolator to be
 \[ q(q_1,q_2)_{\mbox{barycenter}} =  q_{1} \star \left( \bar{q}_{1}\star q_{2} \right)^{1/2} 
  = \mbox{\it slerp} \left(q_1,q_2,\frac{1}{2} \right) \ . \]
 For three or more items, no closed form is currently known.
 
 We start with  a data set of $N$ rotation matrices $R(p_{k})$
 that are represented by the quaternions $p_{k}$, and we want $R(q)$ to be as close as possible 
 to the set of  $R(p_{k})$.  That rotation matrix, or its associated quaternion point, are the 
 orientation frame analogs of the   Euclidean barycenter  for a set of Euclidean points.  As before, it is clear that the mathematically most
 justifiable measure employs the \emph{geodesic arclength} on the quaternion sphere;  but to the best of anyone's knowledge, there is no way to apply linear algebra to find the corresponding
 $R(q_{\opt})$.  Achieving a numerical solution to that problem is the task solved by Buss and Fillmore~\cite{BussFillmore2001}, as well as a number of others, including, e.g., 
 \cite{Moakher2002,MarkleyEtal-AvgingQuats2007,HartleyEtalRotAvg2013}.
 The problem that we can understand algebraically is, once again, the approximate \emph{chord measure},  which we can immediately formulate in a sign-insensitive fashion using the
 Fr\"{o}benius measure $\| M \| ^{2}= \tr(M \cdot M^{\t})$, giving us the following starting point:
\begin{equation} 
\left. \begin{array}{rcl}
 {\mathbf {S}}_{\mbox{barycenter}} (q) &  = &
    \sum_{k=1}^{N} \| R(q )-  R(p_{k})\|^{2} \\ [0.05in]
  & = &     \sum_{k=1}^{N} \tr \left( [R(q )-  R(p_{k})] \cdot  [R^{\t}(q )-  R^{\t}(p_{k})] \right) \\ [0.05in]
 & = &     \sum_{k=1}^{N} \tr \left( 2 \,I_{4} -2  R(q )\cdot  R^{\t} (p_{k})\right) \\[0.05in]
   & = &    \sum_{k=1}^{N} \left( 8 - 2 \tr  R(q)\cdot R( \bar{p}_{k} ) \right) \\
  \end{array} \right\} \ .
\label{qbaryctr.eq}
\end{equation}

Dropping the constants and converting as usual to maximize over the cross-term
instead of minimizing the distance measure, we define a tentative spherical barycenter
as the maximum of the variation over the quaternion $q$ of the following:
\begin{equation} 
\left. \begin{array}{rcl}
{\displaystyle \Delta}_{\mbox{trial barycenter}}(q) & = &   
    {\displaystyle  \frac{1}{4} \sum_{k=1}^{N}  }\tr \left(R(q)\cdot R( \bar{p}_{k}) \right) \\[0.15in]
 & = &  {\displaystyle  \sum_{k=1}^{N}  }q\cdot K_{k}(p) \cdot q 
  \end{array} \right\} \ , 
\label{theqbaryctr.eq}
\end{equation} 
where for each $k=1,\ldots,N$, our first guess at the profile matrix is
\begin{eqnarray}
\lefteqn{K_{k} (p)_{\mbox{trial}} = } \nonumber\\ [0.05in]
&& \hspace{-.3in}\frac{1}{4} \left[
 \begin{array}{cccc}
\textstyle \hspace{-.075in}3 {p_{0}}^2-{p_{1}}^2-{p_{2}}^2-{p_{3}}^2 &
  4 p_{0} p_{1} &
     4 p_{0} p_{2} &  4 p_{0} p_{3} \hspace*{-.1in} \\ 
  4 p_{0} p_{1} &
\textstyle -{p_ {0}}^2 + 3 {p_{1}}^2-{p_{2}}^2-{p_{3}}^2 &
     4 p_{1} p_{2} & 4 p_{1} p_{3} \hspace*{-.1in}\\ 
  4 p_{0} p_{2} & 4 p_{1} p_{2} &
\textstyle -{p_{0}}^2-{p_{1}}^2 + 3 {p_{2}}^2-{p_{3}}^2 &
    4 p_{2} p_{3}  \hspace*{-.1in}\\  
  4 p_{0} p_{3} & 4 p_{1} p_{3} & 4 p_{2} p_{3} &
   \textstyle  -{p_{0}}^2-{p_{1}}^2-{p_{2}}^2 + 3 {p_{3}}^2   \hspace*{-.1in}\\   
\end{array}
\right]  \  .
\label{theBCqmatrix0.eq}
\end{eqnarray}

 But if, as pointed out by \cite{MarkleyEtal-AvgingQuats2007},
 we simply add one copy of the identity matrix in the form 
 $(1/4) I_{4} = (1/4) ( {p_{0}}^2 + {p_{1}}^2 +  {p_{2}}^2 +   {p_{3}}^2 ) I_{4}$ 
 to the matrix $K(p)$, we get a much simpler matrix 
 that we can use instead because constants do not affect the optimization process.
 Our partial profile matrix  
 for each $k=1,\ldots,N$  is now
\begin{eqnarray}
K_{k}(p) & = &  \left[
 \begin{array}{cccc}
  {p_{0}}^2  &
    p_{0} p_{1} &
       p_{0} p_{2} &    p_{0} p_{3}  \\ 
    p_{0} p_{1} &
   {p_{1}}^2  &
       p_{1} p_{2} &   p_{1} p_{3} \\ 
   p_{0} p_{2} &  p_{1} p_{2} &
    {p_{2}}^2 &
      p_{2} p_{3}    \\  
    p_{0} p_{3} &  p_{1} p_{3} &   p_{2} p_{3} &
    {p_{3}}^2   \\   
\end{array}
\right]  \  ,
\label{theBCqmatrix.eq}
\end{eqnarray}
or to be precise, after the sum over $k$, the profile matrix in terms of the
quaternion columns  $[p_{k}]$
becomes of $\Vec{P}$
\begin{equation} 
K(\Vec{P})_{ab} = \sum_{k=1}^{N}   [p_{k}]_{_{\scriptstyle a}} [p_{k}]_{_{\scriptstyle b}}
 \, = \, \left[ \Vec{P} \cdot \Vec{P}^{\,\t} \right]_{ab}
\label{theBCprofile.eq}
\end{equation}
with quaternion indices $(a,b)$ ranging from $0$ to $3$.  Finally, we can write the expression for the chord-based barycentric measure  to be optimized to get $q_{\opt}$ as
 \begin{equation}
 {\displaystyle \Delta}_{\mbox{ barycenter}}(q)
 \, = \,     q\cdot K(\Vec{P})  \cdot q 
    \ .
\label{theqbaryctrFinal.eq}
\end{equation}
We also have another option: if, for some reason, we only have numerical
$3\times 3$ rotation matrices $R_{k}$ and not their associated quaternions
$p_{k}$, we can recast \Eqn{theqbaryctr.eq} in terms of \Eqn{theKRmatrix.eq} for
each matrix $R_{k}$, and use the sum over $k$ of \emph{those} numerical matrices
to extract our optimal quaternion.  This is nontrivial because the simple
eigensystem form of the profile matrix $K$ for the Bar-Itzhack task was valid
only for \emph{one} rotation data matrix, and as soon as we start summing over
additional matrices, all of that simplicity disappears, though the eigensystem
problem remains intact.

The optimizing the approximate chord-measure for the ``average rotation,'' the
``quaternion average,'' or the spherical barycenter of the quaternion
orientation frame data set $\{p_{k}\}$ (or $\{R_{k}\}$) now just reduces, as
before, to finding the (normalized) eigenvector corresponding to the largest
eigenvalue of $K$.  It is also significant that the initial $K_{\mbox{trial}}$
matrix in \Eqn{theBCqmatrix0.eq} is \emph{traceless}, and so the traceless
algebraic eigenvalue methods would apply, while the simpler $K$ matrix in
\Eqn{theBCprofile.eq} is \emph{not traceless}, and thus, in order to apply the
algebraic eigenvalue method, we would have to use the generalization presented
in the Supplementary Material that includes an arbitrary trace term.



\bigskip \bigskip 

\section*{Acknowledgments} 

We thank Sonya M.\ Hanson for
  reacquainting us with this problem, providing much useful information and
  advice, and for motivating us to pursue this challenging investigation
  to its conclusion.  We also express our appreciation to the referees for suggestions
  that added significantly to the completeness and scope of the paper and to Randall Bramley,
  Roger Germundsson, B.\ K.\ P.\ Horn, and Michael Trott for their valuable input.
  
\bigskip \bigskip

 \bibliographystyle{apalike}

  \bibliography{rmsd}


\end{document}






\title{\emph{Supplementary Material:}   The Quaternion-Based \\ Spatial Coordinate  
and Orientation Frame   \\ Alignment   Problems}

\author{Andrew J.\ Hanson\\
Luddy School of Informatics, Computing, and Engineering\\
Indiana University, Bloomington, Indiana, 47405, USA}
\date{}

\maketitle

\begin{abstract}
  Supplementary material for the paper 
  \emph{The Quaternion-Based Spatial Coordinate and Orientation Frame 
      Alignment Problems} is presented here.  The most significant
  additional result is the extension of the 3D treatment in the main
  text to four dimensions.  Following a  review of quaternion properties
  now including the representation of 4D rotations using quaternion pairs,  we give a detailed study
  of  the 4D quaternion-based spatial alignment problem, which is significantly different from
  the 3D problem in the main text.  Next, we use the 4D quaternion rotation method
  to extend our treatment to 4D orientation-frame alignment.  
  The 3D Bar-Itzhak 
  profile-matrix method for extracting a quaternion from a 3D numerical rotation
  matrix is extended to 4D numerical rotation matrices,  followed by a
  look at the algebraic solutions of 2D alignment
  problems, whose deceptive simplicity does not carry over to the 3D and 4D
  cases.  Finally, we supplement the 3D orientation alignment section of
  the main text with careful studies of the properties, limitations,
  and features of our 3D orientation frame alignment methods, followed 
  by an extended exposition and analysis of the strengths and
  weaknesses of the 6DOF combined spatial and orientation frame
  alignment techniques.  The Appendix  provides a comprehensive
  study of the quartic equation solutions to eigenvalue problems, focusing
  on applications to the eigensystems of real symmetric matrices.  
  \end{abstract}



 \section{Foundations of Quaternions for 3D and 4D Problems} 


We begin with a review of quaternion properties used in the 3D 
analysis, folding in some additional details, and then systematically add
the extensions that are exploited to handle the 4D case.
The treatment here is designed to be self-contained, repeating any
relevant material from the main paper,
thus avoiding any confusion involving cross-references to the main
paper for equations and conceptual background.

\qquad

{\bf Quaternions for 3D Analysis.}
We take a quaternion to be a point $q = (q_0, q_1, q_2, q_3)=
(q_0,\,\Vec{q})$ in 4D Euclidean space with 
unit norm, $q \cdot q = 1$  (see, e.g.,
\cite{HansonQuatBook:2006} for further details about
quaternions).  The last three terms, $\Vec{q}$, play
the role of a generalized imaginary number, so  the conjugation operation is
$\bar{q} = (q_0, - \Vec{q})$.  Quaternions obey a multiplication
operation denoted by $\star$ and defined as follows:
\begin{equation} 
 q \star p =    Q(q)   \cdot p =  \left[ \begin{array}{cccc}
   q_0& -q_1& -q_2& -q_3 \\
   q_1& q_0& -q_3& q_2  \\
   q_2& q_3& q_0& -q_1\\
   q_3& -q_2& q_1& q_0
  \end{array} \right] \cdot\left[ \begin{array}{c}
  p_0\\p_1\\p_2\\p_3 \end{array} \right]  = 
( q_0 p_0 - \Vec{q}\cdot \Vec{p},\; q_0 \Vec{p}+ 
   p_0 \Vec{q} + \Vec{q}\times\Vec{p} ) \ ,
\label{qprod.eq} 
\end{equation}
where the orthonormal matrix $Q(q)$ is an alternative form of quaternion multiplication
that explicitly demonstrates that the action of $q$ on $p$ by quaternion multiplication
\emph{literally} rotates the quaternion unit vector $p$ in 4D Euclidean space.
Another non-trivial matrix form of quaternion multiplication that is useful in some
calculations is the left-acting matrix  $ \td{Q}$ producing a \emph{right} multiplication,
\begin{equation}
q \star p =   \td{Q}(p)   \cdot q \ = \ 
\left[
\begin{array}{cccc}
  p_{0} & - p_{1} & - p_{2} & - p_{3} \\
  p_{1} &  p_{0} &  p_{3} & - p_{2} \\
  p_{2} & - p_{3} &  p_{0} &  p_{1} \\
  p_{3} &  p_{2} & - p_{1} &  p_{0} \\
\end{array} \right]\cdot\left[ \begin{array}{c}
  q_0\\q_1\\q_2\\q_3 \end{array} \right]  \ .
\label{qRprod.eq}
\end{equation}

Choosing exactly one of the three imaginary components 
in both $q$ and $p$ to be nonzero 
gives back the classic complex algebra
$(q_0 +{\I} q_1)(p_0 + {\I} p_1)= \left( q_0 p_0 - q_1 p_1\right) + 
   {\I} \left(q_0 p_1  + p_0 q_1  \right)$, 
so there are three copies of the complex numbers embedded in the
quaternion algebra; the difference is that in general the final term
$\Vec{q}\times\Vec{p}$ changes sign if one reverses the order, making
the quaternion product order-dependent, unlike the complex product.
 Quaternions also satisfy the nontrivial ``multiplicative norm'' relation 
\begin{equation}
 \|q\|\,\|p\| = \| q \star p\| \ ,
 \label{multiplicativeNorm.eq}
 \end{equation}
 where  $\|q\|^2 =  q\cdot  q =\Re{(q \star \bar{q} )}$, that
uniquely characterizes the real, complex, quaternion, and octonion number
systems comprising the Hurwitz algebras.  Quaternions also obey a number of interesting scalar triple-product
identities,
 \begin{equation}
 \begin{array}{c@{\ = \ }c@{\ = \ }c}
 r \cdot (q \star p) & q\cdot (r \star \bar{p}) & \bar{r} \cdot (\bar{p} \star \bar{q}) \\
 \end{array}  ,
 \label{quatTriples.eq}
 \end{equation}
 where the complex conjugate entries are the natural consequences of the sign changes
occurring only in the (imaginary) 3D part.

Conjugating a vector $\Vec{x}= (x,y,z)$ written as
a purely ``imaginary'' quaternion
$(0,\Vec{x})$ by quaternion multiplication is isomorphic to
the construction of a 3D Euclidean rotation $R(q)$ generating all
possible elements of the special orthogonal group $\SO{3}$.  If we compute
\begin{equation}
  q \star (c,\, x,\, y,\, z) \star \bar{q} = (c,\, R_{3}(q) \cdot \Vec{x}) 
  \ , \label{qqR3.eq} 
\end{equation} 
we see that only the purely imaginary part is affected, whether or not
the arbitrary real constant $c=0$.
 Collecting coefficients gives this fundamental form of an arbitrary
3D rotation expressed in terms of quaternions,
\begin{equation}
     \left.  \begin{array}{rcl}
R_{ij}(q)    &= & \delta_{ij}\left({q_{0}}^2 - {\Vec{q}}^2\right) + 2 q_{i} q_{j} 
   - 2 \epsilon_{ijk}q_{0} q_{k}\\[0.15in]
  R(q) & =&  \left[ 
 \begin{array}{ccc}
 {q_0}^2+{q_1}^2-{q_2}^2 - {q_3}^2 & 2 q_1 q_2  -2 q_0 q_3 
& 2 q_1 q_3 +2 q_0 q_2   \\
 2 q_1 q_2 + 2 q_0 q_3  &  {q_0}^2-{q_1}^2 + {q_2}^2 - {q_3}^2   
      &  2 q_2 q_3 - 2 q_0 q_1  \\
 2 q_1 q_3 - 2 q_0 q_2  &  2 q_2 q_3 + 2 q_0 q_1 
 & {q_0}^2 - {q_1}^2 - {q_2}^2 + {q_3}^2
\end{array}
   \right] 
   \end{array} \right\}  \ ,
\label{qrot.eq}
\end{equation}
where the mapping from $q$ to $R_{3}(q)$ is two-to-one  because $R_{3}(q) =
R_{3}(-q)$.  Note that  $R(q)$ is a \emph{proper} rotation, with determinant $\det R(q) = (q\cdot q)^3 = +1$,
and that the identity quaternion $q_{\ID} = (1,0,0,0) \equiv q \star \bar{q}$
corresponds to the identity rotation matrix, as does  $ - q_{\ID} = (-1,0,0,0)$. 
The  \emph{columns} of $R(q)$ are exactly the needed 
quaternion representation of the {\it frame triad \/} describing the
orientation of a body in 3D space, i.e., the columns are the vectors
of the frame's local $x$, $y$, and $z$ axes relative to an initial
identity frame.
Choosing the following parameterization preserving $q \cdot
q =1 $ (with $\Hat{n}\cdot \Hat{n}=1$),
\begin{equation}
q = \left(\cos(\theta/2),\,\hat{n}_{1} \sin(\theta/2),\,
     \hat{n}_{2} \sin(\theta/2),\, \hat{n}_{3} \sin(\theta/2) \right)
 \ ,
\label{qqnth.eq}
\end{equation}
gives the ``axis-angle'' form of the rotation matrix,
\begin{equation}
R_{3}(q) = R_{3}(\theta,\Hat{n}) = \!\left[ \!
\begin{array}{ccc}
 \cos \theta +(1-\cos \theta )\, {\hat{n}_1}^{\ 2} & (1-\cos \theta )\, \hat{n}_1 \hat{n}_2-\sin \theta\,  \hat{n}_3 & (1-\cos \theta ) \, \hat{n}_1 \hat{n}_3+\sin \theta\, \hat{n}_2 \\
 (1-\cos \theta )\, \hat{n}_1 \hat{n}_2+\sin \theta \, \hat{n}_3 & \cos \theta +
  (1-\cos \theta )\,  {\hat{n}_2}^{\ 2} &
   (1-\cos \theta ) \, \hat{n}_2 \hat{n}_3-\sin \theta \, \hat{n}_1 \\
 (1-\cos \theta ) \, \hat{n}_1 \hat{n}_3-\sin \theta  \,\hat{n}_2 &
  (1-\cos \theta )\, \hat{n}_2 \hat{n}_3   +\sin \theta\, \hat{n}_1&
   \cos \theta +(1-\cos \theta ) \, {\hat{n}_3}^{\ 2} \\
\end{array}
  \! \right]  .
\label{axangnth.eq}
\end{equation}
This form of the  3D rotation exposes the fact that the direction
$\Hat{n}$ is fixed, so $\Hat{n}$ is the lone real eigenvector of $R_{3}$.
Multiplying a quaternion $p$ by the quaternion $q$ to get a new
quaternion $p' = q\star p$ simply {\it rotates\/} the 3Dframe
corresponding to $p$ by the matrix \Eqn{qrot.eq} written in terms of
$q$, so 
\begin{equation}
R_{3}(q\star p) = R_{3}(q)\cdot R_{3}(p) \ ,
\label{Rqp.eq}
\end{equation}
 and this collapse of repeated rotation matrices into a single rotation
 matrix with a quaternion-product argument can be
 continued indefinitely.

\quad

\noindent{{\bf Remark:} { \it Eigensystem  and properties of $R_{3}$:\/}}
One of our themes is constructing and understanding eigensystems of
interesting matrices, so here, as an aside, we expand the content of
the previous paragraph to include some additional details.  First, note
that we have two ways of writing the 3D rotation, as  $R_{3}(\theta,\Hat{n})$ and as 
$R_{3}(q)$.  Thus there are two ways to write the  eigenvalues, which we
can compute to be  

\comment{  Source for this:
\begin{verbatim}
eigmat3 = {1, Cos[th] + I Sin[th], Cos[th] - I Sin[th]}

In[602]:= ({{nx, ny, nz},
     {-I ny - nx nz, I nx - ny nz, nx^2 + ny^2},
     {+I ny - nx nz, -I nx - ny nz,  nx^2 + ny^2}} .
     Transpose[{{nx, ny, nz},
      {-I ny - nx nz, I nx - ny nz, nx^2 + ny^2},
      {+I ny - nx nz, -I nx - ny nz, nx^2 + ny^2}}] // 
         Simplify) /. {nx^2 + ny^2 + nz^2 -> 1}
Out[602]= {{1, 0, 0}, {0, 0, 2 (nx^2 + ny^2)}, {0, 2 (nx^2 + ny^2),  0}}

From Met2Diag.nb

{{q1, q2, q3}, 
{-q1 q3 - I q2 Sqrt[q1^2 + q2^2 + q3^2], -q2 q3 + 
   I q1 Sqrt[q1^2 + q2^2 + q3^2], 
  q1^2 + q2^2}, 
  {-q1 q3 + I q2 Sqrt[q1^2 + q2^2 + q3^2],
   -q2 q3 -  I q1 Sqrt[q1^2 + q2^2 + q3^2], q1^2 + q2^2}}
   
   eval3qq = {1, (q0^2 - q1^2 - q2^2 - q3^2 + 
    2 I q0 Sqrt[q1^2 + q2^2 + q3^2]), (q0^2 - q1^2 - q2^2 - q3^2 - 
    2 I q0 Sqrt[q1^2 + q2^2 + q3^2])}
\end{verbatim} } 

 \begin{equation}
\begin{array}{c p{.25in} c}  
 \left\{ \begin{array}{c}
    1  \\[0.05in] { e^{\displaystyle \I \theta}} \\ 
    {  e^{\displaystyle - \I \theta} }\\ \end{array}  \right\}
    &&
  \left\{   \begin{array}{c}
    1\\[0.05in]
    ({q_0}^2 - {q_1}^2 - {q_2}^2 - {q_3}^2  
    + 2 \I q_0 \sqrt{{q_1}^2 + {q_2}^2 + {q_3}^2}) \\[0.05in]
    ({q_0}^2 - {q_1}^2 - {q_2}^2 - {q_3}^2   
     - 2 \I q_0 \sqrt{{q_1}^2 + {q_2}^2 + {q_3}^2})
  \\ \end{array} 
  \right\} \ ,
\end{array}   \label{rot3DEigvals.eq}
  \end{equation}
  respectively, where the two columns are of course identical, but we
  have chosen expressions in $q$ (along with an implicit choice of square root
  sign determining $\sin(\theta/2)$) that match exactly with the $R_{3}(q)$ eigenvectors.
  Those eigenvectors (unnormalized for notational clarity)
  can be written as:
 \begin{equation}
\begin{array}{l}  
  \left\{  \begin{array}{ccc}
    \left[ \begin{array}{c}
        n_1 \\ n_2 \\ n_3 \\ \end{array} \right]&
         \left[ \begin{array}{c}
     -\I \, n_2 - n_1 n_3 \\ \I \,n_1 - n_2 n_3\\ {n_1}^2 + {n_2}^2   \\ \end{array} \right]&
        \left[ \begin{array}{c}
         +\I\, n_2 - n_1 n_3\\  -\I\, n_1 - n_2 n_3 \\  {n_1}^2 + {n_2}^2 \\ \end{array} \right] \\
        \end{array} \right\}  \\[0.3in]  
   \hspace*{0.65in}  \left\{    \begin{array}{ccc}
    \left[ \begin{array}{c}
         q_{1}  \\  q_{2}  \\ \ q_{3}  \\ \end{array} \right]&
           \left[ \begin{array}{c}
       \mbox{} -q_1 q_3 - \I\, q_2 \sqrt{ {q_1}^2 + {q_2}^2 + {q_3}^2}\\
        \mbox{}- q_2 q_3 +   \I\, q_1 \sqrt{ {q_1}^2 + {q_2}^2 + {q_3}^2}\\
            {q_1}^2 + {q_2}^2   \\ \end{array} \right]&
      \left[ \begin{array}{c}
       \mbox{} -q_1 q_3 + \I\, q_2 \sqrt{ {q_1}^2 + {q_2}^2 + {q_3}^2}\\
        \mbox{}- q_2 q_3 -  \I\, q_1 \sqrt{ {q_1}^2 + {q_2}^2 + {q_3}^2}\\
         {q_1}^2 + {q_2}^2 \\ \end{array} \right] \\
        \end{array}\right\}
            \\   \end{array}  
  \end{equation}
  where we emphasize that in general $R_{3}$ has only one real eigenvalue (which is unity),
  whose eigenvector is  the direction of the 3D axis $\Hat{n}$  invariant under that particular rotation. 
   Since any quaternion can be written in the form \Eqn{qqnth.eq}, the trace of
  any rotation can be written as 
  \begin{equation}
  \tr   R_{3}   =  3 {q_{0}}^{2} -{q_{1}}^{2}-{q_{2}}^{2}-{q_{3}}^{2} \, = \, 
     4  {q_{0}}^{2} - 1  \, = \,  1 + 2 \cos \theta  \ ,
   \end{equation}
which follows from the half-angle formula.  This means that, in the RMSD formula
maximizing $\tr(R \cdot E)$,
if $E$ is an identity matrix, the rotation giving the maximal trace corresponds
to $R_{3}$ being the identity
matrix, $\theta = 0$, and if $E$ is a rotation matrix, the maximal trace occurs when
the product of the two matrices has vanishing angle $\theta$ for the \emph{composite}
matrix produced by the product of their two quaternions, so the optimal rotation
matrix $R_{3}$ is the inverse of $E$.  This property is exploited in the Bar-Itzhack
algorithm given in Section~\ref{baritzhack.sec}.
    
    \quad
         
\noindent{\bf The Slerp.} Relationships among quaternions can be studied using the \emph{slerp},
or ``spherical linear interpolation'' \cite{shoemake-1985-quat,JuppKent1987}, that
smoothly parameterizes the points on the shortest geodesic quaternion path
between two constant (unit) quaternions, $q_{0}$ and $q_{1}$, as
\begin{eqnarray} 
\mbox{\it slerp\/}(q_{0},q_{1},s) \equiv q(s)[q_{0},q_{1}] & = &
q_{0}\frac{\sin ((1-s)\phi)}{\sin \phi} +  
q_{1}\frac{\sin (s\: \phi)}{\sin \phi} \ .
\label{slerpdef.eq}
\end{eqnarray}
Here $\cos \phi = q_{0} \cdot q_{1}$ defines the angle $\phi$ between
the two given quaternions, while $q(s=0) = q_{0}$ and $q(s=1) =
q_{1}$.  The "long" geodesic can be obtained for $1 \le s \le 2 \pi/\phi$.
For small $\phi$, this reduces to the standard linear
interpolation $(1-s)\, q_{0} + s\, q_{1}$.  The unit norm 
is preserved, $q(s)\cdot q(s) =1$ for all $s$,  
so $q(s)$ is always a valid quaternion and $R(q(s))$ defined by
\Eqn{qrot.eq} is always a valid 3D rotation matrix.
We note that one can formally write \Eqn{slerpdef.eq}   as an exponential of the form
$ q_{0} \star \left(\bar{q}_{0}\star q_{1}\right)^{s}$, but since
this requires computing a logarithm and an exponential whose most efficient
reduction to a practical computer program is \Eqn{slerpdef.eq}, this is 
mostly of pedagogical interest.

\qquad

\noindent{\bf  Double Quaternions and 4D Rotations.}
  We now extend \Eqn{qrot.eq} from three Euclidean dimensions to four
Euclidean dimensions by choosing  two {\it distinct\/} quaternions
and generalizing \Eqn{qqR3.eq} to 4D points $\Vec{x}_{4}= (w,x,y,z)$  as follows:
\begin{equation}
  p \star (w,\, x,\, y,\, z) \star \bar{q} = R_{4}(p,q) \cdot \Vec{x}_{4}
  \ . \label{qqR4.eq} \end{equation} 
Here $R_{4}$ turns out to be an orthonormal 4D rotation matrix that is
quadratic in the {\it pair\/} $(p,q)$ of unit quaternion elements,
which together  
have exactly the six degrees of freedom required for the most general
4D Euclidean rotation in the special orthogonal group $\SO{4}$.  The 
algebraic  form of this 4D rotation matrix is
\begin{equation}  
  \begin{array}{c}
 R_{4}(p,q)  = \left[ 
\begin{array}{c p{0.05in} c}
 p_{0} q_{0}+p_{1} q_{1}+p_{2} q_{2}+p_{3} q_{3} &&
 -p_{1} q_{0}+p_{0} q_{1}+p_{3} q_{2}-p_{2} q_{3} \\
 p_{1} q_{0}-p_{0} q_{1}+p_{3} q_{2}-p_{2} q_{3} &&
 p_{0} q_{0}+p_{1} q_{1}-p_{2} q_{2}-p_{3} q_{3}\\
 p_{2} q_{0}-p_{3} q_{1}-p_{0} q_{2}+p_{1} q_{3} &&
  p_{3} q_{0}+p_{2} q_{1}+p_{1} q_{2}+p_{0} q_{3}\\
 p_{3} q_{0}+p_{2} q_{1}-p_{1} q_{2}-p_{0} q_{3} &&
  -p_{2} q_{0}+p_{3} q_{1}-p_{0} q_{2}+p_{1} q_{3} \\
\end{array} \right. \hspace{.5in}\\[0.05in]
\hspace*{1.5in}
\left. \begin{array}{c p{0.05in} c}
  -p_{2} q_{0}-p_{3} q_{1}+p_{0} q_{2}+p_{1} q_{3} &&
   -p_{3} q_{0}+p_{2} q_{1}-p_{1} q_{2}+p_{0} q_{3} \\
 -p_{3} q_{0}+p_{2} q_{1}+p_{1} q_{2}-p_{0} q_{3} && 
  p_{2} q_{0}+p_{3} q_{1}+p_{0} q_{2}+p_{1} q_{3} \\
 p_{0} q_{0}-p_{1} q_{1}+p_{2} q_{2}-p_{3} q_{3} &&
 -p_{1} q_{0}-p_{0} q_{1}+p_{3} q_{2}+p_{2} q_{3} \\
 p_{1} q_{0}+p_{0} q_{1}+p_{3}q_{2}+p_{2} q_{3} &&
  p_{0} q_{0}-p_{1} q_{1}-p_{2} q_{2}+p_{3} q_{3}\\
\end{array}\right] \  , \\ \end{array} 
\label{doublequat.eq}
\end{equation} 
\comment{  
\begin{equation}  
 R_{4}(p,q) \! = \!\! \left[ \!\!
\begin{array}{c @{\hspace{+.15in}}c @{\hspace{+.15in}}c @{\hspace{+.15in}}c}
 p_{0} q_{0}+p_{1} q_{1}+p_{2} q_{2}+p_{3} q_{3} &
 -p_{1} q_{0}+p_{0} q_{1}+p_{3} q_{2}-p_{2} q_{3} &
 -p_{2} q_{0}-p_{3} q_{1}+p_{0} q_{2}+p_{1} q_{3} &
 -p_{3} q_{0}+p_{2} q_{1}-p_{1} q_{2}+p_{0} q_{3} \\
   p_{1} q_{0}-p_{0} q_{1}+p_{3} q_{2}-p_{2} q_{3} &
  p_{0} q_{0}+p_{1} q_{1}-p_{2} q_{2}-p_{3} q_{3} &
 -p_{3} q_{0}+p_{2} q_{1}+p_{1} q_{2}-p_{0} q_{3} & 
  p_{2} q_{0}+p_{3} q_{1}+p_{0} q_{2}+p_{1} q_{3} \\
   p_{2} q_{0}-p_{3} q_{1}-p_{0} q_{2}+p_{1} q_{3} &
  p_{3} q_{0}+p_{2} q_{1}+p_{1} q_{2}+p_{0} q_{3} &
  p_{0} q_{0}-p_{1} q_{1}+p_{2} q_{2}-p_{3} q_{3} &
 -p_{1} q_{0}-p_{0} q_{1}+p_{3} q_{2}+p_{2} q_{3} \\
   p_{3} q_{0}+p_{2} q_{1}-p_{1} q_{2}-p_{0} q_{3} &
 -p_{2} q_{0}+p_{3} q_{1}-p_{0} q_{2}+p_{1} q_{3} &
  p_{1} q_{0}+p_{0} q_{1}+p_{3}q_{2}+p_{2} q_{3} &
  p_{0} q_{0}-p_{1} q_{1}-p_{2} q_{2}+p_{3} q_{3}\\
\end{array} \!\!\! \right] \!  . 
\label{doublequatx.eq}
\end{equation}   } 
where $\det R_{4}(p,q) = (p \cdot p)^2 (q \cdot q)^2$ and    $\tr R_{4}(p,q)= 4 p_{0} q_{0} $. 
Since this is a quadratic form in $p$ and $q$, the rotation is
unchanged under $(p,q) \to (-p,-q)$, and the quaternions are again a
double covering.  If we set $p=q$, we recover a matrix that leaves the
$w$ component invariant, and is just the rotation \Eqn{qrot.eq} for
the $\Vec{x}_{3}=(x,y,z)$ component.  If we set $p = q_{\ID}$, we find
the interesting result that $R_{4}(q, q_{\ID}) = Q(q)$ from \Eqn{qprod.eq},
and  $R_{4}(q_{\ID},\bar{p}) = \td{Q}(p)$ from \Eqn{qRprod.eq}.

Rotations in 4D can be composed in
quaternion form parallel to the 3D case, with 
\[R_{4}(p,q)\cdot R_{4}(p',q')
= R_{4}(p\star p',\, q\star q')\ . \]
  We observe that the 4D columns of
\Eqn{doublequat.eq} can be used to define 4D Euclidean orientation
frames in the same fashion as the 3D columns of \Eqn{qrot.eq}, and we
will exploit this to treat the 4D orientation-frame alignment problem below.

\qquad

\noindent{\bf Remark:\,} \emph{Eigensystem and properties of $R_{4}$:}
We can also compute the eigenvalues of our 4D rotation matrix $R_{4}(p,q)$ from
\Eqn{doublequat.eq}.  The 3D  form of $R_{3}(q)$ in terms of explicit fixed axes that we used
 does not have an exact analog in 4D because 4D rotations  leave
a \emph{plane} invariant, not an axis.  Nevertheless, we can still find very compact form for
the 4D eigenvalues.  Our exact 4D analog of \Eqn{rot3DEigvals.eq}, after applying the
transformations ${q_1}^2 + {q_2}^2 + {q_3}^2   \rightarrow 1 - {q_{0}}^2$  for $q$ and $p$ 
to simplify the expression, is just
 \begin{equation}
 \left\{ \begin{array}{c}
 p_0 q_0  -  \sigma(p_{0},q_{0}) \left(+ \sqrt{ \left({1 - p_ 0}^2 \right) \left({1 - q_0}^2 \right) } +
    \I \, \sqrt{\left({1 -p_0}^2 \right) {q_ 0}^2} +
     \I \, \sqrt{{p_ 0}^2  \left({1 - q_0}^2\right)} \; \right)  \\[0.15in]
p_0 q_0 -  \sigma(p_{0},q_{0}) \left( +\sqrt{ \left({1 - p_ 0}^2 \right) \left({1 - q_0}^2 \right) } - 
   \I \,  \sqrt{\left({1 - p_0}^2\right) {q_0}^2} - 
    \I \, \sqrt{{p_0}^2 \left({ 1 - q_0}^2 \right)}  \; \right)\\[0.15in]
p_0 q_0 -  \sigma(p_{0},q_{0}) \left(  -\sqrt{ \left({1 - p_ 0}^2 \right) \left({1 - q_0}^2 \right) }+ 
    \I\,  \sqrt{\left({ 1 - p_0}^2\right) {q_0}^2} - 
    \I\,  \sqrt{{p_ 0}^2 \left({1 -q_0}^2 \right)}  \; \right)\\[0.15in]
  p_0 q_0 -  \sigma(p_{0},q_{0}) \left( - \sqrt{ \left({1 - p_ 0}^2 \right) \left({1 - q_0}^2 \right) }  - 
   \I \,  \sqrt{\left({1 - p_0}^2 \right) {q_0}^2} + 
    \I \, \sqrt{{p_ 0}^2 \left({1 -q_0}^2 \right)} \; \right) \\
     \end{array} \right\} \ ,
 \label{rot4DEigvals.eq }
 \end{equation}
 where  the overall sign in the right-hand terms depends on the  sign of $ p_{0} q_{0}=  (1/4)\tr R_{4}(p,q)$,
 \[ \sigma(p_{0},q_{0}) = \sign( p_{0} q_{0}) \ . \]
 This feature is subtle, and arises in the process of removing a spurious apparent  asymmetry
 between $p_{0}$ and $q_{0}$ in the eigenvalue expressions associated with the appearance
 of $\rule{0in}{2.4ex} \sqrt{{q_{0}}^2}$ and $\rule{0in}{2.4ex}\sqrt{{p_{0}}^2}$; 
 incorrect signs arise in removing the square roots
 without $ \sigma(p_{0},q_{0})$,  which is required to make 
 the determinant  equal to the products of the eigenvalues.
 The eigenvectors can be computed in the usual way, but we know of no
 informative simple algebraic form.   Interestingly, the eigensystem
 of the \emph{profile matrix} of $R_{4}(p,q)$, discussed later in Section \ref{baritzhack.sec},
 is much simpler.

\comment{
signPQXYZ [p0_, q0_] := - Sign[p0 q0] (* MINUS *)

{p0 q0 + signPQXYZ[p0, q0] Sqrt[(-1 + p0^2) q0^2] + 
  signPQXYZ[p0, q0] Sqrt[p0^2 (-1 + q0^2)] + 
  signPQXYZ[p0, q0] Sqrt[(-1 + p0^2) (-1 + q0^2)], 
 p0 q0 - signPQXYZ[p0, q0] Sqrt[(-1 + p0^2) q0^2] - 
  signPQXYZ[p0, q0] Sqrt[p0^2 (-1 + q0^2)] + 
  signPQXYZ[p0, q0] Sqrt[(-1 + p0^2) (-1 + q0^2)],
   p0 q0 + signPQXYZ[p0, q0] Sqrt[(-1 + p0^2) q0^2] - 
  signPQXYZ[p0, q0] Sqrt[p0^2 (-1 + q0^2)] - 
  signPQXYZ[p0, q0] Sqrt[(-1 + p0^2) (-1 + q0^2)],
 p0 q0 - signPQXYZ[p0, q0] Sqrt[(-1 + p0^2) q0^2] + 
  signPQXYZ[p0, q0] Sqrt[p0^2 (-1 + q0^2)] - 
  signPQXYZ[p0, q0] Sqrt[(-1 + p0^2) (-1 + q0^2)]}
  }
\section{Double-Quaternion Approach to the 4D RMSD Problem}
\label{4DRMSD.app}


Here we present the nontrivial steps needed to
understand and solve the 4D spatial and orientation-frame RMSD
optimization problems in the quaternion framework.   We extend our
solutions for  $4\times 4$ symmetric, traceless profile matrices
$M_{3}$ arising from 3D Euclidean data 
 to the case of unconstrained $4\times 4$ profile matrices
$M_{4}$, which arise naturally for 4D Euclidean data.  

While we might
expect the quaternion eigensystem of the 4D profile matrix to allow us to 
solve the 4D RMSD problem  in
exactly the same fashion as in 3D, this is, interestingly, false.  We
will need several stages of analysis to actually find the correct way
to exploit quaternions in the 4D RMSD optimization context. 
In this Section, we study the problem by itself, in a way that can be easily
solved using a quaternion approach with the  \emph{numerical} 
methods traditional in the 3D problem.
We devote the Appendix  to a detailed treatment of
 the alternative \emph{algebraic} solutions to the eigensystems of
the $4\times 4$ symmetric real matrices that are relevant to our
quaternion-based spatial and orientation-frame alignment problems in 3D and 4D.

\qquad

\comment{\noindent\emph{Note on enantiomers.}  Mirrored structures often occur in the
alignment problems we are considering (see
\cite{Kabsch1976,KabschErratum1978}).  In 3D,
mirrored geometries cannot be reached by standard rotations, while in 2D
and 4D the completely reflected geometry is \emph{continuously} reachable
by a rotation.  This application of 4D alignment has been considered
by some authors, e.g.,~\cite{ImmelChemEurJ2018,ImmelChirality2019}. 
 We emphasize that in both 3D and 4D obtaining reliable
conclusions when mirror symmetry is present may require special
procedures.  There can also be particular symmetries, or very close approximations to symmetries, that can make some of our natural assumptions about the good behavior of the profile matrix invalid, and many of these issues, including ways to treat degenerate cases, have been carefully studied,
see, e.g.,   \cite{CoutsiasSeokDill2004,CoutsiasWester2019}.   
The latter authors also point out that if a particular data set $M(E)$ produces a negative smallest eigenvalue
$\epsilon_{4}$ such that $| \epsilon_{4} | > \epsilon_{\opt}$, this is a sign of a reflected match,
and the \emph{negative} rotation matrix $R_{\opt} = - R(q(\epsilon_{4}))$ may
 actually produce the best alignment.  These considerations
 can be essential in some applications, and readers are referred to the original literature
 for details.
   }. 
 
 \qquad
 
 \subsection{Review of the Notation for the RMSD Problem}
 \label{3DRMSDalignSupp.sec}
 
Our starting point for all alignment analysis is the minimization of the difference measure
quantifying the rotational alignment of a $D$-dimensional set of point test data
$\{x_{k}\}$ relative to a reference data set $\{y_{k}\}$,
\begin{equation} 
  {\mathbf S}_{D}  = 
   \sum_{k=1}^{N} \| R_{D} \cdot x_{k} -  y_{k}\|^{2} \ ,
\label{RMSD-basic.eq}
\end{equation}
which we replace by a maximization of its cross-term
\begin{equation} 
 \Delta_{D} = \sum_{k=1}^{N} \left(R_{D} \cdot x_{k}\right) \cdot y_{k} =
 \sum_{a=1,b=1}^{D} {R_{D}}^{ba} E_{ab} \ = \  \tr R_{D} \cdot E ,
\label{RMdef.eq}
\end{equation}
   where $E$ is the cross-covariance matrix
 \begin{equation} 
E_{ab} = \sum_{k=1}^{N} [x_{k}]_{\textstyle_a} \:[y_{k}]_{\textstyle_b}  = \left[ \Vec{X}
  \cdot \Vec{Y}^{\t} \right]_{ab} \ ,
\label{Edef.eq}
\end{equation}
and  $[x_{k}]$ denotes the $k$th column of $\Vec{X}$.

For 3D data, we convert this to a quaternion matrix problem by
applying \Eqn{qrot.eq} to get
\begin{equation}
 \Delta(q) = \tr R(q) \cdot E =  (q_0,q_1,q_2,q_3) \cdot M_{3}(E) \cdot
 (q_0,q_1,q_2,q_3)^{\t} \equiv q \cdot M_{3}(E) \cdot q \ , 
\label{qM3q.eq}
\end{equation}
Choosing the traditional 3D indexing $\{x,y,z\}$ for $(a,b)$,    the traceless, symmetric profile matrix
takes the form
\begin{equation} 
     M_{3}(E) \!  = \!
\left[ \begin{array}{cccc}
 \!\!   E_{xx} + E_{yy} + E_{zz} & E_{yz} - E_{zy} & E_{zx} - E_{xz} &
                     E_{xy} - E_{yx} \!  \\
 \!   E_{yz} - E_{zy} &  E_{xx} - E_{yy} - E_{zz} & E_{xy} + E_{yx} & 
                     E_{zx} +E_{xz}  \! \\
 \!   E_{zx} - E_{xz} &  E_{xy} + E_{yx} & - E_{xx} + E_{yy} - E_{zz} & 
          E_{yz} + E_{zy}  \!  \\
  \!   E_{xy} - E_{yx} &  E_{zx} +E_{xz}  &  E_{yz} + E_{zy} & 
              - E_{xx} - E_{yy} + E_{zz}  \!\!
                     \end{array} \right] \ .
       \label{basicHorn.eq}              
\end{equation}
The maximal measure is given by the eigensystem of the maximal
eigenvalue $\epsilon_{\opt}$ of $M_{3}$ and the corresponding 
quaternion eigenvector $q_{\opt}$, with the result
\begin{equation}  \left. \begin{array}{rcl}
 \Delta_{\opt} & = & \tr [R_{3}(q_{\opt}) \cdot E]\\
 & = & q_{\opt} \cdot M_{3} \cdot q_{\opt}\\
  & = & q_{\opt} \cdot \left(\epsilon_{\opt}\, q_{\opt} \right)\\
  & = &  \epsilon_{\opt}   \\ \end{array} \right\} \ .
\label{DeltaOptIsEig.eq}
\end{equation}

 \subsection{Starting Point for the 4D RMSD Problem.}

The 4D double quaternion matrix \Eqn{doublequat.eq} provides the most
general quaternion context that we know of for expressing an RMSD
problem.  We start with the RMSD minimization problem for 4D Euclidean
point data expressed as the maximization problem for the by-now-familiar
cross-term expression
\begin{equation} 
 \Delta_{4} = \sum_{k=1}^{N} \left(R_{4} \cdot x_{k}\right) \cdot y_{k} =
 \sum_{a=0,b=0}^{3} {R_{4}}^{ba}{ E_{4:}}_{ab} \ = \  \tr R_{4} \cdot E_{4} ,
\label{RMSD4Ddef.eq}
\end{equation}
   where  
\begin{equation} 
{E_{4:}}_{ab} = \sum_{k=1}^{N} [x_{k}]_{\textstyle_a} \:[y_{k}]_{\textstyle_b}  = \left[ \Vec{X}
  \cdot \Vec{Y}^{\t} \right]_{ab} \ 
\label{EdefDup.eq}
\end{equation}
is the cross-covariance matrix whose $(a,b)$ indices we will usually write as $(w,x,y,z)$ 
in the manner of \Eqn{basicHorn.eq}.

Using \Eqn{doublequat.eq} in \Eqn{RMSD4Ddef.eq} to perform the
4D version of the rearrangement of the similarity function, we can rewrite our
measure as
\begin{equation}
  \Delta_{4} =  \tr R_{4}(p,q) \cdot E_{4} \, = \, (p_0,p_1,p_2,p_3) \cdot
  M_{4}(E_{4}) \cdot  (q_0,q_1,q_2,q_3)^{\t} \,\equiv \, p \cdot
  M_{4}(E_{4}) \cdot q \ ,  
\label{qM4q2.eq}
\end{equation}
where the profile matrix for the 4D data  now becomes
{\small
\begin{eqnarray} 
 \lefteqn{M_{4} (E_{4}) =} \nonumber \\   
  &&\hspace{-.25in}\left[\!\! \begin{array}{c @{\hspace{0.175in}} c @{\hspace{0.175in}}  c
       @{\hspace{0.175in}} c} 
E_{ww} + E_{xx} + E_{yy} + E_{zz}&   E_{yz} - E_{zy} - E_{wx} + E_{xw} &
   E_{zx} - E_{xz} - E_{wy} + E_{yw} &  E_{xy} - E_{yx}  -E_{wz} + E_{zw} \\
   E_{yz} - E_{zy} + E_{wx} - E_{xw}&   E_{ww} + E_{xx} - E_{yy} - E_{zz}&
  E_{xy} + E_{yx} - E_{wz} - E_{zw} &   E_{zx} + E_{xz} + E_{wy} + E_{yw}\\
  E_{zx} - E_{xz} + E_{wy} - E_{yw}  &   E_{xy} + E_{yx} + E_{wz} + E_{zw} &
   E_{ww} - E_{xx} + E_{yy} - E_{zz} &   E_{yz} + E_{zy} - E_{wx} - E_{xw}\\
  E_{xy} - E_{yx} + E_{wz} - E_{zw} &   E_{zx} + E_{xz} - E_{wy} - E_{yw}&
  E_{yz} + E_{zy} +  E_{wx} + E_{xw} &  E_{ww} - E_{xx} - E_{yy} + E_{zz}\\
                     \end{array}\!\! \right]  \hspace{.2in} 
       \label{basic4DHorn.eq}              
\end{eqnarray} }
and we note that, in contrast to  $M_{3} (E_{3})$, $ M_{4} (E_{4})$ is
\emph{neither} traceless nor symmetric.

\comment{  
 \begin{eqnarray} 
\lefteqn{ M_{4} (E_{4}) =
 \left[ \begin{array}{c p{0.05in} c}
E_{ww} + E_{xx} + E_{yy} + E_{zz}&&  + E_{yz} - E_{zy} - E_{wx} + E_{xw}\\
  + E_{yz} - E_{zy} + E_{wx} - E_{xw}&&   E_{ww} + E_{xx} - E_{yy} - E_{zz}\\
 + E_{zx} - E_{xz} + E_{wy} - E_{yw}  &&  + E_{xy} + E_{yx} + E_{wz} + E_{zw}\\
 + E_{xy} - E_{yx} + E_{wz} - E_{zw} &&  + E_{zx} + E_{xz} - E_{wy} - E_{yw}\\
                     \end{array} \right. \hspace{1in}} \nonumber \\
 &&\hspace*{1in}\left. \begin{array}{c p{0.05in} c}
  + E_{zx} - E_{xz}  - E_{wy} + E_{yw} && + E_{xy} - E_{yx}  -E_{wz}  + E_{zw} \\
 + E_{xy} + E_{yx} - E_{wz} - E_{zw} &&  + E_{zx} + E_{xz} + E_{wy} + E_{yw}\\
 E_{ww} - E_{xx} + E_{yy} - E_{zz} &&  + E_{yz} + E_{zy} - E_{wx} - E_{xw}\\
 + E_{yz} + E_{zy} +  E_{wx} + E_{xw} &&  E_{ww} - E_{xx} - E_{yy} + E_{zz}\\
                     \end{array} \right] \ ,
       \label{basic4DHornAlt.eq}              
\end{eqnarray}    
      } 

\subsection{A Tentative 4D Eigensystem}
\label{4DRMSDEigsys.app}

Our task is now to find an algorithm that allows us to successfully 
compute the quaternion pair $(p_{\opt},q_{\opt})$, or, equivalently, the global
rotation $R_{4}(p_{\opt},q_{\opt})$, that maximizes the measure 
\begin{equation}
  \Delta_{4} =  \tr R_{4}(p,q) \cdot E_{4} \, = \,p \cdot M_{4}(E_{4})\cdot q \ , 
\label{qM4q2A.eq}
\end{equation}
with $M_{4}(E_{4})$ a  general real matrix with a generic trace and no
symmetry conditions.   Note that now we can have \emph{both} 
left and right eigenvectors $p$ and $q$ for a
single eigenvalue of the profile matrix $M_{4}$: 
$q$ would correspond  to the eigenvectors of
$M_{4}$, and $p$ would correspond to the eigenvectors of the transpose
${M_{4}}^{\t}$.  \emph{Warning:}  The eigensystem of $M_{4}$  typically has some complex 
eigenvalues and is furthermore
\emph{insufficient} by itself to solve the 4D RMSD optimization
problem, so additional refinements will be necessary.  We now explore
a path to an optimal solution amenable to quaternion-based numerical evaluation, with
applicable algebraic approaches elaborated in the Appendix.

For some types of calculations, we may find it useful to decompose
$M_{4}$ in a way that isolates particular features using the form
\begin{equation} 
 M_{4}(w,x,y,z,\ldots)  = \left[
\begin{array}{cccc}
 w+x+y+z & a-a_{w} & b-b_{w} & c-c_{w} \\
 a+a_{w} & w+x-y-z & C - C_{w} & B +B_{w} \\
 b+b_{w} & C +C_{w} & w-x+y-z & A - A_{w} \\
 c+c_{w} & B  -B_{w} & A +A_{w} & w-x-y+z \\
\end{array}
\right] \ ,
\label{matA4.eq}
\end{equation} 
where $(w,x,y,z) = (E_{ww},\,E_{xx},\,E_{yy},\,E_{zz})$, $a=E_{yz}-E_{zy}$, cyclic,
$A = E_{yz} +E_{zy}$, cyclic, $a_{w}=E_{wx}-E_{xw}$, cyclic, 
$A _{w}= E_{wx} +E_{xw}$, cyclic, and  $\tr(M_{4}) = 4w$.  This effectively exposes the
structural symmetries of $M_4$.

We next review the properties of the eigenvalue equation $\det [M_{4} - e I_{4} ]=0$, 
where $e$ is the variable we
solve for to obtain the four eigenvalues  $\epsilon_{k}$,  and  $I_{4}$ denotes
the 4D identity matrix;  transposing $M_{4}$ does not change the eigenvalues
but does interchange the distinct left and right eigenvectors.   
While $M_{4}$ itself has new properties,  the corresponding
expressions in terms of $e$ and  $\epsilon_{k}$,  along with the outcome of
eliminating $e$ \cite{AbramowitzStegun1970}, are by now familiar:
\begin{eqnarray} 
\det [M_{4} - e I_{4} ]\, = \, e^4 + e^3 p_1 +  e^2 p_2 + e p_3 + p_4  & = & 0   \label{eigePn.eq}\\
(e - \epsilon_{1})(e - \epsilon_{2})(e - \epsilon_{3})(e - \epsilon_{4})& = & 0 
\label{eigPnEpsk.eq}
\end{eqnarray}
\begin{equation}
\left. \begin{array}{rcl}
p_1 &  = &\left(-\epsilon_{1} - \epsilon_{2} -  \epsilon_{3}
    -  \epsilon_{4}\right)  \\
p_2 & = &\left(\epsilon_{1} \epsilon_{2} + \epsilon_{1}  \epsilon_{3} +
    \epsilon_{2}  \epsilon_{3} + \epsilon_{1}  \epsilon_{4} +
    \epsilon_{2}  \epsilon_{4} +  \epsilon_{3}  \epsilon_{4}\right)
      \\  
p_3 & = & \left(-\epsilon_{1} \epsilon_{2}  \epsilon_{3} -
    \epsilon_{1} \epsilon_{2}  \epsilon_{4} - \epsilon_{1} 
    \epsilon_{3}  \epsilon_{4} - \epsilon_{2}  \epsilon_{3} 
    \epsilon_{4}\right)  \\ 
p_4  & = &   
\epsilon_{1} \epsilon_{2}  \epsilon_{3} \epsilon_{4}  
\end{array}\right\} \ . \label{PofEps.eq}
\end{equation}
We make no assumptions about $M_{4}$, so  its structure
includes a trace term $4 w =-p_1$
as well as the possible antisymmetric components shown in \Eqn{matA4.eq},
yielding the following expressions for the $p_{k}(E_{4})$ following
 from the expansion of  $\det [M_{4} - e I_{4}]$:
\begin{eqnarray} 
p_1(E_{4}) & = & - \tr \left[ M_{4} \right] = -4 w \label{eP1.eq} \\
p_2(E_{4}) & = & \frac{1}{2} \left(\tr\left[ M_{4} \right]\right)^{2} -
 \frac{1}{2} \tr\left[ M_{4} \cdot  M_{4} \right]
 \nonumber \\   
 &=& 6 w^2 - 2 (x^2 + y^2 + z^2) - A^2 - a^2 - B^2 - b^2 - C^2 -c^2
 \nonumber \\ &&
  \mbox{} + {A_w}^2 + {a_w}^2 + {B_w}^2 + {b_w}^2 + {C_w}^2 +{c_w}^2 
 \label{eP2.eq} \\
 p_3(E_{4}) & = &  
    \mbox{} - \frac{1}{6}  \left(\tr\left[ M_{4} \right]\right)^{3} 
  + \frac{1}{2} \tr\left[ M_{4}  \cdot M_{4}\right]
   \tr \left[ M_{4}\right] 
  -  \frac{1}{3} \tr \left[ M_{4} \cdot  M_{4} \cdot M_{4}\right]
\nonumber \\
   &=&  \mbox{} -8 x y z + 4 w (x^2 + y^2 + z^2)   \nonumber \\ 
   && \mbox{}-2 A B C - 2 A b c  - 2 a B c - 2 a b C   \nonumber \\
   && \mbox{}  + 2 A^2 x - 2 a^2 x + 2 B^2 y -2 b^2 y  +2 C^2 z  - 2
   c^2 z \nonumber \\ 
&&  \mbox{}-2 A B_w C_w + 2 A b_w c_w -2 a B_w c_w + 2 a b_w C_w 
           \nonumber \\ &&
  \ \ \ \ \     \mbox{} -2 A_w B C_w + 2 a_w B c_w -2 a_w b C_w + 2 A_w b c_w 
         \nonumber \\ &&
  \ \ \ \ \ \mbox{} - 2 A_w B_w C  + 2 a_w b_w C - 2 A_w b_w c  + 2
                    a_w B_w c  \nonumber  \\  
&&  \mbox{} + 2 a^2 w + 2 A^2 w -2 A_w^2 w-2 A_w^2 x -2 a_w^2 w + 2 a_w^2 x 
        \nonumber \\ &&
  \ \ \ \ \   \mbox{} + 2 b^2 w + 2 B^2 w -2 B_w^2 w-2 B_w^2 y -2
  b_w^2 w + 2 b_w^2 y 
       \nonumber \\ &&
  \ \ \ \ \  \mbox{}+ 2 c^2 w + 2 C^2 w -2 C_w^2 w-2 C_w^2 z -2 c_w^2
  w + 2 c_w^2 z  
            \label{eP3.eq}       \\
p_4(E_{4}) & = & \det \left[ M_{4} \right] \label{eP4.eq}\ .
\end{eqnarray}

\qquad

\subsection{Issues with the Naive 4D Approach}

We previously found that we could maximize $\Delta_{3} =\tr (R_3 \cdot E_{3} )$ 
over the 3D rotation matrices $R_3$ by mapping
$E_{3}$ to the profile matrix $M_{3}$, with $\Delta_{3} = q \cdot
M_{3} \cdot q$, solving for the maximal eigenvalue $\epsilon_{\opt}$ 
of the symmetric matrix $M_{3}$, and choosing $R_{\opt} = R_{3}(q_{\opt})$ with $q_{\opt}$ the
normalized quaternion eigenvector corresponding to $\Delta_{3}(\opt) =
\epsilon_{\opt}$.  The obvious 4D extension of the 3D quaternion RMSD
problem would be to examine $\Delta_{4} =\tr \left(R_4 \cdot
  E_{4}\right) = q_{\lambda} \cdot M_{4} \cdot q_{\rho}$.  This is defined over the 4D
rotation matrices $R_4$, where $M_{4}$ in \Eqn{basic4DHorn.eq} turns out no longer to be
symmetric, so we must split the eigenvector space into a separate
left-quaternion $q_{\lambda}$ and right-quaternion $q_{\rho}$.  We might guess that  
as in the 3D case, $M_{4}$ would have a maximal eigenvalue $\epsilon_{\opt}$
(already a problem -- it may be complex), 
and we could use the ``optimal'' left and right eigenvectors $q_{\lambda:\opt}$ and $q_{\rho:\opt}$
that could be obtained as the corresponding eigenvectors of $M_{4}$ and
${M_{4}}^{\t}$.   Then  the solution to the 4D optimization problem would look
like this:
\begin{equation}
 \Delta_{4}(\opt)\stackrel{\mbox{?}}{ = } q_{\lambda:\opt}\cdot M_{4} \cdot q_{\rho:\opt} =
   (q_{\lambda:\opt}\cdot q_{\rho:\opt})\, \epsilon_{\opt}  \ .
\label{wrongM4.eq}
\end{equation}
Unfortunately, this is wrong.   First, even when this result is
real,  \Eqn{wrongM4.eq} is typically  smaller
 than the actual maximum of $\tr (R_4(q_{\lambda}, q_{\rho}) \cdot
E_{4})$ over the space of 4D rotation matrices (or their equivalent
representations in terms of a search through $q_{\lambda}$ and $q_{\rho}$).  Even a
simple \emph{slerp} through $q_{\ID}$ and just beyond the apparent
optimal eigenvectors $q_{\lambda:\opt}$ and $q_{\rho:\opt}$ from an eigenvalue
of $M_{4}$ can yield
\emph{larger} values of $\Delta_{4}$!  And, to add insult to injury, starting with
those eigenvectors 
$q_{\lambda:\opt}$ and $q_{\rho:\opt}$, one does not in general  even find a
\emph{basis} for some normalized linear combination that yields the true 
optimal result.  What is going wrong, and what is the path to our
hoped-for quaternionic solution to the 4D RMSD problem, which seems so
close to the 3D RMSD problem, but then fails so spectacularly to
correspond to the obvious hypothesis?

%

\subsection{Insights from the Singular Value Decomposition}

We know that the 3D version of \Eqn{wrongM4.eq} is certainly correct with
$\epsilon_{\opt}$ the maximal eigenvalue of $M_{3}(E_{3})$, and we know also
that there is \emph{some}  rotation matrix $R_4(q_{\lambda}, q_{\rho})$ that 
maximizes  $\tr (R_4(q_{\lambda}, q_{\rho}) \cdot E_{4})$ , and therefore
the 4D expression \Eqn{wrongM4.eq} must describe $\Delta_{4}(\opt)$ for
\emph{some} non-trivial pair of quaternions $(q_{\lambda}, q_{\rho})$.
The crucial issue is that the  3D RMSD problem and the 4D RMSD
problem differ, with 3D being a special case due to the symmetry of
the $4\times 4$ profile matrix.  We know also that the SVD form of the
optimal rotation matrix is valid in \emph{any} dimension, so we conjecture that
the key is to look at the commonality of the SVD solutions in 3D and 4D,
and work backwards to see how those non-quaternion-driven equations
might relate to what we know is \emph{in principle} a quaternion approach
to the 4D problem that looks like \Eqn{wrongM4.eq}.

Therefore, we first look at the general singular-value decomposition for
the spatial alignment problem \cite{Schonemann-Procrustes1966,Golub-vanLoan-MatrixComp}
and then analyze the 3D and 4D problems   to understand how we
can recover a quaternion-based construction of the 4D spatial RMSD solution.
For 3D and 4D, the basic SVD construction of  the optimal rotation for a cross-covariance
matrix $E$ takes the form
\begin{eqnarray} 
 \{ U,\, S,\, V\} &=& \mbox{SingularValueDecomposition}\left(
   E  \right) \label{SVDdef.eq}\\
 & \mbox{where }& \nonumber \\
 E (U,S,V) & = & U \cdot S\cdot  V^{\t}   \label{SVDE.eq}\\
 R_{ \opt}(U,D,V) & = &  V \cdot D \cdot U^{\t}  \label{SVDR.eq} \\
  D_{3}  & = &  \mbox{Diagonal}\left( 1,1, \sign \det (V\cdot U^{\t}) \right) \\
  D_{4} & = &  \mbox{Diagonal}\left( 1,1,1, \sign \det (V\cdot U^{\t}) \right)   \  .
\label{SVDD.eq}
\end{eqnarray}
Here $U$ and $V$ are orthogonal matrices that are usually
ordinary rotations, while $D$ is usually the identity matrix but can be nontrivial
in more situations than one might think.  A critical  component for this analysis
is the diagonal matrix   $S$,  whose elements  
are the all-positive  square roots of  the eigenvalues of the symmetric matrix
$\rule{0in}{0.9em}{E_{4}}^{\t}\cdot E_{4}$  (the trace of this matrix is the squared Fr\"{o}benius norm
of $E$).  The first key fact is that in any dimension the RMSD cross-term
obeys the following sequence of transformations following from the SVD
relations of Eqs.~(\ref{SVDE.eq}) --(\ref{SVDD.eq}):
\begin{equation}
\left. \begin{array}{rcl }
    \Delta(\opt) \, = \,  \tr (R_{ \opt} \cdot E )& = &   \tr( R_{ \opt}  \cdot [U \cdot S\cdot  V^{\t} ] ) \\[0.075in]
      &  =  &   \tr \left(   [ V \cdot D \cdot U^{\t}]  \cdot [U \cdot S\cdot  V^{\t} ]  \right)\\[0.1in]
      &  =  &   \tr (D \cdot S)  \\ 
\end{array} \right\} \ .
\label{DeltaAsTrS.eq}
\end{equation}
Note the appearance of $D$ in the SVD formula for the optimal measure; we found 
in numerical experiments that including this term is absolutely essential to guaranteeing
agreement with brute force verification of the optimization results, particularly in 4D.

\qquad

\noindent{\bf 3D Context.}  Thus an alternative to considering the 3D optimization
of $\tr( R \cdot E)$  in the context of $E$  alone   is to look at the $3\times 3$ matrices
\begin{equation}
\left. \begin{array}{rcl}
F & = & E^{\t} \cdot E \\[0.04in]
F' & = & E \cdot E^{\t}
\end{array} \ \right\} 
\label{FFpofE.eq}
\end{equation}
and to note that, although $E$ itself will not in general be symmetric, $F$ and $F'$
are intrinsically symmetric.  Thus they have the same eigenvalues, and like all nonsingular
matrices of this form, and unlike $E$ itself, 
will have real positive eigenvalues \cite{Golub-vanLoan-MatrixComp} that we can write as
  $ \left( \gamma_{1},  \gamma_{2}, \gamma_{3}  \right)$.
  From \Eqn{SVDE.eq}, we can show that $\tr F = \tr F' = \tr( S\cdot S)$,
  and since the trace is the sum of the eigenvalues, the eigensystem of
  $F$ or $F'$ determines $S$.
  The diagonal elements that
 enter naturally into the SVD are therefore  just the square roots
\begin{equation}S(E) = \mbox{Diagonal}\left(\sqrt{\gamma_{1}},\sqrt{\gamma_{2}},
   \sqrt{\gamma_{3}}   \right) \ .
   \label{theSeigs.eq}\end{equation}
      
 So far,  this has no obvious connection to the quaternion system.  For our
 next step, let us now examine how the 3D SVD system relates to
 the profile matrix $M_{3}(E_{3})$ derived from the quaternion decomposition
 to give the form in  \Eqn{basicHorn.eq}.  We define the analogs of \Eqn{FFpofE.eq} for
 a profile matrix as
 \begin{equation}
\left. \begin{array}{rcl}
G & = & M^{\t} \cdot M \\[0.04in]
G' & = & M \cdot M^{\t}
\end{array} \ \right\} \ ,
\label{AApofM3.eq}
\end{equation}
where we recall that in 3D, $\epsilon_{\opt}$ is just the maximal eigenvalue of
$M_{3}(E)$.  Thus if we arrange the eigenvalues  of $M_{3}(E)$ in 
descending order as   $(\epsilon_{1}, \epsilon_{2},  \epsilon_{3}, \epsilon_{4} )$,
we obviously have
\begin{equation}
\mbox{Eigenvalues}(G) \, = \, \mbox{Eigenvalues}(G') \, = \,
 ({\alpha_{1}}, {\alpha_{2}},  {\alpha_{3}}, {\alpha_{4}} )  \, = \
       ({\epsilon_{1}}^2, {\epsilon_{2}}^2,  {\epsilon_{3}}^2, {\epsilon_{4}}^2 ) \ .
\label{AApEigs,eq }
\end{equation}
Therefore, since we already know that $\epsilon_{1}(M)=\Delta(\opt)$,
we have precisely the sought-for connection,
\begin{equation}
\sqrt{\mbox{Max Eigenvalue}(G)}   \, = \,  \sqrt{\alpha_{1}}  \,=\, \tr (D\cdot  S)
   \, = \,   \Delta(\opt)   \,=\, \epsilon_{1}(M) \ .
\label{AEigIsTrS.eq }
\end{equation}
That is, given $E$, compute $M(E)$ from the quaternion decomposition,
and, instead of examining the eigensystem of $M(E)$ itself, take the
square root of the maximal eigenvalue of the manifestly symmetric,
positive-definite  
real matrix $G = M^{\t}\cdot M$.  This is the quaternion-based translation
of the 3D application of the SVD method to obtaining the optimal rotation:
numerical methods in particular do not  care whether you are
computing the maximal eigenvalue of a symmetric quaternion-motivated
matrix $M_{3}$ or of the associated symmetric matrix ${M_{3}}^{\t} \cdot M _{3}$.

\begin{quote}{\bf Note:} In 3D, we can compute \emph{all four} of the eigenvalues of
$G$ from the  \emph{three} elements of $S$~\cite{CoutsiasSeokDill2004}: defining
\begin{equation}
\mbox{Diagonal}(D\cdot S) = \left( \lambda_{1},  \lambda_{2}, \lambda_{3} \right) \ ,
\label{SParts3D.eq }
\end{equation}
then we can write
\begin{equation}
 \left[\begin{array}{c}
{\alpha_{1}}\\ {\alpha_{2}} \\ {\alpha_{3}} \\ {\alpha_{4}} \\
\end{array} \right]
 = 
 \left[ \begin{array}{c}
\left( + \lambda_{1} + \lambda_{2} + \lambda_{3}   \right)^2 \\
\left( - \lambda_{1} - \lambda_{2} + \lambda_{3}  \right)^2 \\
\left( - \lambda_{1} + \lambda_{2} - \lambda_{3}  \right)^2 \\
\left( + \lambda_{1} - \lambda_{2} - \lambda_{3}  \right)^2 \\
\end{array} \right] \ ,
\label{AeigsvsSeigs3D.eq}
\end{equation}
where obviously $\sqrt{\alpha_{1}} = \tr (D\cdot S)$ is maximal.
\end{quote}

The final step is to connect $R_{3}(\opt)$ to a quaternion via $R_{3}(q_{\opt})$
without requiring prior knowledge of the SVD solution \Eqn{SVDR.eq}.   We know that
the square root of the maximal eigenvalue of $G= M^{\t} \cdot M$, which depends only on the quaternion
decomposition, gives us $\tr (D\cdot S) = \Delta(\opt)$ without using the SVD, and we know that
in 3D the profile matrix $M$ is symmetric, so $G$ and $G'$ share a single maximal eigenvector $v$ corresponding
to $\alpha_{1} = (\tr  (D\cdot S))^{2} = \left(\Delta(\opt)\right)^2$.  Using this eigenvector 
we thus have
\[
v \cdot  G \cdot v \,=\, (M\cdot v)^{\t} \cdot (M \cdot v) \,=\,
 v\cdot \left( (\tr  (D\cdot S) )^{2} \cdot v \right)    \,=\, (\Delta(\opt))^2 \ ,
\]
so in this case $v = q_{\opt}$ is itself the optimal eigenvector determining $R_{3}(q_{\opt})$. 

\qquad

\noindent{\bf 4D Context.}  The 4D case, as we are now aware, cannot be solved
using the non-symmetric profile matrix $M_{4}(E_{4})$ directly.  But now we can see
a more general way to exploit the 4D quaternion decomposition of \Eqn{basic4DHorn.eq}
by constructing the \emph{manifestly symmetric products}
\begin{equation}
\left. \begin{array}{rcl}
G & = & {M_{4}}^{\t} \cdot M_{4} \\[0.04in]
G' & = & M_{4} \cdot {M_{4}}^{\t}
\end{array} \ \right\} \ .
\label{AApofM4.eq}
\end{equation}
Although this superficially extends \Eqn{AApofM3.eq} to 4D, it is quite different because $M_{4}$ 
is not itself symmetric (as $M_{3}$ was), and so, while $G$ and $G'$ have the
same eigenvalues, they have \emph{distinct eigenvectors} $q_{\rho}$ and 
$q_{\lambda}$, respectively.  If we use the maximal eigenvalue $\alpha_{1}$ to solve
for  $q_{\rho}$ and $q_{\lambda}$ as follows, these in fact will produce the optimal
quaternion system.  First we solve these equations using the maximal eigenvalue 
$\alpha_{1}$ of $G$,
 \begin{equation}
 \left. \begin{array}{ccccc}
 G \cdot q_{\rho} &= &\alpha_{1} \, q_{\rho}  & = & (\tr  (D\cdot S))^{2} \, q_{\rho}\\
 G' \cdot q_{\lambda} &= &\alpha_{1}\,  q_{\lambda}   & = & (\tr  (D\cdot S))^{2}  \, q_{\lambda}\\
  \end{array}
  \right\} \ .
  \end{equation} 
  At this point, the \emph{signs} of the eigenvectors have to be checked for  a correction,
  since the eigenvector is still correct whatever its sign or scale.  But we know that
  the value of $ q_{\lambda} \cdot  M_{4}(E_{4})\cdot q_{\rho} $ must be positive, so
  we simply check that sign, and change, say, $ q_{\lambda}  \rightarrow - q_{\lambda} $
  if needed to make the sign positive.  There is still an \emph{overall} sign ambiguity,
  but that is natural and an intrinsic part of  the  rotation $R_{4}( q_{\lambda},q_{\rho})$, so
 now we can use these eigenvectors to generate the optimal measure for the 4D
 translational RMSD problem using \emph{only} the quaternion-based data, giving
 finally the whole spectrum of ways to write $\Delta_{4}(\opt)$:
    \begin{equation}
     \Delta_{4}(\opt)  = \tr (R_{4:\opt}( q_{\lambda},q_{\rho}) \cdot E_{4}) =
      q_{\lambda} \cdot  M_{4}(E_{4})\cdot q_{\rho}  =   \sqrt{\alpha_{1}}\ .  
   \label{deltaqlqropt.eq}
    \end{equation}

\begin{quote}{\bf Note:} In 4D, we can compute all the eigenvalues of
$G$ from the  \emph{four} elements of $S$: defining
\begin{equation}
\mbox{Diagonal}(D\cdot S) = \left( \lambda_{1},  \lambda_{2}, \lambda_{3}, \lambda_{4} \right) \ ,
\label{SParts4D.eq }
\end{equation}
then we can write
\begin{equation}
 \left[ \begin{array}{c}
{\alpha_{1}}\\ {\alpha_{2}} \\ {\alpha_{3}} \\ {\alpha_{4}} \\
\end{array} \right]
 = 
 \left[ \begin{array}{c}
\left( +\lambda_{1} + \lambda_{2} + \lambda_{3} + \lambda_{4} \right)^2 \\
\left( +\lambda_{1} + \lambda_{2} -  \lambda_{3} - \lambda_{4} \right)^2 \\
\left( + \lambda_{1} - \lambda_{2} +  \lambda_{3} - \lambda_{4} \right)^2 \\
\left( + \lambda_{1} - \lambda_{2} -  \lambda_{3} + \lambda_{4} \right)^2 \\
\end{array}  \right]\ ,
\label{AeigsvsSeigs4D.eq}
\end{equation}
where again $\sqrt{\alpha_{1}} = \tr (D\cdot S)$ is maximal.
\end{quote}

\qquad

\noindent{\bf Summary:} 
Now we have the entire algorithm for solving the RMSD spatial alignment problem in
4D by exploiting the quaternion decomposition of \Eqn{qM4q2.eq}  and 
\Eqn{basic4DHorn.eq}, based on  \Eqn{doublequat.eq}, inspired by, but in
no way dependent upon knowing, the SVD solution to the problem:
\begin{itemize}
  \item {\bf Compute the profile matrix.}  Using the quaternion
    decomposition \Eqn{doublequat.eq} of the general 4D rotation matrix
    $R_{4}(p,q)$, extract the 4D profile matrix $M_{4}(E_{4})$ of
    \Eqn{basic4DHorn.eq} from the initial proximity measure 
   \begin{equation}
     \Delta_{4}  = \tr (R_{4}(p,q) \cdot E_{4}) = p \cdot  M_{4}(E_{4})\cdot q \ .  
   \label{delta4QPopt.eq}
    \end{equation}
     So far all we know is the numerical value of $M_{4}$ and the fact the $\Delta_{4}$ can be maximized by exploring the entire space of the quaternion pair $(p,\, q)$.
  \item  {\bf  Construct the symmetric matrices and extract the optimal eigenvalue.} 
    The maximal  eigenvalue  $\alpha_{1}$ of the $4\times 4$ symmetric matrix
     $G =  {M_{4}}^{\t}\cdot M_{4}$  is itself easily obtained by numerical means,
     just as one has done  traditionally for $M_{3}$.  If all we need is the optimal value of the proximity 
     measure for comparison, we are done:
     \begin{equation}
       \Delta_{4}(\opt)  \,= \, \sqrt{\mbox{Max Eigenvalue} \left( G =  {M_{4}}^{\t} \cdot M_{4} \right)}
          \,=\, \sqrt{\alpha_{1}}  \  . 
    \label{delta4QPMax.eq}
   \end{equation}
    The alternative algebraic methods for computing the eigenvalues 
    are discussed in the Appendix.
       \item {\bf  If needed, compute the left and right eigenvectors of $G$:} Our two distinct symmetric matrices,  
  $G   =  {M_{4}}^{\t} \cdot M_{4}$ and $G'=  M_{4}\cdot {M_{4}}^{\t}$
  have their own distinct maximal eigenvectors, both corresponding to the maximal
eigenvalue $\alpha_{1}$ shared by $G$ and $G'$, so we can easily
use this common maximal numerical eigenvalue to solve 
\begin{equation}
\left. \begin{array} {rcl}
\left(G  - \alpha_{1} I_{4}\right)\cdot q_{\opt:\rho} & = & 0\\
\left(G' - \alpha_{1} I_{4}\right)\cdot q_{\opt:\lambda}& =& 0 \\
\end {array} \  \right\} 
\label{finalOptqLqR.eq}
\end{equation}
for  the numerical values of $q_{\opt:\lambda}$ and  $q_{\opt:\rho}$. We correct the
signs so that   $q_{\opt:\lambda}\cdot M_{4}(E_{4}) \cdot q_{\opt:\rho} >0 $,
 and then these in turn yield the required 4D rotation matrix
\[  R_{4:\opt}\left(q_{\opt:\lambda},\,q_{\opt:\rho} \right) \]
from \Eqn{doublequat.eq}.  
\end{itemize}

\quad

  If everything is in order, all of the following ways of expressing $\Delta_{4}(\opt)$
  should now be equivalent,
   \begin{equation}
   \begin{array}{ccccccc}
       \Delta_{4}(\opt) &= &\tr ( R_{4:\opt}\left(q_{\opt:\lambda},\,q_{\opt:\rho} \right) \cdot E_{4}) 
          & =  & q_{\opt:\lambda} \cdot  M_{4}(E_{4})\cdot q_{\opt:\rho} & = & \sqrt{\alpha_{1}} 
          \end{array}\ ,
\label{delta4QPoptSummary.eq}
\end{equation}
independently of the fact that one   knows from the SVD decomposition of $E_{4}$  that
$ \Delta_{4}(\opt) = \tr(D\cdot S)=\sqrt{\alpha_{1}}$.


\section{4D Orientation-Frame Alignment}


In this section, we review and slightly expand the details of the 
 3D orientation-frame in the main text.  Then we extend that treatment
to handle the case of 4D orientation-frame alignment to
complete the picture we started in Section \ref{4DRMSD.app}
on the 4D spatial frame alignment problem.
  A  detailed evaluation of
the accuracy of the 3D chord measure compared to the arc-length 
measure, along with other questions, is given 
separately in Section \ref{Evaluate3DOrient.sec}.

\subsection{Details of the 3D Orientation-Frame alignment Problem}

\qquad

We first review the basic structure of our 3D orientation-frame method
and then proceed to present some additional details.
\qquad

\begin{figure}
 \figurecontent{ \centering
 \includegraphics[width=2.5in]{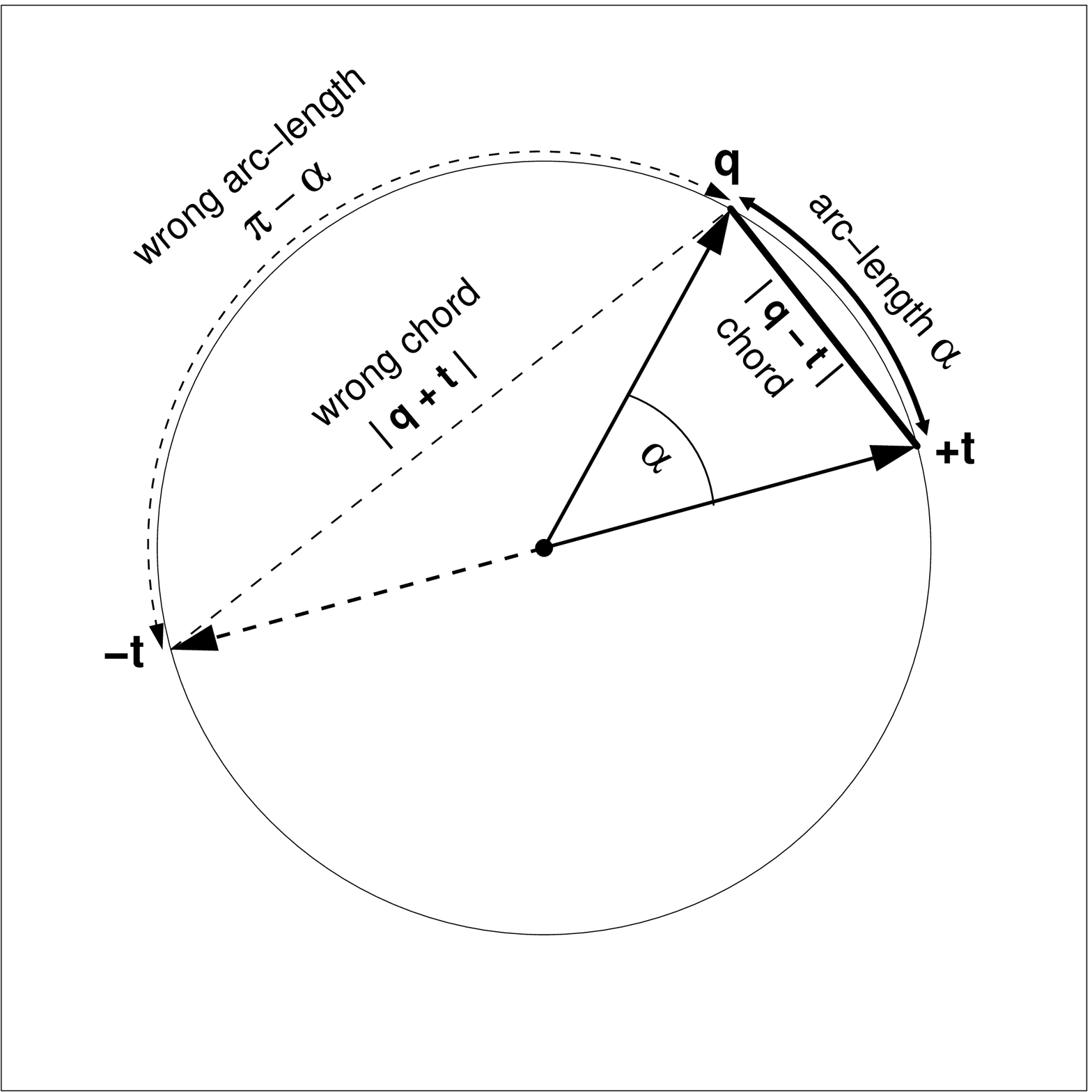} 
 \hspace{0.25in}
  \includegraphics[width=2.5in]{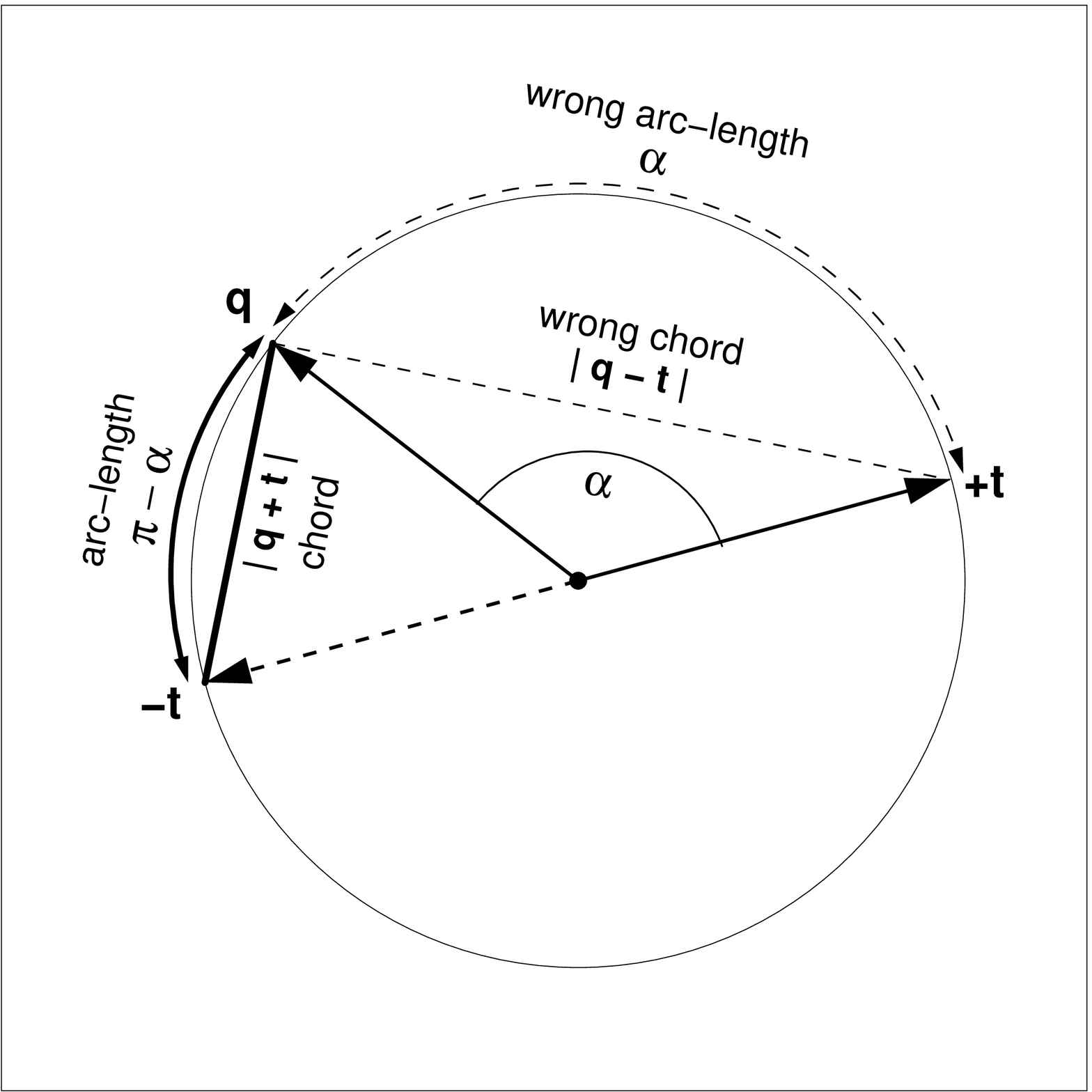} }
  \centerline {\hspace{1.5in} (A) \hfill (B) \hspace{1.5in} }
   \vspace*{.1in}
\caption{\ifnum\ShowFiles=1 {\bf arcvschordA.eps, arcvschordB.eps.} \fi
Geometric context involved in choosing a \emph{quaternion distance} that will result in
the correct \emph{average rotation matrix} when the quaternion
measures are optimized.
 Because  the quaternion vectors represented by $t$ and $-t$
  give the same rotation matrix, one must choose $|\cos \alpha |$
  or the \emph{minima}, that is   $\min\left( \alpha,\, 
  \pi - \alpha \right)$  or $\min\left(\|q-t\|,\,\|q+t \| \right)$ ,
  of the alternative distance measures to get
  the \emph{correct} items in the arc-length or chord measure summations.
  (A) and (B) represent the cases when the first or second choice should be made,
  respectively.}
\label{arc-chord.fig}
\end{figure}

\noindent{\bf Review of Orientation Frames in 3D.}
The  ideal optimization problem for 3D orientation frames
requires a measure 
constructed from the geodesic arc lengths on the quaternion
hypersphere.  Starting with the bare angle between two quaternions on $\Sphere{3}$,
 $\alpha = \arccos(q_{1} \cdot q_{2})$, where we recall that $\alpha\ge 0$,
  we define a \emph{pseudometric} \cite{Huynh:2009:MRC}  for the geodesic arc-length 
 distance as
 \begin{equation}
d_{\mbox{geodesic}}(q_{1},q_{2}) = \min (\alpha,\, \pi - \alpha): 
   \  \ 0 \le  d_{\mbox{geodesic}}(q_{1},q_{2}) \le \frac{\pi}{2} \ ,
\label{dgeo.eq}
\end{equation}
as illustrated in   \Fig{arc-chord.fig}.
An efficient implementation of this is to take 
 \begin{equation}
d_{\mbox{geodesic}}(q_{1},q_{2}) = \arccos( | q_{1} \cdot q_{2} |) \ ,
\label{dgeocos.eq}
\end{equation}
which we now exploit to construct a   measure
from geodesic arc-lengths on the quaternion hypersphere instead of Euclidean distances in space.
Thus to compare  a test quaternion-frame data set $\{p_{k}\}$  to a reference data set $\{ r_{k}\}$,
 we employ the geodesic-based least squares measure
\begin{equation} 
  {\mathbf S}_{\mbox{geodesic}} \  =  \ \sum_{k=1}^{N}
  \left(\arccos{  \left| \left( q \star p_{k}\right) \cdot  r_{k}
      \right| } \right)^2 \  =  \ \sum_{k=1}^{N}
  \left(\arccos{  \left| q \cdot \left( r_{k} \star \bar{p}_{k}\right) \right| } \right)^2 
       \  ,
\label{QRMSD-basic-geodesic.eq}
\end{equation} 
where the alternative second form follows from  \Eqn{quatTriples.eq}.

Since this does not easily fit into a linear algebra approach to construct
optimal solutions to the orientation-frame alignment problem, we choose to
approximate the measure of \Eqn{QRMSD-basic-geodesic.eq} by the
linearizable \emph{chord distance} measure, which does, under certain
conditions, permit a valid closed form solution.  We take as our
approximate measure the chordal pseudometric \cite{Huynh:2009:MRC,HartleyEtalRotAvg2013},
 \begin{equation}
d_{\mbox{chord}}(q_{1},q_{2}) = \min (\|q_{1} - q_{2} \|, \, \|q_{1} + q_{2} \|): 
   \  \ 0 \le  d_{\mbox{chord}}(q_{1},q_{2}) \le \sqrt{2} \ .
\label{dchord.eq}
\end{equation}
 We compare the geometric origins for \Eqn{dgeocos.eq} and \Eqn{dchord.eq} in \Fig{arc-chord.fig}.
 Note that the crossover point between the two expressions in \Eqn{dchord.eq} is at $\pi/2$, so the
 hypotenuse of the right isosceles triangle at that point has length $\sqrt{2}$.
 \quad
 
 The solvable approximate optimization function analogous to $\|R\cdot x - y \|^{2}$ that we will
 now explore for the quaternion-frame alignment problem will thus take the form
 that must be minimized as
 \begin{equation} 
{\mathbf S}_{\mbox{chord}} =  \sum_{k=1}^{N}
\left( \min (\| (q \star  p_{k} ) - r_{k} \|, \, \|(q \star  p_{k} )+ r_{k} \|)\right)^{2} \ .
\label{initchord.eq}
\end{equation} 
 We can convert the sign ambiguity in \Eqn{initchord.eq}
to a deterministic form like \Eqn{dgeocos.eq} by observing, with the help of \Fig{arc-chord.fig}, that 
\begin{equation}
\| q_{1} - q_{2}\|^2 = 2 -2 q_{1} \cdot q_{2}, \hspace{.5in} \| q_{1} +q_{2}\|^2 = 2 +2 q_{1} \cdot q_{2}  \ .
\label{initchordsqrs.eq}
\end{equation}
Clearly $( 2- 2|q_{1} \cdot q_{2}| )$ is always the smallest of the two values.  Thus minimizing 
\Eqn{initchord.eq} amounts to maximizing the now-familiar cross-term form, which we can  write as
\begin{equation}
\left.\begin{array}{rcl}
\Delta_{\mbox{chord}}(q) & = &    \sum_{k=1}^{N}  |(q \star  p_{k} ) \cdot r_{k} | \\[0.1in]
  & = &   
    \sum_{k=1}^{N}  | q  \cdot ( r_{k} \star  \bar{p}_{k} ) | \\[0.1in]
     & = &  \sum_{k=1}^{N}  | q  \cdot t_{k} |
 \end{array} \right\} \ .
 \label{basicQFA.eq}
 \end{equation}
 Here we have used the identity $(q \star  p ) \cdot r  = q  \cdot (r \star  \bar{p}) $ 
 from \Eqn{quatTriples.eq}
 and defined the quaternion displacement or "attitude error" 
 \cite{MarkleyEtal-AvgingQuats2007}
 \begin{equation}\label{deftk.eq}
 t_{k} = r_{k} \star \bar{p}_{k} \ .
 \end{equation}
 Note that we could have derived the same result using \Eqn{multiplicativeNorm.eq} to show that
 $\|q \star p - r\| = \| q \star p - r\| \|p \| = \| q - r \star \bar{p} \|$.

The final step is to   choose the samples of $q$ that include  our expected
 optimal quaternion, and adjust the sign of each data value $t_{k}$ to $\td{t}_{k}$ by
 the transformation
 \begin{equation}
  \td{t}_{k} =  t_{k}\, \sign(q\cdot t_{k})  \ \ \to  |q\cdot t_{k}|  = q \,\cdot \td{t}_{k} \ .
\label {tTotilde.eq}
\end{equation}
The neighborhood of $q$ matters because, as argued by \cite{HartleyEtalRotAvg2013},
even though the allowed range of 3D rotation angles is $\theta \in (-\pi ,  \pi)$  (or 
quaternion sphere angles $\alpha \in (-\pi/2, \pi/2)$), convexity of the optimization
problem cannot be guaranteed for collections outside local regions centered on
some $\theta_{0}$ of size
 $\theta_{0} \in (-\pi/2 ,  \pi/2)$  (or $\alpha_{0}  \in (-\pi/4, \pi/4)$): beyond this range,
local basins may exist that allow the mapping \Eqn{tTotilde.eq} to produce distinct
local variations in the assignments of the $\{\td{t}_{k}\}$ and
in the solutions for $q_{\opt}$.  Within considerations of such constraints, \Eqn{tTotilde.eq}
now allows us to
take the summation outside the absolute value, and write the quaternion-frame optimization
problem in terms of maximizing the cross-term expression
 \begin{equation}
\left.\begin{array}{rcl}
\Delta_{\mbox{chord}}(q) & = & {\displaystyle \sum_{k=1}^{N} }\, q\, \cdot \td{t}_{k}   \\[0.2in]
  & = &   q \cdot V(t)
   \end{array} \right\}
 \label{solnQFA.eq}
 \end{equation}
 where $V=   \sum_{k=1}^{N}   \td{t}_{k}$ is the analog of the Euclidean RMSD profile 
 matrix $M$.   However, since this is \emph{linear} in $q$, we have the remarkable result that,
 as noted in the treatment of \cite{HartleyEtalRotAvg2013} regarding
the quaternion $L_{2}$ chordal-distance norm, the solution is immediate.  We have simply
 \begin{equation}
 q_{\opt} = \frac{V}{\|V \|} \ ,
\label{VsolnQFA.eq}
\end{equation}
since that immediately maximizes the value of $\Delta_{\mbox{chord}}(q)$
in \Eqn{solnQFA.eq}.  This gives the   maximal value of the measure as 
\begin{equation} 
\Delta_{\mbox{chord}}(q_{\opt}) = \|V\| \ ,
\label{DeltaForVlinear.eq}
\end{equation}
and thus $ \|V\|$ is the exact orientation frame analog of the spatial
RMSD maximal eigenvalue $\epsilon_{\opt}$, except it is far
easier to compute.

\qquad

\noindent{\bf Alternative chord-measure approach parallel to the Euclidean case.}  Having understood
the chordal distance approach for the orientation-alignment problem
in terms of the  pseudometric \Eqn{dchord.eq} and the
measure  \Eqn{basicQFA.eq} transformed into the form \Eqn{solnQFA.eq}  involving the corrected 
quaternion displacements $\{\td{t}_{k}\}$, we now observe that we can also express
the problem in a form much closer to our Euclidean RMSD optimization problem.  Returning
to the form
\begin{equation} 
 {\mathbf S}_{\mbox{chord}} 
      =   \sum_{k=1}^{N} \| q \star p_{k} -  r_{k}\|^{2} \ .
\label{QRMSD-basic-chord.eq}
\end{equation}
we see that we can effectively transform the sign of \emph{only} $p_{k}\to \td{p}_{k}$
using the same test as \Eqn{tTotilde.eq} to make \Eqn{QRMSD-basic-chord.eq} valid
as it stands;  we   then proceed,
in the same fashion as the spatial
alignment problem but with the modification required by \Eqn{basicQFA.eq},
 to convert to a cross-term form as follows:
\begin{eqnarray} 
\Delta_{\mbox{chord}}(q) & = &\sum_{k=1}^{N} |(q \star {p}_{k})\cdot r_{k}  |
 \ = \ \sum_{k=1}^{N} (q \star  \td{p}_{k})\cdot r_{k}  \nonumber\\[0.02in]
  & = &\sum_{a=0,b=0}^{3} Q(q)_{ba} \sum_{k=1}^{N} 
 [\td{p}_{k}]_{\textstyle_a} \:[r_{k}]_{\textstyle_b} \nonumber \\[0.05in]
& = & \tr Q(q)\cdot W  \  .
\label{FrameVLinear.eq}
\end{eqnarray}
Here  $W$ is essentially a cross-covariance matrix in 
the quaternion data elements and  $Q(q)$ is the quaternion matrix of \Eqn{qprod.eq}.
Since $Q(q)$ is linear in $q$, we can simply pull out their coefficients, yielding
\begin{equation}
\Delta_{\mbox{chord}}(q)  = q\cdot V(W) \ ,
\label{qV.eq}
\end{equation}
where $V$ is  a four-vector corresponding to the profile matrix in the spatial
problem:
\begin{equation} 
V(W) = \left[ 
\begin{array}{c}
+ W_{00} +W_{11}+W_{22}+W_{33}\\ +W_{01} -W_{10}+ W_{23}-W_{32}\\
+ W_{02} -W_{20}+W_{31}-W_{13}\\ + W_{03} - W_{30} + W_{12} - W_{21}
   \end{array} \right]  \ .
 \label{M4def.eq}
 \end{equation}
 This is of course exactly the same as the quaternion difference transformation
\Eqn{deftk.eq}, expressed as a profile matrix transformation, and 
  \Eqn{qV.eq} leads, assuming consistent data localization,
   to the same  optimal unit quaternion 
\begin{equation} 
q_{\opt} = \frac{V}{\|V\|} \ ,
\label{qForVlinear.eq}    
  \end{equation}
 that  maximizes the value of $\Delta_{\mbox{chord}}$
in \Eqn{solnQFA.eq}, and the   maximal value of the measure is again 
$\Delta_{\mbox{chord}}(q_{\opt}) = \|V\| $.
 
\qquad

\noindent{\bf Matrix Form of the Linear Vector Chord Distance.}
 While \Eqn{solnQFA.eq}  (or \Eqn{qV.eq})  does not immediately fit into the
eigensystem-based  RMSD matrix method used in the spatial problem, it
can in fact be easily transformed from a system linear in $q$ to an
equivalent matrix system \emph{quadratic} in $q$.  Since any power of
the optimization measure will yield the same extremal solution, we can
simply \emph{square} the right-hand side of \Eqn{solnQFA.eq} and write the
result in the form
\begin{eqnarray} 
\Delta_{\mbox{chord-sq}} & = & (q\cdot V)(q \cdot V) \nonumber \\
 & = & \sum_{a=0,b=0}^{3} q_{a}\; V_{a} V_{b}\; q_{b}  \nonumber\\
& = & q \cdot \Omega \cdot q  \ ,
\label{qVsq.eq}
\end{eqnarray}
where $\Omega_{ab} = V_{a} V_{b}$ is a $4\times 4$ symmetric matrix
with $\det \Omega = 0$, and $\tr\Omega = \sum_{a} {V_{a}}^2 \neq 0$.
The eigensystem of $\Omega$  is just defined by the eigenvalue $\|V\|^2$,
and combination with the spatial eigensystem can be achieved either numerically or
algebraically using the $\mbox{trace\ } \neq 0$ case of our quartic solution.
  The process differs dramatically
from what we did with $\Delta_{\mbox{chord}}$, but the forms of the eigenvectors
are necessarily \emph{identical}.  Thus it is in fact possible to
merge  the  QFA system for $\Delta_{\mbox{chord}}$ into the matrix
method of the spatial RMSD using \Eqn{qVsq.eq}.

\qquad

\noindent{\bf Fixing Sign Problem with Quadratic Rotation Matrix Chord
  Distance.}  However, there 
is another approach that has a very natural way to incorporate
manifestly \emph{sign-independent} quaternion chord distances into our general
context, and which has a very interesting close relationship to
$\Delta_{\mbox{chord}}$.  The method begins with the observation that
full 3D rotation matrices like \Eqn{qrot.eq} can be arranged to rotate
the set of frames of the $\{p_{k}\}$ to be as close as possible to the
reference frame $\{r_{k}\}$ by employing a measure that is a
particular product of rotation matrices.  
The essence is to notice that the trace of
any 3D rotation matrix expressed in axis-angle form (rotation about a
fixed axis $\Hat{n}$ by $\theta$) can be expressed in two equivalent forms:
\begin{eqnarray} 
 \tr R(\theta,\Hat{n}) & =&  1 + 2 \cos \theta 
\label{RRRtheta.eq}\\
 \tr R(q) & = & 3 {q_{0}}^{2} - {q_{1}}^{2}- {q_{2}}^{2}- {q_{3}}^{2}
    \; = \; 4 {q_{0}}^2 - 1
 \  ,
 \label{RRRquat.eq}
\end{eqnarray}
and therefore traces of rotation matrices can be turned into maximizable
functions of the angles appearing in the trace.  Noting that the squared
 Fr\"{o}benius norm  of a matrix $M$ is the trace $\tr M\cdot M^{\t}$,
we begin with the goal of minimizing  a   Fr\"{o}benius norm of the form
\[ \|  R(q)\cdot  R(p_{k}) -  R (r_{k}) \|^{2}_{\mbox{Frob.}} \ , \]
and then convert from a minimization problem in this norm to
a maximization of the cross-term as usual.  The result is, remarkably,  
an explicitly symmetric and traceless profile matrix in the
quaternions.
We  thus begin with this form
of the orientation-frame measure (see, e.g.,
\cite{Huynh:2009:MRC,Moakher2002,HartleyEtalRotAvg2013}),
\begin{eqnarray} 
\Delta_{\mbox{\scriptsize RRR}} & = &  \sum_{k=1}^{N} \tr \left[ R(q)\cdot
  R(p_{k})\cdot  {R^{-1}}(r_{k}) \right]   
      \, = \,  \sum_{k=1}^{N} \tr \left[ R(q\star p_{k}\star \bar{r}_{k}) \right]  
  \label{RRRsumdef.eq} \nonumber \\
  & = & \sum_{k=1}^{N} \tr \left[ R(q)\cdot
  R(p_{k}\star \bar{r}_{k}) \right]
  \, = \, \sum_{k=1}^{N} \tr \left[ R(q)\cdot
  R^{-1}(r_{k}\star \bar{p}_{k}) \right]  \label{RRRsumdef2.eq} 
     \ ,
\label{RRRxsumdef3.eq}  
\end{eqnarray}
where $\bar{r}$ denotes the complex conjugate or inverse quaternion.
We note that due to the correspondence of $\Delta_{\mbox{\scriptsize RRR}}$ with a
cosine measure (via  \Eqn{RRRtheta.eq}), this must be \emph{maximized}
to find the optimal $q$, so  both $\Delta_{\mbox{chord}}$ and
$\Delta_{\mbox{\scriptsize RRR}}$
correspond naturally to  the cross-term measure 
we used for Euclidean point data, which we will later refer to as
$\Delta_{x}$ when necessary to distinguish it.

We next observe that the formulas for $\Delta_{\mbox{\scriptsize RRR}}$ and 
the pre-summation arguments of  $\Delta_{\mbox{chord}}$  are related as follows:
\begin{eqnarray} 
  \sum_{k=1}^{N} \tr \left[ R(q)\cdot
  R(p_{k})\cdot  {R}(\bar{r}_{k}) \right] 
  & = &  \sum_{k=1}^{N}\left(4  \left( (q \star p_{k})\cdot r_{k}\right)^2  -
    (q\cdot q)(p_{k}\cdot p_{k})(r_{k}\cdot r_{k})\right)  \ ,
\label{VVeqRRR.eq}
\end{eqnarray}
where of course the last term reduces to a constant since we
apply the unit-length constraint to all the quaternions, but is
algebraically essential to the construction.  The odd
form of \Eqn{VVeqRRR.eq} is not a typographical error: the conjugate
$\bar{r}$ of the reference data must be used in the $R\cdot R\cdot R$
expression, and the ordinary $r$ must be used in both terms on the
right-hand.  We conclude that using the $R\cdot R\cdot R$ measure and
replacing the argument of $\Delta_{\mbox{chord}}$ by its square \emph{before}
summing over $k$ are equivalent maximizing measures that eliminate the
quaternion sign dependence.  Now using  the quaternion triple-term identity
 $(q \star p) \cdot r = q \cdot (r \star \bar{p})$
 of \Eqn{quatTriples.eq}, we see that  each term of $\Delta_{\mbox{\scriptsize RRR}}$ reduces to a quaternion
 product that is a quaternion difference, or a ``quaternion displacement'' 
 $t_{k}= r_{k} \star \bar{p}_{k}$, i.e.,
 the rotation mapping each individual test frame to its corresponding reference frame,
  \begin{equation} 
\left.
\begin{array}{rcl}
  \Delta_{\mbox{\scriptsize RRR}}  = {\displaystyle \sum_{k=1}^{N} }\tr \left[ R(q)\cdot
  R(p_{k})\cdot  {R}(\bar{r}_{k}) \right] 
  & = & {\displaystyle  \sum_{k=1}^{N} } \left(4  \left( (q \star p_{k})\cdot r_{k}\right)^2  -
    (q\cdot q)(p_{k}\cdot p_{k})(r_{k}\cdot r_{k})\right)\\[0.2in]
     & = &  {\displaystyle  \sum_{k=1}^{N}}\left(4   \left(q \cdot (r_{k} \star \bar{p}_{k})\right) ^2   - 1 \right)\\
     & = &  4 \, {\displaystyle \sum_{a,b} } \, q_{a} \left(  {\displaystyle \sum_{k=1}^{N}} [t_{k}]_{\textstyle_a} \:[t_{k}]_{\textstyle_b} 
      \right) q_{b}  - N\\
      & = &  4\,  q \cdot A(t) \cdot q - N  \ . 
     \end{array} \right\}
\label{VVeqRRR2.eq}
\end{equation}
Here the $4\times4$ matrix  $A(t) _{ab} = \sum_{k=1}^{N} [t_{k}]_{\textstyle_a} \:[t_{k}]_{\textstyle_b}$  is the alternative (equivalent) 
profile matrix that was introduced by \cite{MarkleyEtal-AvgingQuats2007,HartleyEtalRotAvg2013}
for the chord-based \emph{quaternion-averaging} problem.  We can therefore use either
the measure $\Delta_{\mbox{\scriptsize RRR}}$ or
\begin{equation}
\Delta_{\mbox{\scriptsize A}} =  q \cdot A(t) \cdot q
\label{qAqmeasure.eq}
\end{equation}
 as our rotation-matrix-based  sign-insensitive chord-distance optimization measure. 
Exactly like our usual spatial measure, these measures must be \emph{maximized}
to find the optimal $q$.  It is, however, important to emphasize that the optimal
quaternion will \emph{differ} for the   $\Delta_{\mbox{\scriptsize chord}}$ , 
 $\Delta_{\mbox{\scriptsize chord-sq}}$ , and 
  $\Delta_{\mbox{\scriptsize RRR}} \sim   \Delta_{\mbox{\scriptsize A}}$   
  measures, though they will normally be very similar.  More details are
  explored in Section \ref{Evaluate3DOrient.sec}.

\qquad

{\bf Details of Rotation Matrix Form.}We now recognize that the sign-insensitive measures are all very closely
 related to our original spatial RMSD problem, and all can be solved by
 finding the optimal quaternion eigenvector $q_{\opt}$ of a $4\times 4$ matrix.
 The procedure for $\Delta_{\mbox{\scriptsize chord-sq}}$ and $\Delta_{\mbox{\scriptsize A}}$
 follows immediately, but it is useful to work out the options for
  $\Delta_{\mbox{\scriptsize RRR}}$ in a little more detail.
  
Choosing \Eqn{RRRsumdef.eq} has the remarkable feature of producing,
via \Eqn{qrot.eq} for $R(q)$, an expression \emph{quadratic} in $q$,
with a symmetric, traceless profile matrix
$U(p,r)$ that is \emph{quartic} in the quaternion elements $p_{k}$ and
$r_{k}$.  This variant of the chord-based QFA problem thus falls
into the same category as the standard RMSD problem, and permits the
application of the same exact solution (or, indeed, the traditional
numerical solution method if that is more efficient).  The profile
matrix equation is unwieldy to write down explicitly in terms of the
quaternion elements quartic in $\{p,r\}$, but we actually have several
options for expressing the content in a simpler form.  One is to write
the matrices in abstract canonical $3\times 3$ form, e.g.,
\begin{equation} 
R(p) =  [P] = \left[\begin{array}{ccc}
  p_{xx} & p_{xy} &  p_{xz} \\
  p_{yx} & p_{yy} &  p_{yz} \\
  p_{zx} & p_{zy} &  p_{zz} \\
\end{array} \right] \label{3b3Rmat.eq} \ ,
\end{equation}
where the \emph{columns} of this matrix are just the three axes of
each data element's frame triad.  This is often exactly what our
original data look like, for example, if the residue orientation
frames of a protein are computed from cross-products of atom-atom
vectors \cite{HansonThakurQuatProt2012}.  Then we can define for each
data element the $3\times 3$ matrix
\[ [T_{k}] =  R(p_{k})\cdot
R(\bar{r}_{k})] = R(p_{k}\star \bar{r}_{k})= R^{-1}(t_{k}) \ , \]
so we can write $T$ either in terms of a $3\times 3$ matrix like
\Eqn{3b3Rmat.eq} derived from the actual frame-column data, or in
terms of \Eqn{qrot.eq} and the quaternion frame data $t_{k} = r_{k}
\star \bar{p}_{k}$.  We then may write the frame measure in general as
\begin{equation} 
\Delta_{\mbox{\scriptsize RRR}} =  \sum_{k=1}^{N} \tr \left( R(q) \cdot
  T_{k} \right) 
 =  \sum_{a=1,b=1}^{3} R_{ba}(q) T_{ab} \ ,
\label{trRS.eq}
\end{equation}
where the frame-based cross-covariance matrix is simply $T_{ab} =
\sum_{k=1}^{N} {[T_{k}]}_{ab}$.  
As before, we can easily expand $R(q)$ using \Eqn{qrot.eq} to
convert the measure to a 4D linear algebra problem of the form
\begin{equation} 
\Delta_{\mbox{\scriptsize RRR}} = \sum_{a=0,b=0}^{3} q_{a} \cdot U_{ab}(p,r) \cdot q_{b} = 
  q \cdot U(p,r) \cdot q \ .
\label{frmCrossCovar.eq}
\end{equation}
Here $ U(p,r) = U(T)$ has the same relation to $T$ as $M(E)$ does to $E$ in
\Eqn{basicHorn.eq}.   We may choose to write the profile matrix $U= \sum_{k} U_{k}$
 appearing in $\Delta_{\mbox{\scriptsize RRR}}$ 
either in terms of the individual $k$-th components of the 
 numerical 3D rotation matrix $\rule{0in}{2.2ex}T =   R^{-1}(t)$ or using the
 composite quaternion $t = r \star \bar{p}\;$: 
\begin{eqnarray} 
\lefteqn{U_{k}(T) \equiv U(t_{k})   }\nonumber \\
\hspace*{-.5in}&\!\! = \!\!\!&  \!\!\!\!\left[
\begin{array}{cccc}
 \!\!  T_{xx}+ T_{yy}+ T_{zz}  &
     T_{yz}- T_{zy}  & 
     T_{zx}- T_{xz}  &
     T_{xy}- T_{yx}  \\
   T_{yz}- T_{zy}  & 
     T_{xx}- T_{yy}- T_{zz}  & 
     T_{xy}+ T_{yx}  & 
     T_{xz}+ T_{zx}  \\
   T_{zx}- T_{xz}  &
     T_{xy}+ T_{yx}  & 
    - T_{xx}+ T_{yy}- T_{zz}  &
    T_{yz}+ T_{zy}  \\
   T_{xy}- T_{yx}  & 
     T_{xz}+ T_{zx}  &
     T_{yz}+ T_{zy}  &
    - T_{xx}- T_{yy}+ T_{zz} \!\!\! \\
\end{array}
\right]_{\textstyle k} \label{UofSmat.eq}\\[.25in]
\hspace*{-.5in}& = & \!\!\!\!\left[  
\begin{array}{cccc}
\textstyle {3 {t_0}^2-{t_1}^2-{t_2}^2-{t_3}^2 }\!\! &
 4 t_0 t_1 & 4  t_0 t_2 & 4 t_0 t_3 \\
 4 t_0 t_1 &
\!\!\textstyle {-{t_0}^2+3 {t_1}^2-{t_2}^2-{t_3}^2}\!\!  &
 4  t_1 t_2 & 4 t_1 t_3 \\
 4 t_0 t_2 & 4 t_1 t_2 & \!\!
\textstyle {-{t_0}^2-{t_1}^2+3 {t_2}^2-{t_3}^2} \!\!&
 4 t_2 t_3 \\
 4 t_0 t_3 & 4 t_1 t_3 & 4 t_2 t_3 & 
 \!\!\textstyle  {-{t_0}^2-{t_1}^2-{t_2}^2+3 {t_3}^2 }\\
\end{array}\!\!
 \right]_{\textstyle k}  .
\label{UofS.eq}
\end{eqnarray}
Both \Eqn{UofSmat.eq} and \Eqn{UofS.eq} are \emph{quartic} (and
identical) when expanded in terms of the quaternion data
$\{p_{k},r_{k}\}$.  To compute the necessary $4\times 4$ numerical
profile matrix $U$, one need only substitute the appropriate 3D frame
triads or their corresponding quaternions for the $k$th frame pair and
sum over $k$ .  Since the orientation-frame profile matrix $U$ is
symmetric and traceless just like the Euclidean profile matrix $M$,
the same solution methods for the optimal quaternion rotation
$q_{\opt}$ will work without alteration in this case, which is
probably the preferable method for the general problem.

\qquad

\noindent{\bf Evaluation.}  The details of evaluating the properties
of our quaternion-frame alignment algorithms, and especially comparing
the chord approximation to the arc-length measure, are tedious and
are available separately in Section \ref{Evaluate3DOrient.sec}.  The top-level result is
that, even for quite large rotational differences, the mean differences
between the arc-length measure's numerical optimal angle and the various chord
approximations are on the order of a few thousandths of a degree.


\comment{   
Even in this compact form, the expression is still quite long,
\begin{equation} 
U_{k}(p,r) = \left[
\begin{array}{cccc}
 \frac{1}{2} \left(p_{xx} r_{xx}+p_{yx}
   r_{xy}+p_{zx} r_{xz}+p_{xy}
   r_{yx}+p_{yy} r_{yy}+p_{zy}
   r_{yz}+p_{xz} r_{zx}+p_{yz}
   r_{zy}+p_{zz} r_{zz}\right) & \frac{1}{2}
   \left(-p_{zx} r_{xy}+p_{yx}
   r_{xz}-p_{zy} r_{yy}+p_{yy}
   r_{yz}-p_{zz} r_{zy}+p_{yz}
   r_{zz}\right) & \frac{1}{2} \left(p_{zx}
   r_{xx}-p_{xx} r_{xz}+p_{zy}
   r_{yx}-p_{xy} r_{yz}+p_{zz}
   r_{zx}-p_{xz} r_{zz}\right) & \frac{1}{2}
   \left(-p_{yx} r_{xx}+p_{xx}
   r_{xy}-p_{yy} r_{yx}+p_{xy}
   r_{yy}-p_{yz} r_{zx}+p_{xz}
   r_{zy}\right) \\
 \frac{1}{2} \left(-p_{zx} r_{xy}+p_{yx}
   r_{xz}-p_{zy} r_{yy}+p_{yy}
   r_{yz}-p_{zz} r_{zy}+p_{yz}
   r_{zz}\right) & \frac{1}{2} \left(p_{xx}
   r_{xx}-p_{yx} r_{xy}-p_{zx}
   r_{xz}+p_{xy} r_{yx}-p_{yy}
   r_{yy}-p_{zy} r_{yz}+p_{xz}
   r_{zx}-p_{yz} r_{zy}-p_{zz}
   r_{zz}\right) & \frac{1}{2} \left(p_{yx}
   r_{xx}+p_{xx} r_{xy}+p_{yy}
   r_{yx}+p_{xy} r_{yy}+p_{yz}
   r_{zx}+p_{xz} r_{zy}\right) & \frac{1}{2}
   \left(p_{zx} r_{xx}+p_{xx}
   r_{xz}+p_{zy} r_{yx}+p_{xy}
   r_{yz}+p_{zz} r_{zx}+p_{xz}
   r_{zz}\right) \\
 \frac{1}{2} \left(p_{zx} r_{xx}-p_{xx}
   r_{xz}+p_{zy} r_{yx}-p_{xy}
   r_{yz}+p_{zz} r_{zx}-p_{xz}
   r_{zz}\right) & \frac{1}{2} \left(p_{yx}
   r_{xx}+p_{xx} r_{xy}+p_{yy}
   r_{yx}+p_{xy} r_{yy}+p_{yz}
   r_{zx}+p_{xz} r_{zy}\right) & \frac{1}{2}
   \left(-p_{xx} r_{xx}+p_{yx}
   r_{xy}-p_{zx} r_{xz}-p_{xy}
   r_{yx}+p_{yy} r_{yy}-p_{zy}
   r_{yz}-p_{xz} r_{zx}+p_{yz}
   r_{zy}-p_{zz} r_{zz}\right) & \frac{1}{2}
   \left(p_{zx} r_{xy}+p_{yx}
   r_{xz}+p_{zy} r_{yy}+p_{yy}
   r_{yz}+p_{zz} r_{zy}+p_{yz}
   r_{zz}\right) \\
 \frac{1}{2} \left(-p_{yx} r_{xx}+p_{xx}
   r_{xy}-p_{yy} r_{yx}+p_{xy}
   r_{yy}-p_{yz} r_{zx}+p_{xz}
   r_{zy}\right) & \frac{1}{2} \left(p_{zx}
   r_{xx}+p_{xx} r_{xz}+p_{zy}
   r_{yx}+p_{xy} r_{yz}+p_{zz}
   r_{zx}+p_{xz} r_{zz}\right) & \frac{1}{2}
   \left(p_{zx} r_{xy}+p_{yx}
   r_{xz}+p_{zy} r_{yy}+p_{yy}
   r_{yz}+p_{zz} r_{zy}+p_{yz}
   r_{zz}\right) & \frac{1}{2} \left(-p_{xx}
   r_{xx}-p_{yx} r_{xy}+p_{zx}
   r_{xz}-p_{xy} r_{yx}-p_{yy}
   r_{yy}+p_{zy} r_{yz}-p_{xz}
   r_{zx}-p_{yz} r_{zy}+p_{zz}
   r_{zz}\right) \\
\end{array}
\right]
\label{MofRRRmatform.eq}
\end{equation}
  }  

\comment{   
\begin{equation} 
\Delta_{f} = q\cdot U(p,r) \cdot q \ ,
\label{RRRDelta.eq}
\end{equation}
where the $4\times 4$ profile matrix is
\begin{equation} 
U = 
\left[
\begin{array}{cccc}
 \frac{1}{2} \left(\left(3 r_0^2-r_1^2-r_2^2-r_3^2\right) p_0^2+8 r_0 \left(p_1 r_1+p_2 r_2+p_3
   r_3\right) p_0-p_2^2 r_0^2-p_3^2 r_0^2-p_2^2 r_1^2-p_3^2 r_1^2+3 p_2^2 r_2^2-p_3^2
   r_2^2-\left(p_2^2-3 p_3^2\right) r_3^2+8 p_2 p_3 r_2 r_3+8 p_1 r_1 \left(p_2 r_2+p_3
   r_3\right)-p_1^2 \left(r_0^2-3 r_1^2+r_2^2+r_3^2\right)\right) & 2 \left(-p_1 r_0+p_0 r_1-p_3
   r_2+p_2 r_3\right) \left(p_0 r_0+p_1 r_1+p_2 r_2+p_3 r_3\right) & 2 \left(-p_2 r_0+p_3 r_1+p_0
   r_2-p_1 r_3\right) \left(p_0 r_0+p_1 r_1+p_2 r_2+p_3 r_3\right) & 2 \left(-p_3 r_0-p_2 r_1+p_1
   r_2+p_0 r_3\right) \left(p_0 r_0+p_1 r_1+p_2 r_2+p_3 r_3\right) \\
 2 \left(-p_1 r_0+p_0 r_1-p_3 r_2+p_2 r_3\right) \left(p_0 r_0+p_1 r_1+p_2 r_2+p_3 r_3\right) &
   \frac{1}{2} \left(-\left(r_0^2-3 r_1^2+r_2^2+r_3^2\right) p_0^2-8 r_1 \left(p_1 r_0+p_3 r_2-p_2
   r_3\right) p_0-p_2^2 r_0^2-p_3^2 r_0^2-p_2^2 r_1^2-p_3^2 r_1^2-p_2^2 r_2^2+3 p_3^2 r_2^2+\left(3
   p_2^2-p_3^2\right) r_3^2-8 p_2 p_3 r_2 r_3+8 p_1 r_0 \left(p_3 r_2-p_2 r_3\right)+p_1^2 \left(3
   r_0^2-r_1^2-r_2^2-r_3^2\right)\right) & 2 \left(p_2 r_0-p_3 r_1-p_0 r_2+p_1 r_3\right) \left(p_1
   r_0-p_0 r_1+p_3 r_2-p_2 r_3\right) & 2 \left(p_3 r_0+p_2 r_1-p_1 r_2-p_0 r_3\right) \left(p_1
   r_0-p_0 r_1+p_3 r_2-p_2 r_3\right) \\
 2 \left(-p_2 r_0+p_3 r_1+p_0 r_2-p_1 r_3\right) \left(p_0 r_0+p_1 r_1+p_2 r_2+p_3 r_3\right) & 2
   \left(p_2 r_0-p_3 r_1-p_0 r_2+p_1 r_3\right) \left(p_1 r_0-p_0 r_1+p_3 r_2-p_2 r_3\right) &
   \frac{1}{2} \left(-\left(r_0^2+r_1^2-3 r_2^2+r_3^2\right) p_0^2-8 r_2 \left(p_2 r_0-p_3 r_1+p_1
   r_3\right) p_0+3 p_2^2 r_0^2-p_3^2 r_0^2-p_2^2 r_1^2+3 p_3^2 r_1^2-p_2^2 r_2^2-p_3^2
   r_2^2-\left(p_2^2+p_3^2\right) r_3^2-8 p_2 p_3 r_0 r_1+8 p_1 \left(p_2 r_0-p_3 r_1\right)
   r_3-p_1^2 \left(r_0^2+r_1^2+r_2^2-3 r_3^2\right)\right) & 2 \left(p_3 r_0+p_2 r_1-p_1 r_2-p_0
   r_3\right) \left(p_2 r_0-p_3 r_1-p_0 r_2+p_1 r_3\right) \\
 2 \left(-p_3 r_0-p_2 r_1+p_1 r_2+p_0 r_3\right) \left(p_0 r_0+p_1 r_1+p_2 r_2+p_3 r_3\right) & 2
   \left(p_3 r_0+p_2 r_1-p_1 r_2-p_0 r_3\right) \left(p_1 r_0-p_0 r_1+p_3 r_2-p_2 r_3\right) & 2
   \left(p_3 r_0+p_2 r_1-p_1 r_2-p_0 r_3\right) \left(p_2 r_0-p_3 r_1-p_0 r_2+p_1 r_3\right) &
   \frac{1}{2} \left(-\left(r_0^2+r_1^2+r_2^2-3 r_3^2\right) p_0^2-8 \left(p_3 r_0+p_2 r_1-p_1
   r_2\right) r_3 p_0-p_2^2 r_0^2+3 p_3^2 r_0^2+3 p_2^2 r_1^2-p_3^2 r_1^2-p_2^2 r_2^2-p_3^2
   r_2^2-\left(p_2^2+p_3^2\right) r_3^2+8 p_2 p_3 r_0 r_1-8 p_1 \left(p_3 r_0+p_2 r_1\right)
   r_2-p_1^2 \left(r_0^2+r_1^2-3 r_2^2+r_3^2\right)\right) \\
\end{array}
\right] \ .
\label{RRRProfile.eq}
\end{equation}
   }  

\subsection{The 4D Orientation-Frame alignment Problem}

Orientation frames in four dimensions have axes that are the columns
of a 4D rotation matrix taking the identity frame to the new
orientation frame.  Therefore, in parallel with the 3D case, such
frames can be represented either as 4D rotation matrices (the action
on a 4D identity frame to get a new set of 4 orthogonal axes), or as
the pair of quaternions $(q,q')$ used in \Eqn{doublequat.eq} to define
$R_{4}(q,q')$.  As in the 3D frame case, we will take advantage of the
chord-distance linearization of the geodesic angular measure, and
we shall present two alternative approaches to the optimization measure.

 \paragraph{Quadratic Form.}  In 3D, with \Eqn{solnQFA.eq}
  having a single quaternion involved in the rotation,
 we were able to write down $\Delta_{\mbox{chord}}$ in
 terms of a simple expression linear in the quaternion $q$ and the
 cumulative data $V$, and we observed that a quadratic expression
 $\left(q \cdot V\right)^2$ would also produce the same optimal eigenvector
 $q = V/\|V\|$.  The optimal frame problem in 4D, in contrast, already
 requires a pair of quaternions, and one strategy is to split the
analogs of the  3D quadratic expression into two parts, yielding
\begin{equation} 
\Delta_{\mbox{4:chord-sq}}(q,q')= \left(q \cdot V\right) \left(q'
  \cdot V'\right) = q_{a} \left(V_{a} V'_{b} \right) q'_{b} = q \cdot
\Omega_{4} \cdot q' \ 
\label{4DpqFrm.eq}
\end{equation}
as the generalization from 3D to 4D.  Here,  each 4D test frame consists of frames denoted by
the quaternion pair $(p,p')$, and each reference frame employs a pair
$(r,r')$, so we build the data coefficients starting from 
\begin{equation}
\left.  \begin{array}{rcccl}
V &= & {\displaystyle  \sum_{k=1}^{N} } \left( r_{k} \star \bar{p}_{k}\right) = \sum_{k=1}^{N} t_{k} \\[0.1in]
V' &= &  {\displaystyle  \sum_{k=1}^{N} } \left( r'_{k} \star \bar{p}'_{k}\right) = \sum_{k=1}^{N} t'_{k}
\end{array}
\right\}
\end{equation}
and then applying the transformation
\begin{equation}
\left.  \begin{array}{rcccl}
 t_{k} &\to & \td{t}_{k}& = &  t_{k} \,\sign(q\cdot t_{k}) \\[0.12in]
 t'_{k} &\to & \td{t'}_{k}& = & t'_{k}  \,\sign(q\cdot t'_{k}) 
 \end{array} \right\}
 \label{ttprimteoTilde.eq}
 \end{equation}
to achieve consistent (local) signs.
According to \Eqn{M4def.eq}, $V$ could also be constructed from $W_{ab}=\sum_{k=1}^{N}  [\td{p}_{k}]_{\textstyle_a} \:[r_{k}]_{\textstyle_b}$,
 and $V'$   from $W'_{ab}=\sum_{k=1}^{N}  
 [\td{p'}_{k}]_{\textstyle_a} \:[r'_{k}]_{\textstyle_b}$, noting that here $p$ is transformed
 by the``tilde'' of \Eqn{ttprimteoTilde.eq}.
Now, for the 4D frame pairs, the solution for the optimal quaternions
must achieve the maximum for \emph{both} elements of the pair, and so
we obtain as a solution maximizing \Eqn{4DpqFrm.eq}
\begin{equation} 
\left. 
\begin{array}{rcl}
 q_{\opt} & = & {\displaystyle \frac{V}{\|V\|}}  \\[.2in]
 q'_{\opt} & = &{\displaystyle \frac{V'}{\|V'\|}}  \\[.2in]
\Delta_{\mbox{4:chord-sq}}(\opt) & = & {\|V\|}{\|V'\|}
\end{array} \right\} \ .
\label{vvprime4Dfrm.eq}
\end{equation}

{\bf Remark:}   There is a particular
reason to prefer \Eqn{4DpqFrm.eq} for the 4D orientation frame problem: in the next section, we will see
that the separate \emph{pre-summation} arguments for $V$ and $V'$, gathered
together, are \emph{exactly} equal to the joint summand of the 4D triple rotation
pre-summation arguments, following the pattern seen in \Eqn{VVeqRRR.eq}
for the 3D orientation-frame analysis.

\paragraph{Quartic Triple Rotation Form.}   
  One can also eliminate the sign choice step altogether by defining 
   a 4D frame similarity measure that is the exact
analog of \Eqn{RRRxsumdef3.eq} in 3D as follows:
\begin{eqnarray} 
\Delta_{\mbox{\tiny RRR4}} & = &  \sum_{k=1}^{N} \tr \left[ R(q,q')\cdot
  R(p_{k},{p'}_{k})\cdot  {R^{-1}}(r_{k},{r'}_{k}) \right] \label{RRR4sumdef.eq}\\
  & = & \sum_{k=1}^{N} \tr \left[ R(q,q')\cdot
  R(p_{k}\star \bar{r}_{k},\;{p'}_{k}\star {\bar{r}'}_{k}) \right] \label{RRR4sumdef2.eq}\\
  & = & \sum_{k=1}^{N} \tr \left[ R(q,q') \cdot R^{-1}(t_{k},\,t'_{k})  
  \right] \label{RRR4sumdef3.eq} \\[0.1in]
& = & q \cdot U(p,{p'};\,r,{r'})\cdot q' \label{RRR4U.eq}\ .
\end{eqnarray}
Remarkably, there is a 4D version of the 3D identity  \Eqn{VVeqRRR.eq}
relating the triple rotation measure to the quadratic realizations of
the linear quaternion rotation measures, namely
\begin{equation} 
\left.
\begin{array}{rcl}
  {\displaystyle \sum_{k=1}^{N} }\tr \left[ R(q,q')\cdot
  R(p_{k},p'_{k})\cdot  {R}(\bar{r}_{k},\bar{r}'_{k}) \right] %
  & = &  
  4 \,{\displaystyle \sum_{k=1}^{N} }  \left( (q \star p_{k})\cdot r_{k}\right)  
                        \left( (q' \star p'_{k})\cdot r'_{k}\right) \nonumber \\[0.1in]
    & = & 4 \, {\displaystyle\sum_{k=1}^{N} }   \left( q \cdot (r_{k}\star \bar{p}_{k})\right)
                             \left( q' \cdot (r'_{k}\star \bar{p}'_{k})\right)\nonumber\\ [0.1in]
    & = & 4 \, {\displaystyle\sum_{k=1}^{N}  }    (q \cdot t_{k}) (q' \cdot t'_{k}   ) \\ [0.1in]
       & = &  4 \, {\displaystyle\sum_{a,b} } q_{a} \left( {\displaystyle \sum_{k=1}^{N}  }\,  [t_{k}]_{\textstyle_a} \:[t'_{k}]_{\textstyle_b} 
      \right) q'_{b} \\ [0.2in]
      & = &  4 \, q \cdot A(t = r  \star \bar{p}, \, t' = r'  \star \bar{p}')  \cdot q'   
     \end{array} \right\}  
                               \ ,
\label{VVeqRRR4.eq}
\end{equation}
Thus the pre-summation version of the arguments in the $(q\cdot V)(q' \cdot
V')$ version of the 4D chord measure turns out to be  \emph{exactly} the
same as the triple-matrix product measure summand without the additional trace
term that is present in 3D.   Furthermore, as long as one follows the rules of changing
\emph{both} the primed and unprimed signs together (the condition for $R_{4}(q,q')$'s invariance),
this measure is sign-independent.  The $4\times4$ matrix $A(t,t')$ is the 4D 
profile matrix equivalent to that of \cite{MarkleyEtal-AvgingQuats2007,HartleyEtalRotAvg2013}
for the 3D chord-based quaternion-averaging problem.  We can therefore use either
the measure $\Delta_{\mbox{\scriptsize RRR4}}$ or
\begin{equation}
\Delta_{\mbox{\scriptsize A4}} =  q \cdot A(t,t') \cdot q'
\label{qAqmeasure2.eq}
\end{equation}
with $A (t, t')_{ab} = \sum_{k=1}^{N} [t_{k}]_{\textstyle_a} \:[t'_{k}]_{\textstyle_b}$ as our rotation-matrix-based  sign-insensitive chord-distance optimization measure.

To get an expression in terms of $R$, we now use
\Eqn{doublequat.eq} for $ R(q,q')$ to decompose the measure
\Eqn{RRR4sumdef2.eq} into the  rotation-averaging form
\begin{eqnarray} 
\Delta_{\mbox{\tiny RRR4}} & = & \tr \left[ R(q,q')\cdot T(p,p';r,r')
\right]\\
 & = & q \cdot U(T) \cdot q' \ ,
\label{deltaRRR4.eq}
\end{eqnarray}
where $T(p,p';r,r') = \sum_{k=1}^{N} R^{-1}( t_{k},\,t'_{k})$ 
 and $U(T)$ has the same
relationship to $T$ as the 4D profile matrix $M(E)$ in
\Eqn{basic4DHorn.eq} does to the cross-correlation matrix $E$. In the
next section, we will see that
the singleton version of this map is unusually degenerate, with rank
one, though that feature does not persist for data sets with $N>1$.

Now, as in the 4D spatial RMSD analysis, we might naturally assume
that we could follow the 3D case by determining the maximal eigenvalue
$\epsilon_{0}$ of $U$ and its left and right eigenvectors
$q_{\lambda}$ and $q_{\rho}$, which would give
\[\Delta_{\mbox{\tiny RRR4}} \stackrel{\mbox{?}}{ = } q_{\lambda}
\cdot U \cdot q_{\rho} = (q_{\lambda} \cdot q_{\rho})\, \epsilon_{0} 
 \ . \] 
As before, this is not a maximal value for the measure
$\Delta_{\mbox{\tiny RRR4}}$ over the possible range of $R(q,q')$.  
To solve the optimization correctly, we must again be very careful,
and work with the maximal eigenvalue $\alpha(\mbox{\small RRR4:opt} ) $
  of $G = U^{\t} \cdot U $ and
$G' = U \cdot  U^{\t}$, which we can get  numerically as usual, or
algebraically from the quartic solution for the eigenvalues 
for symmetric $4\times 4$ matrices with a trace, yielding
\[ \Delta_{\mbox{\tiny RRR4}}(\opt) = 
\sqrt{\mbox{\small max eigenvalue}\; (U^{\t} \cdot U ) } =
\sqrt{ \alpha(\mbox{\small RRR4:opt}) }\ . \]
If we need the actual optimal rotation matrix solving
\[ \Delta_{\mbox{\tiny RRR4}}(\opt) = \tr \left( R_{4}(q_{\opt},q'_{\opt}) \cdot
    S\right) = q_{\opt}  \cdot U \cdot q'_{\opt}=
   \sqrt{  \alpha(\mbox{\small RRR4:opt})} \ , \]
then we just use our optimal eigenvalue to solve
\begin{eqnarray*}
\left(G - \alpha(\mbox{\small RRR4:opt}) I_{4}\right)\cdot q & = & 0\\
\left(G' - \alpha(\mbox{\small RRR4:opt})  I_{4}\right)\cdot q' & =& 0 
\end{eqnarray*}
for $q_{\opt}$ and  $q'_{\opt}$, or use the equivalent adjugate-column method
to extract the eigenvectors.   That gives the desired 4D rotation
matrix $R_{4}(q_{\opt},q'_{\opt})$ explicitly via \Eqn{doublequat.eq}.  
The same approach applies to the solution of $\Delta_{\mbox{\scriptsize A4}} =  q \cdot A(t,t') \cdot q'$,
Note that this can all be accomplished
numerically, directly as above or with Singular Value Decomposition, or using
the quaternion eigenvalue decomposition  on the symmetric matrices
either numerically or algebraically,

\qquad

\section{On Obtaining Quaternions and Quaternion Pairs from 3D and 4D Rotation Matrices}
\label{baritzhack.sec}
 

\subsection{Extracting a Quaternion from 3D Rotation Matrices} 

The quaternion RMSD profile matrix method can be used to implement
 a singularity-free algorithm to 
obtain the (sign-ambiguous) quaternions corresponding to  numerical
3D and 4D rotation matrices.  There are many existing approaches to the
3D problem in the literature (see, e.g., \cite{Shepperd1978},
\cite{ShusterNatanson1993}, or Section 16.1 of
\cite{HansonQuatBook:2006}).  In contrast to these
approaches, Bar-Itzhack~\cite{BarItzhack2000} has observed,
 in essence, that if we simply replace the data matrix $E_{ab}$ by a
numerical 3D orthogonal rotation matrix $R$, the numerical quaternion $q$ that 
corresponds to   $R_{\mbox{numeric}}=R(q)$, as defined by
\Eqn{qrot.eq}, can be found by solving our familiar maximal
quaternion eigenvalue problem.  The initially unknown optimal matrix 
(technically its quaternion) computed by
maximizing the similarity measure turns out to be computable as
a single-element quaternion barycenter problem.
 To see this, take
$S(r)$ to be the sought-for optimal rotation matrix, with its own quaternion $r$,
that must maximize the Bar-Itzhack measure.
We start with the 
Fr\"{o}benius measure describing the match of two rotation matrices corresponding
to the quaternion $r$ for the unknown quaternion and the numeric matrix
$R$ containing  the known
$3\times 3$ rotation matrix data:
\begin{eqnarray*}
  {\mathbf S}_{\mbox{\scriptsize BI}} & =&    \|S(r)-R\|^{2}_{\mbox{Frob}} 
    \ = \  \tr \left( [S(r) - R] \cdot[ S^{\t}(r) - R^{\t}] \right)  \\
  & = & \tr \left( I_{3} + I_{3}  - 2  \left( S(r) \cdot R^{\t} \right) \right)  \\
   & = & \mbox{const} - 2   \tr S(r) \cdot  R^{\t}   \ . 
  \label{FrobBI.eq}
\end{eqnarray*}
Pulling out the cross-term as usual and converting to a maximization
problem over the unknown quaternion $r$, we arrive at
\begin{equation}
\Delta_{\mbox{\scriptsize BI}} = \tr{S(r) \cdot R^{\t} } = r \cdot K(R) \cdot r \ ,
\label{deltaBI.eq}
\end{equation}
where $R$ is (approximately) an orthogonal matrix of numerical data, and 
$K(R)$ is  analogous to the profile matrix $M(E)$.
Since both $S$ and $R$ are $\SO{3}$ rotation matrices, so is their 
product $T=S\cdot R^{\,\t}$, and thus that product itself corresponds to some axis $\Hat{n}$
and angle $\theta$, where 
\[ \tr {S(r)  \cdot R^{\,\t}(q) } = \tr T(r \star \bar{q})= \tr T(\theta,\Hat{n}) = 1 + 2 \cos{\theta} \ . \]
The maximum is obviously close to the ideal value $\theta = 0$, which corresponds to $S \approx R$.
Thus if we find the maximal quaternion eigenvalue  $\epsilon_{\opt}$ of
the profile matrix $K(R)$ in \Eqn{deltaBI.eq},  our closest solution is well-represented
by the corresponding normalized quaternion eigenvector $r_{\opt}$,
\begin{equation}
q =  {r}_{\opt} \ .
\label{deltaBISoln.eq}
\end{equation}
This numerical solution for $q$ will correspond to the targeted numerical rotation
matrix, solving the problem.  To complete the details of the computation,
 we replace the elements
$E_{ab}$ in \Eqn{basicHorn.eq} by a general orthonormal rotation
matrix with columns $\Vec{X}= (x_1,x_2,x_3)$, $\Vec{Y}$, and $\Vec{Z}$,
 scaling by $1/3$, thus obtaining the special $4 \times
4$ profile matrix $K$ whose elements in terms of a known numerical
matrix $R = \left[\Vec{X} | \Vec{Y} | \Vec{Z} \right]$ 
 (transposed in the algebraic
expression for $K$ due to the $R^{\t}$ in $\Delta_{\mbox{\scriptsize BI}}$) are
\begin{equation}
K(R) = \frac{1}{3} \left[
\begin{array}{cccc}
 x_1+y_2+z_3 &  y_3 - z_2 & z_1 - x_3 & x_2 - y_1 \\
  y_3 - z_2 & x_1-y_2-z_3 & x_2+y_1 & x_3+z_1 \\
  z_1 - x_3 & x_2+y_1 & -x_1+y_2-z_3 & y_3+z_2 \\
  x_2 - y_1 & x_3+z_1 & y_3+z_2 & -x_1-y_2+z_3 \\
\end{array}
\right] \ .
 \label{theKRmatrix.eq}
\end{equation}

Determining the algebraic eigensystem of \Eqn{theKRmatrix.eq} is a 
nontrivial task.
However, as we know, any orthogonal 3D rotation matrix $R(q)$, or
equivalently, $R^{\t}(q) = R(\bar{q})$,  can also be ideally
expressed in terms of quaternions via \Eqn{qrot.eq}, and this
yields an alternate useful algebraic form
\begin{eqnarray}
\lefteqn{K(q) = } \nonumber\\
&& \hspace{-.3in}\frac{1}{3} \left[
 \begin{array}{cccc}
\textstyle \hspace{-.075in}3 {q_{0}}^2-{q_{1}}^2-{q_{2}}^2-{q_{3}}^2 &
 4 q_{0} q_{1} &
     4 q_{0} q_{2} & 4 q_{0} q_{3} \hspace*{-.1in} \\ 
 4 q_{0} q_{1} &
\textstyle -{q_ {0}}^2 + 3 {q_{1}}^2-{q_{2}}^2-{q_{3}}^2 &
     4 q_{1} q_{2} & 4 q_{1} q_{3} \hspace*{-.1in}\\ 
 4 q_{0} q_{2} & 4 q_{1} q_{2} &
\textstyle -{q_{0}}^2-{q_{1}}^2 + 3 {q_{2}}^2-{q_{3}}^2 &
    4 q_{2} q_{3}  \hspace*{-.1in}\\  
 4 q_{0} q_{3} & 4 q_{1} q_{3} & 4 q_{2} q_{3} &
   \textstyle  -{q_{0}}^2-{q_{1}}^2-{q_{2}}^2 + 3 {q_{3}}^2   \hspace*{-.1in}\\   
\end{array}
\right]  \ 
\label{theKqmatrix.eq}
\end{eqnarray}
This equation then allows us to quickly prove that $K$ has the correct
properties to solve for the appropriate quaternion corresponding to $R$.
First we note that  the coefficients $p_n$ of the
eigensystem are simply constants,
\[ \begin{array}{c@{\hspace{.25in}}c@{\hspace{.25in}}c@{\hspace{.25in}}c}
p_1 = 0 & p_2 = -\frac{2}{3} & p_3 = - \frac{8}{27} & 
p_4 = - \frac{1}{27} 
\end{array} \ . \]
Computing the eigenvalues and eigenvectors using the symbolic
quaternion form,  we see that the eigenvalues are constant, with
maximal eigenvalue exactly one, and the eigenvectors are almost
trivial, with the maximal  eigenvector being the inverse of the
quaternion $q$ that corresponds to the (numerical) rotation matrix:
\begin{eqnarray}
\epsilon & = & \{ 1, \; -\frac{1}{3},\; -\frac{1}{3},\; -\frac{1}{3}
\} \label{Keigvals.eq} \\
r & = &\left\{ 
  \left[\begin{array}{c} q_0 \\  q_1 \\  q_2 \\   q_3 \end{array}\right], \;
    \left[ \begin{array}{c} -q_1 \\  q_0 \\  0 \\ 0 \end{array}\right], \;
   \left[ \begin{array}{c} -q_2 \\  0  \\  q_0 \\  0 \end{array}\right], \;
   \left[ \begin{array}{c} -q_3 \\  0 \\ 0 \\  q_0 \end{array}\right]
        \right\} \ . 
   \label{Keigvecs.eq}
\end{eqnarray}
The first column is the quaternion $r_{\opt}$, with
 $ \Delta_{\mbox{\scriptsize BI}}  (r_{\opt}) = 1$.  (This would be 3 if we
 had not divided by 3 in the definition of $K$.)

\emph{Alternate version.}  From the quaternion barycenter work of Markley 
 et al.~\cite{MarkleyEtal-AvgingQuats2007}, we know that \Eqn{theKqmatrix.eq} actually
 has a much simpler form with the same unit eigenvalue and natural quaternion eigenvector.
 (This form appears naturally below in the 4D extension of the Bar-Itzhack algorithm.)
 If we simply take \Eqn{theKqmatrix.eq}  multiplied by 3, add the constant term
 $ I_{4} =    ( {q_{0}}^2 + {q_{1}}^2 +  {q_{2}}^2 +   {q_{3}}^2 ) I_{4}$ , and divide by 4,
  we get a more compact quaternion form of the matrix, namely
\begin{eqnarray}
K '(q) & = &  \left[
 \begin{array}{cccc}
  {q_{0}}^2  &
    q_{0} q_{1} &
       q_{0} q_{2} &    q_{0} q_{3}  \\ 
    q_{0} q_{1} &
   {q_{1}}^2  &
       q_{1} q_{2} &   q_{1} q_{3} \\ 
   q_{0} q_{2} &  q_{1} q_{2} &
    {q_{2}}^2 &
      q_{2} q_{3}    \\  
    q_{0} q_{3} &  q_{1} q_{3} &   q_{2} q_{3} &   
    {q_{3}}^2   \\   
\end{array}
\right]  \  .
\label{theBCsimplematrix.eq}
\end{eqnarray}
This has  vanishing determinant and trace $\,  \tr K' =  1=-p_{1}$, with
all other $p_{k}$ coefficients vanishing, and  eigensystem with eigenvalues
identical to \Eqn{theKqmatrix.eq}: 
\begin{eqnarray} \label{Ksimpleeigvals.eq}
\epsilon & = & \{ 1, \;0 ,\;0,\; 0 \}  \\
r& = &\left\{ 
  \left[\begin{array}{c} q_0 \\   q_1 \\   q_2 \\   q_3 \end{array}\right], \;
    \left[ \begin{array}{c} -q_1 \\  q_0 \\  0 \\ 0 \end{array}\right], \;
   \left[ \begin{array}{c} -q_2 \\  0  \\  q_0 \\  0 \end{array}\right], \;
   \left[ \begin{array}{c} -q_3 \\  0 \\ 0 \\  q_0 \end{array}\right]
        \right\} \ . 
   \label{Ksimpleeigvecs.eq}
\end{eqnarray}
As elegant as this is, in practice, our numerical input data are from the
$3\times 3$ matrix $R$ itself, and not the quaternions, so we will almost
always just use those numbers in \Eqn{theKRmatrix.eq} to solve the problem.

\qquad

 {\bf Completing the solution.}  In typical applications, \emph{the solution is immediate, requiring only trivial algebra}.  The maximal eigenvalue is always known in advance to be unity for any valid rotation matrix, so we need only  to compute the eigenvector from the
numerical matrix \Eqn{theKRmatrix.eq} with unit eigenvalue.  We simply
compute  any column
of the adjugate matrix of  $K(R) - I_{4}$,  or  solve the equivalent  linear
equations of the form
\begin{equation}
\begin{array}{c p{0.5in}c}
\left( K(R)- 1  * I_{4} \right) \cdot 
\left[ \begin{array}{c} 1\\v_1\\v_2\\v_3 \\ 
\end{array} \right]  = 0  &&
 q = r_{\opt}  = \mbox{normalize\ }\left[ \begin{array}{c} 1\\v_1\\v_2\\v_3 \\ 
\end{array} \right] \\
\end{array} \ .
\label{TheBIsolnQuat.eq}
\end{equation}
As always, one may need to check for degenerate special cases. 

\qquad

{\bf Non-ideal cases.} It is important to note, as emphasized by Bar-Itzhack, that if there are
\emph{significant errors} in the numerical matrix $R$, then the actual non-unit 
maximal eigenvalue of $K(R)$ 
can be computed numerically or algebraically as usual, and then that eigenvalue's
eigenvector  
determines the \emph{closest}  normalized quaternion to the errorful rotation
matrix, which can be very useful since such a quaternion always produces
a valid rotation matrix.  

 In any case,  \emph{up to an overall sign},  $r_{\opt}$ is the 
desired numerical quaternion $q$  corresponding to the
target numerical rotation matrix $R= R(q)$  .
 In some circumstances, one is looking for a
uniform statistical distribution of quaternions, in which case the
overall sign of $q$ should be chosen \emph{randomly}.  

 The Bar-Itzhack approach solves the
problem of extracting the quaternion of an arbitrary numerical 3D rotation
matrix in a fashion that involves no singularities and only trivial
testing for special cases, thus essentially making the traditional
methods obsolete.  


\comment{ 
\begin{quote}
Below, for completeness, we list the cumbersome unnormalized expressions for $q_{k}$,
 which, though they are 
elementary cubic  expressions in the nine numerical elements of $R$,
have 32 distinct terms in the numerator and 64 in the denominator.  The latter
is used as the $q_0$ term in the following expression,
before normalizing to quaternion-compatible form:\\
 \begin{equation*}
\left. \begin{array}{rcl}
q_{0} & = & \parbox[t]{5.25in}{$
  27 + 2 x_1 y_3 z_2-2 x_1 y_2 z_3+2 x_3 y_2 z_1-2 x_2 y_3 z_1-2 x_3 \
y_1 z_2+2 x_2 y_1 z_3+{x_1}^2 y_2-x_1 {y_1}^2+x_1 {y_2}^2+x_1 \
{y_3}^2-2 x_2 x_1 y_1+6 x_1 y_2-6 x_2 y_1-{x_2}^2 y_2+{x_3}^2 y_2-2 \
x_2 y_1 y_2-2 x_2 x_3 y_3-2 x_3 y_1 y_3+{x_1}^2 z_3-x_1 {z_1}^2+x_1 \
{z_2}^2+x_1 {z_3}^2-2 x_3 x_1 z_1+6 x_1 z_3-6 x_3 z_1-2 x_2 x_3 z_2-2 \
x_2 z_1 z_2+{x_2}^2 z_3-{x_3}^2 z_3-2 x_3 z_1 z_3-{x_1}^3-3 \
{x_1}^2-{x_2}^2 x_1-{x_3}^2 x_1+9 x_1-3 {x_2}^2-3 {x_3}^2+y_2 \
{z_1}^2-y_2 {z_2}^2+y_2 {z_3}^2-2 y_1 y_3 z_1-2 y_2 y_3 z_2-6 y_3 \
z_2-2 y_1 z_1 z_2+{y_1}^2 z_3+{y_2}^2 z_3-{y_3}^2 z_3+6 y_2 z_3-2 y_3 \
z_2 z_3-{y_2}^3-3 {y_1}^2-3 {y_2}^2-y_2 {y_3}^2-3 {y_3}^2-{y_1}^2 \
y_2+9 y_2-{z_3}^3-3 {z_1}^2-3 {z_2}^2-3 {z_3}^2-{z_1}^2 z_3-{z_2}^2 \
z_3+9 z_3 $ }   \vspace{0.1in}\\ 
   q_{1}  &  = &  \parbox[t]{5.25in}{${x_1}^2 y_3+{x_2}^2 y_3-{x_3}^2 y_3+6 x_1 y_3-2 x_1 x_3 y_1-6 
x_3 y_1-2 x_2 x_3 y_2+2 x_1 x_2 z_1+6 x_2 z_1-{x_1}^2 z_2+{x_2}^2  
z_2-{x_3}^2 z_2-6 x_1 z_2+2 x_2 x_3 z_3-{y_3}^2 z_2+y_3 {z_1}^2+y_3 
{z_2}^2-y_3 {z_3}^2+2 y_2 y_3 z_3+2 y_1 y_2 z_1-{y_1}^2 z_2+{y_2}^2 
z_2-2 y_1 z_1 z_3-2 y_2 z_2 z_3-{y_3}^3-{y_1}^2 y_3-{y_2}^2 y_3+9 
y_3+{z_2}^3+z_2 {z_3}^2+{z_1}^2 z_2-9 z_2$}  \\ [0.5in]
   q_{2}  & =  &   \parbox[t]{5.25in}{$   -x_3 {y_1}^2-x_3 {y_2}^2+x_3 {y_3}^2-6 x_3 y_2+6 x_2 y_3+2 
x_1 y_1 y_3+2 x_2 y_2 y_3+{x_3}^2 z_1-x_3 {z_1}^2-x_3 {z_2}^2+x_3 
{z_3}^2-2 x_1 x_3 z_3-{x_1}^2 z_1+{x_2}^2 z_1-2 x_1 x_2 z_2+2 x_1 z_1 
z_3+2 x_2 z_2 z_3+{x_3}^3+{x_1}^2 x_3+{x_2}^2 x_3-9 x_3-{y_1}^2 
z_1+{y_2}^2 z_1+{y_3}^2 z_1+6 y_2 z_1-6 y_1 z_2-2 y_1 y_2 z_2-2 y_1 
y_3 z_3-{z_1}^3-z_1 {z_2}^2-z_1 {z_3}^2+9 z_1     $}  \\ [0.5in]
    q_{3}  & =  &   \parbox[t]{5.25in}{$  -{x_2}^2 y_1+x_2 {y_1}^2-x_2 {y_2}^2+x_2 {y_3}^2+2 x_1 x_2 
y_2+{x_1}^2 y_1-{x_3}^2 y_1-2 x_1 y_1 y_2+2 x_1 x_3 y_3-2 x_3 y_2 
y_3+x_2 {z_1}^2-x_2 {z_2}^2+x_2 {z_3}^2+6 x_2 z_3-6 x_3 z_2-2 x_1 z_1 
z_2-2 x_3 z_2 z_3-{x_2}^3-{x_1}^2 x_2-{x_3}^2 x_2+9 x_2+y_1 
{z_1}^2-y_1 {z_2}^2-y_1 {z_3}^2+6 y_3 z_1+2 y_2 z_1 z_2-6 y_1 z_3+2 
y_3 z_1 z_3+{y_1}^3+y_1 {y_2}^2+y_1 {y_3}^2 - 9 y_1   $}  \\
\end{array} \right\} \ .  \label{IBeigenvect.eq}
\end{equation*}
\end{quote}
  }  
  
\qquad

\subsection{Extracting Quaternion Pairs from 4D Rotation Matrices}  

We know from \Eqn{doublequat.eq} that any 4D orthogonal matrix $R_{4}(p,q)$ can be expressed  
as a quadratic form in two independent unit quaternions.  This is a
consequence of the fact that the 6-parameter orthogonal group $\SO{4}$ is double
covered by the composition of two smaller 3-parameter unitary groups,
that is $\SU{2}\times \SU{2}$;  the group $\SU{2}$ has essentially the
same properties as a single quaternion, so it is not surprising that
$\SO{4}$ should be related to a pair of quaternions.

 We begin our treatment of the 4D case by  extending  \Eqn{deltaBI.eq}
to 4D  with a numerical $\SO{4}$ 
matrix $R_{4}$, giving us a Bar-Itzhack measure to maximize  of the form
\begin{equation}
\Delta_{\mbox{\scriptsize 4:BI}} = \tr{S(\ell,r) \cdot {R_{4}}^{\t}} = \ell\cdot K_{4}(R_{4}) \cdot r 
   = \ell\cdot K_{4}(p,q) \cdot r \ .
\label{delta4DBI.eq}
\end{equation}
Here $(\ell,r)$ are the left and right quaternions over which we are varying
the measure, and $K_{4}(R_{4})$ is the 4D generalization of  \Eqn{theKRmatrix.eq}.
To compute  $K_{4}(R_{4})$, we define   a general 4D orthonormal rotation
matrix with columns $\Vec{W}= (w_0,w_1,w_2,w_3)$, etc., so the
matrix takes the form $R_{4} = \left[\Vec{W} |\Vec{X} | \Vec{Y} |
\Vec{Z} \right]$, producing a numerical profile matrix of the form (taking into
account the transpose in \Eqn{delta4DBI.eq})
\begin{equation}
K_{4}(R_{4}) = \frac{1}{4} 
\left[
\begin{array}{cccc}
 w_0+x_1+y_2+z_3 & -w_1+x_0+y_3-z_2 & -w_2-x_3+y_0+z_1 & -w_3+x_2-y_1+z_0 \\
 w_1-x_0+y_3-z_2  & w_0+x_1-y_2-z_3  & -w_3+x_2+y_1-z_0 &  w_2+x_3+y_0+z_1 \\
 w_2-x_3-y_0+z_1 & w_3+x_2+y_1+z_0 &  w_0-x_1+y_2-z_3  &  -w_1-x_0+y_3+z_2 \\
 w_3+x_2-y_1-z_0 & -w_2+x_3-y_0+z_1 & w_1+x_0+y_3+z_2 &   w_0-x_1-y_2+z_3 \\
\end{array}
\right] \ .
 \label{the4DKRmatrix.eq}
\end{equation}
Now, from \Eqn{doublequat.eq}, we know that we also have an analog to
\Eqn{theKqmatrix.eq}, and for $R_{4}(p,q)$ this takes
the remarkably compact algebraic form
\begin{equation}
K_{4}(p,q) =  \left[
\begin{array}{cccc}
 p_0 q_0 & p_0 q_1 & p_0 q_2 & p_0 q_3 \\
 p_1 q_0 & p_1 q_1 & p_1 q_2 & p_1 q_3 \\
 p_2 q_0 & p_2 q_1 & p_2 q_2 & p_2 q_3 \\
 p_3 q_0 & p_3 q_1 & p_3 q_2 & p_3 q_3 \\
\end{array}
\right] \ .
\label{the4DKpqmatrix.eq}
\end{equation}
This matrix is exactly the outer product of  $p$ and
$q$,  with vanishing determinant, rank 1, and trace $(p
\cdot q)$, which makes it extremely simple. The eigensystem is 
\begin{eqnarray}
\epsilon & = & \{ p\cdot q, \; 0,\; 0,\; 0 \} \label{K4Deigvals.eq} \\
r_{\mbox{right}} & = &\left\{ 
  \left[\begin{array}{c} p_0 \\  p_1 \\ p_2 \\  p_3 \end{array}\right], \;
    \left[ \begin{array}{c} -q_1 \\  q_0 \\  0 \\ 0 \end{array}\right], \;
   \left[ \begin{array}{c} -q_2 \\  0  \\  q_0 \\  0 \end{array}\right], \;
   \left[ \begin{array}{c} -q_3 \\  0 \\ 0 \\  q_0 \end{array}\right]
        \right\}\\
\ell_{\mbox{left}} & = &\left\{ 
  \left[\begin{array}{c} q_0 \\  q_1 \\  q_2 \\ 
      q_3 \end{array}\right], \; 
    \left[ \begin{array}{c}- p_1 \\  p_0 \\  0 \\ 0 \end{array}\right], \;
   \left[ \begin{array}{c}- p_2 \\  0  \\  p_0 \\  0 \end{array}\right], \;
   \left[ \begin{array}{c}- p_3 \\  0 \\ 0 \\  p_0 \end{array}\right]
        \right\} \ , 
   \label{K4Deigvecs.eq}
\end{eqnarray}
with an interesting swap between $p$ and $q$ in the zero eigenvectors,
and the sole non-vanishing eigenvalue is just $\epsilon = \tr
K_{4}(p, q)= p\cdot q$, which is  a convenient  function of the numerical data.
  Thus the left and right eigenvectors can be easily computed
from the numerical data in \Eqn{the4DKRmatrix.eq} using the eigenvalue extracted
from the trace.  Again, if a statistical distribution in the
double quaternion space is desired, the signs can be chosen randomly,
consistent with the sign of $\tr K_{4}(R_{4})$.

Once again, we can simply take the \emph{numerical} value of the eigenvalue
of $K_{4}(R_{4})$, which  is  just the trace, and solve  for the \emph{right} eigenvector
$r_{\mbox{right}}$,
which will be the \emph{left} quaternion $p$, and for the \emph{left} eigenvector $\ell_{\mbox{left}}$
(the eigenvector of the transpose of $K_{4}(R_{4})$), which will be the \emph{right} quaternion $q$.
We can either use any (normalized) adjugate column or just solve some permutation
of the following linear equations  directly for the eigenvectors. \emph{No further
computation is required.}
\begin{equation}
\begin{array}{c p{0.5in}c}
\left( K(R)- \tr K(R)  * I_{4} \right) \cdot 
\left[ \begin{array}{c} 1\\v_1\\v_2\\v_3 \\ 
\end{array} \right]  = 0  &&
p = r_{\opt} =   \mbox{normalize\ }\left[ \begin{array}{c} 1\\v_1\\v_2\\v_3 \\ 
\end{array} \right] \\
\end{array} 
\label{TheBIsolnPQuat1.eq}
\end{equation}
\begin{equation}
\begin{array}{c p{0.5in}c}
\left( (K(R))^{\t}- \tr K(R)  * I_{4} \right) \cdot 
\left[ \begin{array}{c} 1\\v'_1\\v'_2\\v'_3 \\ 
\end{array} \right]  = 0  &&
q = \ell_{\opt}   = \mbox{normalize\ }\left[ \begin{array}{c} 1\\v'_1\\v'_2\\v'_3 \\ 
\end{array} \right] \\
\end{array} \ .
\label{TheBIsolnPQuat2.eq}
\end{equation}
The solution to our problem is thus $R_{4}(p,q) = R_{4}(r_{\opt},\ell_{\opt} )$.
As in 3D, if the numerical matrix $R_{4}$ has some moderate errors and the
maximum eigenvalue \emph{differs} significantly from $\tr K(R)$, we can
solve for the actual maximal eigenvalue and insert that into
Eqs.~(\ref{TheBIsolnPQuat1.eq}) and (\ref{TheBIsolnPQuat2.eq}) to 
find the left and right eigenvectors numerically.

   There is one important caveat:  the 3D quaternion
rotation $R_{3}(q)$ does not care what the sign of $q$ is, but the 4D quaternion rotation $R_{4}(p,q)$ 
is only invariant under \emph{both} $p\to -p$ and $q\to -q$ in tandem.  To ensure that
$R_{4}(p,q)$ is the same matrix, the signs of the quaternions must be adjusted
after the initial computation so that  the sign of $(\ell\cdot r)$ matches
the sign of the numerical input value of ${R_{4}}_{(1,1)} =\tr K_{4}(R_{4})= p\cdot q$. 
That guarantees that the solution describes the same matrix that we used as input, and not its negative. 

\comment{ 
v0Left = 
\left[  27 {w_0}^3+9 x_1 {w_0}^2+9 y_2 {w_0}^2+9 z_3 {w_0}^2+3 \
{w_1}^2 w_0+3 {w_2}^2 w_0+3 {w_3}^2 w_0+3 {x_0}^2 w_0-3 {x_1}^2 w_0-3 \
{x_2}^2 w_0-3 {x_3}^2 w_0+3 {y_0}^2 w_0-3 {y_1}^2 w_0-3 {y_2}^2 w_0-3 \
{y_3}^2 w_0+3 {z_0}^2 w_0-3 {z_1}^2 w_0-3 {z_2}^2 w_0-3 {z_3}^2 w_0+6 \
w_1 x_0 w_0+6 w_2 y_0 w_0-6 x_2 y_1 w_0+6 x_1 y_2 w_0+6 w_3 z_0 w_0-6 \
x_3 z_1 w_0-6 y_3 z_2 w_0+6 x_1 z_3 w_0+6 y_2 z_3 \
w_0-{x_1}^3-{y_2}^3-{z_3}^3-x_1 {x_2}^2-x_1 {x_3}^2+x_1 {y_0}^2-x_1 \
{y_1}^2+x_1 {y_2}^2+x_1 {y_3}^2-y_2 {y_3}^2+x_1 {z_0}^2+y_2 \
{z_0}^2-x_1 {z_1}^2+y_2 {z_1}^2+x_1 {z_2}^2-y_2 {z_2}^2+x_1 \
{z_3}^2+y_2 {z_3}^2-{w_1}^2 x_1+{w_2}^2 x_1+{w_3}^2 x_1-{x_0}^2 x_1-2 \
w_1 x_0 x_1-2 w_1 w_2 x_2-2 w_2 x_0 x_2-2 w_1 w_3 x_3-2 w_3 x_0 x_3+2 \
w_2 x_1 y_0-2 w_1 x_2 y_0-2 x_0 x_2 y_0-2 w_1 w_2 y_1-2 w_2 x_0 y_1-2 \
x_1 x_2 y_1-2 w_1 y_0 y_1-2 x_0 y_0 y_1+{w_1}^2 y_2-{w_2}^2 \
y_2+{w_3}^2 y_2+{x_0}^2 y_2+{x_1}^2 y_2-{x_2}^2 y_2+{x_3}^2 \
y_2-{y_0}^2 y_2-{y_1}^2 y_2+2 w_1 x_0 y_2-2 w_2 y_0 y_2-2 x_2 y_1 \
y_2-2 w_2 w_3 y_3-2 x_2 x_3 y_3-2 w_3 y_0 y_3-2 x_3 y_1 y_3+2 w_3 x_1 \
z_0-2 w_1 x_3 z_0-2 x_0 x_3 z_0+2 w_3 y_2 z_0-2 w_2 y_3 z_0-2 y_0 y_3 \
z_0-2 w_1 w_3 z_1-2 w_3 x_0 z_1-2 x_1 x_3 z_1+2 x_3 y_2 z_1-2 x_2 y_3 \
z_1-2 y_1 y_3 z_1-2 w_1 z_0 z_1-2 x_0 z_0 z_1-2 w_2 w_3 z_2-2 x_2 x_3 \
z_2-2 w_3 y_0 z_2-2 x_3 y_1 z_2+2 x_1 y_3 z_2-2 y_2 y_3 z_2-2 w_2 z_0 \
z_2-2 y_0 z_0 z_2-2 x_2 z_1 z_2-2 y_1 z_1 z_2+{w_1}^2 z_3+{w_2}^2 \
z_3-{w_3}^2 z_3+{x_0}^2 z_3+{x_1}^2 z_3+{x_2}^2 z_3-{x_3}^2 \
z_3+{y_0}^2 z_3+{y_1}^2 z_3+{y_2}^2 z_3-{y_3}^2 z_3-{z_0}^2 \
z_3-{z_1}^2 z_3-{z_2}^2 z_3+2 w_1 x_0 z_3+2 w_2 y_0 z_3+2 x_2 y_1 \
z_3-2 x_1 y_2 z_3-2 w_3 z_0 z_3-2 x_3 z_1 z_3-2 y_3 z_2 z_3  \right]
  }


\section{Two-Dimensional Limit of 3D Problem}
\label{2Dlimit.app}


All rotations of the type we have been trying to optimize reduce to a
rotation in a 2D plane, which in 3D is defined by the plane
perpendicular to the eigenvector $\Hat{n}$ of the rotation matrix
\Eqn{qrot.eq}.  Data sets that are highly linear, determining
a robust straight line from least squares, can even circumvent the
RMSD problem entirely: a very good rotation matrix can be calculated
from the direction $\Hat{x}$ determined by the line fitted to the data
set $\{x_{i}\}$, and the similar direction $\Hat{y}$ corresponding to
the reference data set $\{y_{i}\}$.  An optimal rotation matrix in 3D
is then simply 
\begin{equation}
R(\theta, \Hat{n}) = R(\arccos{(\Hat{x}\cdot \Hat{y})},
\widehat{\Hat{x}\times \Hat{y}})\ ,
\label{Just2DRot.eq}
\end{equation}
which is easily generalized to
any dimension by isolating just the projections of vectors to the
plane determined by $\Hat{x}$ and $\Hat{y}$, and rotating in that 2D
basis.  Thus we conclude that, in general, if we had access to a
prescient preconditioning rotation of the proper form, the entire RMSD
problem would reduce to a very simple rotation in some
  2D plane parameterized by a single angle.  We can simulate
this, giving a massively simpler set of expressions, by assuming the
data are coplanar, all having $z=0$ (or more conditions in higher
dimensions) and thus lying in the canonical $\{\Hat{x},\Hat{y}\}$ plane, for example.
This reduces our fundamental RMSD profile matrix \Eqn{basicHorn.eq} for $M$ to
\begin{equation} 
M_{z=0} = \left[\begin{array}{cccc}
x + y & 0 & 0 & c\\
0 & x - y & C & 0 \\
0 & C & -x + y & 0 \\
c & 0 & 0   & -x - y \\
\end{array} \right] \ ,
\label{Mfor2D.eq}
\end{equation}
where $x=E_{xx}$, $y=E_{yy}$, $c = E_{xy} - E_{yx}$, and $C = E_{xy} +
E_{yx}$.  Then $p_2 = -c^2 - C^2 - 2 (x^2 + y^2)$, $\,p_3 = 0$,
and $\,p_4 = (c^2 + (x + y)^2) (C^2 + (x - y)^2)$, and similarly for
the other cyclic cases, $x=0$ and $y=0$.  The $p_{2}$ and $p_{4}$  are obviously
functions of only two variables, $u = c^2 + (x + y)^2$ and $v=C^2 + (x
- y)^2$, so we can write in general $p_2 = - u - v$ and $p_4 = u v$.
The eigenvalue equation $\det[M - e I_{4}] = e^4 + e^3 p_1 +  e^2 p_2 + e p_3 + p_4  = 0 $
reduces to $ e^4+e^2 p_2 + p_4 =0$ and the
eigenvalues become 
$\epsilon  =  \left( \sqrt{u}, \sqrt{v}, -\sqrt{v}, -\sqrt{u}\, \right)$,
while the normalized (quaternion) eigenvectors become
\begin{equation} 
\begin{array}{rcccc}
q = & \left\{ 
\left[ \!\begin{array}{c} 
\frac{x + y + \sqrt{u}}{\sqrt{c^2+\left(x+y+\sqrt{u}\right)^2}}\\
0\\ 0\\ \frac{c}{\sqrt{c^2+\left(x+y+\sqrt{u}\right)^2}}\\  
\end{array} \! \right] , \right. \! & \!\!
\left[ \!\begin{array}{c} 
0\\ \frac{x - y + \sqrt{v}}{\sqrt{C^2 +\left(x-y+\sqrt{v}\right)^2}}\\ 
\frac{C}{\sqrt{C^2 + \left(x-y+\sqrt{v}\right)^2}}\\ 0\\  
\end{array} \! \right] ,  \!\! & \!\!
\left[\! \begin{array}{c} 
0\\  \frac{x - y - \sqrt{v}}{\sqrt{C^2
    +\left(x-y-\sqrt{v}\right)^2}}\\  \frac{C}{\sqrt{C^2
    +\left(x-y-\sqrt{v}\right)^2}}\\ 0\\  
\end{array} \!\right] , \!\!  & \! \! \left.
\left[ \!\begin{array}{c} \frac{ x + y - \sqrt{u}}{\sqrt{c^2 +\left(x+y
        -\sqrt{u}\right)^2}}\\ 0\\ 0\\ \frac{c}{\sqrt{c^2 +\left(x+y
        -\sqrt{u}\right)^2}}\\  
\end{array} \!\right] \right\}\\ \end{array}  .
\label{eigvec2D.eq}
\end{equation}
The leading eigenvalue and its eigenvector produce this
optimal rotation in the $\{\Hat{x},\Hat{y}\}$ plane:
\begin{equation}
R_{\mbox{\footnotesize 2D}} = \left[
\begin{array}{cc}
  {\displaystyle \frac{\left(x+y+\sqrt{u}\right)^2-c^2}{c^2+\left(x+y+\sqrt{u}\right)^2} } &
  {\displaystyle  -\frac{2 c
   \left(x+y+\sqrt{u}\right)}{c^2+\left(x+y+\sqrt{u}\right)^2}  } \\ [0.2in]
     {\displaystyle \frac{2 c \left(x+y+\sqrt{u}\right)}{c^2+\left(x+y+\sqrt{u}\right)^2}  }&
  {\displaystyle  \frac{\left(x+y+\sqrt{u}\right)^2-c^2}{c^2+\left(x+y+\sqrt{u}\right)^2} }
    \\
\end{array} 
\right] \ . 
\label{rotMat2DZ.eq}
\end{equation}
\paragraph{Yet Another Form.}  However, we have neglected something.
How does this look  if we simply go back to the data matrices for 2D? 
Let us first write down the 2D version of \Eqn{RMSD-basic.eq}, taking
$E_{ab} = \rule{0in}{1.1em}\sum_{k=1}^{N} 
     [x_{k}]_{\textstyle_a} \:[y_{k}]_{\textstyle_b} $ 
 for $a,b = \{1,2\}$, so the raw form for 
the spatial RMSD task is to find the rotation matrix 
\[ R_{2}(\theta) =
\left[\begin{array}{cc} \cos \theta & -\sin\theta \\
\sin \theta & \cos\theta \end{array} \right] \]
 maximizing
\begin{equation} 
 \Delta_{2} = \sum_{k=1}^{N} \left(R_{2} \cdot x_{k}\right) \cdot y_{k} =
 \sum_{a=1,b=1}^{2} {R_{2}}^{ba} E_{ab} \ = \  
    \left( E_{xx} + E_{yy}\right) \cos \theta +
     \left( E_{xy} - E_{yx}\right) \sin \theta  \ .
\label{2DREapp.eq}
\end{equation}
We can either differentiate with respect to $\theta$ and set
$\Delta_{2}'(\theta) =0$, or simply observe directly that
$\Delta_{2}(\theta)$ is largest when the vector
$(\cos\theta,\sin\theta)$  is parallel to its coefficients; both
arguments lead to the solution
\begin{eqnarray} 
\tan\theta & =& \frac{ E_{xy} - E_{yx}}{E_{xx} + E_{yy}}= \frac{N}{M}\\[.2in]
(\cos\theta,\,\sin\theta) &=& 
\left(\frac{M}{\sqrt{M^2+N^2}},\frac{N}{\sqrt{M^2+N^2}}\right)
\ .
\label{2Dxsoln.eq}
\end{eqnarray}
Now we can see that 
\begin{equation} 
\begin{array}{rcccl}
x+ y & = & E_{xx}+E_{yy} &=& M \\
c & = & E_{xy}-E_{yx}& =& N \\[0.05in]
u & = & (E_{xx}+E_{yy})^2 + ( E_{xy}-E_{yx})^2 & = & M^2 +  N^2 \\[0.02in]
\epsilon & = & \lambda & = & \pm \sqrt{M^2 +  N^2}  \ , \end{array}
\label{xycofMN.eq}
\end{equation}
and $ c^2+\left(x+y+\sqrt{u}\right)^2 = 2 \lambda (M + \lambda)$. 
Thus in fact the profile matrix becomes
\begin{eqnarray} 
{\mathbf M}_{2} & = & \left[ \begin{array}{cc}  M & N \\ N &
    -M \end{array} \right] 
\label{M2MN.eq}
\end{eqnarray}
and this has eigenvalues exactly $\epsilon = \pm \sqrt{u} = \pm
\sqrt{M^2 +  N^2}$.  The eigenvectors are the first and last
columns of \Eqn{eigvec2D.eq} expressed in terms of \Eqn{xycofMN.eq},
so the maximal eigenvector is $(a,b)$, where
\begin{equation} 
\begin{array}{rcccccl}
   a & = & \cos (\theta/2) & =&  \displaystyle \frac {\lambda + M}
{\sqrt{2 \lambda(\lambda + M)}}
& = & \displaystyle  \sqrt{\frac {\lambda + M}{2 \lambda}}
  \\
   b & = & \sin(\theta/2) & = & 
 \displaystyle \frac {N}{\sqrt{2 \lambda(\lambda + M)}} 
& = & \displaystyle \sign{N} \sqrt{\frac {\lambda - M}{2 \lambda}} 
 \ . \end{array}
\label{2Dabsln.eq}
\end{equation}
(Note the crucial $(\sign N)$ factor.)
Going back to our original 2D rotation matrix in \Eqn{rotMat2DZ.eq}
and substituting \Eqn{xycofMN.eq}, we recover our optimal result,
 namely
\begin{eqnarray}
 R_{2}(\theta) & = &
\left[ \begin{array}{cc} {\displaystyle \frac{M}{\sqrt{M^2 +  N^2}}} & -
    {\displaystyle \frac{N}{\sqrt{M^2 +  N^2}}  }\\ 
 {\displaystyle \frac{N}{\sqrt{M^2 +  N^2}} }&  {\displaystyle \frac{M}{\sqrt{M^2 +  N^2}} }\end{array} 
\right] \label{rotMat2DX.eq}  \\[0.1in]
 & = & \left[ \begin{array}{cc} \cos \theta & -\sin\theta \\
\sin \theta & \cos\theta \end{array} \right]  \label{rotMat2DY.eq} \ .
\end{eqnarray}

These results are interesting to study because, despite the complexity
of the general solution, 
the intrinsic algebraic
structure of any RMSD problem is entirely characterized by a planar
rotation such as that described by \Eqn{rotMat2DZ.eq} and 
\Eqn{rotMat2DX.eq}.

\section{Evaluating the 3D Orientation Frame Solution.}
\label{Evaluate3DOrient.sec}


 The validity of our approximate
chord-measures for determining the optimal global frame rotation can
be evaluated by comparing their outcomes to the precise geodesic
arc-length measure of \Eqn{QRMSD-basic-geodesic.eq}.  The latter is tricky to
optimize, but choosing appropriate techniques, e.g., in the
Mathematica ${\mathtt {FindMinimum[\ ]}}$ utility, it is possible to
determine good numerical solutions without writing custom code; in our
experiments, fluctuations due to numerical precision limitations were
noticeable, but presumably conventional conditioning techniques, which
we have not attempted to explore, could improve that significantly.
We employed a collection of $1000$  simulated quaternion data
sets of length $100$ for the reference cases, then imposed a normal
distribution of random noise on the reference data, followed by a
global rotation of all those noisy data points distributed around
 45$^{\circ}$ to produce a
corresponding collection of  corresponding quaternion test data
sets to be aligned.  (Observe that we do {\it not\/} expect the
optimal rotation angles to match the exact global rotations, though they will
be nearby.)

 We then collected the optimal quaternions for the following cases:
\begin{itemize} 
\item[(a)] {\bf Arc-Length (numerical).} This is the ``gold
  standard,'' modulo the occasional data pair that seems to challenge
  the numerical stability of the computation (which was to be
  expected).  We obtained the data set {(a)} of quaternions that
  numerically minimized the nonlinear geodesic arc-length-squared
  measure of \Eqn{QRMSD-basic-geodesic.eq}; this is in principle the best
  estimate one can possibly get for the optimal quaternion rotations
  to align a set of 3D test-frame triads with a corresponding set
  of reference-frame triads.  There is no known way to find this set of
  optimal quaternions using our linear algebra methods.
\item[(b)] {\bf Chord-Length (numerical and algebraic).}  This
  approach, designated as the data set {(b)}, is based on the approximation to
  \Eqn{RMSD-basic.eq} illustrated in Fig~\ref{arc-chord.fig}\,,
  replacing the arc-length by the chord-length, which amounts to
  removing the arccosine and using the effective maximal cosines
  ($t \to \td{t}$) to define the
  measure.  The form given in \Eqn{QRMSD-basic-chord.eq} is a
  minimization problem that is exactly the quaternion analog of the
   RMSD problem definition in \Eqn{RMSD-basic.eq} for spatial data, with the
  additional constraint that all the spatial data must be unit-length
  4-vectors (which have only 3 degrees of freedom) instead of
  arbitrary 3-vectors.  In addition, the convergence condition for
  clustering of the data within ball should in principle be satisfied
  for the optimal solution of \Eqn{QRMSD-basic-chord.eq}  to be
  global; our data simulation pushes these limits, but in practice
  the convergence is typically satisfied.   Just as
  \Eqn{RMSD-basic.eq} and its cross-term form \Eqn{RMdef.eq} give
  exactly the same results for spatial data when the measures are
  minimized and maximized, respectively, the orientation-problem
  equations \Eqn{QRMSD-basic-chord.eq} and \Eqn{FrameVLinear.eq} do
  the same for the quaternion measure.  Finally, the two cross-term forms
  \Eqn{qV.eq} and \Eqn{qVsq.eq} give the same optimal quaternions,
  with the interesting fact that \Eqn{qV.eq} yields the optimal
  quaternion from a linear equation, and \Eqn{qVsq.eq} gives an
  identical result from a quadratic matrix equation that works the
  same way as the RMSD matrix optimization, except that the symmetric
  profile matrix is no longer traceless.

   Thus there are in fact four ways of looking at the chord-length
   measure and obtaining exactly the
  same optimal quaternions, and we have checked these using two
  numerical optimizations and two algebraic optimizations.  These
  options are:
\begin{itemize} 
\item {\bf Minimizing Euclidean Chord-Length Squared.}  Here we write
  the chord-approximation to the QFA problem using
  \Eqn{QRMSD-basic-chord.eq}, which is exactly parallel to the RMSD
  problem employing \Eqn{RMSD-basic.eq}, modulo the sign ambiguity
  issue.  We test this by performing a numerical minimization.
\item {\bf Maximizing Chord-Length Cross-Term.} Just as the RMSD
  cross-term maximization problem  \Eqn{RMdef.eq} is equivalent to the RMSD
  minimization problem of \Eqn{RMSD-basic.eq}, we can use
  maximization of the quaternion cross-term  \Eqn{FrameVLinear.eq}
  equivalently with the minimization of the chord-length
  \Eqn{QRMSD-basic-chord.eq}.  We test this by performing a numerical
  maximization.
\item {\bf Linear Reduction of Chord-Length Cross-Term.} Pulling out
  the linear coefficients of the each quaternion component in
  \Eqn{FrameVLinear.eq} generates \Eqn{qV.eq}, where the 4-vector $V_{a}(W)$ of 
 \Eqn{M4def.eq} plays the role of the RMSD profile matrix $M_{ab}(E)$
 in  \Eqn{qM3q.eq}.   Here we test the
 optimization by algebraically solving the linear expression
 \Eqn{qV.eq}.
\item {\bf Quadratic Equivalent Matrix Form of the Chord-Length
    Cross-Term.}  Finally, there is in fact a maximal matrix
  eigenvalue problem \Eqn{qVsq.eq} that works like \Eqn{qM3q.eq} by
  squaring \Eqn{qV.eq} to get a matrix problem $q \cdot \Omega \cdot
  q$ with $\Omega_{ab}= V_{a} V_{b}$.  Despite the presence of a
  nonvanishing trace, the maximal quaternion eigenvectors are the same
  as the other three cases above.  This produces the same optimal
  quaternion solutions as solving the (much, much simpler) linear
  problem of \Eqn{qV.eq}.  This can also be checked algebraically.
\end{itemize}
\item[(c)] {\bf {$\mathbf{(\tr R(q)\cdot R(p)\cdot R(\bar{r}))}$}
    Chord-Length (algebraic).} Finally, the most rigorous method if
  consistency of quaternion signs cannot be guaranteed is to use a
  measure in which algebraic squares occur throughout and enforce
  rigorous sign-independence.  This is our (c) data set.
  Such measures must of necessity be
  \emph{quartic} in the quaternion test and reference data, and thus are distinct
  from the measures of {(b)} that are \emph{quadratic} in the data
  elements.  This $(\tr R(q)\cdot R(p)\cdot R(\bar{r}))$ measure is
  the form that is most easily integrated into the combined
  rotational-translational problem treated in the next section,
  because the combined matrices are both symmetric and traceless like
  the original RMSD profile matrices.  Furthermore, it is obvious from
  \Eqn{VVeqRRR.eq} that this measure is exactly the same as the one
  obtained from \Eqn{FrameVLinear.eq} if we squared \emph{each term in
    $k$} before summing the cross-term data elements in option {(b)}.
  Thus, whichever actual formula we choose, we appear to have
  exhausted the options for quaternion-sign-independent quartic
  measures for the orientation data problem.
\end{itemize}

The task now is simply to evaluate how close the optimal quaternion
solutions for the arc-length measure {(a)} are to the quadratic
chord-length measures {(b)} and the quartic chord-length measures
{(c)}.  In addition, we would like to know how close the fragile but
very elegant quadratic measures {(b)} are to the rigorously
sign-insensitive quartic measures {(c)}; we expect them to be 
similar, but we do not expect them to be identical.

\begin{figure}
 \figurecontent{\vspace{.1in}  \centering
 \includegraphics[width=1.75in]{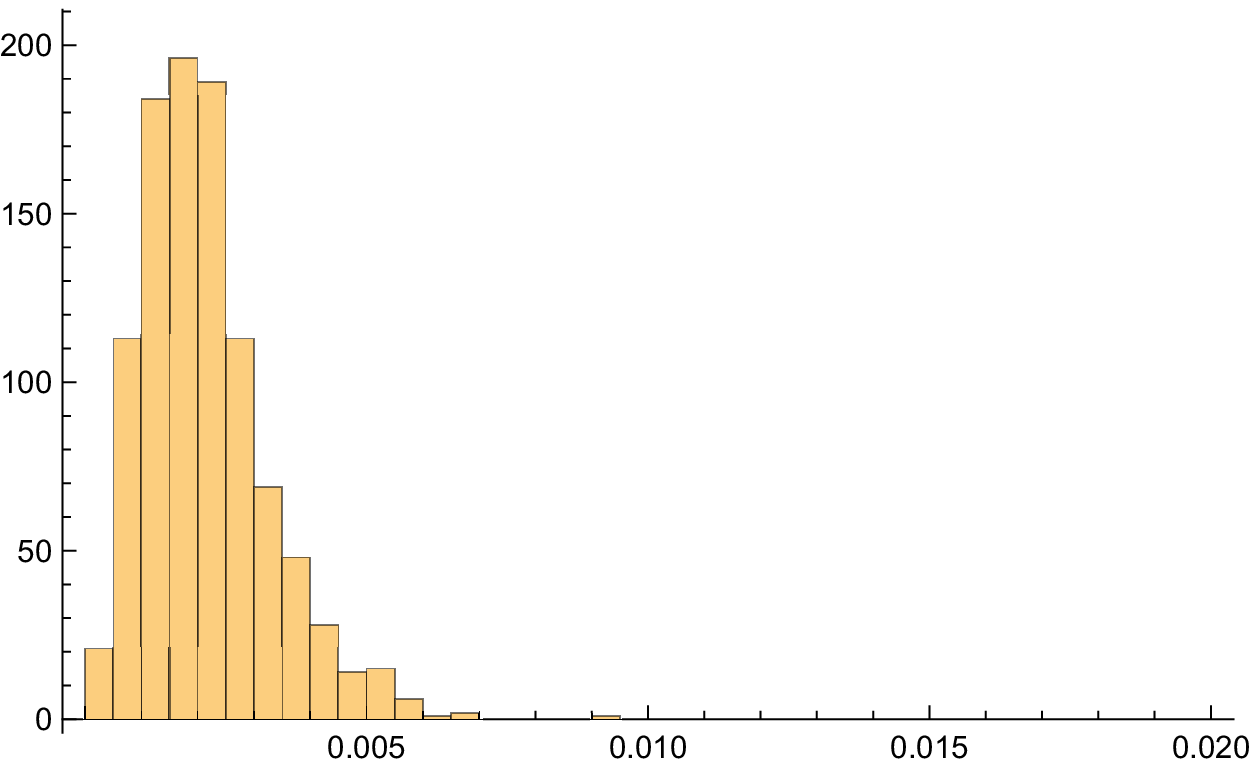} 
 \includegraphics[width=1.75in]{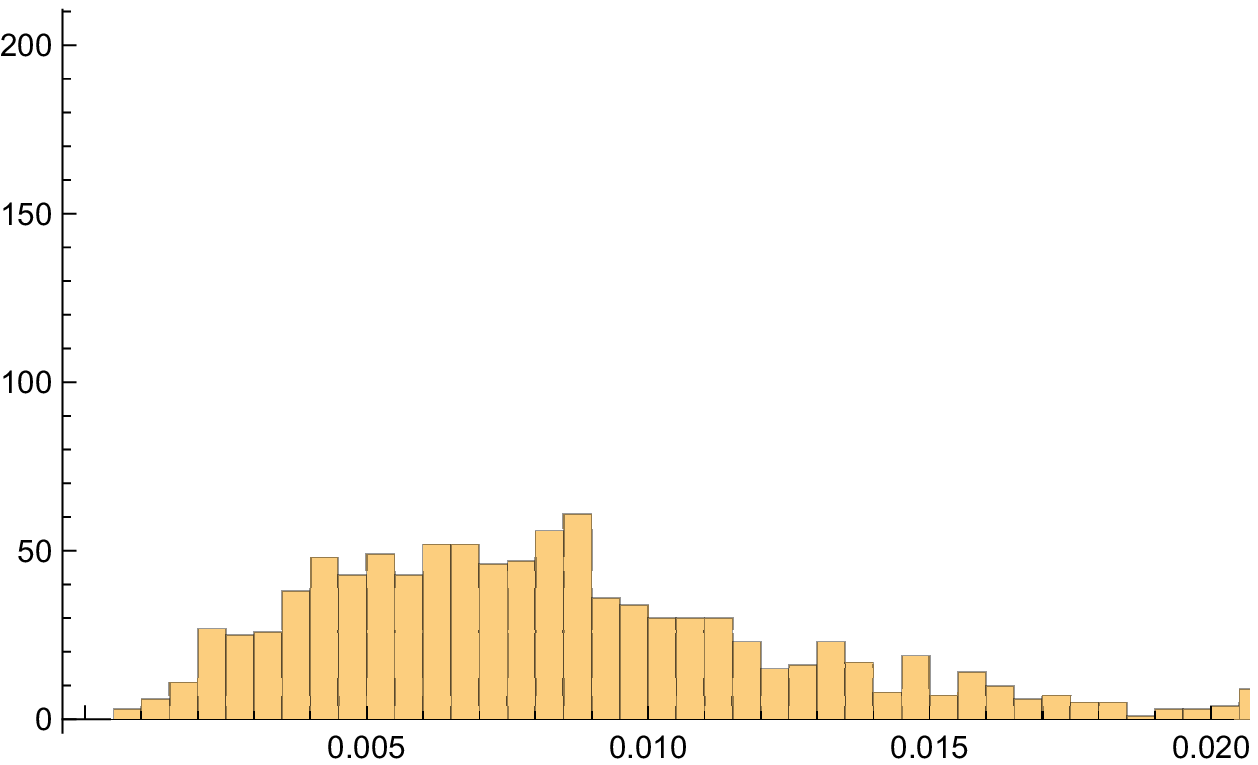}
 \includegraphics[width=1.75in]{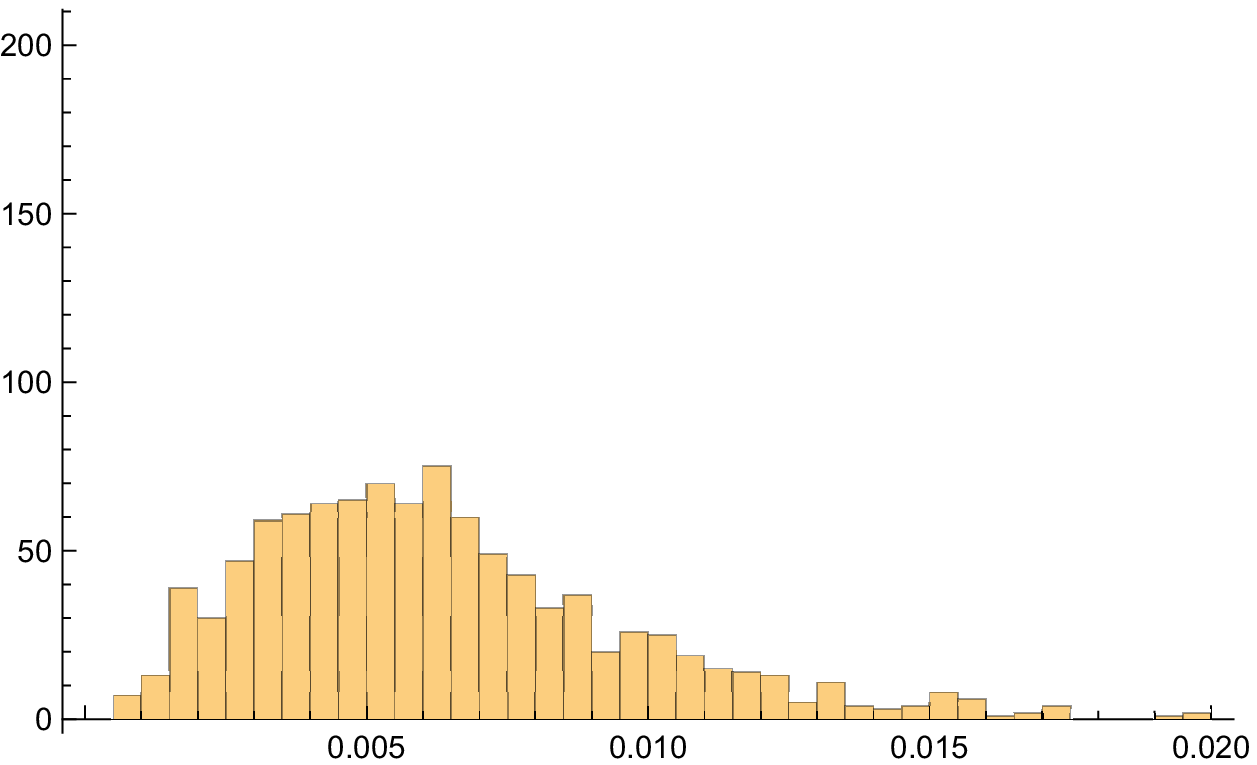}
 \\
\centerline{\hspace{0.9in} (a:b) \hfill (a:c) \hfill (b:c) \hspace{0.9in} } }
\vspace*{0.05in}
\caption{\ifnum\ShowFiles=1 {\bf aab1Histogram.eps, aac3Histogram.eps,
    b1c3Histogram.eps.} \fi 
 Spectrum in degrees of angular differences
  between optimal quaternion alignment rotations for quaternion
  frames.
(a:b): (a) vs (b), true arc-length vs approximate quadratic
chord-length measure. 
(a:c): (a) vs (c), true arc-length vs approximate quartic
chord-length measure. 
(b:c): (b) vs (c),  approximate quadratic vs approximate quartic
chord-length measure.
 }
\label{arc-histo-abc.fig}
\end{figure}

To quantify the closeness of the measures, we took the magnitude of
the inner products between competing optimal quaternions for the same
data set, which is essentially a cosine measure, took the arccosines,
and converted to degrees.  The results were histogrammed for 1000
random samples consisting of $N=100$ data points, and are presented in
Fig~\ref{arc-histo-abc.fig}.
The means and standard deviations of the optimal total rotations relative to
the identity frame for the three cases are:\\
\begin{center}
\begin{tabular}{|l||c|c|}
\hline
{\bf Measure Type}& Mean(deg) & Std Dev(deg) \\ \hline
{\bf (a) arc-length } & 44.8062 & 11.2307  \\
{\bf (b) chord quadratic }& 44.8063 & 11.2308  \\
{\bf (c) chord quartic } &44.8065  & 11.2310 \\ \hline
\end{tabular}\ .\end{center}
One can see that our simulated data set involved a large range of
global rotations, and that all three methods produced a set of
rotations back to the optimal alignment that are not significantly
different statistically.  We thus expect very little difference in the
histograms of the case-by-case optimal quaternions produced by the
three methods.  The mean differences illustrated in the Figures are
summarized as follows:\\
\begin{center}
\begin{tabular}{|r||c|c|c|}
\hline
{\bf Figure:(Pair) }& Mean(deg) & Std Dev(deg) \\
\hline  
{\bf  Figure~\ref{arc-histo-abc.fig} ({a:b})} & 0.0021268& 0.0011284 \\
{\bf  Figure~\ref{arc-histo-abc.fig} ({a:c})} & 0.0084807& 0.0044809 \\
{\bf  Figure~\ref{arc-histo-abc.fig} ({b:c})} & 0.0063539& 0.0033526\\
\hline \end{tabular}\ .\end{center} 
We emphasize that these numbers are in degrees for 1000 simulated
samples with a distribution of global angles having a standard
deviation of $11^{\circ}$.  Thus we should have no issues using the chord
approximation, though it does seem that the $q\cdot V$ measure is
significantly better both in accuracy and simplicity of computation.



\section{The 3D Combined Point+Frame Alignment Problem.}


The 3D combined alignment problem for both spatial data and
orientation-frame data involves a number of issues and subtleties that
we were able to treat only superficially in the main text. In this
section, we explore various options and evaluate their performance.
This is necessary for anyone who might think of trying to attempt a
combined alignment problem, so we have attempted to anticipate the
questions and alternatives that might be explored and check their
properties.  The overall result is that it seems difficult to obtain
significant additional information from the combined alignment
strategies that we examined, so potential exploiters of this paradigm
are forewarned.

From the main text, we are in possession of precise alignment
procedures for both 3D spatial coordinates and 3D frame triad data
(using the exact measure for the former and one of the approximate
chord measures for the latter), and thus we can consider the full 6
degree-of-freedom alignment problem for combined data from a single
structure.  In fact this problem can also be solved in closed
algebraic form given the our existing eigensystem formulation of the
orientation alignment problem.  While there are clearly appropriate
domains of this type, e.g., any protein structure in the PDB database
can be converted to a list of residue centers and their local frame
triads~\cite{HansonThakurQuatProt2012}, little is known at this time
about the potential value of combined alignment.  To establish the most
complete possible picture, we now proceed to describe the details of
our solution to the alignment problem for combined translational and
rotational data.

In our treatment, we will assume the $\Delta_{\mbox{\tiny RRR}}$
measure since its profile matrix is traceless and manifestly
independent of   the quaternion signs, but there
is no obstacle to using $\Delta_{\mbox{frame-sq}}$ if the data are
properly prepared and one prefers the simpler measure.  For notational
simplicity, we will let $\Delta_{f}$ stand for whatever orientation
frame measure we have chosen, corresponding to $\Delta_{x}$ for the
spatial measure, and thus we will denote the combined measure by
$\Delta_{xf}$.

\qquad

{\bf The Combined Optimization Measure.}
A  significant aspect of establishing a combined measure
including the point measure $\Delta_{x}$ and the frame orientation
measure  $\Delta_{f}$ is the fact that the
measures are \emph{dimensionally incompatible}.  We {\it cannot\/}
directly combine the corresponding data minimization measures
$\Delta_{x}(q_{x}) = \epsilon_{\scriptstyle{x\mbox{\tiny :max}}}$ and $
\Delta_{f}(q_{f}) = \epsilon_{\scriptstyle{f\mbox{\tiny :max}}}$ because the
spatial measure has dimensions of $\mbox{(length)}^2$ and the frame
measure is essentially a dimensionless trigonometric function (the
arc-distance measure produces $\mbox{(radians)}^2$, which is still
incompatible).

While it should be obvious that a combined measure requires an
arbitrary, problem-specific, interpolating constant with dimensions of
length to produce a compatible measure, there has been some confusion
in the molecular entropy literature, where such measures seem first to
have been employed.  These issues were resolved and dimensionful
constants introduced, e.g., in the work of Fogolari, et
al.~\cite{FogolariEtAl:2016,Hugginsdj2014}.  Our approach
to defining a valid heuristic combined measure has three components:
\begin{itemize} 
\item {\bf Normalize the Profiles.}  The numerical sizes of the
  maximal eigenvalues of the $\Delta_{x}$ and the
  $\Delta_{f}$ systems
  can easily differ by orders of magnitude.  Since scaling the profile
  matrices changes the eigenvalues \emph{but not the eigenvectors}, it
  is perfectly legitimate to start by dividing the profiles by their
  maximal eigenvalues before beginning the combined optimization,
  since this accomplishes the  sensible effect of assigning
 maximal eigenvalues of exactly unity to both of our scaled profile
 matrices. 
\item {\bf Interpolate between the Profiles.} To allow an arbitrary
  sensible weighting distinguishing between a location-dominated
  measure and an orientation-dominated measure, we simply incorporate
  a linear interpolation parameter $t \in [0,1]$, with $t=0$ singling
  out $\Delta_{x}$ and the pure (unit eigenvalue) location-based RMSD,
  and $t=1$ singling out $\Delta_{f}$ and the pure
  orientation (unit eigenvalue) QFA solution.
\item {\bf Scale the Frame Profile.}  Finally, we incorporate the
  mandatory dimensional scaling adjustment 
 by incorporating one additional (nominally dimensional) parameter
  $\sigma$ that scales the orientation parameter space described by
$\Delta_{f}$ to be more or
  less important than the ``canonical'' spatial dimension component
  $\Delta_{x}$, which we leave unscaled.  That is, with $\sigma=0$ 
   only the spatial measure survives, with $\sigma=1$, the normalized
   measures have equal contributions, and with $\sigma>1$, the
   orientation measure dominates (this effectively undoes the original
   frame profile eigenvalue scaling).
\end{itemize}

We thus start with a combined spatial-rotational measure of the
form 
\begin{eqnarray} 
 \Delta_{\mbox{\small initial}} & = & (1 - t) \sum_{a=1,b=1}^{3} {R^{ba}(q) E_{ab}} +
     t \, \sigma \sum_{a=1,b=1}^{3} {R^{ba}(q) S_{ab}}  \nonumber \\
  & = & (1-t) \tr \left( R(q) \cdot E\right)  + t\, \sigma  \tr \left(
    R(q) \cdot S \right) \nonumber \\
 & = & \sum_{a=0,b=0}^{3} q_{a} \left[ (1-t) M_{ab}(E) + t\, \sigma\, U_{ab}(S)
   \right] q_{b} \nonumber \\
 & = & q \cdot \left[ (1-t) M(E) + t \,\sigma\,  U(S) \right] \cdot q \ ,
\label{XQMdef.eq}
\end{eqnarray}
and then impose the unit-eigenvalue normalization on $M(E)$ and
$U(S)$, giving our final measure as
\begin{eqnarray} 
 \Delta_{xf}(t,\sigma) & = & 
 q \cdot \left[ (1-t) \frac{M(E)}{\epsilon_{x}} +  t \; \sigma
 \frac{U(S)}{\epsilon_{f}} \right] \cdot q \ .
\label{ScaledXQMdef.eq}
\end{eqnarray}
  Because of the dimensional incompatibility of
$\Delta_{x}$ and $\Delta_{f}$,  we  have to treat the ratio
\[ \lambda^2 = \frac{t \sigma}{1-t} \]
as a dimensional constant such as that adopted by Fogolari et
al.~\cite{FogolariEtAl:2016} in their entropy calculations, so if $t$
is dimensionless, then $\sigma$ carries the dimensional scale
information. 

From the profile matrix of \Eqn{ScaledXQMdef.eq}, we now 
extract our optimal rotation solution  using the same equations as always,
solving for the maximal eigenvalue and its eigenvector either numerically
or algebraically, leading to the equivalent of
\Eqn{DeltaOptIsEig.eq}, as we have solved the standard RMSD maximal
eigenvalue problem.   The result is a parameterized eigensystem
\begin{equation} 
\left. \begin{array}{c}
\epsilon_{\opt}(t,\sigma) \\
q_{\opt}(t,\sigma) \end{array} \right\} 
\label{SDNeValVec.eq}
\end{equation}
yielding the optimal values $R(q_{\opt}(t,\sigma))$, 
$\Delta_{xf}=\epsilon_{\opt}(t,\sigma)$ based on the data
$\{E,S\}$ no matter what we 
take as the values of the two variables $(t,\sigma)$.

\begin{figure}
 \figurecontent{ \centering
 \includegraphics[width=3.5in]{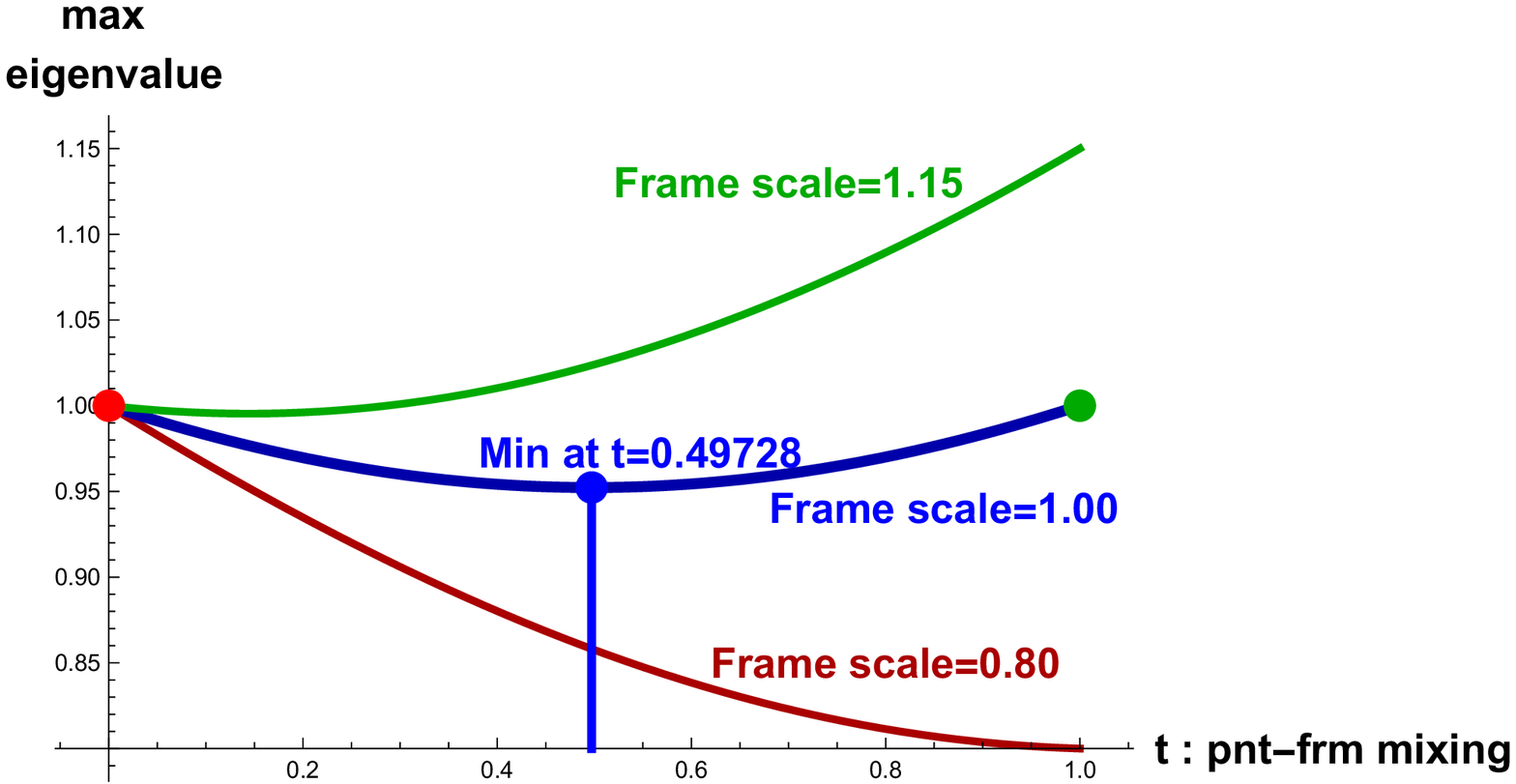} }
\vspace*{.1in}
\caption{\ifnum\ShowFiles=1 {\bf combinedSimpleTcurves1.eps.} \fi
  The blue curve is the
  path of the composite eigenvalue for Data Set 1 (the value of the
  similarity measure $\Delta_{xf}(t,1)$) in the interpolation variable
  $t$ with equally weighted space and orientation data, i.e.,
  $\sigma=1$. It has maxima only at the ``pure'' extremes at $t=0,1$,
  but there is a minimum that occurs, for these data, not at $t=1/2$,
  but very nearby at $t=0.49728$.  Increasing the influence of the
  spatial data by taking $\sigma=0.8$ gives the red curve, and
  increasing the influence of the orientation data by taking
  $\sigma=1.15$ gives the green curve.  }
\label{combinedSimpleTcurves1.fig}
\end{figure}

\begin{figure}
 \figurecontent{ \centering
 \includegraphics[width=3.5in]{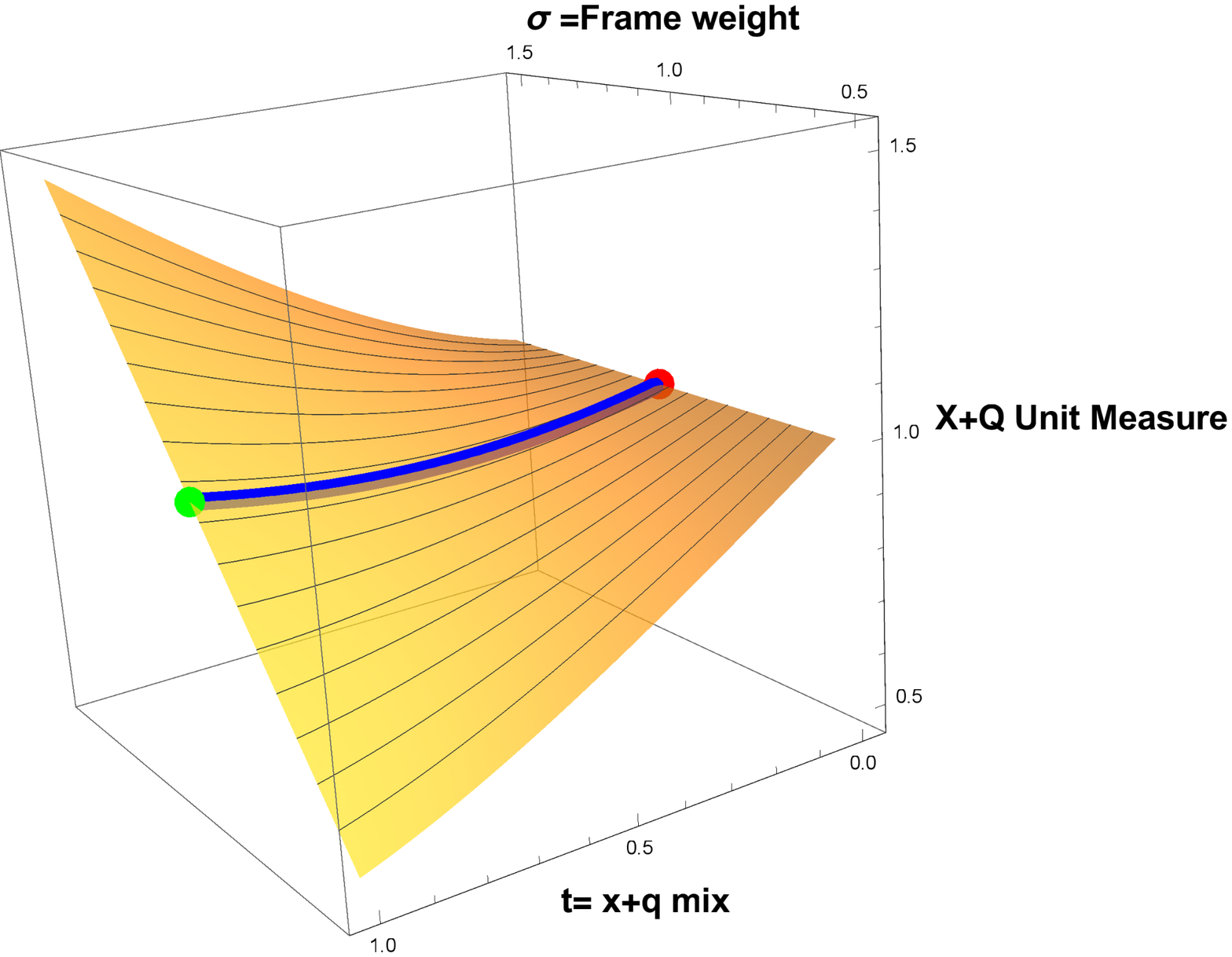} }
\vspace*{.1in}
\caption{\ifnum\ShowFiles=1 {\bf combinedTSigsurf1.eps.} \fi
  The $\Delta(t,\sigma)$
  similarity-measure surface for Data Set 1 as a function of the interpolation
  parameter $t$ and the relative scaling of the orientation term with
  $\sigma$, with the slightly concave curve at $\sigma=1$ in the
  middle.  The other data sets look very much like this one. }
\label{combinedTSigsurf1.fig}
\end{figure}

\paragraph{Properties of the Combined Optimization.}  Substantially
different features arise in the solutions depending on how close the
optimal rotations were for the initial, separate, systems $\Delta_{x}$
and $\Delta_{f}$.  We now choose a selection of 
simulated data sets with the following choices of approximate initial
global rotations of the
test data sets relative to the reference data: 
      \begin{table} 
\caption{Offsets of sample data for the spatial vs  orientation data
 used in exploring the properties of combined   measures.}
\label{datasets.eq} 
     \end{table}
\begin{center}
\begin{tabular}{|l|c|c|}
\hline
{\bf DATA ID} & {\bf \parbox[t]{1.8in}{\ \ \ (Space, Orientation)}} %
 & \parbox[t]{1.2in}{\ \ \  Measured  Offset} \\ 
 \hline
Data Set 1 &  $\mathbf{(22^{\circ} , -22^{\circ})}$ & 44.60  \\
Data Set 2 & $\mathbf{(22^{\circ} , -11^{\circ}) }$ & 21.98  \\
Data Set 3 & $\mathbf{(22^{\circ} , 0^{\circ}) }$ & 11.15 \\
Data Set 4 & $\mathbf{(22^{\circ} , 11^{\circ}) }$ & 11.15 \\
Data Set 5 & $\mathbf{(22^{\circ} , 21^{\circ}) }$ & 1.20 \\ \hline
\end{tabular} 
\end{center} 
In Fig~\ref{combinedSimpleTcurves1.fig}, we plot the trajectory of the
maximal combined similarity measure for Data Set 1 as a function of
$t$, showing the behavior for $\sigma=1.0, 0.80,\,\mbox{and}\
1.15$. Figure \ref{combinedTSigsurf1.fig} shows a more comprehensive
representation of the continuous behavior with $\sigma$, and in both
figures, we see that the true optima are \emph{at the end points},
$t=0,1$, the locations associated with the pure profile eigenvector
solutions $q_{x}({\opt})$ and $q_{f}({\opt})$.  There is no
\emph{better} optimal eigenvector (i.e., global rotation) for any
intermediate value of $t$.  In some circumstances, however, it might
be argued that it is appropriate to choose the \emph{distinguished
value} of $t$ at the minimum of the curve $\Delta_{xf}(t,\sigma=1)$.
As we shall see in a moment, just as in
Fig~\ref{combinedSimpleTcurves1.fig} for $\sigma=1$, this point is
generally within a few percent of $t=0.5$.  As the spatial and
orientation optima get closer and closer, the curves in $t$ become
much flatter and less distinguished, while the variation in $\sigma$
is qualitatively the same as in Fig~\ref{combinedTSigsurf1.fig}\,.

\begin{figure}
 \figurecontent{ \centering
 \includegraphics[width=5.5in]{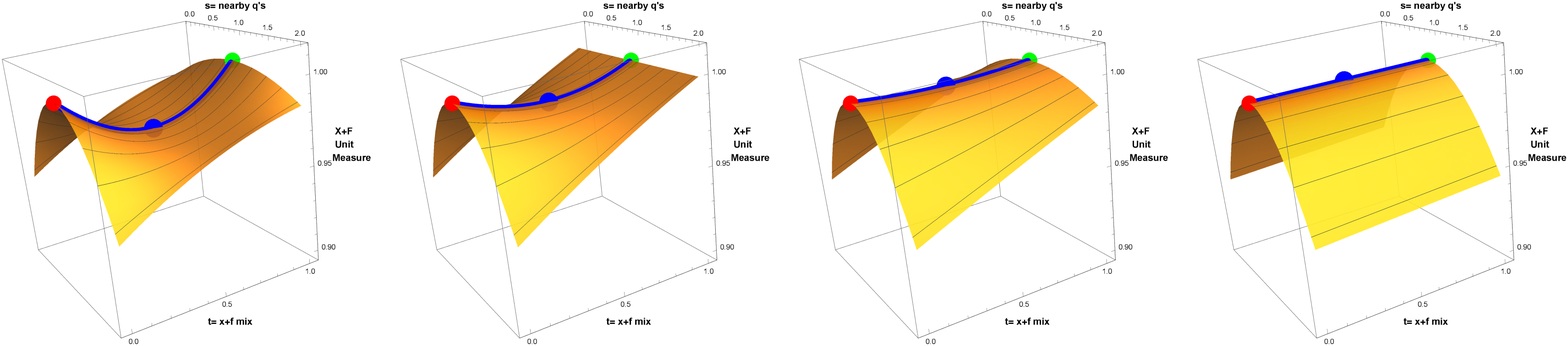} }
\centerline {\hspace{0.5in}(Set 2)\hfill (Set 3) \hfill (Set 4) \hfill
(Set 5)\hspace{0.7in}}
\vspace*{.05in}
\caption{\ifnum\ShowFiles=1 {\bf combinedTS2345.eps .} \fi
  The $\Delta_{xf}(t,1)$
  similarity-measures with $q(s)$ interpolated from the identity
  through the optimum for $\Delta_{xf}$ and past to the
  identity-mirror point,   for Data Sets  2, 3, 4, and 5, where
  Data Set 5 has the x-angle and the f-angle only one degree apart, as
  we might have for real experimental data. 
}
\label{combinedTS2345.fig}
\end{figure}

Finally, we examine one more amusing visualization of the properties
of the composite solutions, restricting ourselves to $\sigma=1$ for
simplicity, and examining the ``sideways warp'' in the quaternion
eigenvector $q_{\opt}(t,\sigma=1)$ in \Eqn{SDNeValVec.eq}.  We
examine what happens to the combined similarity measure
\Eqn{ScaledXQMdef.eq} if we smoothly interpolate from the identity
matrix (that is, the quaternion $q_{\ID} =(1,0,0,0)$) through the optimal
solution for each $t$ and beyond the optimum by the same amount, using
the \emph{slerp} interpolation defined in \Eqn{slerpdef.eq}, i.e.,
$q(s) = \mbox{\it slerp}(q_{\ID},q_{\opt}(t,\sigma=1),s)$.   Figure   
\ref{combinedTSlerp1.fig} shows Data Set 1, with the
largest relative spatial vs orientation angular
differences,  Figure~\ref{combinedTS2345.fig} corresponds to the
intervening Data Sets 2, 3, 4, and 5, with the Data Set parameters
given above in Table \ref{datasets.eq}.   Data Set 5 in particular is perhaps
the most realistic example, having nearly identical spatial and
angular rotations, and we see negligible differences between the spatial and
angular structures.  These graphics also show how the local,
non-optimal, neighboring quaternion values peak in $s$ at the optimal ridge
going from $t=0$ to $t=1$.  The red dot is the maximum of $\Delta_{x}$
at $t=0$, the green dot is the maximum of $\Delta_{f}$ at $t=1$, and
the blue dot, specific to each data set, is the distinguished point at
the \emph{minimum} of $\Delta_{xf}(t,\sigma=1)$ in $t$, which for our data
sets are always within $1\%$ of $t = 0.5$.
We observe that for equal and opposite rotations, the midpoint
coincides almost exactly with the identity quaternion that occurs at
the left and right boundaries of the plot.  In other respects, the
data in these figures show that we do not have \emph{maxima} in the
middle of the interpolation in $t$, but we do have a distinguished
value, always very near $t=0.5$, that could be used as a baseline for
a hybrid translational-rotational rotation choice.

\begin{figure}
 \figurecontent{ \centering
 \includegraphics[width=3.0in]{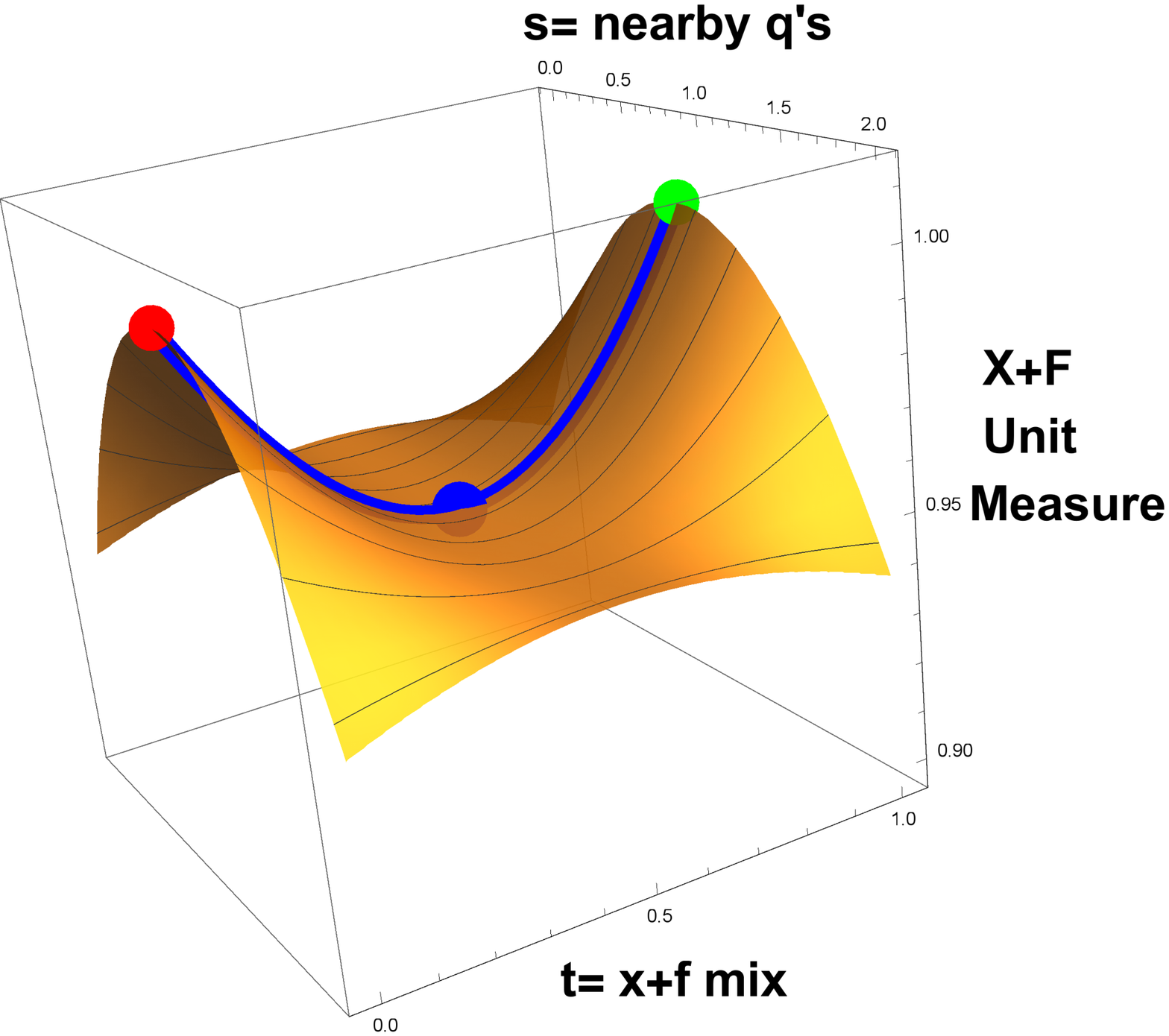} }
\centerline {\hfill(Set 1) \hfill}
\vspace*{.05in}
\caption{\ifnum\ShowFiles=1 {\bf combinedTSlerp1.eps .} \fi
 The $\Delta_{xf}(t,1)$
  similarity-measure 
  surface for Data Set 1, x-angle $22^{\circ}$, f-angle $-22^{\circ}$,
  and fixed $\sigma=1$ showing the deviation with the
  quaternion varying perpendicularly around the solution $q(t)$,
  starting at the identity quaternion at $s=0$, as a function of the
  interpolation parameter $t$.  Since $q(t)$ is the maximal
  eigenvector, all variations in $q$ peak there.
  Both have distinguished central points 
  at $t\approx 0.5$.
}
\label{combinedTSlerp1.fig}
\end{figure}

\comment{  
\begin{figure}[h!]
 \figurecontent{ \centering
 \includegraphics[width=3.0in]{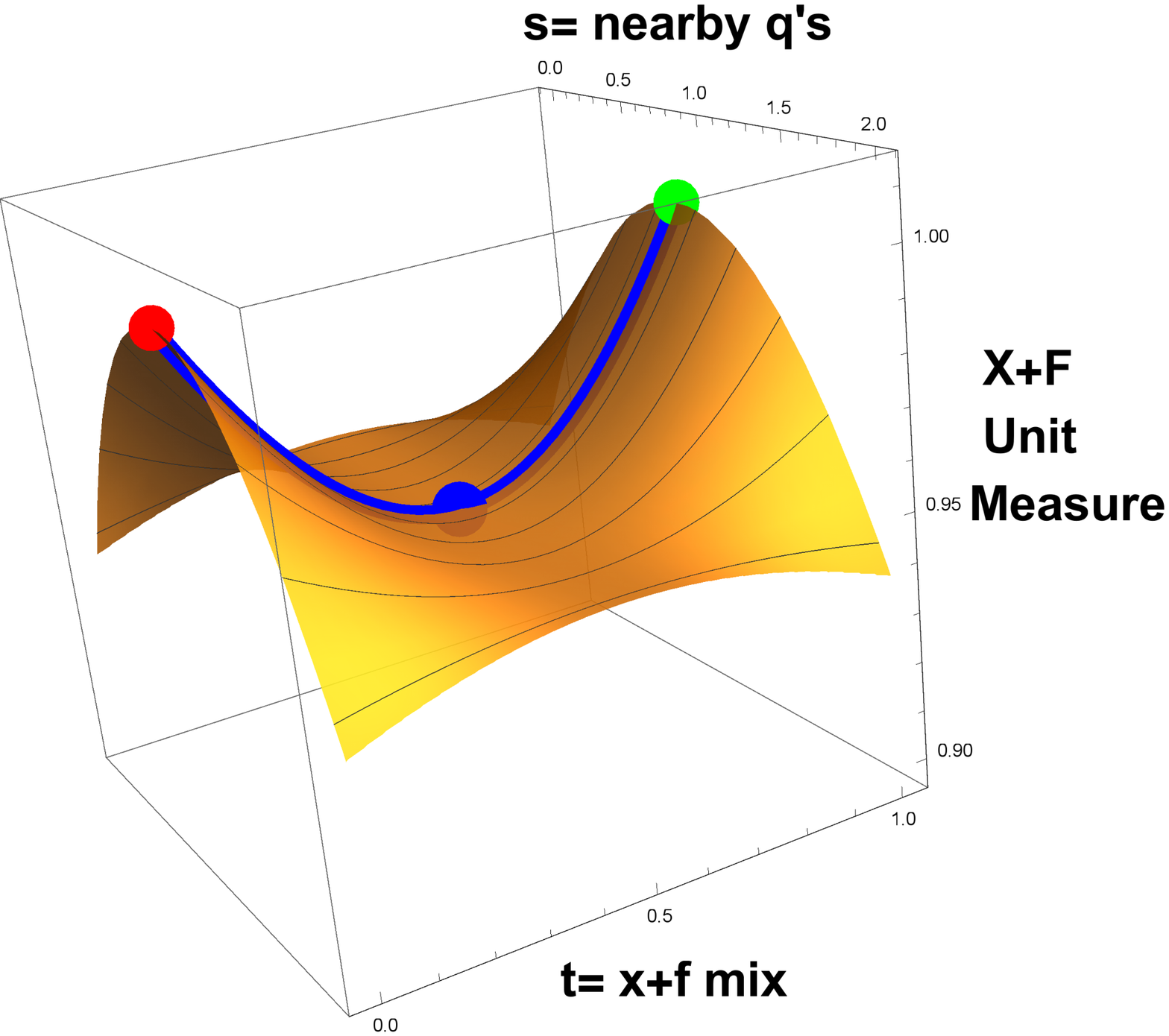} }
\vspace*{.1in}
\caption{\ifnum\ShowFiles=1 {\bf combinedTSlerp1.eps .} \fi
 The $\Delta_{xf}(t,1)$
  similarity-measure 
  surface for Data Set 1 and fixed $\sigma=1$ showing the deviation with the
  quaternion varying perpendicularly around the solution $q(t)$,
  starting at the identity quaternion at $s=0$, as a function of the
  interpolation parameter $t$.  Since $q(t)$ is the maximal
  eigenvector, all variations in $q$ peak there.
  Both have distinguished central points 
  at $t\approx 0.5$.
}
\label{combinedTSlerp1.fig}
\end{figure}

\begin{figure}[h]
 \figurecontent{ \centering
 \includegraphics[width=3.0in]{combinedTSlerp2.eps} }
\vspace*{.1in}
\caption{\ifnum\ShowFiles=1 {\bf combinedTSlerp2.eps .} \fi
 The $\Delta_{xf}(t,1)$
  similarity-measure   surface for Data Set 2 and fixed $\sigma=1$
  showing the deviation with the 
  quaternion varying perpendicularly around the solution $q(t)$.
  There is a distinguished point on the central curve at
  $t\approx 0.5$.
}
\label{combinedTSlerp2.fig}
\end{figure}

\begin{figure}[h]
 \figurecontent{ \centering
 \includegraphics[width=3.0in]{combinedTSlerp3.eps} }
\vspace*{.1in}
\caption{\ifnum\ShowFiles=1 {\bf combinedTSlerp3.eps .} \fi
 The $\Delta_{xf}(t,1)$
  similarity-measure   surface for Data Set 3 and fixed $\sigma=1$
  showing the deviation with the 
  quaternion varying perpendicularly around the solution $q(t)$.
  There is a distinguished point on the central curve at
  $t\approx 0.5$.
}
\label{combinedTSlerp3.fig}
\end{figure}

\begin{figure}[h]
 \figurecontent{ \centering
 \includegraphics[width=3.0in]{combinedTSlerp4.eps} }
\vspace*{.1in}
\caption{\ifnum\ShowFiles=1 {\bf combinedTSlerp4.eps .} \fi
 The $\Delta_{xf}(t,1)$
  similarity-measure   surface for Data Set 4 and fixed $\sigma=1$
  showing the deviation with the 
  quaternion varying perpendicularly around the solution $q(t)$.
  There is a distinguished point on the central curve at
  $t\approx 0.5$.  As one can see, with the global rotations for
  spatial and orientation data being essentially the same for Data Set
  4, there is no significant structure as $t$ varies between the two
  limits. 
}
\label{combinedTSlerp4.fig}
\end{figure}

\begin{figure}[h]
 \figurecontent{ \centering
 \includegraphics[width=3.0in]{combinedTSlerp5.eps} }
\vspace*{.1in}
\caption{\ifnum\ShowFiles=1 {\bf combinedTSlerp5.eps .} \fi
 The $\Delta_{xf}(t,1)$
  similarity-measure   surface for Data Set 5 and fixed $\sigma=1$
  showing the deviation with the 
  quaternion varying perpendicularly around the solution $q(t)$.
  There is a distinguished point on the central curve at
  $t\approx 0.5$.  As one can see, with the global rotations for
  spatial and orientation data being essentially the same for Data Set
  5, there is no significant structure as $t$ varies between the two
  limits. 
}
\label{combinedTSlerp5.fig}
\end{figure}

   } 

\begin{figure}
 \figurecontent{ \centering
 \includegraphics[width=3.5in]{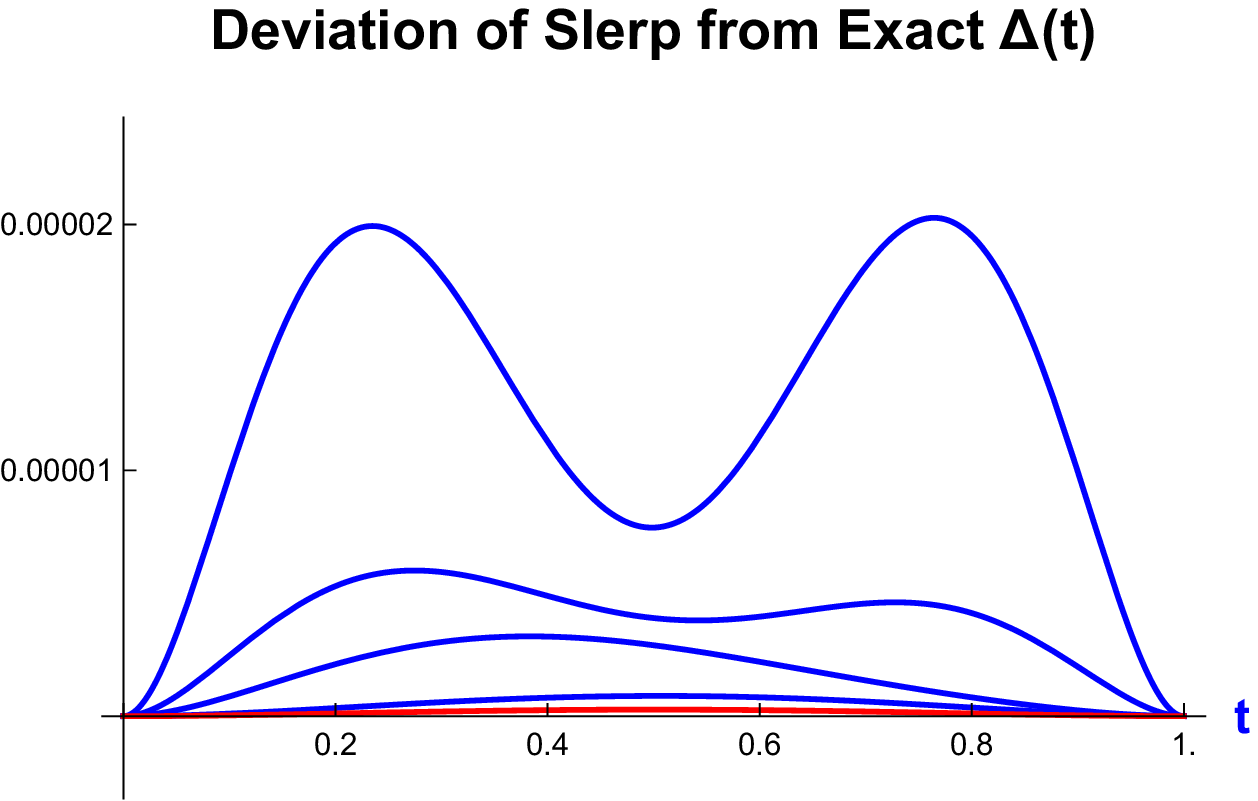} }
\caption{\ifnum\ShowFiles=1 {\bf combinedQQSlerpOffset.eps.} \fi
  Here we see how close a
  simple \emph{slerp}$(t)$ between the extremal optimal eigenvectors 
$q_{\opt}(t=0,\sigma=1)=q_{x}({\opt})$ and
$q_{\opt}(t=1,\sigma=1)=q_{f}({\opt})$ is to the rigorous
result where we optimized $q_{\opt}(t,\sigma=1)$ \emph{for all
  $t$}.
The differences are \emph{relative to the unit eigenvalue}, and thus
are of order thousandths of a percent, decreasing significantly as the
global rotations applied to the space and orientation data approach
one another.  The largest deviation is for Data Set 1, which
interestingly has a third minimum near the center in $t$; for the
highly similar data in Data Set 5, the difference shown in red had to
be magnified by 100 even to show up on the graph.
 }
\label{combinedQQSlerpOffset.fig}
\end{figure}

\bigskip \bigskip

\paragraph{The Simple Approximation.} Having now observed that it is
possible to construct and solve a rigorous combined RMSD-QFA
problem (with the chord-distance approximation in the angular measure), 
one might ask how that compares to the very simplest idea
one might use to interpolate between the measures:  what if we take
the rigorous combined profile matrix defined by \Eqn{ScaledXQMdef.eq},
compared to the \emph{slerp} relating the two optimal eigenvectors of
the independent spatial and orientation frame problems, that is 
\begin{equation}
q(t) = \mbox{\it slerp} (q_{x:{\opt}},q_{f:{\opt}},t) \ .
\label{simple6DOFslerp.eq}
\end{equation}
Given the individual optimal eigenvectors, if we  compare this
simple $q(t)$ to  \Eqn{ScaledXQMdef.eq} for any $t$ (and
$\sigma=1$), we find  that the differences are essentially 
negligible.   In Fig~\ref{combinedQQSlerpOffset.fig}, we plot the continuous
differences of the similarity functions, which we recall are scaled to
have a maximal eigenvalue equal to unity.  These scaled
differences are on the order of one thousandth of a percent or less as
the global rotations applied to the spatial and rotational data become
close to one another.  We conclude that, for all practical purposes, we
might as well use \Eqn{simple6DOFslerp.eq} to estimate the combined similarities.


\clearpage

\appendix

\section{\vspace{.25in}Details of the Algebraic Solutions to the \\[-.2in] Quartic Eigenvalue Problem}
\label{QuarticSolutions.app} 


  \quad

Given the data for the 3D  or 4D test and reference structures, we can numerically
solve for the maximal eigenvalue of $M_{3}(E_{3})$ and its eigenvector in 3D, or the 
  maximal eigenvalue of $G =  {M_{4}}^{\t}(E_{4}) \cdot  M_{4}(E_{4})$ and the 
left and right eigenvectors of $G$  in 4D.  Alternatively,
we can apply the numerical  SVD method directly to $E_{3}$ or
$E_{4}$ to determine  the optimal rotation matrix.

\emph{However}, we can also work out the properties of the 
eigensystems of the various matrices that have come up in our
treatment \emph{algebraically}, using classic methods  
\cite{AbramowitzStegun1970} for solving quartic polynomial equations
for the eigenvalues,
 to provide deeper insights into the structure of the problem.  We now study  some features of these
results in more detail, and in particular we consider  real symmetric matrices, with
and without a trace, since essentially every problem we have encountered reduces
to finding the maximal eigenvalues of a matrix in that category.

\qquad 

\noindent{\bf The Eigenvalue Expansions.}
We begin by writing down the eigenvalue expansion of an arbitrary real 4D matrix $M$ as
\begin{equation}
 \det [M - e I_{4}]=0 \ ,
 \label{EigvalEqnS.eq}
 \end{equation}
 where $e$ denotes a generic eigenvalue and $I_{4}$ is the 4D
identity matrix.   Our task is to express these eigenvalues, particularly the maximal
eigenvalue, in terms of the elements of the matrix $M$, and also to find their eigenvectors.

  By expanding \Eqn{EigvalEqnS.eq} in powers of $e$, we see how
the four eigenvalues   $e = \epsilon_{k=1,\ldots,4}$
depend on the known components of
the matrix $M$ and correspond to the solutions of the quartic equations that 
we can express in two useful forms, 
\begin{eqnarray} 
e^4 + e^3 p_1 + e^2 p_2 + e p_3 + p_4 &=& 0 \label{3DeigEqneS.eq}\\
(e - \epsilon_{1})(e - \epsilon_{2})(e - \epsilon_{3})(e - \epsilon_{4}) &
= & 0 \ .
\label{3DeigEqnepsS.eq}
\end{eqnarray}
Here the $p_{k}$ are homogeneous polynomials of order $k$  that can 
be expressed alternatively employing elements of $M$ or elements of $E$ for
the  3D amd 4D spatial data, or with the corresponding orientation-frame data.  
At this point we want to be as  general as possible,
and so we note the form valid for all $4\times 4$ matrices $M$ in the expansion
of  \Eqn{EigvalEqnS.eq} and \Eqn{3DeigEqneS.eq}:
\begin{eqnarray} 
p_1(M) & = & - \tr \left[ M \right]   \label{eP1M.eq} \\
p_2(M) & = & -   \frac{1}{2} \tr\left[ M \cdot  M \right]   
   +  \frac{1}{2} \left(\tr\left[ M \right]\right)^{2}   \label{eP2.Meq} \\
 p_3(M) & = &   -  \frac{1}{3} \tr \left[ M  \cdot   M 
  \cdot   M\right]  
  + \frac{1}{2}  \tr\left[ M  \cdot  M\right] \tr \left[ M\right]
   - \frac{1}{6} \left( \tr\left[ M \right]\right)^{3} 
   \label{eP3M.eq}       \\
p_4(M) & = & -\frac{1}{4} \tr \left[M \cdot M \cdot M \cdot M\right]
      + \frac{1}{3} \tr \left[M \cdot M \cdot M\right] \tr \left[M\right]
     + \frac{1}{8} \tr \left( \left[M \cdot M\right] \right)^2 
       - \frac{1}{4} \tr \left[M \cdot M\right]  \left(\tr \left[M\right] \right)^2 
        + \frac{1}{24}  \left( \tr \left[M\right] \right)^4   \nonumber \\
     & = & \det \left[ M \right] \label{eP4M.eq}\ .
\end{eqnarray}
Remember that for our problem, $M$ is just a real symmetric numerical matrix,
and the four expressions $p_{k}(M)$ are also just a list of real numbers.

Matching the coefficients of powers of $e$ in Eqs.~(\ref{3DeigEqneS.eq})
and (\ref{3DeigEqnepsS.eq}), we 
 can also eliminate $e$ to express the 
 the matrix data expressions $p_{k}$ in terms of the symmetric polynomials
of the eigenvalues $\epsilon_{k}$  as   \cite{AbramowitzStegun1970}
\begin{equation} 
\left. \begin{array}{rcl}
p_1  &  = &-\epsilon_{1} - \epsilon_{2} -  \epsilon_{3}
    -  \epsilon_{4}  \\
p_2 & = &\epsilon_{1} \epsilon_{2} + \epsilon_{1}  \epsilon_{3} +
    \epsilon_{2}  \epsilon_{3} + \epsilon_{1}  \epsilon_{4} +
    \epsilon_{2}  \epsilon_{4} +  \epsilon_{3}  \epsilon_{4}
     \\
p_3  & = & -\epsilon_{1} \epsilon_{2}  \epsilon_{3} -
    \epsilon_{1} \epsilon_{2}  \epsilon_{4} - \epsilon_{1} 
    \epsilon_{3}  \epsilon_{4} - \epsilon_{2}  \epsilon_{3} 
    \epsilon_{4}    \\
p_4    & = &   \epsilon_{1} \epsilon_{2}  \epsilon_{3} \epsilon_{4} 
\end{array} \right\} \ .
\label{pNasEpsS.eq}
\end{equation}

\quad

Both  \Eqn{3DeigEqneS.eq} and  \Eqn{pNasEpsS.eq}  can in principle be solved directly 
for the eigenvalues in terms of the matrix data using 
the solution of the quartic published by Cardano in 1545 and investigated further
by Euler \cite{Euler-onRoots-1733,Euler-Bell-1733,NIckalls2009} (see also \cite{AbramowitzStegun1970,mathworld:QuarticEquation,NIckalls1993,ArsMagnaCardano1545}).
Applying, e.g., the  Mathematica function  
\begin{equation}
\mathtt{Solve[myQuarticEqn[e] == 0, e, Quartics}\, \rightarrow\, \mathtt{True ]}   
 \label{SolveQuarticS.eq}
\end{equation}
 to \Eqn{3DeigEqneS.eq} 
 immediately returns a usable algebraic formula.
However, applying $\mathtt{Solve[\ ]}$ to \Eqn{pNasEpsS.eq} is in fact unsuccessful, although 
invoking
$\mathtt{ Reduce[pkofepsEqns, }
     \{ \epsilon_{1}, \epsilon_{2},  \epsilon_{3}, \epsilon_{4}\},  \mathtt{Quartics}\, \rightarrow\,
  \mathtt{True,  Cubics} \rightarrow  \mathtt{True ]  }$
 can solve \Eqn{pNasEpsS.eq} iteratively and
produces the same final answer that we obtain from \Eqn{3DeigEqneS.eq}, as does using 
a  Gr\"{o}bner basis based on \Eqn{pNasEpsS.eq}.  

In the main paper, we presented 
a robust algebraic solution  that could be evaluated numerically 
for the quaternion eigenvalues in the special case of a symmetric
 traceless  $4\times 4$ profile matrix $M_{3}(E_{3})$ based on the 3D cross-covariance
  matrix $E_{3}$; we will complete the steps deriving that solution below.
But first we will   study   general $4\times 4$ real  matrices,  and then specialize
to symmetric matrices with and
without a trace, as all of our cases of interest are of this type.  We note
  \cite{Golub-vanLoan-MatrixComp} that any nonsingular real matrix that can be written
  in the form $[S^{\t} \cdot S]$  is itself symmetric and has only positive real eigenvalues;
  in general, the symmetric matrices
   $[S^{\t} \cdot S]$ and $[S\cdot S^{\t}]$ share one set of eigenvalues, 
  but have distinct eigenvectors.  
  Thus, even if we study only symmetric matrices, we can get significant information
  about \emph{any}  matrix $S$  as long as we can recast our investigation
  to exploit the associated symmetric matrices  $[S^{\t} \cdot S]$  and $[S \cdot S^{\t} ]$.
 
 \qquad

\noindent{\bf  The Basic Structure: Standard Algebraic Solutions for 4D Eigenvalues.}   
When we solve  \Eqn{3DeigEqneS.eq} directly using the  textbook quartic solution without 
explicitly imposing restrictions, we find that 
the general structure for the eigenvalues $e= \epsilon_{k}(p_1,p_2,p_3,p_4)$ 
takes the form
\begin{equation}
\left. \begin{array}{rclp{0.25in}rcl}
\epsilon_{1}(p) & =&  -{\displaystyle\frac{p_1}{4} } + F (p) + G_{+}(p)&&
\epsilon_{2}(p) & =&   -{\displaystyle\frac{p_1}{4} }+ F (p) - G_{+}(p)\\[0.25in]
\epsilon_{3}(p) & =&   -{\displaystyle\frac{p_1}{4} } - F (p) + G_{-}(p)&&
\epsilon_{4}(p) & =&   -{\displaystyle\frac{p_1}{4} } - F (p)  - G_{-}(p)\\
\end{array}  \ \right\}  \ . \label{longsolnEps4.eq} 
\end{equation}
 Here  $-p_{1}=  (\epsilon_{1} + \epsilon_{2} +  \epsilon_{3}
    + \epsilon_{4})$ is the trace, and we can see that ``canonical form'' for the quartic
    \Eqn{3DeigEqneS.eq},
    with a missing cubic term in $e$, results from simply changing variables from
    $e \to e + (\epsilon_{1} + \epsilon_{2} +  \epsilon_{3}
    + \epsilon_{4})/4$ to effectively add $1/4$ of the trace to each eigenvalue.  
The other two types of terms have the following explicit expressions in terms of the four
independent coefficients $p_{k}$:
\begin{equation}
\left. \begin{array}{rcl}
   F(p_1,p_2,p_3,p_4) & =&
  \sqrt{ {\displaystyle \frac{{p_1}^2}{16}  - \frac{p_2}{6}  + 
  \frac{1}{12} 
\left(\sqrt[{\textstyle 3}]{a+\sqrt{-b^2}}
              + \frac{r^{2}}{ \sqrt[{\textstyle 3}]{a + \sqrt{-b^2}}} \right) } }   \\[.25in]
   G_{\pm}(p_1,p_2,p_3,p_4)       & =&      \sqrt{ \displaystyle 
             \frac{3{p_1}^2}{16} - \frac{p_2}{2} - F^{2}(p)
             \pm  \frac{  s(p) }  {32 \; F(p) } }  \\[0.20in]
                 & =& 
    \\[0.25in]
 \multicolumn{3}{l}{  \sqrt{ \displaystyle
  \frac{{p_1}^2}{8}-\frac{p_2}{3} -
         \frac{1}{12 }\left( \sqrt[{\textstyle 3}]{a+\sqrt{-b^2}} 
                 + \frac{r^{2}}{ \sqrt[{\textstyle 3}]{a + \sqrt{-b^2}}} \right)
          \pm \frac{ s(p)  }
          {32 \; \sqrt {\displaystyle \frac{{p_1}^2}{16}  - \frac{p_2}{6}   +  
      \frac{1}{12}\left( \sqrt[{\textstyle 3}]{a+\sqrt{-b^2}} 
                 + \frac{r^{2}}{ \sqrt[{\textstyle 3}]{a + \sqrt{-b^2}}}   \right) }}}}  \\[0.25in]
          \end{array}   \\  \right\} \label{longsolnFGH4.eq}  
  \end{equation}
\centerline{with} \vspace{.0in}
\begin{equation} 
\left. \begin{array}{rcl}
r^{2}(p_1,p_2,p_3,p_4) &=&  
   {p_{2}}^{2} - 3 p_{1} p_{3}   + 12 p_{4}  \; = \; \sqrt[\scriptstyle 3]{a^2 + b^2}    \\[0.07in]
       a(p_1,p_2,p_3,p_4) & = &  {p_{2}}^{3}  
   +  { \frac{  \textstyle 9}{ \textstyle 2} } \left(  3  {p_3}^2 + 3 {p_1}^2 {p_4}
          -   {p_1} {p_2}{p_3} - 8  {p_2}{p_4} \right)   \\[0.1in]
b^{2}(p_1,p_2,p_3,p_4) & = &  r^{6}(p) - a^{2}(p) \\[0.07in]
s(p_1,p_2,p_3,p_4) & = &  4 p_1 p_2  - {p_1}^3  -  8 p_3 
\end{array} \right\} \ . 
\label{4DxyzsolnabrDefs.eq}
\end{equation}
For general real matrices, which may have complex conjugate pairs of eigenvalues,
the sign of $r^2$ can play a critical role, so giving in to the temptation to write
\[  \frac{r^{2}}{ \sqrt[{\textstyle 3}]{a + \sqrt{-b^2}}} \; \rightarrow  \;
 \sqrt[{\textstyle 3}]{a - \sqrt{-b^2}} \]
leads to anomalies; in addition, 
$b^2$  can take on any value, so evaluating this algebraic expression numerically
while getting the phases of all the roots right
can be problematic.   So far as we can confirm, setting aside matrices with individual
peculiarities, the formula \Eqn{longsolnEps4.eq} yields correct complex eigenvalues for
all real matrices, though the numerical order of the  eigenvalues  can be irregular.  
When  we restrict
our attention to real symmetric matrices, a number of special constraints come into
play that significantly improve the numerical behavior of the algebraic solutions, as well
as allowing us to simplify the algebraic expression itself.  The real
symmetric matrices are all that concern us for any of the alignment problems.

\qquad

\noindent{\bf Symmetric Matrices.}
We restrict our attention from here on to general symmetric $4\times 4$ real matrices, 
for which the eigenvalues
must be real,  and so the roots of the  matrix's quartic characteristic polynomial must be real.  
 A critical piece of information
comes from the fact that the quartic roots are based on an underlying cube
root solution (a careful examination of how this works can be found, for example,
in \cite{CoutsiasSeokDill2004,CoutsiasWester2019,NIckalls2009}).  
As noted. e.g.,  in \cite{AbramowitzStegun1970}, the roots of this
cubic are \emph{real} provided that a particular discriminant is \emph{negative}. 
This expression takes the form
\[ {q_{\mbox{\small AS }}}^{3} + {r_{\mbox{\small AS }}}^{2}  \le 0 \ , \]
where $\{ \mbox{\small AS}\}$ disambiguates the Abramowitz-Stegun variable names,
and the relationship to our parameterization in terms of the eigenequation coefficients
$p_{k}$ is simply
\begin{equation} \begin{array}{rclp{.25in}rcl}
q_{\mbox{\small AS }} & = & -  {\displaystyle \frac{1}{9}} \, r^{2}(p_1,p_2,p_3,p_4) \ , & &
r_{\mbox{\small AS }} & = & {\displaystyle \frac{1}{27}}\, a(p_1,p_2,p_3,p_4) \end{array} \ .
\label{qrDefs.eq}
\end{equation}
Thus we can see  from \Eqn{4DxyzsolnabrDefs.eq} that
\begin{equation}
\label{cubicDiscrim.eq}
b^{2}(p_1,p_2,p_3,p_4) \, = \, r^{6}(p) -  a^{2}(p) = 
     - \,{9}^{3} \,\left({q_{\mbox{\small AS}}}^{3} + {r_{\mbox{\small AS}}}^{2} \right) \ ,
\end{equation}
and hence  for symmetric real matrices 
we must have $b^2(p)\ge 0$.  Therefore for this case  we can always write
\begin{equation}
 \left(a(p) + \sqrt{-b(p)^{2}}\right)\longrightarrow \left(a + \I \,b \right)  \label{rtbsqToIbsq.eq} \ , 
\end{equation}
and then 
we can rephrase our  general  solution   
 from Eqs.~(\ref{longsolnEps4.eq}), (\ref{longsolnFGH4.eq}), and (\ref{4DxyzsolnabrDefs.eq})
 as
\begin{equation}
\left. \begin{array}{rcl}
   F(p_1,p_2,p_3,p_4) & =&
      \sqrt {\displaystyle \frac{{p_1}^2}{16}  - \frac{p_2}{6}   +  
            \frac{1}{6} \, r(p) \, c(a,b)  } \\[.20in]
   G_{\pm}(p_1,p_2,p_3,p_4) & =&    \sqrt{ \displaystyle 
      \frac{{p_1}^2}{8}-\frac{p_2}{3} -
         \frac{1}{6}\, r(p)\,c(a,b) 
              \pm  \frac{ s(p)  }
                 {32 \; \sqrt {\displaystyle \frac{{p_1}^2}{16}  - \frac{p_2}{6}   +  
                     \frac{1}{6} \, r(p) \, c(a,b)  } }} \\[0.40in]
                     & =&    \sqrt{ \displaystyle 
             \frac{3{p_1}^2}{16} - \frac{p_2}{2} - F^{2}(p)
             \pm  \frac{  s(p)   }
                 {32 \; F(p) } }  \\
\end{array}   \\  \right\} \ ,\label{longsolnFGHSym.eq}
\end{equation}
 where the cube root terms can now be reduced to real-valued trigonometry:
\begin{equation} 
\left. \begin{array}{rcl}
r(p) \,c(a,b) &  = & r(p)\,\cos\left( {\displaystyle\frac{\arg{(a + \I b)}}{3} }   \right)\, = \,
     {\displaystyle  \frac{1}{2} }\left(  \rule{0in}{1.2em}(a+\I \,b)^{1/3}+(a-\I \,b)^{1/3}  \right) \\[0.12in]
r^{2}(p) &=&  
   {p_{2}}^{2} - 3 p_{1} p_{3}   + 12 p_{4} \; = \;\sqrt[\textstyle3]{a^2 + b^2} \;= \;(a+\I \,b)^{1/3}  (a-\I \,b)^{1/3}  \\[0.1in]
 r^{6}(p)     & = & a^2(p) + b^2(p) \\
 s(p) & = &  4 p_1 p_2  - {p_1}^3  -  8 p_3 
     \end{array}  \right\}   \ .
     \label{realquarticTrig.eq}
\end{equation}

\qquad

\noindent{\bf  Alternative Method: The Cube Root Triples Method and Its Properties.}
Our first general method above corresponds directly to \cite{AbramowitzStegun1970},
and consists of   combinations of signs in  two blocks of expressions.
  The second method that we are about to explore uses
sums of three expressions in all four eigenvalues, with each term having 
a square root ambiguity;  this is fundamentally Euler's solution, discussed,
for example, in \cite{CoutsiasSeokDill2004,CoutsiasWester2019}  and \cite{NIckalls2009}.
The correspondence between this triplet and the four expressions in \Eqn{longsolnFGHSym.eq}
is delicate, but deterministic, and we will show the argument leading to the equations
we introduced in the main text.

The ``Cube Root Triple'' method follows from the observation that if we break up the general
form of the four quartic eigenvalues into a trace part and a sum of three identical parts
whose signs are arranged to be traceless, we find an equation that can be easily
solved, and which (under some conditions that we will remove) evaluates numerically 
to the same eigenvalues as \Eqn{longsolnFGHSym.eq}, but
can be expressed in terms of a one-line formula for the eigenvalue system.
The Ansatz that we start with is the following:
\begin{equation}  
\left. \begin{array}{rcl}
\epsilon_{1} & \stackrel{\mbox{?}}{ = }  &  - {\displaystyle  \frac{p_{1}}{4}}+ \sqrt{X} + \sqrt{Y} + \sqrt{Z} \\[.1in]
\epsilon_{2} & \stackrel{\mbox{?}}{ = }  &  - {\displaystyle \frac{p_{1}}{4}} + \sqrt{X} -  \sqrt{Y} - \sqrt{Z} \\[.1in]
\epsilon_{3} & \stackrel{\mbox{?}}{ = } &  - {\displaystyle \frac{p_{1}}{4}} - \sqrt{X} + \sqrt{Y}  - \sqrt{Z}  \\[.1in]
\epsilon_{4} &  \stackrel{\mbox{?}}{ = }  & - {\displaystyle \frac{p_{1}}{4} } - \sqrt{X} -  \sqrt{Y} + \sqrt{Z}  \\ \end{array} \right\} 
\label{eps1to4S.eq} 
  \ .
\end{equation}

If we now insert   our expressions for $\epsilon_{k}(p_{1},X,Y,Z)$ from
\Eqn{eps1to4S.eq} into \Eqn{pNasEpsS.eq}, we see 
that the  $p_{k}$  equations are transformed into a quartic system of equations
that can in principle be solved for the components of the eigenvalues,
\begin{equation} 
\left. \begin{array}{rcl}
p_{1} & = & p_{1}   \\[0.1in]
p_2 & = & { \displaystyle \frac{ 3 {p_1}^2}{8} }\, - 2\,(X + Y + Z) \\[0.15in]
p_3  & = & {\displaystyle \frac{  {p_1}^3}{16}}\, -8\,  \sqrt{X\, Y\, Z} 
        - p_1 (X + Y + Z)  \\ [0.15in]
p_4 & = &  {\displaystyle     \frac{ {p_1}^4}{256} }\! + \!  X^2 + Y^2 + Z^2 \!
   -2 \left( Y Z  + Z X  + X Y \right)  
      - p_{1}  \sqrt{X\, Y\, Z}  -    {\displaystyle \frac{  {p_1}^{2}}{8}} (X + Y + Z)  \\
 \end{array} \hspace{0.1in}\right\} \ .
\label{pkOfXYZS.eq}
\end{equation}
While our original equation \Eqn{pNasEpsS.eq} does not
respond to $\rule{0in}{1.1em} \mathtt{Solve[... , \,\{\epsilon_{1},\epsilon_{2},\epsilon_{3},\epsilon_{4}\}, ... ]}$,  and \Eqn{pkOfXYZS.eq} with $X\to u^2$, $Y\to v^2$,
$Z \to w^2$ does not respond to $\rule{0in}{1.1em} \mathtt{Solve[... ,\,\{u,v,w\}, ... ]}$,   for some reason
\Eqn{pkOfXYZS.eq} with $X,Y,Z$ as the free variables responds immediately
to $\rule{0in}{1.1em} \mathtt{Solve[ pkEqnList\,, \{X,Y,Z\}, Quartics } \,\rightarrow\, \mathtt{True \,]}$,
and produces a solution for $X(p)$, $Y(p)$, and $Z(p)$ that we can manipulate into the  following form,
\begin{equation}
F_{f}(p) = \frac{{p_{1}}^2}{16} - \frac{p_{2}}{6}
 - \frac{1}{12} \left(\phi(f) \left(a(p) + \sqrt{-b^{2}(p)}\right)^{1/3}+
   \frac{r^{2}(p) }{\phi(f)\left(a(p) + \sqrt{-b^{2}(p)}\right)^{1/3}} \right) \  .
\label{4DeigSolnS.eq}
\end{equation}
Here   $F_{f}(p)$ with $f=(x,y,z)$ represents $X(p)$, $Y(p)$, or $Z(p)$
corresponding to one of the three values of the cube roots $\phi(f)$
of $(-1)$ given by
\begin{equation}
\begin{array}{l@{\hspace{.3in}}l@{\hspace{.3in}}l}
 \phi(x) \; = \;  -1 \; ,&
 \phi(y) \; = \; \frac{1}{2}\left(1 + \I \sqrt{3}\right)  \; ,&
 \phi(z) \; =\; \frac{1}{2}\left(1 - \I \sqrt{3}\right)  \  , \end{array}
\label{cubicrts.eq}
\end{equation}
and the utility functions are defined as above in \Eqn{4DxyzsolnabrDefs.eq}.
Once again, because we have symmetric real
matrices with real eigenvalues, we know that the discriminant condition for real 
solutions requires $b^2(p)\ge 0$, so we can again apply \Eqn{rtbsqToIbsq.eq} to transform each
$\left(a(p) \pm \sqrt{-b^{2}(p)}\right)$ term into the form $\left(a(p) \pm \I \, b(p) \right)$.
This time we get a slightly different formula because there is a different $\sqrt[\textstyle 3]{-1\,}$ phase
incorporated into each of the $X,Y,Z$ terms, and we obtain the following  intermediate result:
\begin{eqnarray}
F_{f}(p)  & =  &\frac{{p_{1}}^2}{16} - \frac{p_{2}}{6}
 - \frac{1}{12} \left(\phi(f) \left(a + \I b \right)^{1/3}+
   r^{2}(p) \frac{1}{\phi(f)\left(a + \I b\right)^{1/3}} \right) \nonumber \\
  & =  &\frac{{p_{1}}^2}{16} - \frac{p_{2}}{6}
 - \frac{1}{6}  \left( \phi(f)(a+\I b)^{1/3}+  \overline{\phi(f)}\, (a-\I b)^{1/3}  \right) \\
   & =  &\frac{{p_{1}}^2}{16} - \frac{p_{2}}{6}
 - \frac{1}{6}  \left( \phi(f)(a+\I b)^{1/3}+  \overline{\phi(f) (a+\I b)^{1/3} } \right) \ ,
\label{4DeigSolnSym.eq}
\end{eqnarray}
where  $ \overline{\phi(f)}$, etc., denotes the complex conjugate,
and we took advantage of the relation $\sqrt[\textstyle{3}]{a^2 + b^2} = r^2(p)$.
  The cube root terms again reduce to real trigonometry,
giving our final result (remember that $\phi(x)=-1$, changing the sign)
\begin{equation} 
\begin{array}{rcl}
F_{f}(p_1,p_2,p_3,p_4) & =& {\displaystyle 
\frac{{p_{1}}^2}{16} - \frac{p_{2}}{6} +
\frac{1}{6}}
\left(\rule{0in}{1.2em} r(p) \cos_{f}(p)  \right) \ ,\\[0.05in]  
\end{array}\\[0.1in] \label{4DeigSolnSS.eq} \end{equation}
but now with the direct incorporation of the three phases of $\sqrt[\textstyle{3}]{-1\,}$
 from \Eqn{cubicrts.eq} (see, e.g.,\ \cite{NIckalls1993}), we get nothing but phase-shifted real 
 cosines,
\begin{equation}
\begin{array}{c@{\hspace{.07in}}c@{\hspace{.07in}}c}
  \cos_{\textstyle x}(p) \! = \! \cos\left( {\displaystyle \frac{\arg(a + \I b)}{3} } \right), & 
  \cos_{\textstyle y}(p) \! = \! \cos\left( {\displaystyle \frac{\arg(a + \I b)}{3} -
    \frac{2 \pi}{3} } \right), &
  \cos_{\textstyle z}(p) \! = \! \cos\left( {\displaystyle \frac{\arg(a + \I b)}{3} +
    \frac{2 \pi}{3} } \right)\\  \end{array} .
   \label{x3soln1.eq}   
\end{equation}
The needed subset of the utility functions now reduces to
\begin{equation} 
\left. \begin{array}{rcl}
r^{2}(p_1,p_2,p_3,p_4) &=&  
   {p_{2}}^{2} - 3 p_{1} p_{3}   + 12 p_{4} \; = \;\sqrt[\textstyle{3}]{a^2 + b^2} \;= 
          \;(a+\I b)^{1/3}  (a-\I b)^{1/3}  \\[0.07in]
     a(p_1,p_2,p_3,p_4) & = &  {p_{2}}^{3}  
   +  { \frac{  \textstyle 9}{ \textstyle 2} } \left(  3  {p_3}^2 + 3 {p_1}^2 {p_4}
          -   {p_1} {p_2}{p_3} - 8  {p_2}{p_4} \right)   \\[0.07in]   
     b^{2}(p_1,p_2,p_3,p_4) & = &  r^{6}(p) - a^{2}(p) \\ 
\end{array} \right\} \ .
\label{4DxyzsolnabrFunsSS.eq}
\end{equation}

\qquad

{\bf Repairing Anomalies in  the Cube Root Triple Form.}
We are not quite finished, as  our $X,Y,Z$ triplets acquire  an
ambiguity due to possible alternate sign choices when
we take the square roots of $X,Y,Z$  to construct the
eigenvalues themselves using the Ansatz of \Eqn{eps1to4S.eq}.
As long as all the terms of one part change sign together, the
tracelessness of the $X,Y,Z$ segment of the eigenvalue system
is maintained, so there are a number of things that could  happen with
the signs without invalidating the general properties of  \Eqn{eps1to4S.eq}.
We can check that, with random symmetric matrix data,
  \Eqn{eps1to4S.eq}  with \Eqn{4DeigSolnSS.eq} will yield the correct eigenvalues about half the time,
while  \Eqn{longsolnEps4.eq} with \Eqn{longsolnFGHSym.eq} always works.
Inspecting  \Eqn{longsolnFGHSym.eq}  and   \Eqn{4DeigSolnSS.eq} with \Eqn{x3soln1.eq},
we observe
that  $F(p_{1}, p_2, p_3, p_4)=\sqrt{\rule{0in}{.85em} F_{\textstyle x}(p_{1},p_2,p_3,p_4)} = \sqrt{X}$;
we can also see that  \Eqn{longsolnFGHSym.eq}  suggests that a relation of the following form
should hold, 
\[G_{\pm}(p_{1}, p_2, p_3, p_4) \sim   \sqrt{Y}  \pm \sqrt{Z}  \ , \]  
so we can immediately conjecture that something is going wrong with the sign choice of 
the root $\sqrt{Z}$.
It turns out that $G_{+}(p)$  changes its algebraic structure to
essentially that of $G_{-}(p)$ when the numerator $\,s(p) =  (4 p_{1}p_{2} -{p_{1}}^{3} - 8 p_{3})\,$ 
inside the square root  in \Eqn{longsolnFGHSym.eq} changes  sign.  That tells us exactly where there is a discrepancy with the choice 
 ${\rule{0in}{1.1em}\sqrt{Y}} +{\rule{0in}{1.1em}\sqrt{Z}}$.  If  we define the following sign test,
\begin{equation}
\sigma(p_1,p_2,p_3,p_4) = \sign\left( 4 p_{1}p_{2} -{p_{1}}^{3} - 8 p_{3} \right) \ ,
\label {xyzSign.eq}
\end{equation}
we discover that we can make \Eqn{eps1to4S.eq} agree exactly with the robust $G_{\pm}(p)$ 
from  \Eqn{longsolnFGHSym.eq}  
for all the random symmetric numerical matrices we were able to test,
 provided we make the following simple change to the final form of the $X,Y,Z$
 formula for the eigenvalue solutions:
\begin{equation} 
\left. \begin{array}{rcl}
\epsilon_{1} & = &  - {\displaystyle  \frac{p_{1}}{4}}
  + \sqrt{X} + \sqrt{Y} + \sigma(p) \sqrt{Z} \\[.1in]
\epsilon_{2} & = &  - {\displaystyle \frac{p_{1}}{4}} + \sqrt{X} - \sqrt{Y} - \sigma(p)\sqrt{Z} \\[.1in]
\epsilon_{3} & = &  - {\displaystyle \frac{p_{1}}{4}} - \sqrt{X} + \sqrt{Y} - \sigma(p) \sqrt{Z}  \\[.1in]
\epsilon_{4} & = & - {\displaystyle \frac{p_{1}}{4} }- \sqrt{X} - \sqrt{Y} + \sigma(p) \sqrt{Z}  \\
 \end{array} \right\} 
\label{Signeps1to4Sym.eq}   \ .
\end{equation}

\qquad

{\bf Algebraic Equivalence of Standard and Cube Root Triple Form.}  With the benefit
of hindsight, we now complete the picture by working out the algebraic properties of
\Eqn{longsolnEps4.eq} and \Eqn{longsolnFGH4.eq} that confirm our heuristic derivation of \Eqn{Signeps1to4Sym.eq}.   First, we look back at \Eqn{pkOfXYZS.eq} 
and discover that, using the relations for $p_{2}$ and $p_{3}$, we can 
incorporate $X+Y+Z = 3 {p_{1}}^{3}/16 - p_{2}/2$ into $p_{3}$
to get a very suggestive form for our expression $s(p)$  from \Eqn{4DxyzsolnabrDefs.eq}
in terms of the only square-root ambiguity in our original equations that we used to solve
for $\left( X(p), Y(p), Z(p) \right)$, which is
\begin{equation}
s(p_1,p_2,p_3,p_4) \, = \,   4 p_1 p_2  - {p_1}^3  -  8 p_3 
  \,  =  \,  64 \sqrt{ X(p) Y(p) Z(p) }  \ . 
\label{SFunOfXYZ.eq}
\end{equation}
Already we see that this is potentially nontrivial because $s(p)$  does not
have a deterministic sign, but $\sqrt{ X(p) Y(p) Z(p) } $  will always be positive
unless we have a deterministic reason to choose the negative root.

Next, using \Eqn{4DeigSolnS.eq}, we  recast  \Eqn{longsolnFGH4.eq} in a form
that uses $F(p) \equiv \sqrt{X(p)} \equiv \sqrt{F_{x}(p)}$, as well as \Eqn{SFunOfXYZ.eq},
to give
\begin{equation}
\left. \begin{array}{rcl}
   F(p_1,p_2,p_3,p_4) & =&  \sqrt{X(p_1,p_2,p_3,p_4) } \\[0.15in]
    & = & \sqrt{ {\displaystyle \frac{{p_1}^2}{16}  - \frac{p_2}{6}  + 
  \frac{1}{12} 
\left(\sqrt[{\textstyle 3}]{a-\sqrt{-b^2}}+\sqrt[{\textstyle 3}]{a+\sqrt{-b^2}} \right) }}   \\[.25in]
   G_{\pm}(p_1,p_2,p_3,p_4) & =& 
                      \sqrt{ \displaystyle 
             \frac{3{p_1}^2}{16} - \frac{p_2}{2} - F^{2}(p)
             \pm  \frac{  s(p) }
                 {32 \; F(p) } }  \\[0.2in]
                    & =&   
                      \sqrt{  A(p_1,p_2,p_3,p_4)  \pm \ B(p_1,p_2,p_3,p_4) } \\[0.15in]
                              \end{array}   \\  \right\} \label{lFGGtoAB.eq}  
  \end{equation}
  where in fact we know  a bit about how $B(p)$ should look:
\begin{equation}
B(p)  =    \displaystyle { \frac{  s(p) }  {32 \; \sqrt{X(p)}}}
                   \ . \label{DefsOfB.eq}  
  \end{equation}
Now we solve the equations
\begin{equation}
 \sqrt{ A(p) \pm B(p) } =  \sqrt{Y}  \pm  \sigma(p) \sqrt{Z}
         \label{SolveAB.eq}  
  \end{equation}
for $A(p)$ and $B(p)$, to discover
\begin{eqnarray}
A (p)& = & Y(p)  + \sigma^2 Z(p) \nonumber  \\
       & = & Y (p) +   Z(p)  \label{SolnofA.eq}\\[0.05in]
B(p) & = & 2 \sigma \sqrt{Y(p) Z(p)} \  , \label{SolnofB.eq} 
\end{eqnarray}
where we note that these useful relations are nontrivial to discover \emph{directly} from
our original expressions for $F(p)$ and $G_{\pm}(p)$.
Finally, using  \Eqn{DefsOfB.eq},
we conclude that
\begin{equation}
 s(p)   =   64\, \sigma(p) \sqrt{X(p) Y(p) Z(p)} \ ,
 \label{SfunWsigma.eq}
 \end{equation}
which confirms that the appearance of 
\begin{equation}
 \sigma(p) = \sign(s(p)) = \sign( 4 p_1 p_2  - {p_1}^3  -  8 p_3 ) \label{finalSfun.eq}
 \end{equation}
in the $(X,Y,Z)$ expression of \Eqn{Signeps1to4Sym.eq}  is rigorous and inevitable,
as it can  be deduced directly from its appearance in  $B(p)$.

\qquad

{\bf  Alternative Reduction of the Quartic Solution.}   Perhaps a more explicit
way to connect the $(F,G_{\pm})$ and $(X,Y,Z)$ forms, and one we might have used
from the beginning with further insight, is to observe that $G_{\pm}$ is actually
the square root of a perfect square,
\begin{equation}
\left. \begin{array}{rcl} 
G_{\pm}  & =&    \sqrt{ \left( \sqrt{Y}  \pm  \sigma \sqrt{Z}\right) ^2 }  \\[0.2in]
      & =&    \sqrt{  Y + Z  \pm 2 \sigma \sqrt{Y Z}} \\[0.2in]
        & =&    \sqrt{  Y + Z  \pm 2 \sigma \displaystyle { \frac{\sqrt{X Y Z}}{\sqrt{X}}}  } \\[0.2in]
         & =&    \sqrt{  Y + Z  \pm 2 \sigma \displaystyle { \frac{64 \sqrt{X Y Z} }{64\sqrt{X}}} }\\[0.2in]
            & =&    \sqrt{  Y + Z  \pm 2 \sigma\displaystyle {  \frac{| s(p)|} {64\sqrt{X}}} }\\[0.2in]
             & = &   \sqrt{  Y + Z  \pm \displaystyle { \frac{  s(p) }  {32 \; \sqrt{X(p)}}} } \\[0.25in]
                \end{array}   \\  \right\}  \ , \label{GpmOfXYZ.eq}  
  \end{equation}
  where we used the fact that $ \sigma(p)  { |{s(p)}|} = s(p)$.  As long as the sign with 
  which $G_{\pm}$ enters into the solution is consistent, the alternative overall signs of the
  radicals in \Eqn{GpmOfXYZ.eq} will be included correctly.
  

\qquad

\comment{
{\bf  Repairing Anomalies in the Equations for General Nonsymmetric Matrices.}
The case of  non-symmetric $4\times 4$ matrices is not required for any of
our applications because, using the insights from SVD, we can recast any
eigenvalue problem of interest to us in terms of strictly symmetric matrices.
However, for completeness, let us examine the general matrix case, which
presents some challenges when we try, e.g., to evaluate  \Eqn{longsolnFGH4.eq} 
numerically for nonsymmetric matrices.
So far as we can determine,  while the canonical solution 
   \Eqn{longsolnEps4.eq} with \Eqn{longsolnFGH4.eq} should in principle yield the
   eigenvalues of a non-symmetric matrix when evaluated numerically, it succeeds in
   selecting correct phases only half the time, and the correct cases occur for both
   all-real and some-complex eigenvalue systems.  On the other hand, the $(X,Y,Z)$ 
   system defined by \Eqn{Signeps1to4Sym.eq}  with the bare, not-necessarily-real
   conditions in \Eqn{4DeigSolnS.eq} appears to obtain \emph{almost} all  eigenvalues of
   non-symmetric matrices correctly.  
   
   To complete the analysis, we consider the ambiguity of the overall sign of $G_{\pm}$.
   We might expect   phase  corrections for the nonsymmetric case
   similar to the one used with $\sigma(p)$ in \Eqn{Signeps1to4Sym.eq} that would
   also yield the correct numerical general eigenvalue solutions using   minor modifications of
    \Eqn{longsolnEps4.eq} and  \Eqn{longsolnFGH4.eq}, or 
    \Eqn{Signeps1to4Sym.eq}   and \Eqn{4DeigSolnS.eq}.  If we define another sign
    function
    \begin{equation}
    \left. \begin{array}{rcl}
     \tau(p)  & = & \sign \left(\displaystyle 
             \frac{3{p_1}^2}{16} - \frac{p_2}{2} - F^{2}(p)
             \pm  \frac{  s(p) }
                 {32 \; F(p) } \right) \\[0.15in]
       & = & \sign{\left( Y + Z  \pm 2 \sigma(p) \sqrt{Y Z}\, \right)} \\[0.05in]
       & = & \sign{\left( Y + Z  \pm \displaystyle { \frac{  s(p) }  {32 \; \sqrt{X}}}  \right)}\\[0.1in]
       & \!\!\! \mbox{where}\!\!\!&\\[0.05in]
       F^{2}(p) & = &  {\displaystyle \frac{{p_1}^2}{16}  - \frac{p_2}{6}  + 
  \frac{1}{12} 
\left(\sqrt[{\textstyle 3}]{a-\sqrt{-b^2}}+\sqrt[{\textstyle 3}]{a+\sqrt{-b^2}} \right) }\\
\end{array} \right\}  \label{signOfGpm.eq}
  \ ,
\end{equation}
 we can consider replacing the ambiguous sign of the square root of the square in \Eqn{GpmOfXYZ.eq}
 by
 \begin{equation}
 \left.    \begin{array}{rcl}
     G_{\pm}(p) & = & \tau(p)  \left(\displaystyle 
             \frac{3{p_1}^2}{16} - \frac{p_2}{2} - F^{2}(p)
             \pm  \frac{  s(p) }
                 {32 \; F(p) } \right) \\[0.04in]
 \end{array} \right\}  \label{signedGpmVal.eq}
  \ .
\end{equation}
  }   

 \qquad 

\noindent{\bf The Traceless Triple Form.}  The explicitly  traceless $X,Y,Z$ triplet form that 
corresponds to a set of eigenvalues in descending magnitude order that we introduced
for the 3D RMSD problem in the main text is is obtained by imposing the traceless
condition, $p_{1}=0$, obeyed by the 3D profile matrix $M_{3}(E_{3})$:
\begin{equation}
\left. \begin{array}{rcl}
\epsilon_{1} & = &  + \sqrt{X} + \sqrt{Y} +  \sigma(p) \sqrt{Z}  \\
 \epsilon_{2} & = &   + \sqrt{X} - \sqrt{Y} -  \sigma(p) \sqrt{Z}  \\
  \epsilon_{3} & = &  - \sqrt{X} + \sqrt{Y} -   \sigma(p) \sqrt{Z} \\
  \epsilon_{4} & = &  - \sqrt{X} - \sqrt{Y} +  \sigma(p) \sqrt{Z} 
\end{array} \right\}\label{3DepsilonT0.eq}
  \ . 
\end{equation}
Then \Eqn{pkOfXYZS.eq} simplifies to
\begin{eqnarray} 
p_{1} & = & 0 \label{xyz1T0.eq} \\
p_2 & = &  - 2\,(X + Y + Z) \label{xyz2T0.eq}\\
p_3  & = &  -8\,  \sigma(p)\, \sqrt{X\, Y\, Z} \label{xyz3T0.eq} \\ 
p_4 & = &  X^2 + Y^2 + Z^2   -2 \left( Y Z  + Z X  + X Y \right) \ ,
 \label{xyz4T0.eq}
\end{eqnarray}
and the  solutions
for  $X(p)$, $Y(p)$, and $Z(p)$ (and thus for $\epsilon_{k}(p)$) reduce to:
\begin{equation}
 \begin{array}{rcl}   
F_{ f}(p_2,p_3,p_4) & =& {\displaystyle+ \frac{1}{6}}
\left(\rule{0in}{1.2em} r(p) \cos_{ f}(p) -p_{2}  \right) \\[0.05in]  
\end{array} \ , \label{xyzSolnApp.eq}  \end{equation}
  where the phased cosine terms retain their form
\begin{equation}
 \hspace{-.095in}  \begin{array}{c@{\hspace{.07in}}c@{\hspace{.07in}}c}
  \cos_{\textstyle x}(p) \! = \! \cos\left( { \displaystyle \frac{\arg{(a + \I b)}}{3}} \right), & 
  \cos_{\textstyle y}(p) \! = \! \cos\left(  {\displaystyle \frac{\arg{(a + \I b)}}{3} -  \frac{2 \pi}{3} }\right), &
  \cos_{\textstyle z}(p) \! = \! \cos\left(  {\displaystyle\frac{\arg{(a + \I b)}}{3} +
    \frac{2 \pi}{3} }\right) \ . \end{array}  \label{xyzcosApp.eq} 
\end{equation}
Here $F_{f}(p)$ with $f=(x,y,z)$ as always represents $X(p)$, $Y(p)$, or $Z(p)$
and the utility functions simplify to
\begin{equation} 
\left. \begin{array}{rcl}
  \sigma(p_3) &  = &  \sign(-   p_{3} )\\[0.075in]
r^{2}(p_2,p_3,p_4) &=&  
   {p_{2}}^{2}   + 12 p_{4} \; = \;\sqrt[3]{a^2 + b^2} \;= \;(a+\I b)^{1/3}  (a-\I b)^{1/3}  \\[0.075in]
a(p_2,p_3,p_4) & = &  {p_{2}}^{3} + 
  {\textstyle \frac{\textstyle 9}{\textstyle 2} } \left(  3   {p_3}^2   - 8 {p_2}   {p_4} \right)   \\[0.075in]
b^{2}(p_2,p_3,p_4) & = &  r^{6}(p) - a^{2}(p) \\[0.07in]
 & = &  {\displaystyle \frac{27}{4} }\left(16 p_4 {p_2}^4 - 4 {p_3}^2 {p_2}^3 - 
128 {p_4}^2 {p_2}^2 + 144 {p_3}^2 p_4 p_2 - 27 {p_3}^4 + 
   256 {p_4}^3 \right)  \\
\end{array} \right\} \ . 
\label{4DxyzsolnabrFuns2A.eq}
\end{equation}  

\qquad 
 
\begin{quote} \fbox{\parbox{5.5in}{{\bf  Summary:} We therefore have two alternate robust expressions, 
  \Eqn{longsolnEps4.eq} with \Eqn{longsolnFGHSym.eq}  and 
  \Eqn{Signeps1to4Sym.eq}  with \Eqn{4DeigSolnSS.eq},
  for the entire eigenvalue spectrum  of any real, symmetric $4\times 4$ matrix $M$
  characterized by its four intrinsic eigenequation coefficients $(p_{1}, p_{2},p_{3},p_{4})$. 
  For the simpler traceless case, we can take advantage of \Eqn{3DepsilonT0.eq} with \Eqn{xyzSolnApp.eq}. }}
  \end{quote}

%
%


\bibliographystyle{apalike}

\bibliography{rmsd}